\pdfoutput=1
\documentclass[a4paper,11pt,openright]{report} 

\newcommand{\previouslypublished}[1]{\vspace{-0.5cm}\emph{\small #1}}

\makeatletter
\newcommand{\object}[2][\simbadname]{%
\hypersetup{urlbordercolor=\objectcolor}%
\def\simbadname{#2}%
\StrSubstitute{#1}{ }{+}[\parsedname]%
\href{http://simbad.u-strasbg.fr/simbad/sim-id?Ident=\parsedname}{#2}%
\hypersetup{urlbordercolor=\urlcolor}}%
\makeatother

\newcommand{\arcsec}{\hbox{$^{\prime\prime}$}}

\long\def\symbolfootnote[#1]#2{\begingroup%
\def\thefootnote{\fnsymbol{footnote}}\footnote[#1]{#2}\endgroup}
\let\oldsqrt\sqrt
\def\sqrt{\mathpalette\DHLhksqrt}
\def\DHLhksqrt#1#2{%
\setbox0=\hbox{$#1\oldsqrt{#2\,}$}\dimen0=\ht0
\advance\dimen0-0.2\ht0
\setbox2=\hbox{\vrule height\ht0 depth -\dimen0}%
{\box0\lower0.4pt\box2}}

\newcommand{\myurl}[1]{\href{#1}{#1}}

\newcommand{\jnlref}[1]{\textrm{#1}}

\newcommand{\aj}{\jnlref{AJ}}
\newcommand{\araa}{\jnlref{ARA\&A}}
\newcommand{\apj}{\jnlref{ApJ}}
\newcommand{\apjl}{\jnlref{ApJ}}
\newcommand{\apjs}{\jnlref{ApJS}}

\newcommand{\aap}{\jnlref{A\&A}}
\newcommand{\aapr}{\jnlref{A\&A~Rev.}}
\newcommand{\aaps}{\jnlref{A\&AS}}

\newcommand{\mnras}{\jnlref{MNRAS}}

\newcommand{\prd}{\jnlref{Phys.~Rev.~D}}

\newcommand{\pasp}{\jnlref{PASP}}
\newcommand{\pasj}{\jnlref{PASJ}}

\newcommand{\ssr}{\jnlref{Space~Sci.~Rev.}}

\newcommand{\nat}{\jnlref{Nature}}

\newcommand{\thesistitle}{A Tale of Tidal Tails in the Milky Way}
\newcommand{\fullname}{Andrew Raithby Casey}
\newcommand{\shortname}{A. R. Casey}
\newcommand{\thesissubjects}{astronomy,astrophysics}

\newcommand{\fullthesisdate}{\today}

\usepackage[british]{babel}
\usepackage[pdftex]{geometry}
\usepackage{graphicx}
\usepackage{captcont}
\usepackage{multirow}
\usepackage{array} 
\usepackage{epigraph}
\DeclareGraphicsExtensions{.pdf}
\usepackage[twoside]{rotating}
\usepackage{ctable} 
\usepackage{xstring}
\usepackage[bottom]{footmisc}
\usepackage{deluxetable}
\usepackage[T1]{fontenc}
\usepackage{textcomp}
\usepackage{ucs}
\usepackage[utf8x]{inputenc}
\usepackage{amsmath,amssymb}
\usepackage{pxfonts}
\usepackage{tgheros,tgpagella}

\usepackage{amssymb}
\usepackage{longtable}
\usepackage{lscape}

\usepackage[small,bf,singlelinecheck=off]{caption}

\newcommand\phn{\phantom{0}}%

\usepackage{microtype}
\usepackage[]{caption,subfig}
\DeclareCaptionLabelSeparator{mysep}{\ \ \ }
\usepackage[round]{natbib}
\newcommand{\citecolor}{blue}
\newcommand{\linkcolor}{blue}
\newcommand{\urlcolor}{blue}
\newcommand{\objectcolor}{gray}
\usepackage[hyperref,x11names]{xcolor}
\usepackage[unicode,pageanchor,colorlinks=true,plainpages=false,pdfpagelabels,pdftex]{hyperref}
\hypersetup{pdftitle=\thesistitle,%
  pdfauthor=\shortname,%
  pdfsubject={\thesissubjects},%
  citebordercolor=\citecolor,%
  linkbordercolor=\linkcolor,%
  urlbordercolor=\urlcolor,%
  citecolor=\citecolor,%
  linkcolor=\linkcolor,%
  urlcolor=\urlcolor%
}
\usepackage[section]{placeins}
\usepackage[titletoc]{appendix}

\usepackage{ifthen}
\usepackage{xifthen}
\usepackage{fncychap}
\makeatletter
\ChNameUpperCase
\ChNameVar{\centering\Huge\usefont{OT1}{qbk}{m}{n}\selectfont}
\ChNumVar{\Huge}
\ChTitleVar{\centering\Huge\usefont{OT1}{qbk}{m}{n}\selectfont}
\ChRuleWidth{2pt}
\renewcommand{\DOCH}{%
  \CNV\FmN{\@chapapp}\space \CNoV\thechapter
  \par\nobreak
  \vskip -0.5\baselineskip
}
\renewcommand{\DOTI}[1]{%
  \mghrulefill{\RW}\par\nobreak
  \CTV\FmTi{#1}\par\nobreak
  \vskip 60\p@
}
\renewcommand{\DOTIS}[1]{%
  \mghrulefill{\RW}\par\nobreak
  \CTV\FmTi{#1}\par\nobreak
  \vskip 60\p@
}

\usepackage{fancyhdr}
\fancypagestyle{plain}{
\fancyhf{}
\fancyhead[LE]{\usefont{OT1}{qbk}{m}{n}\selectfont \thepage}
\fancyhead[RO]{\usefont{OT1}{qbk}{m}{n}\selectfont \thepage}
\renewcommand{\headrulewidth}{0pt}
\renewcommand{\footrulewidth}{0pt}
}

\makeatletter
\def\cleardoublepage{\clearpage\if@twoside \ifodd\c@page\else
	\hbox{}
	\vspace*{\fill}
	\thispagestyle{empty}
	\newpage
	\if@twocolumn\hbox{}\newpage\fi\fi\fi}
\makeatother

\usepackage[calcwidth]{titlesec}
\titlelabel{\thetitle.\quad}

\begin{document}
\setlength{\epigraphwidth}{10cm}
\setlength{\afterepigraphskip}{1.5cm}
\renewcommand{\epigraphsize}{\small}
\renewcommand{\epigraphrule}{0pt}
\renewcommand{\epigraphflush}{center}
\renewcommand{\textflush}{flushleft}


\pagestyle{empty}

%
%
%
%
\begin{titlepage}\usefont{OT1}{qbk}{m}{n}\selectfont
  \begin{center}
  \phantom{}
  \vspace{0cm}
    \huge{\thesistitle}
    \\[2.5cm]
    \huge{\fullname}
    \\[2.5cm]
    \LARGE{A thesis submitted for the degree of}\\[0.5cm]
      \LARGE{Doctor of Philosophy}\\[0.5cm]
      \LARGE{of the Australian National University}
    \\[3.cm]
     \includegraphics[width=0.5\textwidth]{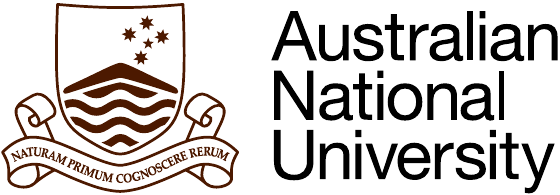}
     \\[10mm]
        \LARGE{Research School of Astronomy \& Astrophysics}
        \vfill
        \Large{Submitted 27th September 2013}\\[1mm]
        \Large{Accepted 14th January 2014}
  \end{center}
\end{titlepage}

\setlength{\parindent}{0pt}
\setlength{\parskip}{1ex plus 0.5ex minus 0.2ex}

\setcounter{page}{0} 
\phantomsection\addcontentsline{toc}{section}{Notes on the digital copy}
\setlength\fboxsep{3mm}
\setlength\fboxrule{1pt}
\begin{center}
\framebox{\huge{\textsc{Notes on the Digital Copy}}}
\end{center}
\vspace{5mm}
This document makes extensive use of the hyperlinking features of \LaTeX. References to figures, tables, sections, chapters and the literature can be navigated from within the PDF by clicking on the reference. Internet addresses will be displayed in a browser. The bibliography at the end of this thesis has been hyperlinked to the NASA Astrophysics Data Service (\myurl{http://www.adsabs.harvard.edu/}). Further information on each of the cited works is available through the link to ADS.




%
\cleardoublepage
\normalsize
%
%
\pagestyle{plain}
\pagenumbering{roman}
%
%
%
\section*{Disclaimer}

I hereby declare that the work in this thesis is that of the candidate alone, except where indicated below or in the text of the thesis. The work was undertaken between February 2010 and August 2013. During the time from July 2012 and December 2012, this work was undertaken at the Massachusetts Institute of Technology, Boston. With the exception of this period, all other work was undertaken at the Australian National University, Canberra. It has not been submitted in whole or in part for any other degree at any other university. 

Chapter \ref{chap:sgr} is an updated version of the paper \textit{`Kinematics and Chemistry of Halo Substructures: The Vicinity of the Virgo Overdensity'}. Casey, A. R., Keller, S. C., Da Costa, G. S., 2011, AJ, 143, 88. 

Chapter \ref{chap:orphan_I} is an updated version of the paper \textit{`Hunting the Parent of the Orphan Stream: Identifying Stream Members from Low-Resolution Spectroscopy'}. Casey, A. R., Da Costa, G. S., Keller, S. C., Maunder, E. 2012, ApJ, 764, 39. Stefan Keller, Gary Da Costa, and Elizabeth Maunder obtained the AAT/AAOmega spectra presented in Chapters \ref{chap:sgr} and \ref{chap:orphan_I}.

Chapter \ref{chap:orphan_II} is an updated version of the paper \textit{`Hunting the Parent of the Orphan Stream II: The First High-Resolution Spectroscopic Study'}. Casey, A. R., Keller, S. C., Da Costa, G. S., Frebel, A., Maunder, E., submitted to the Astrophysical Journal on September 4th, 2013. The Orphan stream candidates presented in Chapter \ref{chap:orphan_II} were initially recognised by Elizabeth Maunder. These targets (and others from the same sample) were later verified by the candidate, and a qualitative probability of stream membership was determined, as discussed in Chapter \ref{chap:orphan_I}.

Chapter \ref{chap:aquarius} is an updated version of the paper \textit{`The Aquarius Co-Moving Group is Not a Disrupted Classical Globular Cluster'}. Casey, A. R., Keller, S. C., Alves-Brito, A., Frebel, A., Da Costa, G. S., Karakas, A., Yong, D., Schlaufman, K., Jacobson, H. R., Yu, Q., Fishlock, C., submitted to the Monthly Notices of the Royal Astronomical Society on September 11th, 2013. Alan Alves-Brito obtained the Magellan/MIKE spectra for the program stars analysed in Chapter \ref{chap:aquarius}. David Nidever and Steven Majewski kindly produced Figure \ref{fig:omega-cen-tidal-debris}, which is a modified figure from \citet{majewski;et-al_2012} that helps place the star C222531-145437 in context of $\omega$-Centauri tidal debris.

The balance of the observations, data reduction and analysis presented in these papers was performed solely by the candidate, who also wrote the text in its entirety. The co-authors provided valuable discussions and commentary on the text.

\vspace{1cm}
\begin{flushright}
\includegraphics[width=0.5\linewidth]{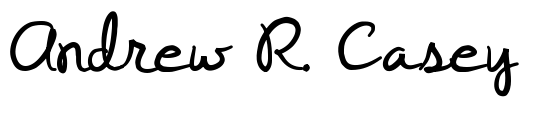}\\[2mm]
\fullname

\fullthesisdate
\end{flushright}

\section*{Supplementary Research}

In addition to the work presented in this thesis, the candidate has contributed to a number of other research projects. Some of these collaborations have resulted in peer-reviewed publications. All papers authored or co-authored by the candidate throughout the 3.5 year program are listed here in reverse chronological order:

\begin{enumerate}

\item{\textbf{Spectroscopy Made Hard: Doing Stellar Spectroscopy `The Right Way\texttrademark'}\\
{\it \underline{\textbf{Casey, Andrew R.}}; et al.\\
\\ Ready for submission.}}

\item{\textbf{The Aquarius Co-Moving Group is Not a Disrupted Classical Globular Cluster}\\
{\it \underline{\textbf{Casey, Andrew R.}}, Keller, Stefan C; Alves-Brito, Alan; Frebel, Anna; Da Costa, Gary; Karakas, Amanda; Yong, David; Schlaufman, Kevin; Jacobson, Heather R.; Yu, Qinsi; and Fishlock, Cherie, 2013.\\
\\ Submitted to the the Monthly Notices of the Royal Astronomical Society \href{http://arxiv.org/abs/1309.3562}{\rm{(arXiv entry)}}}}

\item{\textbf{Hunting the Parent of the Orphan Stream II: The First High-Resolution Spectroscopic Study}\\
{\it \underline{\textbf{Casey, Andrew R.}}; Keller, Stefan C.; Da Costa, Gary S.; Frebel, Anna; and Maunder, Elizabeth, 2013.\\
\\ Accepted to the the Astrophysical Journal \href{http://arxiv.org/abs/1309.3563}{\rm{(arXiv entry)}}}}

\item{\textbf{A single low-energy, iron-poor supernova as the source of metals in the star SMSS J031300.36-670839.3}\\
{\it Keller, Stefan C.; Bessell, Michael S.; Frebel, Anna; \underline{\textbf{Casey, Andrew R.}}; Asplund, Martin; Jacobson, Heather R.; Lind, Karin; Norris, John E.; Yong, David; Heger, Alexander; Magic, Zazralt; Da Costa, Gary S; Schmidt, Brian P.; and Tisserand, Patrick, 2013.\\
\\ Accepted to Nature}}

\item{\textbf{Astropy: A Community Python Package for Astronomy}\\
{\it The Astropy Collaboration; Robitaille, Thomas P.; Tollerud, Erik J.; Greenfield, Perry; Droettboom, Michael; Bray, Erik; Aldcroft, Tom; Davis, Matt; Ginsburg, Adam; Price-Whelan, Adrian M.; Kerzendorf, Wolfgang E.; Conley, Alexander; Crighton, Neil; Barbary, Kyle; Muna, Demitri; Ferguson, Henry; Grollier, Fr\'{e}d\'{e}ric; Parikh, Madhura M.; Nair, Prasanth H.; G\"{u}nther, Hans M.; Deil, Christoph; Woillez, Julien; Conseil, Simon; Kramer, Roban; Turner, James E. H.; Singer, Leo; Fox, Ryan; Weaver, Benjamin A.; Zabalza, Victor; Edwards, Zachary I.; Azalee Bostroem, K.; Burke, D. J.; \underline{\textbf{Casey, Andrew R.}}; Crawford, Steven M.; Dencheva, Nadia; Ely, Justin; Jenness, Tim; Labrie, Kathleen; Lian Lim, Pey; Pierfederici, Francesco; Pontzen, Andrew; Ptak, Andy; Refsdal, Brian; Servillat, Mathieu; and Streicher, Ole, 2013.\\ 
\\
Published in Astronomy and Astrophysics \href{http://adsabs.harvard.edu/abs/2013arXiv1307.6212T}{\rm{(ADS entry)}}}}

\newpage

\item{\textbf{The 300\,km s$^{-1}$ Stellar Stream near Segue 1: Insights from High-resolution Spectroscopy of Its Brightest Star}\\
{\it Frebel, Anna; Lunnan, Ragnhild; \underline{\textbf{Casey, Andrew R.}}; Norris, John E.; Wyse, Rosemary F. G.; and Gilmore, Gerard, 2013. \\ \\ Published in the Astrophysical Journal, 771, 39 \href{http://adsabs.harvard.edu/abs/2013ApJ...771...39F}{\rm{(ADS entry)}}}}

\item{\textbf{Deriving Stellar Effective Temperatures of Metal-poor Stars with the Excitation Potential Method}\\
{\it Frebel, Anna; \underline{\textbf{Casey, Andrew R.}}; Jacobson, Heather R.; and Yu, Qinsi, 2013. \\ \\ Published in the Astrophysical Journal, 769, 57 \href{http://adsabs.harvard.edu/abs/2013ApJ...769...57F}{\rm{(ADS entry)}}}}

\item{\textbf{Hunting the Parent of the Orphan Stream: Identifying Stream Members from Low-resolution Spectroscopy}\\
{\it \underline{\textbf{Casey, Andrew R.}}; Da Costa, Gary; Keller, Stefan C.; and Maunder, Elizabeth, 2013. \\ \\ Published in the Astrophysical Journal, 764, 39 \href{http://adsabs.harvard.edu/abs/2013ApJ...764...39C}{\rm{(ADS entry)}}}}

\item{\textbf{Kinematics and Chemistry of Halo Substructures: The Vicinity of the Virgo Overdensity}\\
{\it \underline{\textbf{Casey, Andrew R.}}; Keller, Stefan C.; and Da Costa, Gary, 2012. \\ \\ Published in the Astronomical Journal, 143, 88 \href{http://adsabs.harvard.edu/abs/2012AJ....143...88C}{\rm{(ADS entry)}}}}

\item{\textbf{The Extragalactic Distance Scale without Cepheids. IV.}\\
{\it Hislop, Lachlan; Mould, Jeremy; Schmidt, Brian; Bessell, Michael S.; Da Costa, Gary; Francis, Paul; Keller, Stefan; Tisserand, Patrick; Rapoport, Sharon; and \underline{\textbf{Casey, Andrew R.}}, 2011. \\ \\ Published in the Astrophysical Journal, 733, 75 \href{http://adsabs.harvard.edu/abs/2011ApJ...733...75H}{\rm{(ADS entry)}}}}

\end{enumerate}
%
\cleardoublepage
\section*{Acknowledgments}

When I started my doctorate, I had no formal teaching in astrophysics. All I knew was what I had learned from leisurely reading, which gradually transitioned from popular science magazines towards highly specialised textbooks. After I first met my prospective supervisors I explained that if I did start a research doctorate, I might have to either suspend from the program, or go part-time on a whim. Even with this outlandish condition -- after just a five minute conversation -- Stefan Keller, Gary Da Costa, and Brian Schmidt expressly decided they'd take me on as a student. I couldn't be more grateful for this. I have always thought to some degree, my supervisors \emph{really} took a gamble on me. I might have turned out to be completely hopeless. At some points near the start of my candidature I'm sure they thought I was. To that end, I've spent the last 3.5 years motivated by the hope that some day they wouldn't want to tell others that they should have put their money on black instead. Whether they knew it or not, they've inspired me to always do better.

I honestly can't thank the people at Mount Stromlo enough. They -- particularly Harvey Butcher, Brian Schmidt, Martin Asplund, David Yong, Stefan Keller, and Gary Da Costa -- have given me wonderful opportunities throughout my candidature, which I appreciate enormously. Because of these opportunities I've met dozens of Nobel laureates, presidents of nations, royalty, Fields medal winners, pillars of science, as well as incredible people working on cutting-edge research in widely varied fields. I've been to conferences, observatories and institutes all over the world. I could never have imagined any of this three years ago, and it exemplifies the generous nature of Stromloneans, as well as the opportunities available at the Australian National University. I've really grown to treasure the wonderfully friendly and unique culture at Stromlo, which you just don't find elsewhere in the world. It's a place I'll forever revere.

It is a pleasure to thank the people who have provided invaluable advice and assistance to me over the years, particularly Martin Asplund, Michael Bessell, Gary Da Costa, Anna Frebel, Amanda Karakas, Stefan Keller, Andreas Korn, John Norris, Brian Schmidt, and David Yong. If I ever end up doing anything that others find to be particularly noteworthy or important, it's largely thanks to these people. I've been extremely fortunate to be surrounded by such fantastic and friendly people, who also happen to be incredibly remarkable physicists. Without their scientific and political advice, I certainly wouldn't be continuing in astrophysics. Special thanks also go to Melissa Ness for helping to keep me (relatively) sane. I already miss our office banter.

Lastly, I'm pleased to thank my loving and supportive fianc\'e, Alana Shaw. Even though Alana has always been orders of magnitude more important to me than all else, this doctorate has forced us apart far more often than when we were together. Thankfully, she's been foolish enough to stay with me.

%
\cleardoublepage
\section*{Abstract}

Hundreds of globular clusters and dwarf galaxies encircle the Milky Way. Many of these systems have undergone partial disruption due to tidal forces, littering the halo with stellar streams. These tidal tails are sensitive to the Galactic potential, facilitating an excellent laboratory to investigate galaxy formation and evolution. To better understand the emergence of the Milky Way, this thesis examines the dynamics and chemistry of a number of known stellar streams. In particular the Sagittarius, Orphan and Aquarius streams are investigated.

Low-resolution spectra for hundreds of stars in the direction of the Virgo Over-Density and the Sagittarius northern leading arms have been obtained. Multiple significant kinematic groups are recovered in this accretion-dominated region, confirming detections by previous studies. A metal-poor population ([Fe/H] = $-1.7$) in the Sagittarius stream is discovered due to a photometric selection that was inadvertently biased towards more metal-poor stars. Positions and kinematics of Sagittarius stream members are compared with existing best-fitting dark matter models, and a tri-axial dark matter halo distribution is favoured.

The Orphan stream is appropriately named, as no parent system has yet been identified. The stream has an extremely low surface brightness, which makes distinguishing stream members from field stars particularly challenging. From low-resolution spectra obtained for hundreds of stars, we identify likely Orphan stream red giant branch stars on the basis of velocity, metallicity, surface gravity, and proper motions. A negligible intrinsic velocity dispersion is found, and a wide spread in metallicities is observed, which suggests the undiscovered parent is similar to the present-day dwarf galaxies in the Milky Way.

High-resolution spectra were obtained for five Orphan stream candidates, and the intrinsic chemical dispersion found from low-resolution spectra is confirmed from these data. Detailed chemical abundances for high-probability Orphan stream candidates further indicates a dwarf galaxy host. Low [$\alpha$/Fe] abundance ratios are observed, and lower limits for [Ba/Y] are found, which sit well above the observed chemical evolution in the Milky Way. This thesis provides the first detailed chemical evidence for a dwarf galaxy origin, allowing us to rule out any association between the Orphan stream and the globular cluster NGC 2419.

High-resolution, high $S/N$ spectra for one third of the Aquarius stream have also been obtained. Contrary to previous work, there is no evidence that the Aquarius stream has resulted from a disrupted globular cluster. Detailed chemistry suggests that the Aquarius stars are galactic in origin, and not disrupted members from either a globular cluster or a dwarf galaxy. In the absence of compelling dynamic and/or chemical evidence to suggest otherwise, we advocate the `Aquarius Group' as a more appropriate description, and hypothesise that the moving group has resulted from a disk-satellite interaction.

The high-resolution spectra presented in this thesis has been analysed using custom written software, ironically named `Spectroscopy Made Hard'. A detailed description of the software, capabilities and algorithms are presented. Spectroscopy Made Hard includes an intuitive graphical user interface, allowing the spectroscopist to interactively modify any aspect of their analysis. The software is designed to facilitate the transition between small and massive sample sizes, while ensuring data provenance, tangibility, and reproducibility. Future applications for this software are outlined, with a particular focus on the large scale high-resolution spectroscopic surveys being planned or currently undertaken.

%
\setcounter{tocdepth}{1} 
\tableofcontents\cleardoublepage
\phantomsection\addcontentsline{toc}{section}{List of Figures}
\listoffigures\cleardoublepage
\phantomsection\addcontentsline{toc}{section}{List of Tables}
\listoftables\cleardoublepage

\newcommand\urllinklabel{View Online}
%

%
%
\pagenumbering{arabic}
\pagestyle{fancy}
\fancyhead[LO]{\usefont{OT1}{qbk}{m}{n}\selectfont \rightmark}
\fancyhead[RE]{\usefont{OT1}{qbk}{m}{n}\selectfont \leftmark}
\fancyhead[LE]{\usefont{OT1}{qbk}{m}{n}\selectfont \thepage}
\fancyhead[RO]{\usefont{OT1}{qbk}{m}{n}\selectfont \thepage}
\fancyfoot{}
\renewcommand{\headrulewidth}{0.5pt}
\renewcommand{\footrulewidth}{0pt}

\chapter{Introduction}\label{sec:introduction}

\epigraph{Given enough time, hydrogen starts to wonder where it came from, and where it is going.}{-- Edward R. Harrison}

The creation of the Universe gave birth to space-time and the laws of nature. A rapid exponential expansion transpired a mere picosecond after the big bang, smoothing out all previous structure in the Universe with the exception of quantum-scale fluctuations \citep[e.g.,][]{guth_1997}. Over time, these fluctuations resulted in cold dark matter (CDM) density perturbations, causing minuscule gravitational potential wells. This process furnished the Universe with the seeds required to build cosmological structure \citep{liddle;lyth_2000,springel;et-al_2005}. Baryonic matter followed the influence of the gravitational field  \citep[$z \sim 200$;][]{rees_1986}, and subsequently began to condense into primordial gas clouds of Hydrogen, Helium, as well as trace elements of Deuterium and Lithium \citep{2004PhLB..592....1P}.

A prolonged, quiescent period followed. The first stars didn't start to form until after $\sim$200 Myr \citep{bromm;et-al_1999,abel;et-al_2002,barkana;loeb_2007}, when primordial gas had sufficiently cooled. The first (so-called Population III) stars had limited cooling capability, and were subsequently were massive \citep[$\sim$10$-100 M_\odot$;][]{bromm;yoshida_2011} and  short-lived \citep{yoshida;et-al_2004,gao;theuns_2007}. They were metal-free\footnote{Following the inexplicable tradition of astronomers, the term `metals' describes elements heavier than He.}, and extremely turbulent \citep{bromm;yoshida_2011}. Given their size, it is unlikely any such objects have survived to the present day \citep{loeb_2010}, and no such object has been found. These objects filled the interstellar medium with the first nucleosynthetic products through their spectacular deaths. The chemically-enriched gas ultimately condensed to initiate the next cycle of star formation \citep[for a review, see][]{bromm;yoshida_2011}. Unlike most stars observable today, which are the product of many hundreds of supernovae, these stars formed from the nucleosynthetic products of a single star. Since the explosive deaths of the Population III stars, successive cycles of star formation have continued to enrich the interstellar medium until the present day. Initially, this chemical enrichment occurred largely independently in each initial potential wells. As time passed, baryonic matter within seed formation sites conglomerated together to form larger stellar systems, and cosmological structure has continued to build hierarchically in a bottom-up fashion.

While Population III stars are no longer expected to exist, a fraction of their progeny are considered to have stellar masses low enough ($\sim$0.8$M_\odot$) to survive to the present epoch. These stars are particularly rare, and provide us a view of the early Universe. With the exception of C, N, O, and Li, the observed stellar abundances (i.e., those in the upper photosphere) remain largely unchanged during a star's lifetime. Regardless of their age, stars are fossilised records containing the chemical composition of their natal gas cloud. While the dynamical information of stars dissipates over time, chemical signatures remain. This permits the use of chemical and dynamical information in stellar groups to be effectively used for the study of galaxy formation and evolution. This was first demonstrated by \citet{eggen;et-al_1962}, which has grown to become the most influential paper of galactic archaeology\footnote{The study of galaxy formation and evolution from observations of fossilised records (e.g., stars).}.

Understanding galaxy formation remains a key goal of astrophysics. This thesis seeks to further our understanding of galaxy formation through the observed motions and chemistry of nearby stars. Most of the stars studied in this thesis have been enriched in different star formation environments before later being accreted onto the Galaxy through hierarchical merging. Thus, these stars posses important information about the merger history of the Milky Way. Throughout this thesis we\footnote{For stylistic reasons I use the plural form throughout this thesis to refer to work led by myself, but naturally performed in collaboration with my supervisors and colleagues. The text however is solely my own.} seek to identify signatures of galaxy formation from these stars, and to chronicle these events in the context of the Milky Way's history. An introduction to hierarchical structure within a context of galaxy formation is provided, and the distinguishable structural components of our Galaxy are discussed in detail. Observational evidence for signatures of galaxy formation is presented, and the accreted stellar groups examined in each chapter are outlined.

\section{History of Hierarchical Structure}

Hierarchical structure formation has grown to become the dominant paradigm since the work of \citet{de_vaucouleurs_1970a}. The preferred $\Lambda$CDM paradigm  proposes structure formation is driven by the effect of the large-scale distribution of CDM. The $\Lambda$CDM paradigm provides simple models of galaxy formation and produces hierarchical structure across cosmic scales as a natural consequence \citep{peebles_1974,white;rees_1978,blumenthal;et-al_1984}. From an observational perspective, the efficiency of multiplexing has enabled redshift surveys to explore the properties of galaxies across cosmic time in great detail \citep[for a review, see][]{mo;et-al_2010}. Within the $\Lambda$CDM paradigm, N-body and semi-analytic models simulating cosmic growth since the early universe have been quite successful at replicating these surveys. Simulations include star formation, gas pressure, radiative cooling and heating, as well as metal production. However an accurate prescription for how baryonic matter produces realistic galaxies within a hierarchy of dark matter is required to fully comprehend galaxy formation.

Observations of Milky Way stars have further aided our understanding of galaxy formation and cosmological structure. \citet{eggen;et-al_1962} was instrumental in demonstrating this link. \citet{eggen;et-al_1962} observed the motions of high velocity stars and found while orbital energies and eccentricities increased, the metal abundance and orbital angular momenta decreased. This implied a metallicity gradient with galactocentric radii. From these data they inferred metal-poor halo stars were created from the rapid collapse of a mostly uniform protogalactic cloud. 

This framework was challenged by \citet{searle;zinn_1978}. Following their observations of globular clusters at a range of galactic radii, \citet{searle;zinn_1978} concluded that globular clusters were found at all metallicities, regardless of their distance to the Galactic center. They suggested the halo could emerge over a long period of time from coalescing gas fragments, each with their own level of independent chemical evolution. Smaller systems would fall into the gravitational potential of larger ones, and become tidally disrupted during this process. This Searle \& Zinn view of galaxy formation was certainly influential, but not entirely new. \citet{searle;zinn_1978} made a noteworthy link with theoretical studies of cosmology and galaxy formation ongoing at the time, where hierarchical aggregation was becoming the preferred interpretation of the time \citep{peebles_1971,press;schechter_1974}. This paradigm has become the accepted dominant formation mechanism for the stellar halo (see Section \ref{sec:stellar-halo}).
 
Observationally, evidence of hierarchical formation remains observable today through the accretion of satellite systems onto the Milky Way. Hundreds of globular clusters and dwarf galaxies orbit our Galaxy \citep{harris_1996}, and many are observed to undergo tidal disruption (c.f. Table \ref{tab:tidal-tails}). This has provided the strongest evidence for the \citet{searle;zinn_1978} paradigm in the stellar halo. Disrupted systems serve as tracers of the Milky Way's merger history, allowing us to reconstruct the chemical and dynamical evolution of the Galaxy, probe the Galactic potential, and explore galaxy formation more generally.

\section{Components of The Milky Way}
Humans are unable to observe the Milky Way with the same context or perspective that we are afforded for other galaxies, making it difficult to distinguish the overall breadth and structure of our Galaxy. This is a natural consequence of our location, yet astronomers remain undeterred. \citet{de_vaucouleurs_1970b} classified the Milky Way as a SAB(rs)bc galaxy: a spiral weakly barred and ringed galaxy. However there is ongoing \citep{urquhart;et-al_2014} debate about whether our Galaxy has 2 or 4 spiral arms, and whether these arms are symmetric. The existence of a central bar has been debated, as well as the interacting effect it would have on the disk. The existence of a thick disk has been brought into question \citep{bovy_2011}, and the warp/flare of the disk has been discussed at length \citep[e.g., see][]{momany_2006}. Generally accepted structural components of the Milky Way include a central bulge/bar, a thin disk, thick disk, a dark matter halo, and a stellar halo. While this thesis focusses on accreted substructure in the Milky Way halo, there is interplay between the structural components of our Galaxy, making it important to introduce these pieces.

\subsection{Bulge}
A concentrated region of stars exists at the very heart of our Galaxy. The Milky Way bulge has a somewhat boxy shape, which are generally seen to have an exponential surface brightness distribution in other galaxies. Bulges in spiral galaxies -- including the Milky Way -- are usually assumed to be old. Color-magnitude diagrams for populations in the Milky Way bulge suggest it is old. The presence of RR Lyrae and metal-poor RGB stars also indicate that at least some fraction of the Milky Way bulge is old. Although a large metallicity range is observed, most stars in the bulge are also predominantly metal-rich. This is seemingly counterintuitive with the expectation that metal-poor stars are fossil remnants of an older stellar population. The old, metal-rich population in the bulge is explainable by rapid metal enrichment at early times, facilitated by extremely short dynamical timescales. Following this intense star formation, mean metallicities around [Fe/H] $\sim -1$ at $z \sim 1$ are not unreasonable.

Cold dark matter simulations suggest the bulge formed via mergers, whereas the boxy/peanut shape of the bulge suggests it has arisen from disk instabilities. A combination of the two mechanisms is also possible. The formation mechanisms that led to the present-day bulge are represented in the dynamics and chemistry of stellar populations still present in the bulge. Recent large-scale spectroscopic surveys targeting the bulge (e.g., the ARGOS survey) have drastically improved our understanding of these formation processes \citep{argosI,argosII,argosIII}. \citet{ness;et-al_2012} find distinct stellar components in the galactic bulge, identifiable by their velocities and metallicities: stars with [Fe/H] $> -0.5$ are claimed to be part of the boxy/peanut bar/bulge, and stars with [Fe/H] $< -0.5$ are associated with the thick disk. Distinct sub-components are also present, and their population gradients across the bulge suggest an instability-driven bar/bulge formation originating from the thin disk. Thus, there is clear interplay between the structural components in the Milky Way since the earliest times of formation, whose signatures remain observable today.

\subsection{The Thin Disk}
The defining component for disk galaxies is the existence of an exponential thin disk. The thin disk is considered as the final product of quiescent dissipation of baryons within a virialized dark matter halo, and contains most of the baryonic angular momenta in the Galaxy \citep{read;et-al_2008}. In the Milky Way, the vertical scale height of the thin disk is $\sim$300\,pc \citep{gilmore;reid_1983}, and the oldest disk stars are observed to be between 10 to 12 gigayears old \citep{oswalt_1996, phelps;et-al_1994}. This is consistent with the disk being almost entirely formed by $z \sim 1$. 

The metallicity of stars in the thin disk varies between [Fe/H] $\sim -0.5$ to $+0.3$ \citep{bensby;et-al_2003}. While the bulge underwent rapid star formation at early times, chromospheric ages of thin disk stars indicate star formation in the thin disk near the Sun has been roughly uniform over the last 10 gigayears, although episodic variations by a factor of 2 are evident \citep{rocha-pinto;et-al_2000}. Five independent surveys aiming to measure the age of the thin disk population find an average of 8.7 $\pm$ 0.4 Gyr \citep{lineweaver_1999}. The presence of an old thin disk suggests that the Galaxy has experienced most of its accretion in early times. The accretion of satellite systems at later times would dynamically heat the thin disk. In fact, the presence of the thick disk (Section \ref{sec:thick-disk}) is perhaps an example of such a thin disk-subhalo  (baryonic or dark) interaction that occurred at early times, dynamically heating the thin disk before it had fully settled.

\subsection{The Thick Disk}
\label{sec:thick-disk}
In addition to the thin disk, the Milky Way has a second fainter disk component, referred to as the thick disk \citep{gilmore;reid_1983}. The thick disk is much more diffuse than the thin disk. It has a scale height of $\sim$1-1.5\,kpc \citep{reynolds;et-al_1989} -- approximately three times that of the thin disk -- and a surface brightness only 10\% of the thin disk \citep[e.g.,][]{fuhrmann;et-al_2011}. Metallicities for thick disk stars range from [Fe/H] = $-2.2$ to $-0.5$ \citep{reddy;et-al_2006}, considerably more metal-poor than the thin disk.

A thick disk is not an essential element for the formation of disk galaxies. Given the diffusivity of the thick disk compared to the thin disk, detailed chemical abundances are usually necessary to discern the two populations. Stars in the thick disk are typically more metal-poor than stars in the thin disk.


\subsection{Stellar Halo}
\label{sec:stellar-halo}
The stellar halo is the largest baryonic component of the Milky Way, extending out to at least 100\,kpc. Although the extent of the stellar halo is large, it is extremely diffuse. The mass of the stellar halo is $\sim$1\% of the total mass, and as such it is not readily observed in other galaxies, as the surface brightness is too low to detect from diffuse light. Most objects in the halo were probably some of the first objects to form, and they remain relics of galaxy formation in the present day.

Many groups have examined the stellar density distribution of the Milky Way, finding that it roughly follows a power law density distribution $\rho \propto r^{-3.5}$ \citep[e.g.,][]{Juric;et-al_2008}. There are substantial offsets in the magnitude of this exponent, which appear to depend on the stellar population chosen as a tracer. Even between groups using the same stellar tracer, the results are varied \citep{sluis;arnold_1998,de_propris;et-al_2010}. This is because the contamination, recoverability and fractional volume of the halo observed at a given galactocentric radii can be non-trivial to accurately quantify. Nevertheless the stellar halo diminishes quickly with increasing galactocentric radii, where the presence of accreted substructure becomes evermore apparent.

In recent decades it has become clear that at least part of the stellar halo has been formed through the accretion of small metal-poor satellite systems \citep{ibata;et-al_1994,belokurov;et-al_2006,grillmair_2006,bell;et-al_2008}. Stars from a disrupted parent system have coherent velocities that persist over long timescales. This allows us to identify individual stars that were once disrupted, and chronicle the merger tree of the halo from events that occurred long ago. Groups of stars can be identified by their position, kinematics and chemistry, and associated to a progenitor satellite that matches the phase space signature. It's possible that the entire stellar halo has formed through accretion \citep{bell;et-al_2008}, implying that every halo star is be a remnant from a fossil object accreted by the proto-Milky Way, each with completely independent -- and perhaps, distinguishable -- star formation histories.

\subsection{Dark Halo}
Kinematic galaxy studies as far back as \citet{freeman_1970} and \citet{rubin;et-al_1970} have repeatedly found galaxies to rotate faster than expected \citep[see also][]{rubin;et-al_1980}. This has led to a flurry of research, with the general acceptance that the dynamics are driven by non-luminous (dark) matter. The actual nature of dark matter remains elusive, but its effects are observable through the gravitational field. Exploring the properties of dark matter has shifted away from astrophysics towards particle physics. However, the quantity and distribution of dark matter remains quantifiable on astrophysical scales, providing important constraints for direct dark matter detection techniques.

The Milky Way is dark matter-dominated, with dark matter contributing $>$90\% of the total galactic mass \citep[$1-1.5 \times 10^{12} M_\odot$;][]{smith;et-al_2007,nichols;et-al_2012,kafle;et-al_2012}. The dark halo follows a $\rho \sim r^{-2}$ distribution out to at least 100\,kpc \citep[e.g.,][]{kochanek_1996,battaglia;et-al_2005}, enveloping most of the baryonic matter in the Milky Way. The distribution of dark matter is believed to be spheroidal rather than disk-like \citep{creze;et-al_1998,ibata;et-al_2001}. Although non-luminous, the gravitational forces of the dark halo significantly contribute to the structure of the luminous Galactic components. The misalignment of angular momentum between the dark and baryonic components can result in warps in the thin/thick disk. This implies that a triaxial dark halo distribution has strong repercussions for the tidal disruption of luminous systems in the stellar halo. Interestingly, from their models of the Milky Way--Sagittarius interaction, \citet{Law;et-al_2009} finds a tri-axial dark matter halo to best represent the observed positions and kinematics of stars in the Sagittarius (Sgr) stream (c.f. Chapter \ref{chap:sgr}).

Cosmological simulations \citep{moore;et-al_1999} of cold dark matter suggests that the dark halo is strongly substructured (i.e., clumpy). This has important consequences for the formation of the Milky Way, and the dissipation of the disk. While significant work has been placed on quantifying the complete mass distribution of dark matter sub-halos \citep[e.g., ][]{madau;et-al_2008}, some groups have attempted to identify individual clumps of dark matter via measurable dips in the surface brightness along a stream \citep{diemand;et-al_2008,carlberg;et-al_2012,carlberg;et-al_2013}. However it is not clear whether these observations can be used to identify individual dark matter sub-halos, or they are simply observing the presence of epicyclic stripping. 

\section{Substructure in the Milky Way}

The term substructure is used to describe a group of stars that demonstrate coherent and distinguishable signature in phase space. This coherence may be in any number of observables, including their position, kinematics, or possibly through individual elemental abundances. In the last two decades astronomers have discovered that the Milky Way is rich with substructure, particularly in the outer halo. This has been observed in the presence of globular clusters, open clusters, diffuse over-densities, and the tidal tails of disrupted systems -- also referred to as stellar streams. 

\begin{figure}[h]
	\includegraphics[width=\columnwidth,height=0.5\columnwidth]{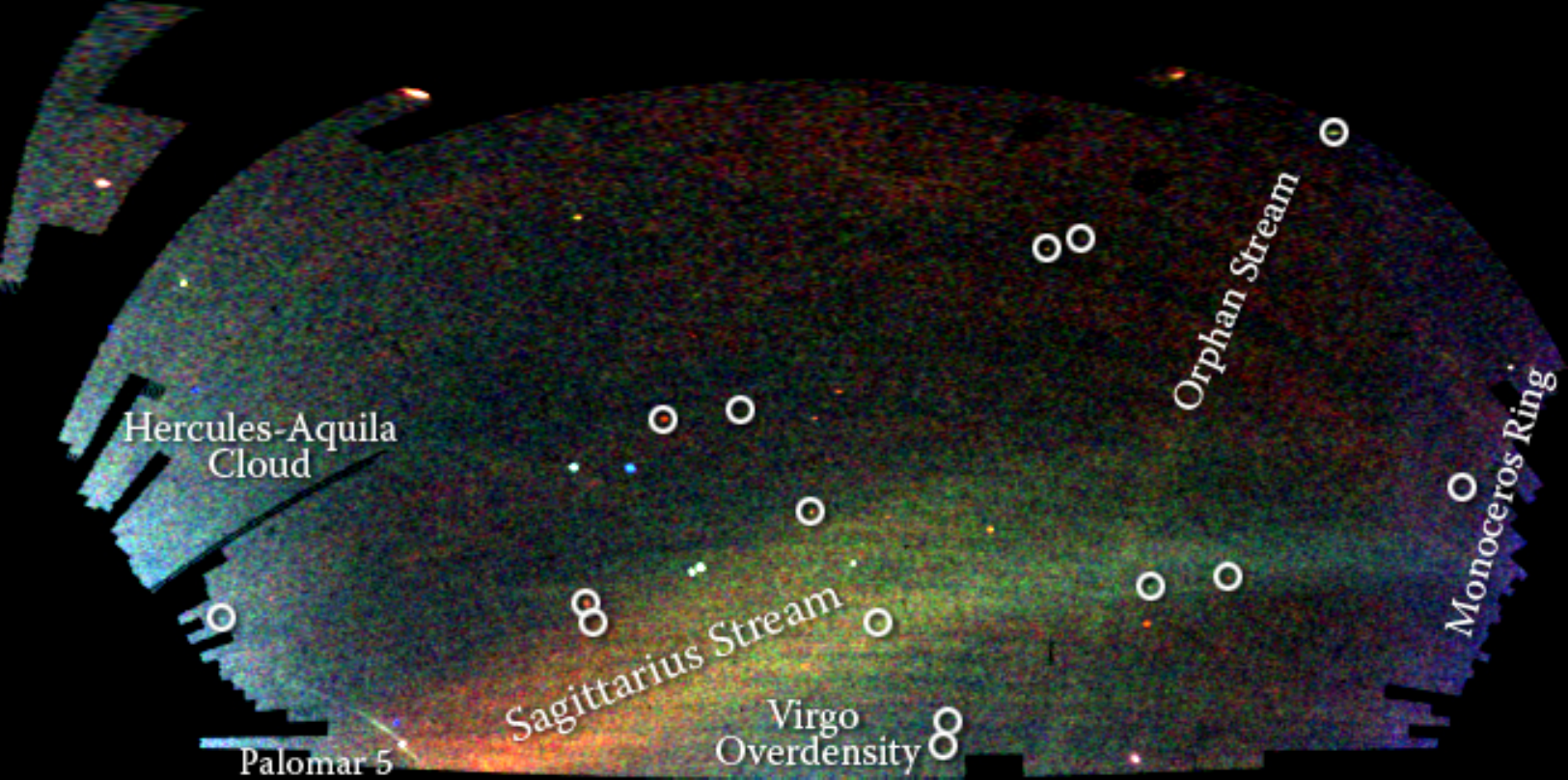}
	\caption[A mercator projection of the ``Field of Streams'']{A mercator projection of the ``Field of Streams'' \citet{belokurov;et-al_2006} assembled from SDSS images of the northern sky. Color indicates the approximate distance to the stars. Structures visible in this region of the sky are marked, including the prominent Sgr stream, the Orphan stream, the `Monoceros Ring', as well as a number of diffuse overdensities and globular clusters.\\Credit: Vasily Belokurov (Institute of Astronomy, Cambridge) and the Sloan Digital Sky Survey collaboration.}
	\label{fig:field-of-streams}
\end{figure}

The presence of substructure in the Milky Way is evident by numerous photometric and spectroscopic studies \citep[for a review, see][]{helmi_2008}. Both diffuse and coherent over-densities (e.g. tidal tails) are observable in the Sloan Digital Sky Survey \citep[hereafter SDSS;][]{York;et-al_2000} with extremely simple colour cuts (e.g., see Figure \ref{fig:field-of-streams}). Radial velocities obtained by follow-up spectroscopy of individual stars in tidal tails reveals a composite velocity distribution: an expected Galactic velocity dispersion, and a significant peak of stars with similar, or indistinguishable, velocities. This signature is usually evidence of a co-moving group of stars which cannot be explained by a smooth halo distribution, radial mixing, or disk heating. Table \ref{tab:tidal-tails} outlines a number of known tidal tails in the Milky Way\footnote{Table \ref{tab:tidal-tails} is an updated and extended version of the database compiled by \href{http://homepages.rpi.edu/~newbeh/mwstructure/MilkyWaySpheroidSubstructure.html}{Heidi Newberg}.}. Diffuse photometric over-densities often have more elaborate explanations as the velocity distribution may be nearly homologous with the expected Galactic contribution. In these cases -- if the over-density is the result of a disrupted satellite -- it suggests considerable mixing has occurred since accretion, and at least some of the dynamical information has already been lost. Nevertheless, any stellar over-density in a focussed direction of the halo is a signature of a significant Galactic formation event. While dynamical information may be lost, distinguishing between accreted and in-situ scenarios may be discernible from detailed chemical abundances (see \ref{sec:intro-chemical-tagging}). Through successive generations of star formation, the detailed chemical abundances represent the integrated star formation history of the local environment. Thus, accreted and in-situ stars will exhibit distinguishable chemical signatures that are representative of their formation environment. With increasing observational data astronomers may be able to identify all substructure in the Milky Way, even those that have undergone nearly complete disruption. This will allow us to identify the inflexion point from in situ-dominated and accretion-dominated formation, build up the Galactic merger history, and would drastically increase the level of known accretion in the Milky Way.

\newcommand{\vat}[1]{\begin{tabular}{@{}l@{}}#1\end{tabular}}
\newcommand{\vatc}[1]{\begin{tabular}{@{}c@{}}#1\end{tabular}}

\begin{table}
\caption{Tidal tails in the Milky Way\label{tab:tidal-tails}}
\resizebox{\textwidth}{!}{%
\begin{tabular}{lccccl}
\hline
\vat{\\Name\\} & \vatc{Right\\Ascension} & \vatc{Declination\\} & \vatc{Position\\Angle} & \vatc{Tail\\Length} & \vat{\\Literature\\ } \\
& (J2000) & (J2000) & ($^\circ$) & (kpc) & \\
\hline
\\

\multicolumn{6}{c}{\textbf{Disrupting Dwarf Galaxies}} \\
\hline
\object[NAME FORNAX dSPH]{Fornax} 			& 02:39:59.30	& $-$34:26:57.0	& --42		& 2.5	& \citet{coleman_fornax} \\
\vat{\object[NAME LMC]{LMC}\\\\ }	& \vat{05:23:33.61\\\\ }&\vat{$-$69:45:36.2\\\\ } & \vat{\nodata\\\\ }	& \vat{22\\\\ }& \begin{tabular}{@{}l@{}}\citet{putman_1998};\\\citet{munoz_2006}\end{tabular} \\
\vat{\object[CARINA dSPH]{Carina}\\\\ } & \vat{06:41:36.70\\\\ } & \vat{$-$50:57:58.0\\\\ } & \vat{65\\\\ } & \vat{\nodata\\\\ } & \begin{tabular}{@{}l@{}}\citet{irwin_1995};\\\citet{munoz_2006}\end{tabular} \\
\object[LEO I]{Leo I}	& 10:08:28.12 & $+$12:18:23.4 & 84 & $\sim$0.4 & \citet{sohn;et-al_2007} \\
\object[URSA MAJOR dSPH]{Ursa Major} & 10:34:52.80 & $+$51:55:12.0 & 78 & $\sim$29 & \citet{okamoto;et-al_2008} \\
\vat{\object[URSA MINOR dSPH]{Ursa Minor}\\\\ } & \vat{15:09:11.34\\\\ } & \vat{$+$67:12:51.7\\\\ } & \vat{53\\\\ } & \vat{3\\\\ } & \vat{\citet{irwin_1995};\\\citet{palma;et-al_2003}} \\
\vat{\object[NAME HER DWARF GALAXY]{Hercules}\\\\ } & \vat{16:31:02.40\\\\ } & \vat{$+$12:47:30.0\\\\ } & \vat{$-$73\\\\ } & \vat{$\sim$21\\\\ } & \vat{\citet{belokurov;et-al_2007_quintet};\\
\citet{coleman;et-al_2007}} \\ 
\object[NAME SDG]{Sagittarius} & 18:55:03.10 & $-$30:28:42.0 & 100 & $>$500 &
\citet{pakzad;et-al_2004}
\\
\object[NGC 6822]{NGC 6822} & 19:44:56.20 & $-$14:47:51.3 & $\sim$170 & $\sim$1.4 & \citet{de_block;et-al_2000} \\
\hline

\\
\multicolumn{6}{c}{\textbf{Disrupting Globular Clusters}} \\
\hline

\object[NGC 362]{NGC 362} & 01:03:14.26	& $-$70:50:55.6	& $\sim$135 & $\sim$2.5 & \citet{grillmair;et-al_1995} \\
\object[NGC 1261]{NGC 1261} & 03:12:16.21 & $-$55:12:58.4	& $\sim$135 & $\sim$2.8 & \citet{leon;et-al_2000} \\
\object[NGC 1851]{NGC 1851} & 05:14:06.76 & $-$40:02:47.6  & $\sim$140 & $\sim$2 & \citet{leon;et-al_2000} \\
\vat{\object[NGC 1904]{NGC 1904} (M 79)\\\\ }& \vat{05:24:10.59\\\\ } & \vat{$-$24:31:27.3\\\\ } & \vat{$\sim$40\\\\ } & \vat{$\sim$2.4\\\\ } & \vat{\citet{grillmair;et-al_1995}\\\citet{leon;et-al_2000}} \\
\object[NGC 2298]{NGC 2298} & 06:48:59.41 & $-$36:00:19.1 & $\sim$150 & $\sim$1.5 & \citet{leon;et-al_2000} \\
\object[NGC 2808]{NGC 2808} & 09:12:03.10 & $-$64:51:48.6 & $\sim$160 & $\sim$3 & \citet{grillmair;et-al_1995} \\
\object[NGC 3201]{NGC 3201} & 10:17:36.82 & $-$46:24:44.9 & $\sim$140 & $\sim$4.4 & \citet{grillmair;et-al_1995} \\
\object[NGC 4372]{NGC 4372} & 12:25:45.43 & $-$72:39:32.7 & \nodata & \nodata & \citet{leon;et-al_2000} \\
\object[NGC 4590]{NGC 4590} (M 68) & 12:39:27.98 & $-$26:44:38.6 & $\sim$90 & $\sim$3.7 & \citet{grillmair;et-al_1995} \\
\vat{\object[NGC 5139]{NGC 5139} ($\omega$Cen)\\\\ } & \vat{13:26:47.28\\\\ } & \vat{$-$47:28:46.1\\\\ } & \vat{$\sim$165\\\\ } & \vat{$\gtrsim$5.3\\\\ } & \vat{\citet{leon;et-al_2000}; \\ \citet{majewski;et-al_2012}\\} \\

\object[NGC 5272]{NGC 5272} (M 3) & 13:42:11.62 & $+$28:22:38.2 & $\sim$85 & $\sim$3.5 & \citet{leon;et-al_2000} \\
\vat{\object[NAME NGC 5466]{NGC 5466}\\\\ } & \vat{14:05:27.29\\\\ } & \vat{$+$28:32:04.0\\\\ } & \vat{130\\\\ } & \vat{45\\\\ } & 
\vat{\citet{belokurov;et-al_2006_5466};\\\citet{grillmair_johnson_2006}} \\
\object[NGC 5694]{NGC 5694} & 14:39:36.52 & $-$26:32:18.0 & $\sim$135 & $\sim$1.3 & \citet{leon;et-al_2000} \\
\object[NGC 5824]{NGC 5824} & 15:03:58.61 & $-$33:04:06.7 & $\sim$175 & $\sim$3.8 & \citet{leon;et-al_2000} \\
\object[PAL 5]{Palomar 5} & 15:16:05.30 & $-$00:06:41.0 & 45 & 22.5 & \citet{grillmair_dionatos_2006a} \\

\object[NGC 5904]{NGC 5904} (M 5) & 15:18:33.22 & $+$02:04:51.7 & $\sim$140 & $\sim$1.3 & \citet{leon;et-al_2000} \\
\object[NGC 6205]{NGC 6205} (M 13) & 16:41:41.63 & $+$36:27:40.8 & $\sim$150 & $\sim$1 & \citet{leon;et-al_2000} \\
\object[NGC 6254]{NGC 6254} (M 10) & 16:57:09.05 & $-$04:06:01.1 & $\sim$5 & $\sim$4 & \citet{leon;et-al_2000} \\
\object[NGC 6397]{NGC 6397}  & 17:40:42.09 & $-$53:40:27.6 & $\sim$100 & $\sim$3.5 & \citet{leon;et-al_2000} \\
\object[NGC 6535]{NGC 6535}  & 18:03:50.51 & $-$00:17:51.5 & $\sim$160 & $\sim$0.8 & \citet{leon;et-al_2000} \\
\object[NGC 6712]{NGC 6712}  & 18:53:04.32 & $-$08:42:21.5 & \nodata & \nodata & \citet{de_marchi;et-al_1999} \\
\object[NGC 6864]{NGC 6864} (M 75) & 20:06:04.84 & $-$21:55:20.1 & $\sim$125 & $\sim$3 & \citet{grillmair;et-al_1995} \\
\object[NGC 6934]{NGC 6934} & 20:34:11.37 & $+$07:24:16.1 & $\sim$135 & $\sim$3 & \citet{grillmair;et-al_1995} \\
\object[NGC 6981]{NGC 6981} (M 72) & 20:53:27.70 & $-$12:32:14.3 & $\sim$100 & $\sim$2.8 & \citet{grillmair;et-al_1995} \\
\object[NGC 7078]{NGC 7078} (M 15) & 21:29:58.33 & $+$12:10:01.2 & $\sim$160 & $\sim$3.6 & \citet{grillmair;et-al_1995} \\
\object[NGC 7089]{NGC 7089} (M 2) & 21:32:35.27 & $-$44:04:02.9 & $\sim$55 & $\sim$3.3 & \citet{grillmair;et-al_1995} \\
\object[PAL 12]{Palomar 12} & 21:46:38.84 & $-$21:15:09.4 & $\sim$0 & $\sim$2.3 & \citet{leon;et-al_2000} \\
\object[NGC 7492]{NGC 7492} & 23:08:26.68 & $-$15:36:41.3 & $\sim$70 & $\sim$0.6 & \citet{leon;et-al_2000} \\
\hline
\end{tabular}}
\end{table}

Not all Galactic accretion has occurred long ago. The discovery that the Sgr dwarf spheroidal (dSph) galaxy was experiencing ongoing accretion is a prominent example for recent and substantial accretion onto the Milky Way. \citet{ibata;et-al_1994} detected the Sgr stream by a clear peak in their velocity distribution of stars in the direction of the bulge: a significant fraction of stars in their field all had nearly the same velocity, and that signature was distinct from the background Galactic contribution. Stars from the disrupting dwarf remained kinematically cold, and a coherent velocity histogram is observable for long timescales. Since its discovery, the tidal arms of the Sgr dwarf have been seen to encircle around the Milky Way with overlapping tidal tails, littering debris throughout the Galaxy. The Sgr stream remains the most prominent evidence of ongoing accretion in the Milky Way, inspiring extensive observational and theoretical work in the last two decades. 

Sgr is by far not the only example of recent accretion in the Milky Way. The epoch of large scale digital imaging surveys has facilitated significant leaps forward in galaxy formation and substructure identification. The SDSS and 2-Micron All Sky Survey \citep[hereafter 2MASS;][]{skrutskie;et-al_2006} have been particularly useful in this regard. Using these data, a wealth of structure has been identified across a broad range of cosmological scales. In fact, perhaps the most iconic example of substructure in the Local Group has  resulted from the SDSS catalogue. The aptly named `Field of Streams' \citep[Figure \ref{fig:field-of-streams};][]{belokurov;et-al_2006} contains a wealth of substructure in a single image. Multiple disrupting tidal tails are present including two arms of Sgr, as well as numerous globular clusters and diffuse over-densities. The coming decade will be equally exciting for substructure discovery, as the SkyMapper \citep{keller;et-al_2007} and Pan-STARRS \citep{hodapp;et-al_2004} telescopes begin their deep imaging surveys of the southern sky.

There is a wide distribution of masses and structure sizes that have been accreted onto the stellar halo across all cosmic times. Around $\sim$10 dwarf galaxies are known to be have undergone tidal disruption, and $\sim{}$30 of the 150 Galactic globular clusters present evidence of tidal tails. In addition to these systems, at least another dozen or so stellar streams have been identified \emph{without} the presence of any disrupting host (see Table \ref{tab:tidal-tails} for a referenced overview). A detailed account of the discovery and properties of these substructures is beyond the scope of this thesis\footnote{A full examination of known substructures in the Milky Way would serve as the subject of a (presumably) excellent review article in the future, years after a thorough deep photometric survey of the southern sky has concluded.}. A table highlighting some of the disrupting systems in the Milky Way without known parent systems is shown in Table \ref{tab:milky-way-substructure}.

Some tension exists between observations and theory of substructure in the Milky Way. Cold dark matter simulations suggest the level of substructure observed in the Milky Way is at least an order of magnitude less than what ought to be present. Simulations indicated there should be approximately $\sim{}$10-100 times more globular clusters and dwarf galaxies surrounding the Galaxy \citep{klypin;et-al_1999}. Although recent work \citep{kravtsov;et-al_2004,simon;et-al_2007,brooks;et-al_2012} has found the discrepancy between observations and theory is not as substantial as first proposed \citep[$\sim$4 too many;][]{simon;et-al_2007}, this disparity is commonly referred to as the `Missing Satellites Problem'. Additional Galactic satellites may be found through deep multi-band imaging of the Southern Sky, which would further alleviate or resolve this issue.

\subsection{Quantifying the Level of Substructure}

Many studies have tried to quantify the level of substructure in the halo, and the results are varied. Some groups have placed lower limits \citep{starkenburg;et-al_2009}, whereas others claim the entire halo ($D>20$\,kpc) to have been formed through accretion \citep{bell;et-al_2008}. There are many difficulties in measuring the substructure contribution, including the stellar tracer used \citep{preston;et-al_1991,green;morrison_1993,vivas;zinn_2006}, the definition employed to identify substructure, as well as recoverability and contamination, to name a few. The stellar tracer must also be intrinsically bright -- such that they can probe far distances in the halo -- and be relatively free of contamination.

When quantifying the level of substructure in the halo, one must be particularly careful when specifying what qualifies as a substructure. In the simplest scenario, a substructure may be definable as a cohort of stars with a common observable that is statistically significant above the smooth halo background. Unfortunately, even this abstract description is plagued. Defining statistical significance usually requires some comparison to a smooth model of the Galaxy. Even a statistical measure that is non-parameterized will require a comparison with a model of the smooth Galactic background. Thus, some model of the Galaxy has historically been employed to quantitatively measure substructure in the Milky Way. If the model employed is oversimplified, then the warp of the disk may be qualitatively recognised as statistically significant accreted substructure. In reality our over-simplification of the Galaxy is resulting in an over-estimation of the substructure present. These considerations become important in Chapter \ref{chap:aquarius}, where we examine the existence of the recently discovered Aquarius stream \citep{williams;et-al_2011}.

\subsection{Weighing the Milky Way}

In addition to probing the merger history of the Galaxy, there are key measurements that can be made with stellar streams. One of particular interest is measuring the mass of the Milky Way. Since the positions and velocities of stars in stellar streams are sensitive to the gravitational potential, tidal tails serve as powerful probes to measure the galactic potential \citep[e.g.,][]{adrn_2013}. Through this process, one can also infer the shape, extent and distribution of dark matter \citep[e.g., ][]{Law;et-al_2009}. These are critical measurements for cosmological simulations. 

Several groups have used the prominent Sgr stream for inferring the Milky Way potential \citep{edelsohn_1997,Law;et-al_2005,Law;et-al_2009}. The more diffuse, less luminous streams have also been used \citep{newberg;et-al_2010}, but this is reliant on a wealth of accurate positional and kinematic data \citep{adrn_2013}. As more exquisite distances are obtained from the Gaia satellite mission \citep{perryman}, more representative models of the Milky Way can be devised.

\begin{table}[h!]
\caption{Stellar streams in the Milky Way without associated systems\label{tab:milky-way-substructure}}
\resizebox{\textwidth}{!}{%
\begin{tabular}{lccccl}
\hline
\vat{\\Name\\} & \vatc{Right\\Ascension} & \vatc{Declination\\} & \vatc{Position\\Angle} & \vatc{Tail\\Length} & \vat{\\Literature\\ } \\
& (J2000) & (J2000) & ($^\circ$) & (kpc) & \\
\hline
\object[Cetus Polar Stream]{Cetus Polar Stream} & 01:33:24 & $+$03:13:48 & \nodata & \nodata & \citet{newberg;et-al_2009} \\
\object[Triangulum Stream]{Triangulum Stream} & 01:35:54 & $+$23:12:04 & $\sim$160 & 5.5 & \citet{bonaca;et-al_2012} \\
\vat{\object[Orphan Stream]{Orphan Stream}\\\\ } & \vat{10:40:00\\\\ } & \vat{$+$05:00:00\\\\ } & \vat{159\\\\} & \vat{60\\\\ } & \vat{\citet{grillmair_2006};\\\citet{belokurov;et-al_2006}} \\
\object[Styx Stream]{Styx Stream} & 13:56:24 & $+$26:48:00 & $\sim$45 & 65 & \citet{grillmair_2009} \\
\object[Acheron Stream]{Acheron Stream} & 15:50:24 & $+$09:48:36 & $\sim$70 & 37 & \citet{grillmair_2009} \\
\object[Lethe Stream]{Lethe Stream} & 16:15:45 & $+$29:57:00 & $\sim$30 & 87 & \citet{grillmair_2009} \\
\object[Cocytos Stream]{Cocytos Stream} & 16:29:22 & $+$26:40:12 & $\sim$40 & 80 & \citet{grillmair_2009} \\
\hline
\end{tabular}}
\end{table}

\section{Chemical Evolution of the Milky Way}

Astrophysics is limited by observations; we cannot easily replicate the conditions of stellar interiors in any laboratory. Additionally, we cannot directly observe the interiors of stars. The interior must be inferred with models employing observables as boundary constraints. Only the outer layers of the photosphere are accessible for the inference of stellar structure and elemental abundances. Modulo surface abundance variations due to internal mixing, atomic diffusion and mass transfer from a binary companion, the abundances in the photosphere of a star retain the chemical signature of the proto-stellar gas cloud from which they formed. Thus by measuring the elemental abundances of a given star, we are effectively observing the gas cloud composition where that star was born. 

This facilitates extraordinarily research. Since each star forms from gas chemically enriched by its predecessors, the observed chemical abundances for any star are the cumulative sum of elements produced throughout its star formation history. Through measuring the elemental abundances in stellar photospheres we can effectively trace the changing gas composition throughout the Milky Way history, as well as the yields and frequency of supernovae throughout throughout the Galaxy. Even as dynamical signatures dissipate over time, the chemical signature remains largely fossilised in the photospheres of stars.

High-resolution spectra are required to accurately capture the full extent of chemical abundances for a given system. While overall metallicities can be reliably obtained from low-resolution spectra ($\mathcal{R} < 2000$), the evolution of abundance groups (e.g., $\alpha$-elements, Fe-peak elements, and neutron-capture elements) is critically important to infer the environment where those stars have formed, and can only be accurately ascertained with high-resolution spectra ($\mathcal{R} \sim 20,000+$). With high-resolution spectra from enough stars, the chemical enrichment of a system can be fully characterised. Following this idea, large samples of high-resolution spectra allow us to identify stars that have formed from the same gas composition, but are now dispersed in Milky Way \citep[e.g., `chemical tagging';][]{freeman;bland-hawthorn_2002}.

For tidal tails in the Milky Way that do not have a known parent system, this allows us to infer properties of the undiscovered host before it is discovered. With high-resolution spectra of disrupted members, one can reveal the chemical enrichment of the stream system, and compare its chemical signatures to known satellites in the Galaxy. This approach becomes relevant in Chapter \ref{chap:orphan_II}, where we undertake the first high-resolution spectroscopic study on the Orphan stream.

\subsection{Chemically Tagging the Galaxy}
\label{sec:intro-chemical-tagging}

In the coming years a number of high-resolution spectroscopic surveys will be completed. The APOGEE \citep{apogee} and the Gaia-ESO \citep{gilmore} surveys are already underway, and the GALAH survey \citep{hermes} is commencing in 2014A. These surveys complement each other in their scientific objectives, target selection, and spectral coverage. Together, they will obtain high-resolution spectra for over a million stars -- an unprecedented volume of spectral data in astronomy.

With these data, the chemical evolution of the Galaxy can be unravelled. Moreover it may be possible to chemically tag individual stars as remnants of individual proto-galaxy fragments. The merger history of the Milky Way can be exposed with unprecedented detail, allowing for the exploration of stellar interiors, nucleosynthesis, and galactic structure formation. However the crux of these interpretations will rely on the homogeneity of their spectral analyses. As we will outline in Chapters \ref{chap:smh} and \ref{chap:aquarius}, non-negligible offsets in results are found between spectroscopists, even with high-resolution spectra. To infer minute chemical differences throughout the Galaxy, a homogenous analysis is essential. For the high-resolution spectra presented in Chapters \ref{chap:orphan_II} and \ref{chap:aquarius}, we have strived for an easily reproducible, homogenous analysis through the use of the analysis software outlined in Chapter \ref{chap:smh}.

%


\section{Thesis Outline}
Although the formation history of the Milky Way is tangled and chaotic, it serves as an excellent -- and more importantly, accessible -- laboratory to investigate the evolution of the Universe since the earliest times. To understand the formation history of the Milky Way, this thesis investigates known substructures in our Galaxy. Specifically, we focus on the Virgo Over-Density (VOD), and three previously discovered tidal tails; the Sgr stream, and the Orphan and Aquarius streams. We aim to place these systems in context with the rest of the Galaxy, and to further chronicle the Milky Way's evolution. Thus far we have presented an introduction to cosmological structure and galaxy formation and the components of the Milky Way have been outlined, providing context for the following chapters.

In Chapter \ref{chap:sgr} we present an analysis of low-resolution spectra taken in the vicinity of the VOD and the Sgr stream. Although debris from the Sgr stream is littered throughout the Galaxy, it is distinctive from the galactic background and coherent in position and velocity. In contrast the VOD is a diffuse cloud of stars, showing an excess of stellar material from its polar opposite. Both of these substructures -- and others -- reside in a complex, accretion-dominated region of the sky. We explore the metallicity distribution function of each system, and use phase-space information on the Sgr debris to constrain the shape and extent of dark matter in the Milky Way.

The structure of Chapter \ref{chap:smh} is distinguishable from the rest of this thesis. Chapter \ref{chap:smh} outlines a software package that was written during the course of this doctoral candidature. This software, named `Spectroscopy Made Hard', has been written with the intent to streamline the analysis of high-resolution stellar spectra. The motivation for this software is outlined, and the steps involved in a classical curve-of-growth analysis are discussed in the context of presenting the software. This software has been used for a number of projects by expert spectroscopists, and has been used for the analysis of spectra in Chapters \ref{chap:orphan_II} and \ref{chap:aquarius}.

Chapter \ref{chap:orphan_I} presents the first of two chapters where we investigate the Orphan stream. The Orphan stream was independently discovered by \citet{grillmair_2006} and \citet{belokurov;et-al_2006} in the SDSS catalogue, stretching over 60$^\circ$ of arc. In Chapter \ref{chap:orphan_I} we present our method to identify sparse red giant branch stars in the stream from low-resolution spectra.

High-resolution spectra were obtained for a number of targets presented in Chapter \ref{chap:orphan_I}. Strong conclusions about the undiscovered host system can be drawn from the detailed chemical abundances of the disrupted stream stars. We present our analysis of these spectra in Chapter \ref{chap:orphan_II}, and discuss our chemical abundances in context of  known systems that have been suggested as the parent to the Orphan stream.

A detailed examination of the Aquarius stream is chronicled in Chapter \ref{chap:aquarius}. The Aquarius stream was discovered by \citet{williams;et-al_2011} from the Radial Velocity Experiment \citep[hereafter RAVE;][]{steinmetz;et-al_2006} survey catalogue. A follow-up spectroscopic analysis by \citet{wylie-de-boer;et-al_2012} proposed the Aquarius stream to have resulted from a disrupted globular cluster. We have obtained high-resolution, high signal-to-noise ($S/N$) ratio spectra for a third of the known members in the stream. Detailed chemical abundances suggests the stream is actually Galactic in origin, ruling out any accretion origin.

Concluding remarks are summarised in Chapter \ref{chap:conclusions}, which includes a discussion of future work planned for the analysis software presented in Chapter \ref{chap:smh}.


\chapter{Halo Substructures in the Vicinity of the Virgo Over-Density}
\label{chap:sgr}

\previouslypublished{Parts of this chapter have been previously published as \href{http://adsabs.harvard.edu/abs/2012AJ....143...88C}{`Kinematics \& Chemistry of Halo Substructures: The Vicinity of the Virgo Over-Density', {Casey}, A.~R., {Keller}, S.~C., {Da Costa}, G., 2012, AJ, 143, 88C}. The work is presented here in expanded and updated form.} \\

We present observations obtained with the AAT's 2dF wide field spectrograph AAOmega of K-type stars located within a region of the sky which contains the VOD and the leading arm of the Sgr Stream.  On the basis of the resulting velocity histogram we isolate halo substructures in these overlapping regions including Sgr and previously discovered Virgo groups. Through comparisons with \textit{N}-body models of the Galaxy-Sgr interaction, we find a tri-axial dark matter halo is favoured and we exclude a prolate shape. This result is contradictory with other observations along the Sagittarius leading arm, which typically favour prolate models.  We have also uncovered K-giant members of Sgr that are notably more metal poor ($\langle[$Fe/H$]\rangle = -1.7\,\pm\,0.3$ dex) than previous studies. This suggests a significantly wider metallicity distribution exists in the Sagittarius Stream than formerly considered. We also present data on five carbon stars which were discovered in our sample.

\section{Introduction}
	
	
The most prominent ongoing accretion event in the Milky Way is the Sgr dSph galaxy. The tidal tails of the disrupting satellite were originally discovered by \citet{ibata;et-al_1994} as a co-moving group of K- and M-type giants. Almost two decades since this discovery, stream debris is seen to circle our Galaxy. As such, the Sgr tidal tails have been extensively traced with red-clump stars \citep{Majewski;et-al_1999}, carbon stars \citep{Totten;Irwin_1998, ibata;et-al_2001}, RR Lyrae stars \citep{Ivezic;et-al_2000, Vivas;et-al_2005, keller;et-al_2008, Watkins;et-al_2009, Prior;et-al_2009b}, A-type stars \citep{Newberg;et-al_2003}, BHB stars \citep{Ruhland;et-al_2011} and K/M-giants \citep{Majewski;et-al_2003,  Yanny;et-al_2009, keller;et-al_2010}. Tracers originating from the host system can be identified with spatial and kinematical information because they remain dynamically cold, and are identifiable as kinematic substructures long after they are stripped from their progenitor \citep[e.g.,][]{Ibata;Lewis_1998, Helmi;White_1999}. 
	
Stars within these tidal tails are kinematically sensitive to the galactic potential. This has led various groups to model the Sgr interaction with different dark matter profiles. \citet{Martinez-Delgado;et-al_2004} traced the Northern leading arm and found a near spherical or oblate ($q \approx 0.85$) dark matter halo best represented the observed debris, coinciding with the findings of \citet{ibata;et-al_2001}. In contrast, \citet{Helmi_2004} found evidence in the Sgr leading debris that most favoured a prolate ($q = 1.25$) halo. \citet{Vivas;et-al_2005} found that either a prolate or spherical model of \citet{Helmi_2004} would fit their RR Lyrae observations, rather than those of an oblate model. \citet{Johnston;et-al_2005} later pointed out that no prolate model can reproduce the orbital pole precession of the Sgr debris, but an oblate potential could. \citet[hereafter LJM05]{Law;et-al_2005} performed simulations using data of the Sgr debris from the 2MASS catalogue and found that the kinematics of leading debris was best fit by prolate halos, whereas the trailing debris typically favoured oblate halos. \citet{Prior;et-al_2009a} reached a similar conclusion.
	
\citet{belokurov;et-al_2006} found an apparent bifurcation within the Sgr debris which \citet{Fellhauer;et-al_2006} argued can only result from a dark halo having a near spherical shape. \citet{Law;et-al_2009} introduced a tri-axial model with a varying flattening profile $q$, which replicates the orbital precession seen and matches kinematic observations of the Sgr debris. However \citet{Law;Majewski_2010} (hereafter LM10) concede this may be a purely numerical solution as tri-axial halos are dynamically unstable. They emphasize that more kinematic measurements in other regions of the Sgr stream are required.

After the Sgr stellar stream, the VOD is arguably the next most significant substructure within our Galaxy. The first over-density in the vicinity of the VOD was observed as a group of RR Lyrae stars by the QUEST survey \citep{Vivas;et-al_2001}. The collaboration later named this the ``$12{\mbox{$.\!\!^{\mathrm h}$}}4$ clump'' \citep{Zinn;et-al_2004}. The broad nature of the VOD was uncovered from the SDSS catalogue as a diffuse over-density of main-sequence turnoff stars centered at $r_\odot \sim18$ kpc \citep[which ][dubbed as S297+63-20.5]{Newberg;et-al_2002}. The nomenclature on the substructure names within this region is varied. In this chapter when referring to the VOD we are discussing the spatial over-density of stars within the region, separate from any detected co-moving groups.
	
It is difficult to accurately distinguish stars in the VOD, as there are multiple substructures along this line-of-sight. \citet{Duffau;et-al_2006} took observations of BHB and RR Lyrae within the ``$12{\mbox{$.\!\!^{\mathrm h}$}}4$ clump'' and found a common velocity of $V_{GSR} = 99.8$\,km s$^{-1}$ with $\sigma_{v} = 17.3$\,km s$^{-1}$. It should be noted that the dispersion in kinematics measured by \citet{Duffau;et-al_2006} is essentially their velocity precision, so the substructure may possess a much smaller kinematic dispersion. This co-moving group was coined the `Virgo Stellar Stream' (VSS) to differentiate it from the broad spatial over-density. This distinction from the VOD was strengthened with new distance measurements that placed the VOD centroid at $r_\odot = 16$\,kpc \citep{Juric;et-al_2008, keller;et-al_2010}, and the VSS 3\,kpc further away \citep{Duffau;et-al_2006}. Although, when considered in the light of systematic and observational uncertainties, this is of marginal significance. Additionally, \citet{Juric;et-al_2008} suggests the VOD may extend between $r_\odot = 6$ to 20\,kpc, which further complicates the matter of distance separation.
 
The relationship between the VSS and the S297+63-20.5 over-density is still unclear. \citet{Newberg;et-al_2007} found a kinematic signature at $V_{GSR} = 130 \pm 10$\,km s$^{-1}$ for members of the VOD/S297+63-20.5, which is extremely close to the VSS peak. The VSS and S297+63-20.5 are co-incident in space, but their velocity difference has not yet been reconciled. The distance measurements between S297+63-20.5 and the VSS are similar enough ($\sim$1\,kpc) for \citet{Newberg;et-al_2007} and \citet{Prior;et-al_2009a} to infer they are part of the same structure. \citet{Newberg;et-al_2007} estimate a distance to S297+63-20.5 of $r_\odot = 18$\,kpc from $g_0 = 20.5$ turnoff stars, but they concede the structure is likely dispersed along the line-of-sight as the Color-Magnitude Diagram (CMD) for this region does not demonstrate a tight sequence.  Certainly this region of sky, aptly coined the 'Field of Streams' by \citet{belokurov;et-al_2006}, is complex territory.
	
Photometric studies are inadequate to fully untangle this region. Kinematics are essential to identify co-moving groups that are distinct from the field. Chemical information is vital to accurately distinguish these substructures and understand their origins. However, very few studies have directly investigated metallicities for these stars. In this chapter we present spectroscopic observations of K-giants in this region.  Sgr stream kinematics are used to probe the shape of the dark matter halo in the Galaxy. We report both the velocities and metallicities of these giants in an effort to help untangle this accretion-dominated region.

Target selection methodology is outlined in the next section which is followed with details regarding the observations. Techniques used to separate K-giants from dwarfs are discussed in Section \ref{sec:ch1-dwarf-giant-separation}, and our analysis procedure for kinematics (Section \ref{sec:ch1-kinematics}) and metallicities (Section \ref{sec:ch1-metallicities}) follows. A discussion of substructures is outlined in Section \ref{sec:ch1-discussion}, and in Section \ref{sec:ch1-carbon-stars} we report the carbon stars discovered in our sample. In Section \ref{sec:ch1-conclusions} we conclude with some final remarks and critical interpretations. 

\section{Target Selection}
\label{sec:ch1-target-selection}

\begin{figure}[h]
	\includegraphics[width=\columnwidth]{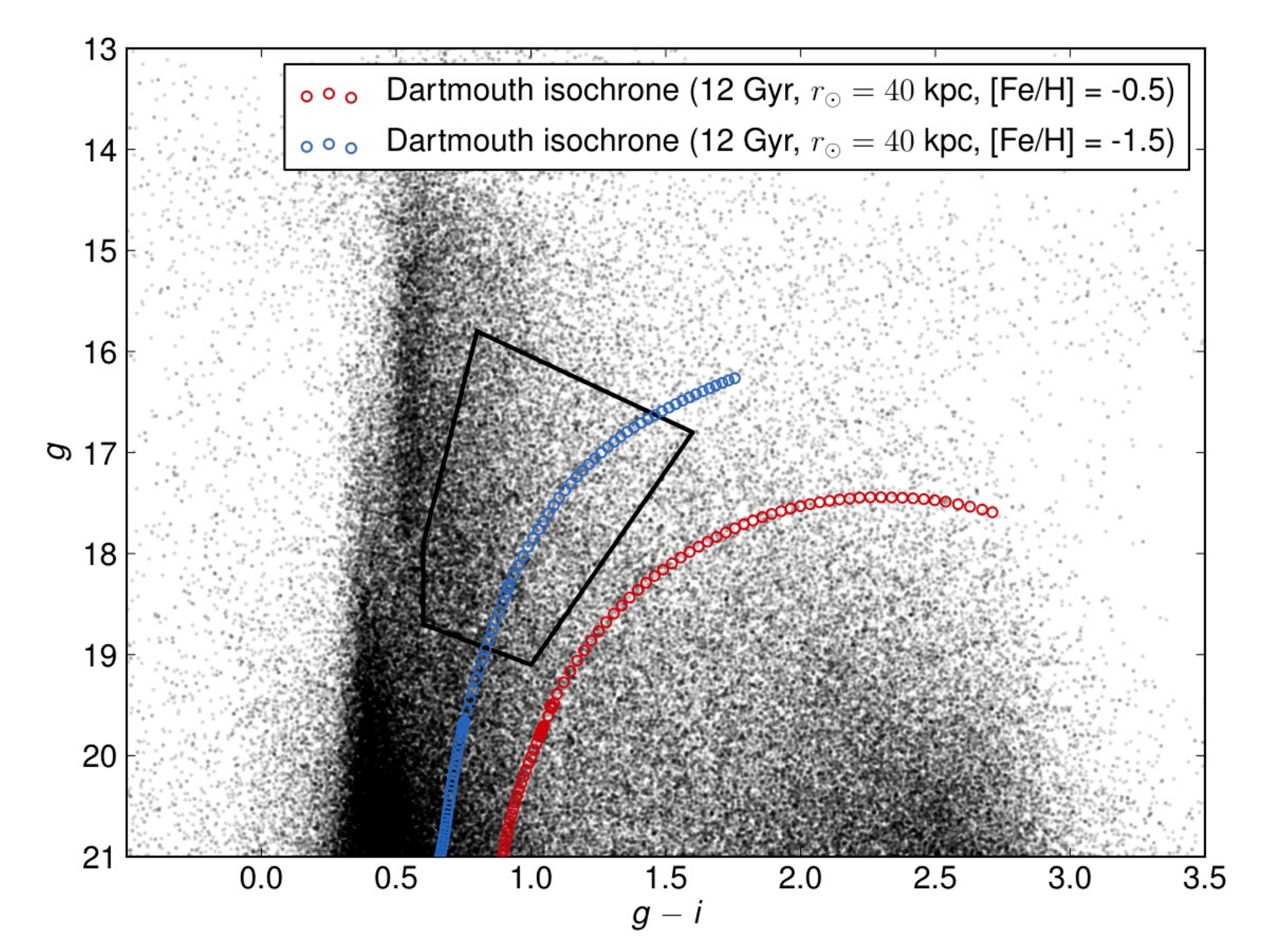}
	\caption[Photometric target selection for the Virgo Over-Density]{CMD of our observed regions from SDSS DR7, overlaid with the color selection criterion used to target K-giants. Appropriate Dartmouth isochrones \citep{Dotter;et-al_2008} are shown for Sgr debris at a distance of $r_\odot =40$ kpc \citep{belokurov;et-al_2006}}
	\label{fig:cmd-target-selection}
\end{figure}

When the presence of a stellar substructure is uncovered, K-giants provide excellent candidates for spectroscopic follow-up. Their spectra permit precise radial velocities and chemical abundances. In order to target K-giants we have targeted candidates within the color selection box shown in Figure \ref{fig:cmd-target-selection}, taken from the SDSS DR7 catalogue. This selection box favors a metal-poor population at Sgr stream-like distances ($\sim$40 kpc) but will also contain similarly metal-poor stars at VOD/VSS distances ($\sim$20 kpc) without being significantly biased. Some field dwarfs are expected to contaminate the sample due to their similarity in colors. Although K-dwarfs are difficult to distinguish photometrically, we can spectroscopically separate these through the equivalent width (EW) of the  gravity-sensitive Mg I triplet lines at 516.7\,nm, 517.3\,nm, and 518.4\,nm (see Section \ref{sec:ch1-dwarf-giant-separation}).

\section{AAT Observations with AAOmega}
\label{sec:ch1-observations}

Low-resolution spectra were obtained for targets over two runs using AAOmega on the 3.9-m Anglo-Astronomical Telescope at Siding Springs Observatory in New South Wales, Australia. AAOmega is a double-beam, multi-object (392) fibre-fed spectrograph covering a two degree field of view. The targets were observed in normal visitor mode in April 2009. Throughout all observations, sufficient sky fibres ($\sim$30) were allocated to ensure optimal sky subtraction. In total 3,453 science targets were observed across 4 fields within the VOD/Sgr region, as outlined in Table \ref{tab:ch1-observations}. Multiple configurations were observed for most fields to permit measurements of bright ($i < 16$) and faint ($i > 16$) stars, as well as repeat observations on a subset of stars.

The beam was split into red and blue arms using the 570\,nm dichroic. The 580V grating in the blue arm yields spectra between 370-580\,nm, with a resolution of $\mathcal{R} = 1300$. In the red arm we used the 1000I grating ($\mathcal{R} = 4400$) which spans the spectral range from 800-950\,nm. This coverage includes the Ca II near infrared triplet (CaT) lines, which are used for radial velocities and metallicities. Science targets on each configuration were limited to 1.5 magnitudes in range to minimise scattered-light cross talk between fibres. Globular clusters NGC 5024, 5053 and 5904 were observed as radial velocity and metallicity standards.

\begin{table}[h!]
\centering
\caption{Observed field centers and number of configurations\label{tab:ch1-observations}}
\begin{tabular}{cccc}
\hline
\hline
Field & $\alpha$ & $\delta$ & Field \\
& (J2000) & (J2000) & Configurations \\
\hline
A & 12 00 00 & $+00$ 00 00 & 1 \\ 
B & 12 20 00 & $-01$ 00 00 & 2 \\ 
C & 12 40 00 & $-02$ 00 00 & 3 \\ 
D & 12 56 00 & $-02$ 42 00 & 6 \\ 
\hline
\end{tabular}
\end{table}

The data were reduced using the 2\textsc{DFDR}\footnote{http://www.aao.gov.au/AAO/2df/aaomega} pipeline. After being flat-fielded, the fibres were throughput calibrated and the sky spectrum was subtracted using the median flux of the dedicated sky fibres. Wavelength calibration was achieved from Th-Ar arc lamp exposures taken between each set of science fields. Multiple object frames were combined to assist with cosmic ray removal.

\section{Dwarf / Giant separation}
\label{sec:ch1-dwarf-giant-separation}

When discussing our data with respect to halo substructures in this chapter, we are referring only to K-type giants.  Dwarfs that fall within our apparent magnitude limit are not sufficiently distant to probe halo substructures. Here we utilise the equivalent width (EW) of the Mg I triplet lines to identify and exclude dwarfs from our analyses. 	

A grid of synthetic spectra have been generated to quantitively establish a suitable giant/dwarf separation criterion. The grid was generated using \citet{castelli;kurucz_2004} model atmospheres with the MOOG\footnote{http://www.as.utexas.edu/$\sim$chris/moog.html} spectral synthesis code \citep{sneden_1973} and the line list of \citet{Kurucz;Bell_1995}. Spectra were also generated using stellar parameters for the Sun and Arcturus. The strength of the Mg I lines were tuned to match both the Solar and Arcturus atlases of \citet{Hinkle;et-al_2003}. \citet{girardi;et-al_2004} isochrones have been used to translate our de-reddened $g - i$ color range to effective temperature. Reddening is accounted for using the \citet{Schlafly;Finkbeiner_2011} corrected dust maps of \citet{schlegel;et-al_1998} assuming a \citet{fitzpatrick_1999} dust profile where $R_V = 3.1$. The corresponding effective temperature region ranges from 3900\,K to 5200\,K, and is stepped at 25\,K intervals. We have assumed a surface gravity of $\log{g} = 2$ for giants and $\log{g} = 4.5$ for dwarfs. Metallicities of [Fe/H] $= -0.5$, $-1.5$, and $-2.5$ were considered for both surface gravities.

\begin{figure}[h]
	\includegraphics[width=\columnwidth]{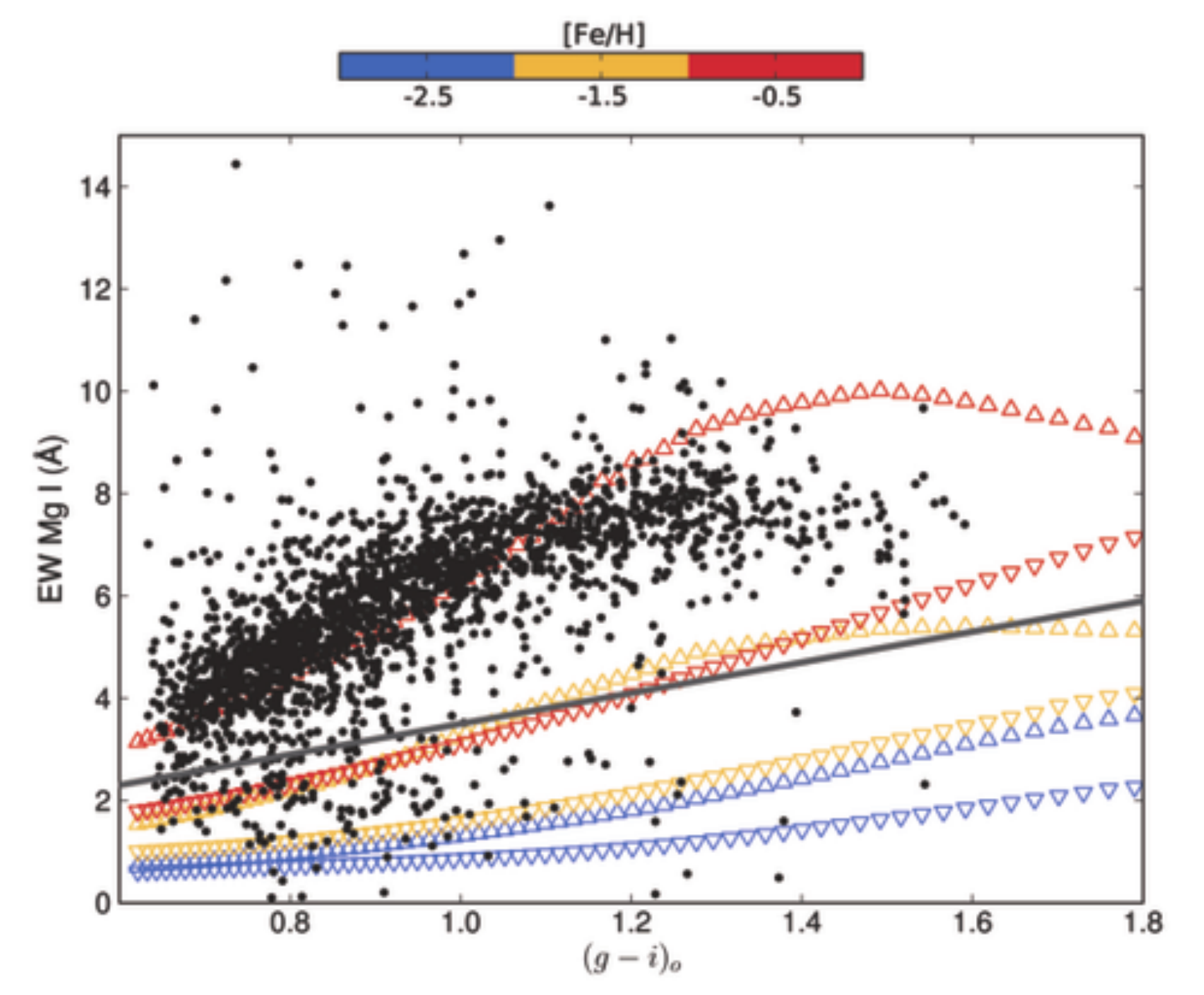}
	\caption[Dwarf/giant discriminant for Virgo Over-Density and Sgr stars]{The sum of the Mg I triplet EWs for observations and synthetic spectra shown against the Sloan $g\,-\,i$ color. Synthetic spectra are shown for $\log{g} = 2$ ($\bigtriangledown$) and $\log{g} = 4.5$ ($\bigtriangleup$), and points are colored by metallicity. Our dwarf/giant separation line is shown (solid). }
	\label{fig:dwarf-giant-separation}
\end{figure}

All synthetic spectra were convolved with a Gaussian kernel of 3.03\,\AA{} to match the resolution of our 580V spectra, and interpolated onto the observed wavelength dispersion map. Mg I line strengths for our observations and synthetic spectra are shown against $g - i$ in Figure \ref{fig:dwarf-giant-separation}. As expected, there is an overlap of Mg I strengths between metal-rich ``giants'' ($\log{g} = 2$) and metal-poor ``dwarfs'' ($\log{g} = 4.5$). However we do not expect metal-poor dwarfs to be a principle contaminant due to their intrinsically low luminosities and comparative rarity.

We have adopted a linear separation line to distinguish dwarf and giant stars. The slope and offset of this line were varied to assess the effectiveness in differentiating giants from dwarfs. A balance must be made between the recoverability of giant stars, and the contamination of dwarf stars. Using the rule,
\begin{equation}
	EW_{Mg\,I} < 3(g-i)_o + 0.5
\end{equation}
\noindent{}we identify 185 giants in our observed fields. An analysis of our 185 giant candidates revealed three stars with high proper motions \citep[PPMXL Catalog, ][]{roser;et-al_2010}, all of which lay close to our dwarf/giant separation line. This suggests they are dwarfs, and they have been excluded. Some possible giant targets were also discarded due to insufficient $S/N$, or because the Mg I triplet fell on bad columns of the detector. All efforts were made to minimise these losses. The distilled giant sample size is 178 stars. All identified K-giant stars in our distilled sample had no proper motion detectable above measurement errors.

\section{Radial Velocities}
\label{sec:ch1-kinematics}

The Ca II triplet absorption lines at 849.8\,nm, 854.2\,nm and 866.2\,nm have been used to measure radial velocities. These lines are strong, and easily identifiable in red giant branch (RGB) stars even at low resolution.  Observed spectra have been cross-correlated with a synthetic spectrum of a metal-poor giant to measure radial velocities. Heliocentric corrections were also applied. Radial velocity measurements made of the standard stars in our globular clusters agree (within $\pm$3\,km s$^{-1}$) with the catalogue of \citet{harris_1996} (2011 edition). Furthermore, a number of our targets were observed on multiple fields, allowing us to calculate the internal measurement error. The differences between multiple measurements of the same target were calculated and they form a half-normal distribution with a {HWHM $= 3.58$\,km s$^{-1}$}. 
	
In order to compare our kinematic results in a homogenous manner, we have translated our heliocentric velocities to a galactocentric frame. We have adopted the circular velocity of the Local Standard of Rest (LSR) at the Sun as 220\,km s$^{-1}$ \citep{kerr;lynden-bell_1986} and accounted for the Sun's peculiar velocity to the LSR by using 16.5\,km s$^{-1}$ towards $l = 53^\circ, b = 25^\circ$ \citep{mihalas;binney_1981}. The corrected line-of-sight velocity is then given by,

\begin{equation}
	V_{GSR} =  V_{OBS} + 220\sin{l}\cos{b} + 16.5\times[\sin{b}\sin{25^\circ} + \cos{b}\cos{25^\circ}\cos{(l - 53^\circ)}]
\end{equation}

\noindent where $V_{OBS}$ is the heliocentric-corrected observed line-of-sight velocity. A caveat to this reference transformation is that other authors in the literature have used slightly different formulae to transpose their kinematics to a galactocentric frame. This will result in possible systematic shifts in velocities between authors of up to $\sim$11\,km s$^{-1}$. 

\section{Metallicities}
\label{sec:ch1-metallicities}

We have measured metallicities for our giants using the strength of the Ca II triplet lines. This technique was first empirically described for individual stars in globular clusters \citep{armandroff;da_costa_1991}. A spectroscopic analysis using VLT/FLAMES observations of RGB stars from composite populations led \citet{battaglia;et-al_2008} to conclude that a calibrated CaT--[Fe/H] relationship can be confidently used in composite stellar populations \citep[see also][]{Rutledge;Hesser;Stetson_1997, starkenburg;et-al_2010}. The caveat to this technique is that a luminosity (specifically $V - V_{HB}$) is required for calibration, and we have to assume a $V_{HB}$ luminosity here. Johnson $V$-band magnitudes for our giants were calculated from SDSS \textit{ugriz} photometry using \citet{Jester;et-al_2005} transformations. The weaker third Ca II triplet line is more susceptible to noise and residual sky-line contamination \citep{Tolstoy;et-al_2001,battaglia;et-al_2008}. Consequently, only the strongest two CaT lines (854.2\,nm and 866.2\,nm) have been used to form a total EW ($W'$) such that,

\begin{eqnarray}
	\textstyle\sum{W}\, &=& EW_{8542} + EW_{8662} \\
	W' &=&\textstyle\sum{W} + 0.64\left(\pm 0.02\right)\left(V-V_{HB}\right)
\end{eqnarray}
\noindent and the metallicity linearly varies with W' where,
\begin{equation}
	\mbox{[Fe/H]}_{\mbox{\small{CaT}}} = (-2.81\pm0.16) + (0.44\pm0.04) W'
	\label{eq:feh-cat}
\end{equation}

Using this calibration our globular cluster standard stars (with known $V_{HB}$ magnitudes) have metallicities that match well with the \citet{harris_1996} catalogue (2011 edition). Only K-giants within the valid calibration range ($0 > V-V_{HB} > -3$) were considered for metallicities. We assume that we have two dominant substructures present in our observations; the leading arm of Sgr and the Virgo Over-Density (see Section \ref{sec:ch1-substructure-identification}). The Sgr stream dominates our negative $V_{GSR}$ population, and our positive kinematic space primarily comprises VOD and VSS members. As such we have separated our population at $V_{GSR} = 0$\,km s$^{-1}$ into two samples with assumed $V_{HB}$ magnitudes. Although this introduces a (known) systemic effect, it is a required assumption to estimate the metallicity distribution of each population.

The horizontal branch magnitude assumed for the VSS/VOD has been ascertained from previous RR Lyrae studies. As RR Lyrae stars sit on the horizontal branch, the $V_{HB}$ is taken as the V-band median of four \citet{Prior;et-al_2009a} and three  \citet{Duffau;et-al_2006} VSS members to yield $\langle{}V\rangle{} = 17.09$. This combined value matches well with the mean horizontal branch magnitude reported by \citet{Duffau;et-al_2006}. In our K-giant sample with positive galactocentric velocities, we assume $V_{HB} = 17.09$, implying these stars are at a distance of $\sim$20 kpc.

The distance to the Sgr debris varies greatly throughout the Milky Way. Through CMD comparisons with the Sgr core \citep{Bellazzini;et-al_2006}, \citet{belokurov;et-al_2006} determined distances along the two branches of the Sgr bifurcation.  Revised distance measurements by \citet{Siegel;et-al_2007} (as tabulated in Table 1 of LM10) have been crucial in constraining dark halo models. Our observations lay on the edge of Branch A (Figure \ref{fig:sgr-field-of-streams}).

We have assumed distances to these stars by interpolating their position along a cubic spline fitted to published distances \citep{Siegel;et-al_2007}. Recent distance measurements of the stream by \citet{Ruhland;et-al_2011} match the stream distances used here within the uncertainties. The horizontal branch magnitude $V_{HB}$ is calculated using this assumed distance, and the known luminosity of RR Lyrae stars \citep[$M_V = +0.69;$][]{Tsujimoto;et-al_1998}. The typical $V_{HB}$ magnitude derived for the Sgr members is $18.7$, corresponding to a distance of $\sim{}$41 kpc. Reddening is accounted for using the \citet{schlegel;et-al_1998} maps with \citet{Schlafly;Finkbeiner_2011} corrections as described in Section \ref{sec:ch1-dwarf-giant-separation}. Uncertainties in distance are not interpolated; they are taken as the largest published uncertainty of the closest neighbouring data points, and this is propagated through to our reported metallicity uncertainties. The kinematics and metallicities derived for our giant sample are discussed in the following section.

\section{Discussion}
\label{sec:ch1-discussion}
		
	\subsection{Substructure Identification}
	\label{sec:ch1-substructure-identification}
		
\begin{figure}
		\includegraphics[width=\textwidth]{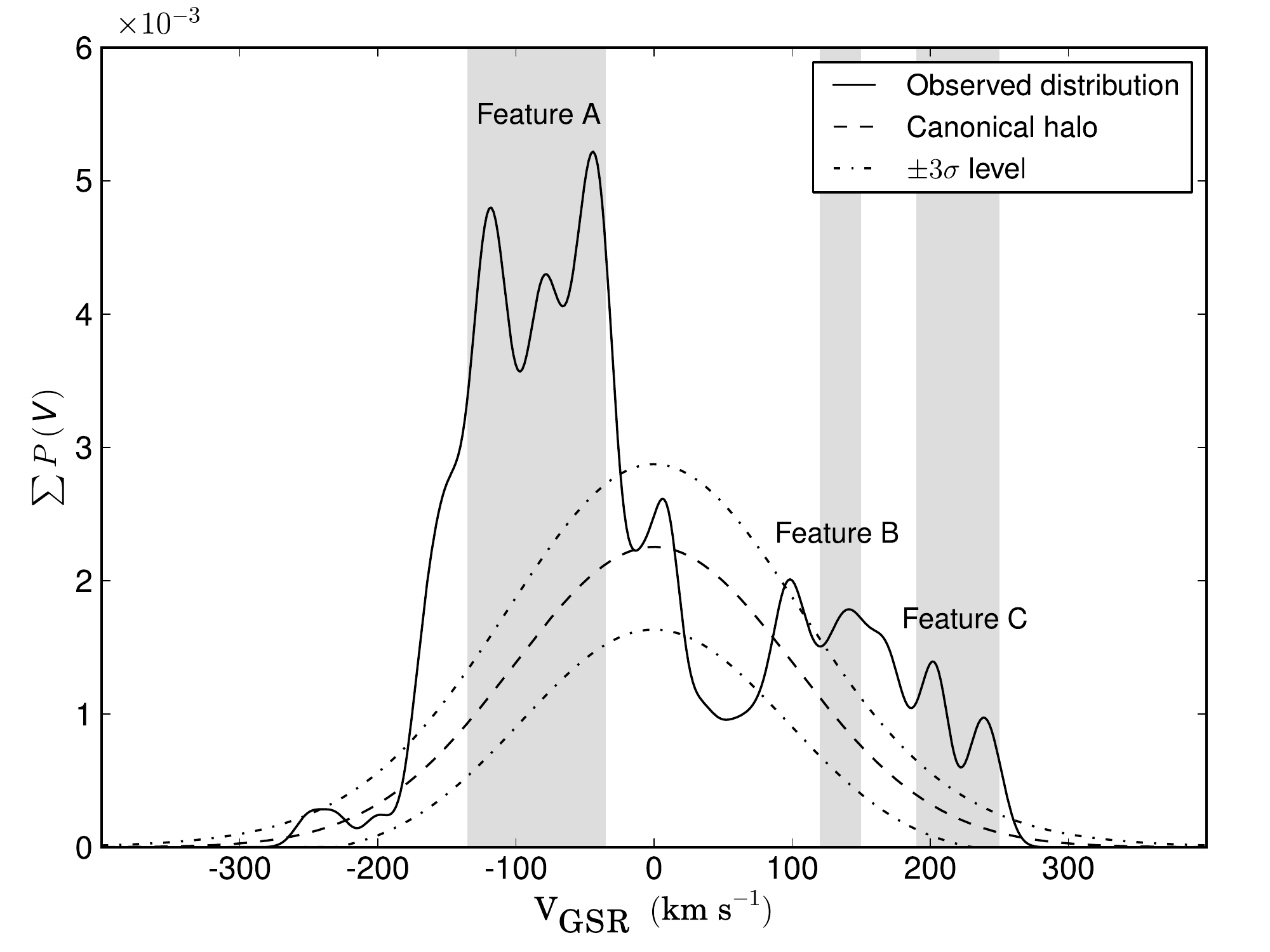}
		\caption[Generalised histogram of galactocentric velocities, illustrating the substructure present]{Generalised histogram of $V_{GSR}$ for our 178 K-type giants, highlighting the substructure present. The Gaussian distribution represents the halo contribution; it is normalised such that the integral equals the number of observed stars excluding those outside a 2.5-$\sigma$ excess. Significant ($>3\sigma$) kinematic deviations from the smooth distribution are appropriately grouped and labelled.}
		\label{fig:velocity-histogram}
	\end{figure}

A significant kinematic deviation from a canonical halo population signifies a co-moving group. There are multiple substructures along our line-of-sight. These features are identified first and discussed separately. We have represented our galactocentric velocities with a generalised histogram (Figure \ref{fig:velocity-histogram}) to quantitatively compare the observed stellar kinematics with the hypothesis that the parent distribution can be described by a canonical halo where $\mu = 0$ km s$^{-1}$, and $\sigma = 101.6$\,km s$^{-1}$ \citep{Sirko;et-al_2004}.

The generalised histogram represents each data point with a Gaussian kernel of an equal deviation. As our internal kinematic errors between multiple measurements summate to a half-normal distribution with a FWHM of 7.16\,km s$^{-1}$, we have opted for a sigma value of 10 km s$^{-1}$ for the generalised histogram in Figure \ref{fig:velocity-histogram} to avoid under-smoothing. The most significant ($>3\sigma$) kinematic peaks have been labeled Features A, B, and C. Features A and C are our most significant structures, and the importance of Feature B becomes more evident through Monte-Carlo simulations.

\begin{figure}
		\includegraphics[width=\textwidth]{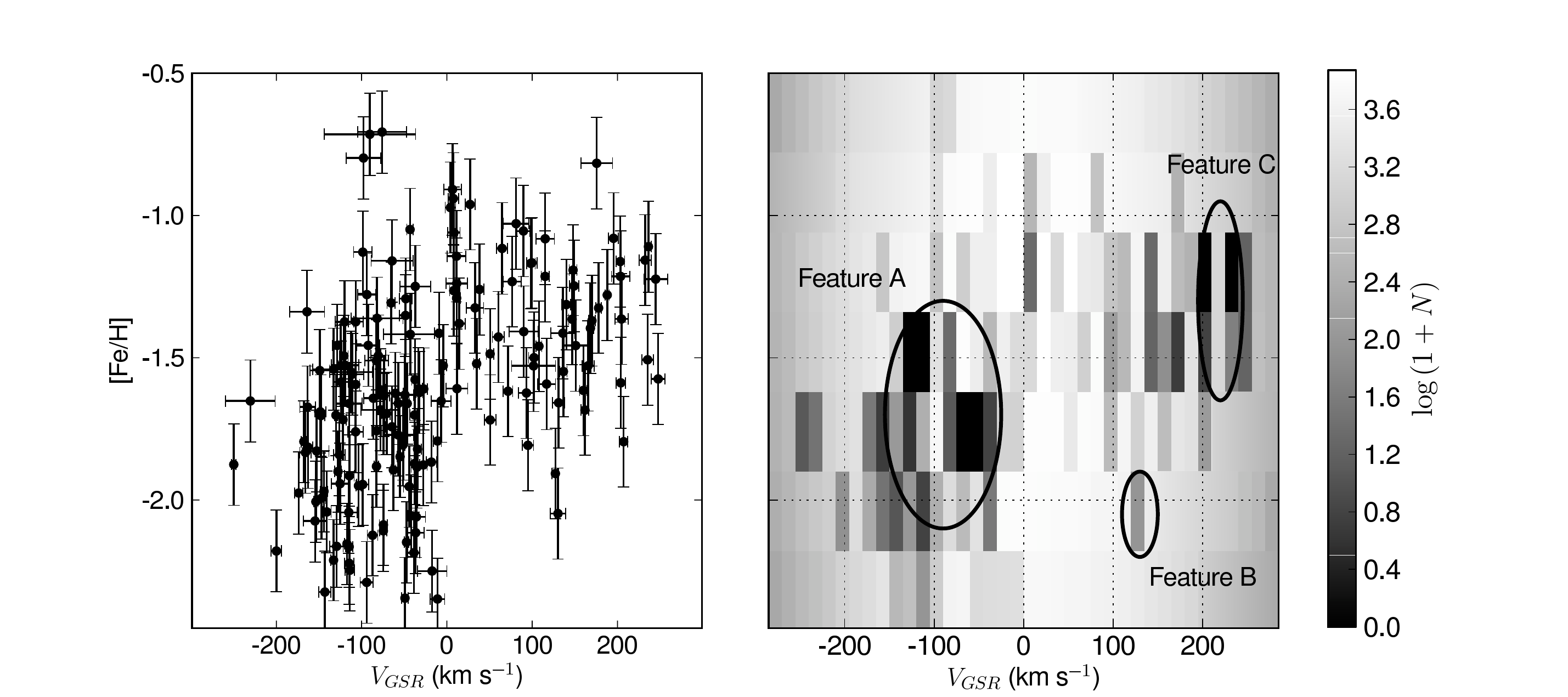}
		\caption[Monte-Carlo simulations to quantify the level of kinematic substructure present]{The observed data for the 178 giant star sample (left panel). Monte-Carlo simulation results illustrating the number of times simulations could reproduce our observed data in the equivalent multi-dimensional bin (right panel). Significant features discussed in the text are labelled.}
		\label{fig:monte-carlo}
	\end{figure}

	Velocities and metallicities of our K-giants, shown in the left panel of Figure \ref{fig:monte-carlo}, were binned into grid blocks of 15\,km s$^{-1}$ and 0.3\,dex \--- roughly twice the error in each dimension. To interpret these data a population of 178 stars were randomly drawn from a simulated halo with canonical kinematics. Metallicities were randomly assigned using the observed halo metallicity distribution function (MDF) of \citet{Ryan;Norris_1991}. Each simulated population was binned identically to our observed sample. Simulation grid blocks with counts equal to or exceeding stars in the equivalent observed grid block were noted, and summed after 10,000 simulations. Results from our Monte-Carlo simulations are illustrated in Figure \ref{fig:monte-carlo} (right). The identified Features A, B, and C in our results were consistently significant when the grid was midpoint offset in both dimensions, except for the grid-block centered at $\sim185$\,km s$^{-1}$ and $\sim-1.4$\,dex, which has consequently been left unlabelled.
	
	 Substructure becomes statistically significant when the observed grid blocks are rarely replicated in Monte-Carlo simulations. Specifically, the number of members in Feature A were never reproduced in some grid blocks. This feature is well in excess of the halo, and has a wide spread in kinematics. \citet{Chou;et-al_2007} mapped the Sgr debris across the sky using K/M-giants and found galactocentric velocity signatures between $-205$\,km s$^{-1}$ to $-31$\,km s$^{-1}$ in this region. As such we attribute this wide, significant kinematic peak as the leading arm of the Sgr tidal tail. Feature A is discussed in the next section. Another feature where grid blocks were never reproduced in our simulations was Feature C. This feature may also be attributed to Sgr debris and is discussed further in Section \ref{sec:ch1-feature-c}.
	 	
	 The clump of stars in the velocity bin centered on $V_{GSR} \approx 130$ km s$^{-1}$ and [Fe/H] $\approx -2$ was replicated only $\sim1$\% in 10,000 simulations. We attribute these stars to the VSS, as they are coincident in spatial position, velocity, and metallicity with previously reported values of the VSS obtained by F turnoff/BHB stars \citep{Newberg;et-al_2007} and RR Lyrae stars \citep{Prior;et-al_2009a}. Feature B is discussed in Section \ref{sec:ch1-the-vss}.

	\begin{figure}
		\includegraphics[width=\textwidth]{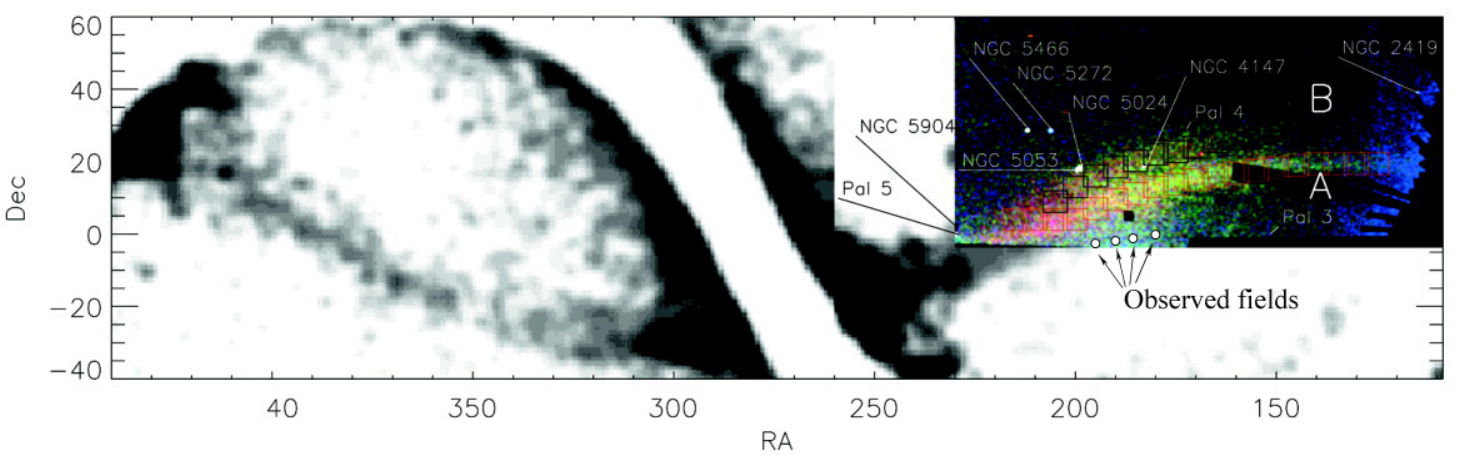}
		\caption[Observed fields upon a panoramic view of the Sagittarius stream]{Observed fields are outlined upon a panoramic view of the Sgr stream, to demonstrate our field locations in context with the Sgr stream. This plot is an adaptation of Figure 2 in \citet{belokurov;et-al_2006}, which uses the 2MASS M-giant sample of \citet{Majewski;et-al_2003}.}
		\label{fig:sgr-field-of-streams}
\end{figure}
	
	\subsection{Feature A \--- Sagittarius Debris}
	\label{sec:ch1-sgr-debris}
	
	  The locations of observed fields are overlaid upon a panoramic projection of the Sgr stream and the  `Field of Streams'  \citep{belokurov;et-al_2006} in Figure \ref{fig:sgr-field-of-streams}. This is a crowded region of globular clusters, substructures and overlapping stellar streams, primarily populated by the Sgr northern leading arm. Simply from a spatial perspective, we expect the Sgr debris to dominate our data. Although the VOD is present, it is much more diffuse. 
	
	\subsubsection{Comparing Sagittarius Debris to Dark Matter Halo Models}	

	In order to investigate the spatial coverage and kinematics of our Sgr members, we have compared our data with the constant flattening spheroidal models of LJM05, and the more recent tri-axial model of LM10. The simulated data output from these models is readily available online\footnote{http://www.astro.virginia.edu/$\sim$srm4n/Sgr/}, and the released models have the best-fitting parameters for each dark halo shape (prolate, spherical, oblate and tri-axial). These were the only simulations available at the time which made use of an all-sky data set; the 2MASS M-giant sample. We have not considered particles farther than 60\,kpc to match realistic observation limitations. The $10^5$ model Sgr debris particles from their simulations are represented along the best-fit great-circle ($\Lambda_\odot$, $B_\odot$) where the longitudinal coordinate $\Lambda_\odot$ is zero in the direction of the Sgr core and increases along the trailing debris. For comprehensive details of the simulations the reader is referred to the papers of \citet{Law;et-al_2005} and \citet{Law;Majewski_2010}. 
		
\begin{figure}
	\centering
		\includegraphics[width=0.7\textwidth]{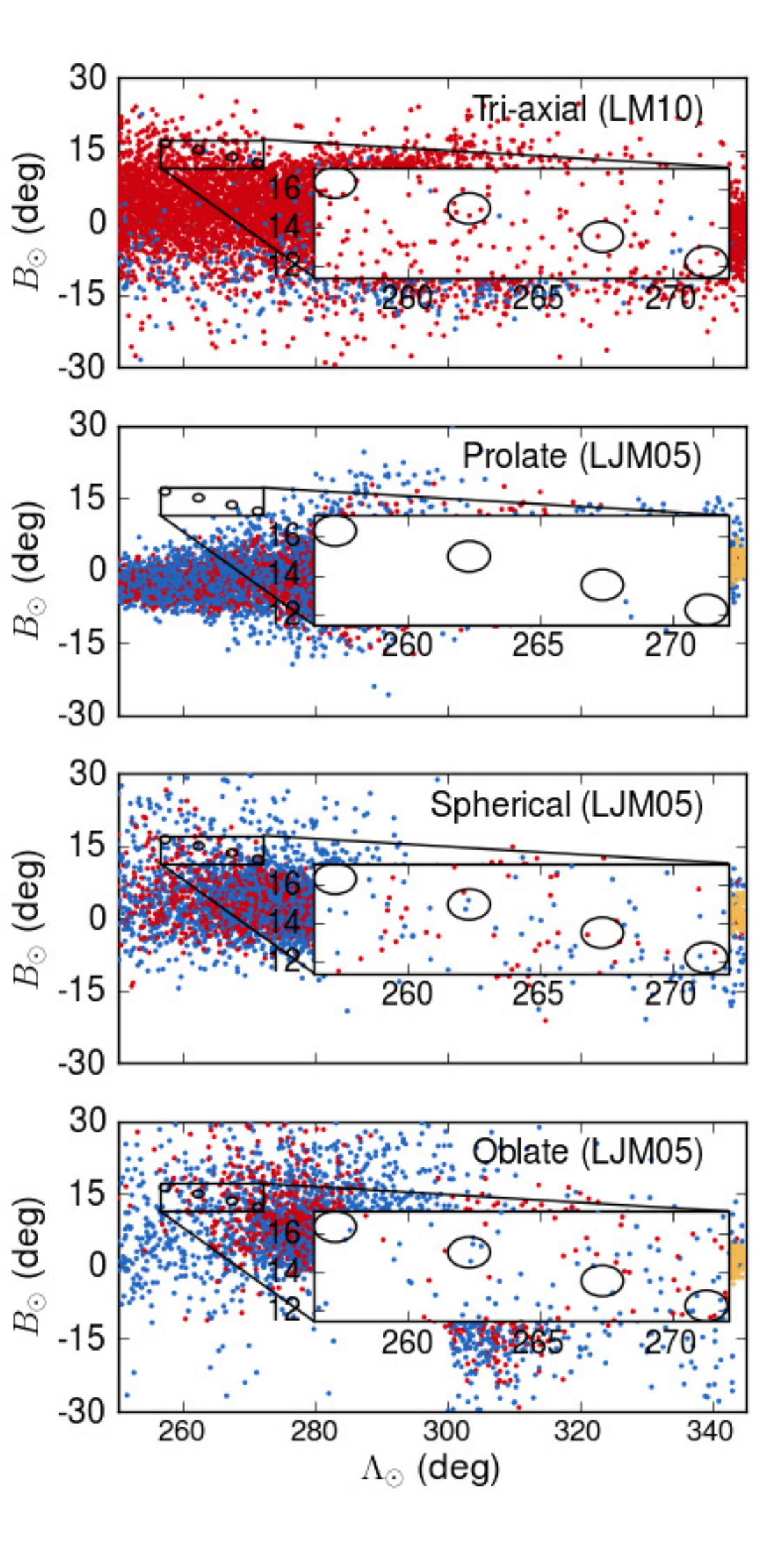}
		\caption[Generalised histograms showing predicted galactocentric velocities from best-fitting dark matter halos compared with observations]{Our fields and observed bounds are shown with the simulated particles from multiple \textit{N}-body simulations by LJM05 and LM10. These particles are distance restricted (up to 60 kpc), and color-coded by their peri-centric passage (yellow; most recent passage, red; previous passage, blue; oldest observable passage).}
\label{fig:law-spatial}
\end{figure}

	The northern leading arm of Sgr is particularly sensitive both kinematically and spatially to the shape of the Milky Way dark halo (Figure \ref{fig:law-spatial}). The spherical, oblate and tri-axial models predict the Sgr stream to pass directly through our fields, whereas the prolate model predicts only the edge of the stream near this region, and no particles directly in our fields. Spatially, the prolate model predicts the Sgr stream to pass much lower in latitude ($B_\odot$) than our observed fields. When we extend a rectangle bounded by the edges of our fields (as per the zoomed insets in Figure \ref{fig:law-spatial}) a mere two model particles are found. The $10^5$ model particles are proportional to the stream density so the lack of model particles implies a negligible stream concentration in our observed region. When the number of field configurations is accounted for, our observed Sgr K-giants are uniformly distributed across the fields. If this prolate halo model is a true representation of the Sgr debris, this does not exclude the potential of finding some Sgr debris in our fields. However, it does imply that if the LJM05 prolate model is an accurate representation of the dark halo then Sgr would not be the dominant population \--- contrary to our preferred interpretation of the observations.
	
	In comparison, the tri-axial, spherical and oblate models predict varying amounts of debris from previous peri-centric passages in our fields. A kinematic comparison against our observations is necessary to evaluate model predictions. As the prolate model has only two simulated particles within our extended bounds there is no qualitative kinematic comparison to be made for this model, therefore the prolate model is excluded from further kinematic analysis.

	\begin{figure}
		\includegraphics[width=\textwidth]{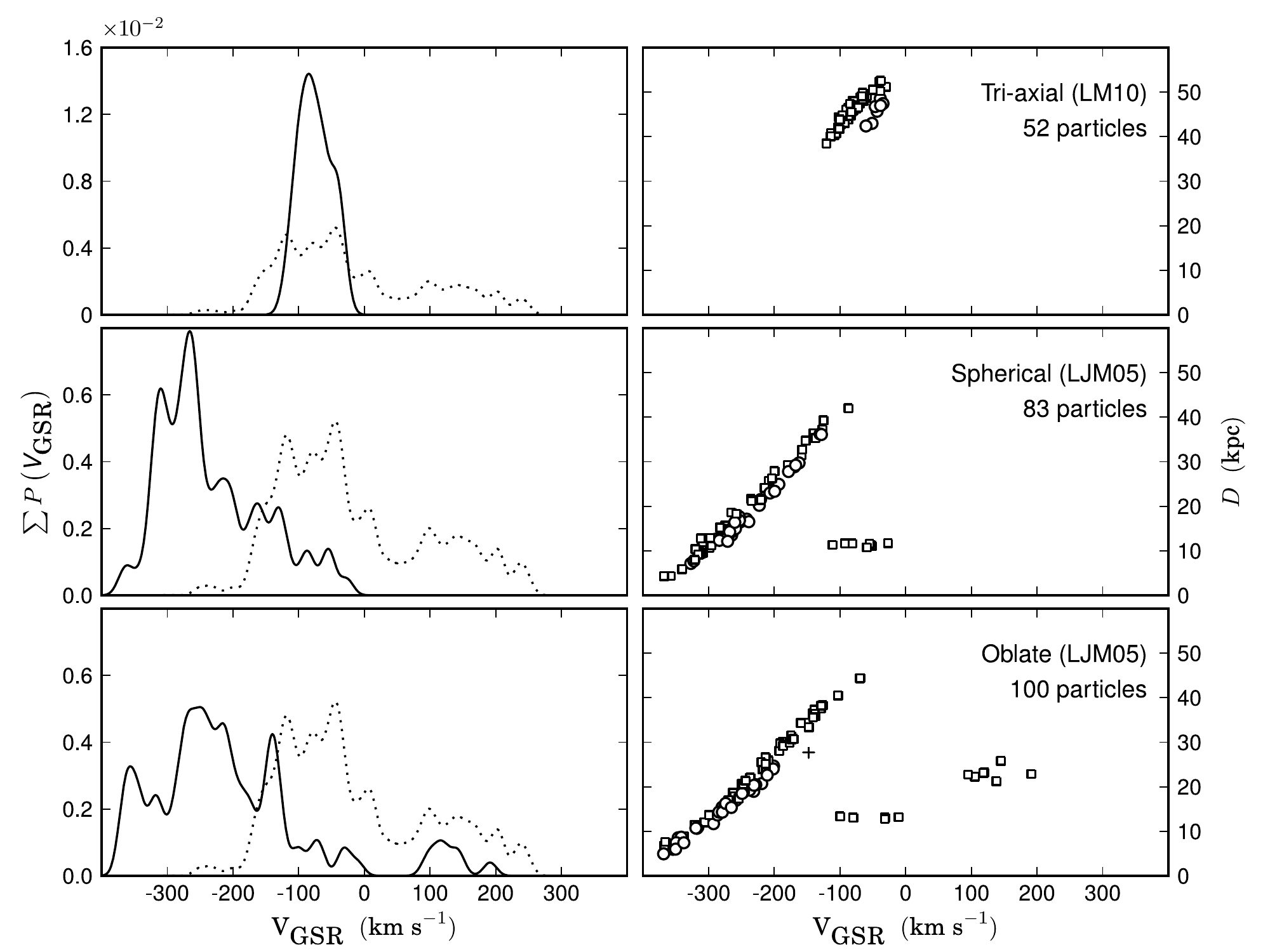}
		\caption[Generalised histogram of galactocentric velocities and heliocentric distances for simulated particles]{A generalised histogram (left) of the galactocentric rest frame velocities in our sample (dotted line) compared to the velocities (solid line) from the \textit{N}-body models of LJM05, LM10 of particles within the same spatial coverage and distance range (up to $r_\odot \sim$60 kpc) of our observed sample. Heliocentric distances for particles used to generate each velocity histogram are shown on the right. Particles are marked by their peri-centric passage. A plus ($+$) denotes debris from the current peri-centric passage by Sgr, circles ($\circ$) mark the previous passage, and squares ($\square$) represent debris from two previous passages. The prolate model from LJM05 is not considered in this plot (see text).}
		\label{fig:law-vel-compare}
	\end{figure}
	
	The number of predicted particles observable differs between models. Consequently, the kinematics for each model have been represented as a generalised histogram (Figure \ref{fig:law-vel-compare}). For consistency all simulation output and observed data has been convolved with a Gaussian kernel of 10\,km s$^{-1}$; the sigma used for the previous generalised histogram in Figure \ref{fig:velocity-histogram}. Predicted distances and peri-centric ages are shown, demonstrating that debris from multiple passages are predicted along the line-of-sight. A much tighter velocity sequence is predicted by the tri-axial model than the constant flattening models. This is because the tri-axial model predicts a peri-center almost precisely in our observable region, whereas the spherical and oblate models predict a large wrap along the line-of-sight, resulting in a wide spread of distances and velocities.
	
	Predicted distances are consistent with what we would expect from our observations. The median luminosities of our giants is $g\sim$17.5, so for the tri-axial model particles at 45\,kpc this corresponds to an absolute magnitude of $M_g\sim-0.8$; quite reasonable for a K-giant. Similarly for the spherical and oblate models, typical distances of $\sim$25\,kpc will yield luminosities of $M_g\sim+0.5$ which is also reasonable.  
	
	The spherical and oblate models also predict close-by debris with extremely negative galactocentric velocities. This signature is not represented in our data. The lowest observed velocity is $\sim{}$$-$250\,km s$^{-1}$, and we have only two observations less than $V_{GSR} < -200$\,km\,s$^{-1}$. In contrast, the spherical/oblate models predict velocities well below $V_{GSR} < -300$\,km\,s$^{-1}$. These predicted stars with highly negative velocities were also not found in Sgr RR Lyrae study of \citet{Prior;et-al_2009b}. If the dark matter potential is well-represented by either an oblate or spherical halo, then this discrepancy must be reconciled. 
	
	The oblate and spherical model also illustrate a similar signature at $r_\odot \sim$12\,kpc, with particle velocities ranging between $-100 < V_{GSR} < 0$\,km s$^{-1}$. This is the edge of a predicted crossing-point between different wraps of the stream, which occurs at $\sim$12\,kpc. These particles --and the positive kinematic signature around $\sim$20\,kpc in the oblate model -- are relatively minor signatures when compared to the northern leading arm. If these signatures are present in our observed fields, their relatively low density compared to the Northern arm would prevent them from appearing as significant. 
	
	Although the predicted velocity distribution is narrower than what we observe, the LM10 tri-axial dark halo model best fits our observations. The observed sample broadens most prominently towards more negative galactocentric velocities at $\sim-200$\,km s$^{-1}$, considerably higher than oblate/spherical model predictions. Halo contamination at highly negative velocities is likely to be small. One reconciliation for this discrepancy between our observations and the tri-axial model may lie in the workings of the tri-axial model itself. Unlike other models considered, the tri-axial model does not reproduce the observed bifurcation in the Sgr stream \citep{belokurov;et-al_2006}. Although observationally untested, it is reasonable to suggest a bifurcation may result from a significant kinematic disruption. Such an effect could result in a broader kinematic distribution \--- similar to what we have observed \--- which is most notable at the stream edges.
	
	There is further work required through observations and simulations to reconcile kinematic discrepancies. Typical examinations of the leading arm debris usually favour prolate halos and evidence along the trailing arm typically favours oblate halos \citep{Helmi_2004, Martinez-Delgado;et-al_2004, Law;et-al_2005}. Kinematic predictions of the LM10 tri-axial model reasonably match our observations, whereas the prolate model has been excluded as significant debris is not predicted in this edge of the stream. These observations along the northern leading arm are the first which are not reproducible with the current prolate model of LJM05, contrary to previous groups who have surveyed closer to the leading arm debris.

	\subsubsection{A Metal-Poor Population Uncovered in Sagittarius Debris}
	\label{sec:ch1-sgr-metal-poor}
		
	Many groups have found the population of the Sgr core to possess a mean [Fe/H] $\sim -0.5$\,dex \citep{Cacciari;et-al_2002, Bonifacio;et-al_2004, Monaco;et-al_2005}. The observed metallicity gradient along the stream suggests that as the host circles the Milky Way the older, more metal-poor stars are preferentially stripped from the progenitor \citep{Chou;et-al_2007, keller;et-al_2010}. In this region of the stream very few Sgr member metallicities have been reported. \citet{Vivas;et-al_2005} found a mean metallicity of $\langle$[Fe/H]$\rangle = -1.77$ from spectra of 16 RR Lyrae stars along a nearby region of the Sgr leading arm. Similarly, \citet{Prior;et-al_2009b} found $\langle$[Fe/H]$\rangle = -1.79 \pm 0.08$\,dex for 21 type \textit{ab} RR Lyrae stars in the region. This is somewhat expected since only the oldest, most metal-poor stars can form RR Lyraes. 
	
\begin{figure}[h]
		\includegraphics[width=\textwidth]{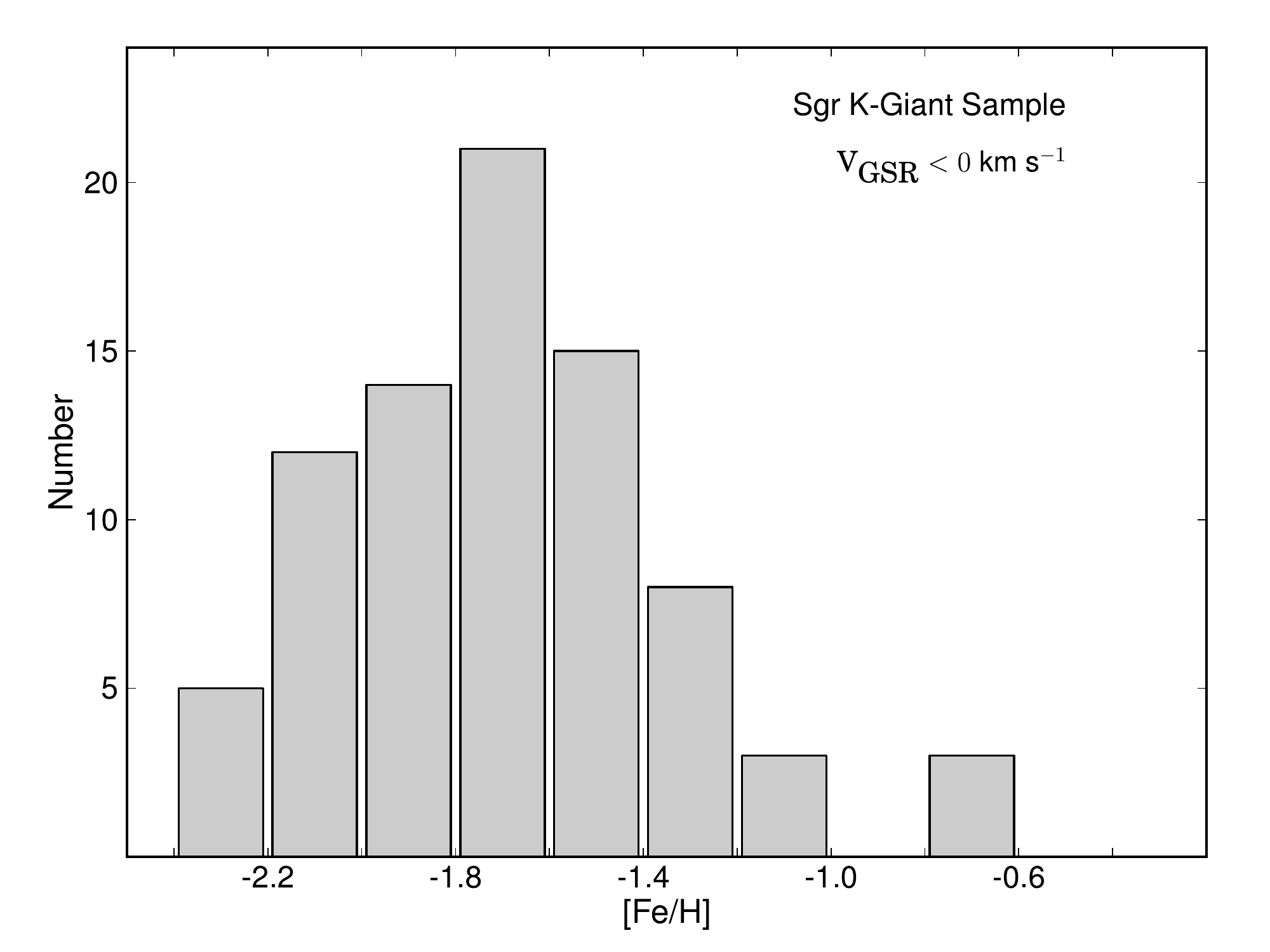}
		\caption[Observed metallicity histogram for giant stars with negative galactocentric velocities]{Metallicity histograms for K-giant members in our negative galactocentric velocity sample, which is largely populated by Sgr debris from the Northern leading arm.}
		\label{fig:sgr-metallicity-hist}
	\end{figure}

	Investigating the MDF of the Sgr debris requires an unbiased sample. Generally K-giants are excellent stellar candidates for such studies, as all stars go through this evolutionary phase, whereas M-giants are consistently metal-rich. If we apply the metallicity gradient found by \citet[][with M-giants]{keller;et-al_2010} to these observed K-giants, we would expect an abundance mean near $\sim{}-1.2$ dex in this observed region. The metallicity distribution for our entire negative $V_{GSR}$ sample is shown in Figure \ref{fig:sgr-metallicity-hist}, and illustrates a metal-poor population. The mean of our distribution is $\langle$[Fe/H]$\rangle = -1.7\,\pm\,0.3$\,dex. If we include only the stars attributed to Feature A (as defined by the shaded region in Figure \ref{fig:velocity-histogram}, between $-140 < V_{gsr} < -30$\,km s$^{-1}$) this value increases by only 0.04\,dex; well within observational uncertainties. As a comparison, in a sample of metal-rich biased M-giants from the 2MASS data \citet{Chou;et-al_2007} found a mean metallicity of $\langle$[Fe/H]$\rangle = -0.72$\,dex for their best subsample in the northern leading arm of Sgr. Although there may be some halo contamination in our sample, in general our more metal-poor Feature A members have higher negative velocities, which is expected for Sgr stream members in this region.
		
	It is likely that our observed K-giant sample is biased towards more metal-poor members. Given a typical distance of $r_\odot\sim 40$\,kpc for the Sgr debris in this region \citep{belokurov;et-al_2006}, a 12\,Gyr old Dartmouth \citep{Dotter;et-al_2008} isochrone with a metallicity of [Fe/H] $= -1.5$ falls directly within our target selection window. The more metal-rich isochrone with [Fe/H] $= -0.5$ does not pass through our selection range. This is consistent with our observed metallicities for Sgr. Although this does not prohibit the possibility of more metal-rich members within our sample, it implies that our distribution is biased towards the metal-poor end of the MDF. The tail of our Sgr distribution extends from [Fe/H] = $-0.71$ to $-2.35$\,dex.

	\subsection{Feature B \--- The Virgo Stellar Stream}
	\label{sec:ch1-the-vss}
	
	Another substructure which was distinguished from the VOD is the Virgo Stellar Stream \citep{Duffau;et-al_2006}. Several RR Lyrae stars have been observed with a common velocity of $V_{GSR} = 130$\,km s$^{-1}$ \citep{Newberg;et-al_2007, Prior;et-al_2009a}. \citet{Prior;et-al_2009a} also measured metallicities of stars with this kinematic signature and found a mean [Fe/H] $= -1.95\,\pm\,0.1$ for the VSS, although this is based on RR Lyrae stars, and therefore is potentially biased to the oldest, most metal-poor populations. The RR Lyrae sample from \citet{Duffau;et-al_2006} found a $\langle$[Fe/H]$\rangle = -1.86$ with a much larger abundance spread of $\sigma = 0.40$, several times larger than the relative uncertainty in individual values ($\sigma_{{\rm [Fe/H]}} = 0.08$\,dex). This led \citet{Duffau;et-al_2006} to infer the progenitor of the VSS was likely a dSph galaxy, as a large dispersion in $\mbox{[m/H]}$ indicates self-enrichment, implying the structure is a galaxy and not an isolated mono-metallic cluster. Although some globular clusters are also now known to contain multiple populations and sizeable $\mbox{[m/H]}$ dispersions, globular clusters in general have a much smaller dispersion than that observed here. Similar to \citet{Duffau;et-al_2006}, we  also observe an abundance spread within this kinematic bin, including two K-giants with mean kinematics and metallicities ($\langle{}V_{GSR}\rangle{} = 130 \pm 9$ km s$^{-1}, \langle\mbox{[Fe/H]}\rangle = -2.0 \pm 0.16$) matching those found by \citet{Prior;et-al_2009a}. Monte-Carlo simulations reproduced these targets a mere $\sim1$\% in 10,000 simulations \--- highlighting their significance.

	High resolution follow-up spectroscopy on these three targets and other VSS candidates at higher metallicities in this kinematic range will provide crucial information about the origin of the VSS.  Investigating [$\alpha$/Fe] ratios in these candidates can constrain the mass of the VSS progenitor and discern whether the host was indeed a dSph \citep{venn;et-al_2004, Casetti-Dinescu;et-al_2009}. Future observations are planned.

	\subsection{Feature C \--- Sagittarius debris?}
	\label{sec:ch1-feature-c}

	The last significant kinematic substructure we identify in our data is Feature C. There are two peaks identified at $V_{GSR} = 200$\,km s$^{-1}$ and $240$\,km s$^{-1}$. These features cannot be explained by a smooth halo distribution. \citet{Newberg;et-al_2007} identified a 4$\sigma$ peak at $V_{GSR} = 200$\,km\,s$^{-1}$ in their sample of F turnoff stars centered on $(l, b) = (288^\circ, 62^\circ)$. Our nearest field to their plate (Field B) hosts only one star associated with this peak; most of our stars are located within our most populated Field D. \citet{Prior;et-al_2009b} also noted stars in this kinematic range and compiled a list of authors who have similarly observed such peaks (see \citet{Sirko;et-al_2004, Duffau;et-al_2006, starkenburg;et-al_2009}). \citet{Prior;et-al_2009b} argue these stars may be associated with trailing debris of Sgr, as suggested by models (LJM05, LM10). Although this trailing debris would be much closer than the nearby visible leading arm, our observations support the interpretation by \citet{Prior;et-al_2009b} as the metallicities are quite similar to the nearby Sgr debris.
	
	This substructure has a range of metallicities. These metallicities were derived assuming a single distance modulus to the VOD/VSS. The abundance dispersion we observe is either representative, or these stars have a common metallicity and are dispersed along the line-of-sight. Unfortunately neither scenario can be positively excluded without further observations. If these peaks are associated with the trailing debris of Sgr debris, they are likely to be further away than the VOD. On our scale that would make these stars more metal-rich than shown in Figure \ref{fig:monte-carlo}. However, the nearest trailing debris particles predicted from any best-fitting dark halo potential that could explain this kinematic peak occur at $\sim\Lambda_\odot = 310^\circ$; our observations range in longitude between $\Lambda_\odot = 256-273^\circ$. Whether this substructure is separate from Sgr debris or not remains equivocal. Nevertheless, the apparent metallicities of the Feature C stars are not inconsistent with those of a Sgr population given our results for Feature A and those of \citet{keller;et-al_2010}.
	
\section{Carbon Stars}
\label{sec:ch1-carbon-stars}	

	Five carbon stars have been identified in this sample. The colour selection shown in Figure \ref{fig:cmd-target-selection} overlaps with the colours of carbon stars. These stars were not specifically targeted, and their surface densities are so low \citep[$\approx$ 1 per 50 deg$^2$;][]{Green;et-al_1994} that we would not expect them as a substantial contaminant.  Although our sky coverage is small ($\sim$12.5 deg$^2$), the observed carbon star spatial density is $\sim$20x higher than expected. All five carbon stars are recognisable by the presence of distinctive 473.7\,nm and 516.5\,nm Swan C$_2$ bands.

\begin{figure}
	\includegraphics[width=\textwidth]{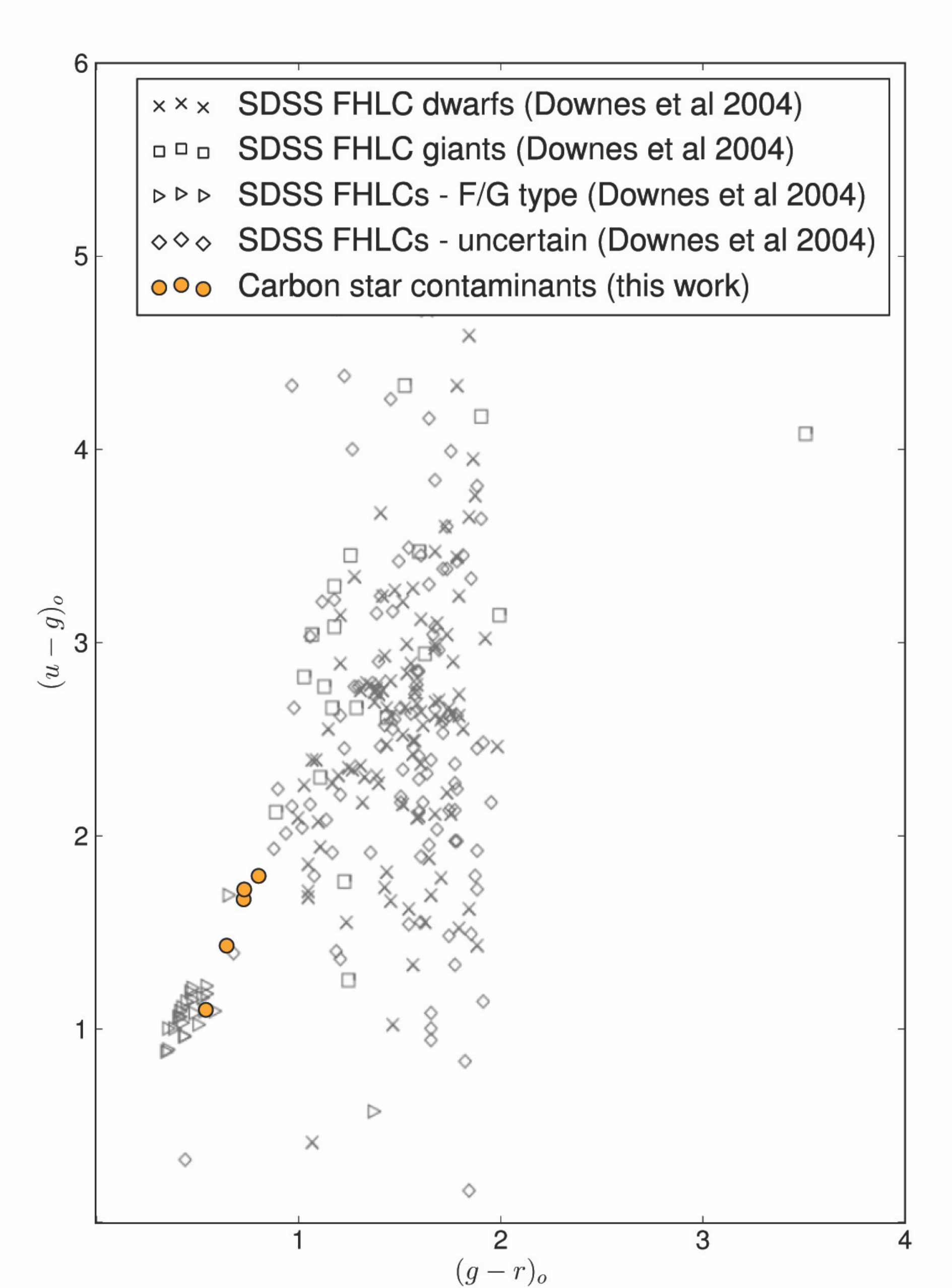}
	\caption[Carbon stars identified in this study]{Sloan \textit{u \--- g} and \textit{g \--- r} colors for our carbon stars, and the identified carbon stars and classifications by \citet{Downes;et-al_2004}. Carbon stars identified in this work have been shaded for clarity.}
	\label{fig:carbon-sdss}
\end{figure}

	 Dwarf carbon stars exhibit a spectral signature which mimics that of a typical CH-type giant carbon star, however they have anomalous JHK colors \citep{Green;et-al_1992} and high proper motions. The existence of the 430\,nm CH G-band is representative of a CH-type carbon star, and is found in all five of our carbon stars. SDSS photometry for our stars match well with the F/G-type CH carbon stars identified by \citet{Downes;et-al_2004}, as seen in Figure \ref{fig:carbon-sdss}. However these targets were not spectroscopically observed in the SDSS follow-up survey SEGUE. Through comparisons with previous carbon-type star catalogues \citep{Totten;Irwin_1998, Downes;et-al_2004, Goswami;et-al_2010}, the stars tabulated in Table \ref{tab:carbon-stars} are previously unclassified carbon stars. This is largely because our objects are too faint to have been observed by previous spectroscopic carbon star surveys.
	
\begin{table}[h!]
\centering
\caption{Properties of carbon stars found in our sample\label{tab:carbon-stars}}
\resizebox{\textwidth}{!}{%
\begin{tabular}{ccccccccccrc}
\hline
\hline
SDSS Name J+\tablenotemark{a} &
$M_g$ & 
$u - g$ & $g - r$ & $r - i$ & $i - z$ & $H-K$ & $J-H$ & $\mu_{\alpha\cos{\delta}}\tablenotemark{c}$ & $\mu_\delta\tablenotemark{c}$ & $V_{GSR}$ & Likely \\
& & & & & & & & (mas yr$^{-1}$) & (mas yr$^{-1}$) & (km s$^{-1}$) & dwarf? \\
\hline

121740.94-001839.5 & 18.66 & 1.10 & 0.54 & 0.16 & 0.05 & ...\tablenotemark{b}& ...\tablenotemark{b} & $-3.0 \pm 5.4$  & $-18.0 \pm 5.4$ & $-71 \pm 34$ & Yes\\
121853.18-004628.4 & 17.71 & 1.67 & 0.73 & 0.22 & 0.11 & 0.24 & 0.36 &  $-3.7 \pm 4.5$ & $-0.5 \pm 4.5$ & $176 \pm 9$ & - \\
122053.71-011709.5 & 17.65 & 1.79 & 0.80 & 0.26 & 0.13 & 0.16 & 0.56 &  $-7.3 \pm 4.3$ & $-3.9 \pm 4.3$ & $31 \pm 11$ & - \\
125410.80-032744.0 & 18.59 & 1.43 & 0.64 & 0.22 & 0.05 & 0.56 & 0.32 &$-21.1 \pm 5.0 $ & $-7.7 \pm 5.0$ & $-41 \pm 9$ & Yes \\ 
125416.52-031437.6 & 16.85 & 1.72 & 0.73 & 0.25 & 0.11 & -0.11 & 0.50 & $-19.8 \pm 4.2$ & $+8.7 \pm 4.2$& $15 \pm 4$ & Yes \\
\hline
\end{tabular}}
\tablenotetext{a}{To keep with convention, positional information has been truncated. Full information available through the SDSS archive.}
\tablenotetext{b}{No \textit{JHK} photometry is available for this object as it is fainter than the 2MASS survey limit.}
\tablenotetext{c}{Proper motions taken from the PPMXL Catalog \citep{roser;et-al_2010}.}
\end{table}

	There is 2MASS photometry available for four of these stars. Of those, two stars (SDSS J125410.80-032744.0 and J125416.52-031437.6) exhibit anomalous JHK colors and significant ($>3\sigma$) proper motions \citep[PPMXL;][]{roser;et-al_2010}; characteristics of a dwarf carbon star. A third star also exhibits significant proper motion (SDSS J121740.94-001839.5), however JHK photometry is unavailable. The two remaining stars in our sample have JHK photometry characteristic of the F/G type CH carbon stars found by \citet{Downes;et-al_2004} in their Faint High-Latitude Carbon (FHLC) star survey (Figure \ref{fig:carbon-sdss}), and do not exhibit significant proper motion - which strongly suggests they are not dwarfs \citep{Green;et-al_1994, Deutsch_1994}.

\section{Conclusions}
\label{sec:ch1-conclusions}

	We present spectroscopic observations of K-type stars in a region of the 'Field of Streams', where significant substructure is present. We utilise the gravity-sensitive Mg I triplet lines to separate giant and dwarf stars. Radial velocities and metallicities of 178 K-giants have been determined. We have recovered kinematic substructure found by other authors, which cannot be explained by a smooth halo distribution. Highly negative velocity signatures match those expected by the Sgr stream debris, and we also identify a group of K-giants with metallicities and kinematics which make them highly probable members of the VSS.

	Stars in this region of the Sgr stream are kinematically sensitive to the shape of the Galactic dark halo. As such, we have compared our velocity distribution to the Sgr-Milky Way dark matter models of \citet{Law;et-al_2005, Law;Majewski_2010}. Typically, leading arm debris favour prolate models and trailing arm debris favour oblate models. A prolate dark halo predicts relatively little debris in our observed fields. If the spheroid is prolate and LJM05 presents an accurate representation, Sgr debris would not be expected to dominate our sample even if we were $\Delta{}B_\odot\sim10^\circ$ closer to the great circle best-fit. However, Sgr debris is our most significant kinematic structure observed. No single model perfectly represents our data, although we find the more recent tri-axial model (LM10) best represents our observed kinematics. 
	
	Observed metallicities for K-giants in this region of the Sgr stream are notably lower than expected based on other Sgr samples, demonstrating the presence of a metal-poor population in the  Sgr debris. Although isochrones indicate we are biased towards metal-poor members at these distances, these stars are unequivocally Sgr in origin. Metallicity gradients reported suggest a mean of $\langle[$Fe/H$]\rangle\sim{}-1.2$, whereas our population has $\langle[$Fe/H$]\rangle = -1.7 \pm 0.3$\,dex. Previously reported abundances in this region have been ascertained from observing RR Lyrae stars (where the mean metallicity found was $\langle\mbox{[Fe/H]}\rangle\sim{}-1.7$\,dex) or from observing M-giants (where $\langle\mbox{[Fe/H]}\rangle\sim{}-0.7$\,dex was reported). In both scenarios, the measurements are reasonable with what we would expect from each particular stellar population. Thus the sample presented here is more representative of the complete MDF, and less susceptible to age or metallicity biases. These observations also demonstrate that the Sgr stream likely hosts a substantially larger abundance range than previously suggested by RR Lyrae and M-giant studies, including a previously un-detected metal-poor population of K-giants.


\newcommand\smh{SMH}
\newcommand\starname{HD 76932}

\chapter{Spectroscopy Made Hard}
\label{chap:smh}


\previouslypublished{This chapter will be submitted to the Publications of the Astronomical Society of Australia.}\\

In this chapter we present innovative astronomical software to fulfil the inevitable transition from manual curve-of-growth analyses of high-resolution stellar spectra, to completely automatic approaches. The software, named `Spectroscopy Made Hard'\footnote{Ironically so.} (SMH), is written in Python and wraps the well-tested MOOG spectral synthesis code \citep{sneden_1973}. This software has been employed for the analyses presented in Chapter \ref{chap:orphan_II} and Chapter \ref{chap:aquarius}, as well as a number of other projects. We describe a number of built-in analysis tools: normalisation of apertures, inverse variance-weighted stitching of overlapping apertures and/or sequential exposures, doppler measurement and correction, automatic measurement of EWs, multiple methods for inferring stellar parameters, measuring elemental abundances from EWs or spectral synthesis, as well as a rigorous uncertainty analysis. Atomic and molecular data are included, as are existing 1D and state-of-the-art $\langle$3D$\rangle$\footnote{$\langle$3D$\rangle$ refers to 3D model atmospheres that have been temporarily and spatially averaged for use with existing 1D spectral synthesis codes.} model atmospheres. All aspects of the analysis can be performed either automatically or interactively through an intuitive graphical user interface. SMH allows scientists to save their analysis to a single file for distribution to other spectroscopists, or to release them with their publication. Anyone with the SMH software can load these files and easily reproduce -- or adjust -- every aspect of the analysis. This ensures scientific accuracy, reproducibility and data provenance. An overview of the software is presented, and detailed explanations are provided for the algorithms employed.

\section{Introduction}

The analysis of high-resolution stellar spectroscopic data has historically been a careful and rigorous observational subfield. It has often colloquially been considered a `dark art' because of the subjectivity and caveats involved in the analysis. This subjectivity can raise a number of discrepancies in the analysis of a single star, which leads to significant offsets in results between experienced spectroscopists. A plethora of input atomic and molecular data are also available, as well as a numerous model atmospheres or analysis options, which complicate any attempt of a fair comparison between different studies. Moreover, the subjectivity in subtle choices like continuum placement will result in non-negligible measurement offsets \citep[e.g., see ][for a recent and controversial example]{kerzendorf}. A wealth of items need to be considered when untangling differences between spectroscopists.

In the last decade the landscape of stellar astrophysics has irrevocably changed. Sample sizes of high-resolution spectroscopic data have increased by orders of magnitude for a single project. This trend is certain to continue as the number of large-scale spectroscopic surveys increases \citep[e.g.,][]{hermes,apogee,gaia-eso}. These spectroscopic surveys have enormous scientific potential. The chemical evolution of the Milky Way can be untangled, and we will be capable of probing stellar interiors and nucleosynthetic pathways to an unprecedented detail. If the scientific potential of these spectroscopic surveys is to be achieved these samples must be analysed homogeneously, with careful consideration given to subtle systematic effects.

However, it has become an insurmountable task for a single spectroscopist to homogeneously analyse such sample sizes in any reasonable lifetime. Even multiple observers trained in the same manner may find non-negligible offsets in stellar parameters and chemical abundances. With any human approach, the results will inevitably be heterogeneous. A need has arisen to transition from manual analysis techniques to completely automatic processes in a graceful manner. However any such transition should not occur at the expense of data accuracy, provenance, or tangibility with the data.

This problem has been met with solutions from a number of groups with markedly different approaches. Some researchers have automated the most time-consuming portion of a classical curve-of-growth analysis: the measurement of EWs \citep{ewdet,zhao,daospec,ares,tame,fama,robospect}. In addition to these software, a number of authors have their own scripts to automate this process. Indeed, even IRAF can be easily scripted to automate EW measurements. Most of these codes produce diagnostic plots for the user to either inspect or ignore. Typically a comparison is shown made between the automatic approach and a trusted source of manual measurement, then the remainder of the analysis is performed by the user in the classical, manual way.

\begin{figure*}[t!]
	\includegraphics[width=\textwidth]{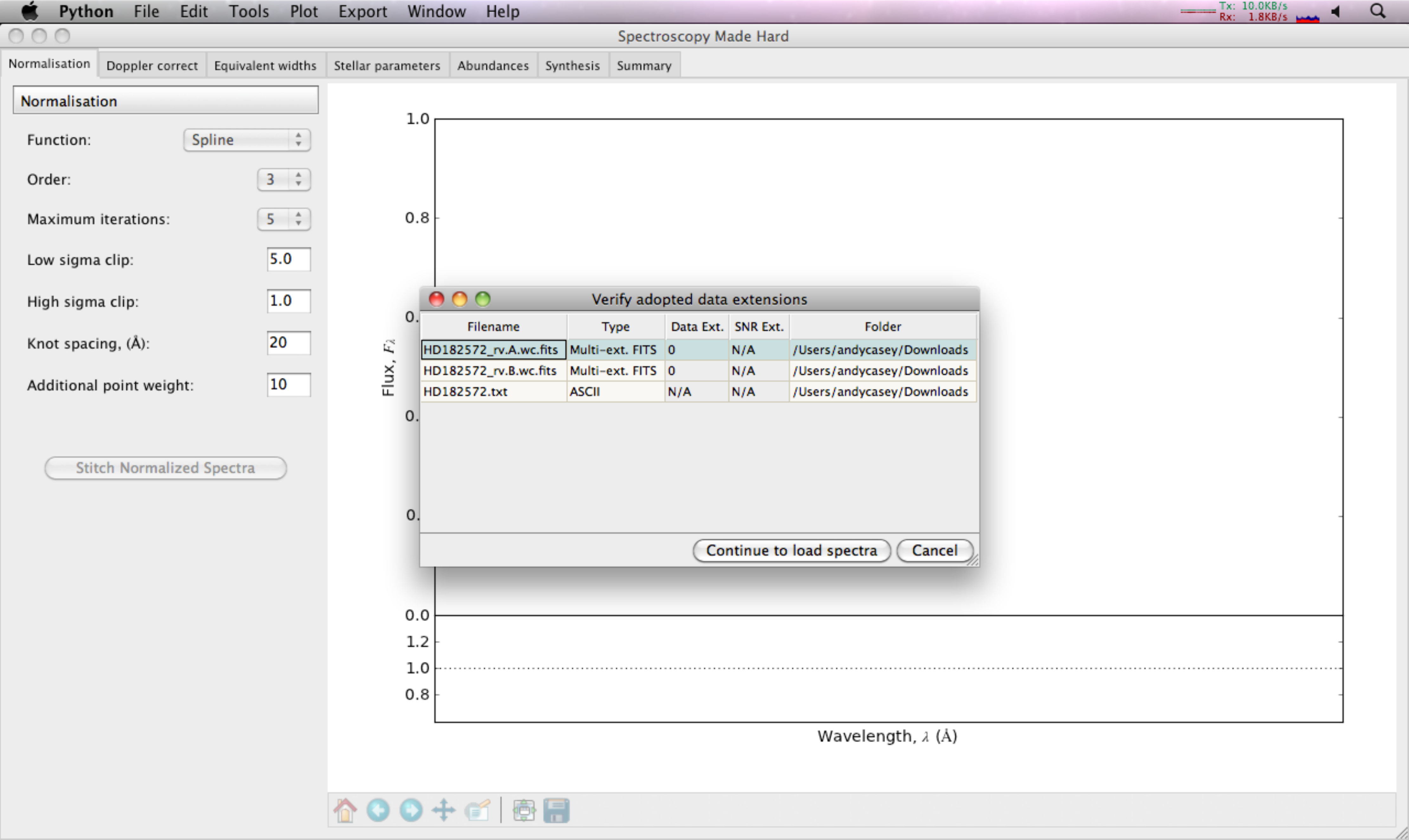}
	\caption[Screenshot of Spectroscopy Made Hard illustrating the different types of acceptable input file formats]{The initial view of Spectroscopy Made Hard when the software is loaded without any input spectra. A new session is being created with multiple spectra of different formats. Two input files are multi-extension FITS images, where the data index was not immediately discernible.}
\label{fig:new-session-ambiguity}
\end{figure*}

Other approaches to solve this problem have sidestepped curve-of-growth approaches entirely and opted for a global $\chi^2$-minimisation technique \citep{matisse,sspp,sme}. In this paradigm, the observed spectrum is compared against a grid of spectra produced by synthesis. Careful spectral masks and/or weighting functions are often required to provide accurate and consistent results \citep[e.g.,][]{matisse,kordopatis}. Even with such convolutions, post-analysis corrections may be required to a subset of the results \citep[e.g.,][]{apogee_dr10}.

These kind of $\chi^2$-fitting approaches are entirely automated: reduced spectra is provided as input, stellar parameters and some chemical abundances are provided as output. The careful observer will no doubt inspect all of their diagnostic plots for ``common sense''. However as sample sizes continue to increase, it will become an insurmountable task for even a single observer to inspect the resultant diagnostic plots for every target. Additionally, none of the aforementioned approaches easily accommodate an retrospective interactive analysis of the data. A manual analysis would be required to investigate subtle effects of continuum placement, atomic/molecular transition data or the rejection of certain atomic lines.

In this chapter we present a new approach to facilitate analysing high-resolution stellar spectra. The software has two primary modes: a graphical user interface and a batch processing mode. In this contribution, we focus on the interactive mode. This chapter is organised as follows. In Section \ref{sec:smh-background} we introduce background knowledge that is relevant for the following sections. Section \ref{sec:smh-analysis} chronologically outlines the usual steps involved in a curve-of-growth analysis for high-resolution stellar spectra, in the context of this software. We conclude in Section \ref{sec:smh-conclusions}, and some future work relevant to \smh{} is outlined in the final chapter of this thesis.




\section{Background to the Software}
\label{sec:smh-background}

\smh{} is written in Python\footnote{http://www.python.org} and works on most operating systems. The main analysis steps are compartmentalised by tabs in the graphical user interface (GUI), and ordered in the standard way most high-resolution analyses are performed. Some functionality (e.g., stellar parameter inference) is intentionally restricted or disabled in the GUI until it becomes relevant. As an example, stellar parameter inference becomes available once EWs have been measured. The behaviour of disabling and enabling functionality in this manner is designed to encourage the user to visually inspect the data for quality. The user must verify the results make sense. In the end the spectroscopist publishes the research, not a piece of software.


\smh{} utilizes a concept of `sessions' to ensure data provenance and scientific reproducibility. Sessions can be created and manipulated either in an interactive mode or through batch processing. Each session can be compressed and saved to a single file (with a `smh' file name extension), then distributed and opened by anyone with the \smh{} software. Only one star can be analysed per session. All measurements stored in a \smh{} session file can be exported to external databases, or to external ASCII-type tables (e.g., comma separated files or \LaTeX{} tables). Additionally, results from multiple sessions can be easily combined to produce extensive \LaTeX{} tables or to interact with databases. 

New sessions can be created in the GUI from the standard application `File' menu. The saving and loading of sessions is also performed through the `File' menu.  At least one wavelength-calibrated spectrum is necessary to initialize a session. \smh{} is ambivalent about file formats: one dimensional spectra in FITS or ASCII format are seamlessly read, as are multi-extension FITS images\footnote{Orders within a multi-extension FITS image are referred to in \smh{} as `apertures'. The spectrum contained in a single ASCII file is also referred to as an individual aperture.} with complex non-linear wavelength-pixel dispersion maps. In the case of multi-extension FITS images, if the data extension is not obvious, then a new window will allow the user to explicitly set the data extension for each file (Figure \ref{fig:new-session-ambiguity}). Multiple spectra with input formats of mixed-type are acceptable for a single session. Pixels with non-finite or negative flux are considered unphysical and ignored. Consequently, individual apertures or spectra that \emph{only} contain non-finite or negative flux are also ignored.

\section{Analysis Steps}
\label{sec:smh-analysis}
This section is ordered in the typical way a curve-of-growth analysis would be carried out. Although there may be minor variations in these steps between spectroscopists (e.g., performing a radial velocity correction prior to normalisation), a logical analysis flow is presented. This is intended to provide context to the reader whilst introducing the software.

\subsection{Normalization}
\begin{figure}[h]
	\includegraphics[width=\columnwidth]{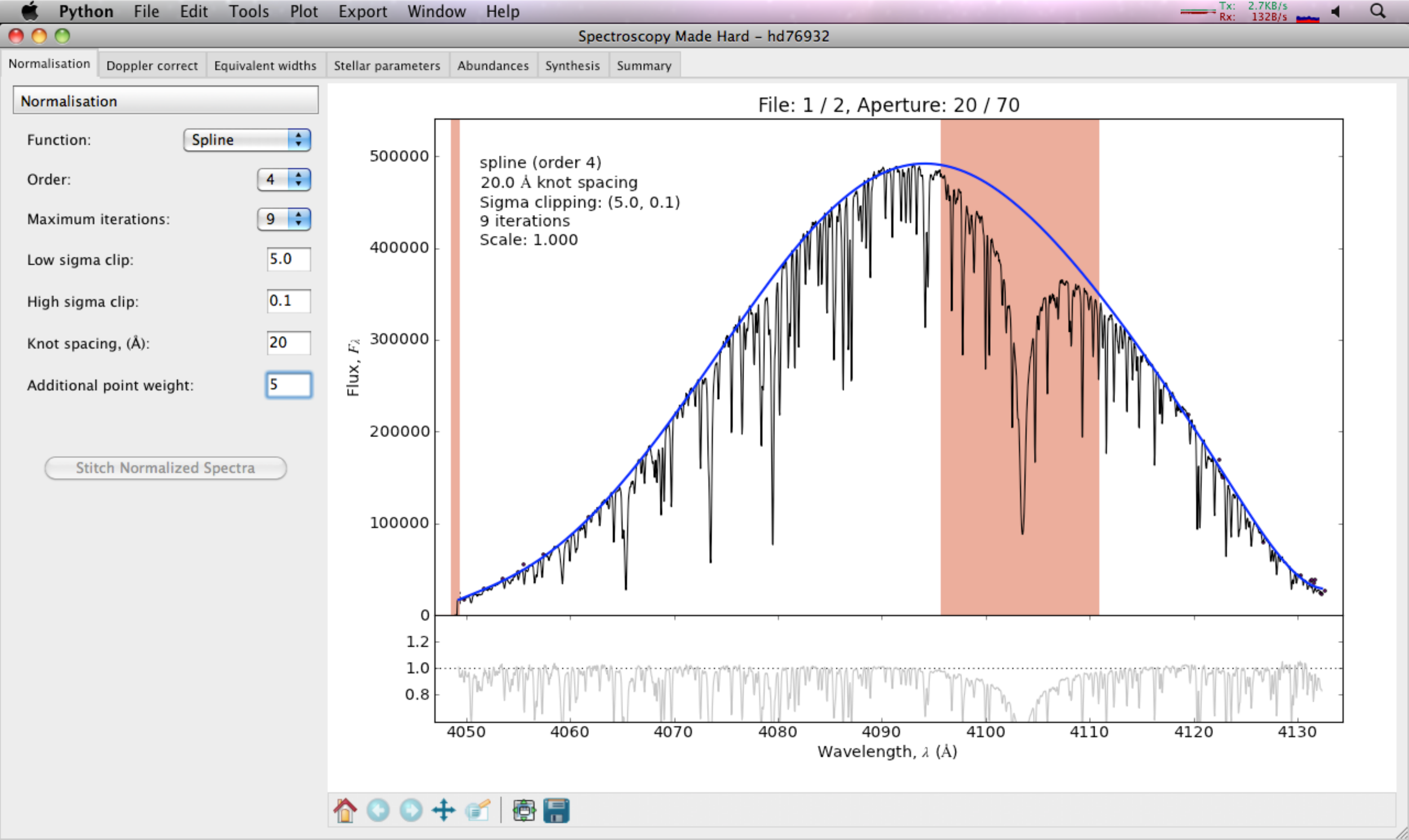}
	\caption[`Normalisation' tab in Spectroscopy Made Hard, showing an example surrounding the H-$\delta$ line]{Normalisation for a single aperture of \starname{} surrounding the H-$\delta$ line, observed with the MIKE spectrograph \citep{bernstein;et-al_2003} on the Magellan Clay telescope. The exclusion regions are highlighted in red, and additional weighted points are shown. The continuum fit to the data is shown in blue. A preview of the normalisation (prior to aperture stitching) is shown in the lower panel.}
\label{fig:normalisation}
\end{figure}

In order to accurately measure EWs, a nearly-flat spectrum is recommended in all apertures. As such, each aperture is normalised separately. Overlapping echelle orders or separate apertures from independent beams are treated the same way in \smh{}. Spectra can be fit with polynomials (order 1-9) or cubic Beizer splines with iterative sigma-clipping of outlier flux points (Figure \ref{fig:normalisation}). In the case of cubic splines, the knot spacing between splines is a free parameter. Entire spectral regions can be excluded to avoid molecular band heads or telluric absorption, and additional points can be placed with positive relative weights. A screenshot of \starname{} open in \smh{} is shown in Figure \ref{fig:normalisation}, where the normalisation tab is in focus.

There are a number of keyboard shortcuts to speed up the normalisation of apertures. Left/right directional arrows will move between previous and following apertures, and up/down keys will scale the flux by $\pm$0.5\% increments. The continuum is re-fitted and re-drawn when any changes are made in the (i) normalisation parameters, (ii) flux scaling, or by any (iii) addition of faux weighted-flux points (by moving the mouse and pushing `a') or (iv) the exclusion of spectral regions (by pushing `e' and selecting two points). In this sense, any change to the continuum fitting and normalisation preview is always live. Whenever a new aperture is displayed -- which does not have a continuum determined already -- a fit is automatically performed using the current normalisation settings, and instantly shown in the \smh{} window. Thus, holding down the right directional key in the normalisation tab will sequentially display all apertures, automatically fitting and displaying the continuum for each aperture in the process. The continuum fits for every aperture remain available in the session for later inspection or manipulation by toggling through with the left and right directional keys.

Every aperture must have continuum fitted before the apertures can be stitched, and before the radial velocity can be measured. In some cases spectra may already be normalised outside of \smh{}. \smh{} assumes any aperture with a mean flux less $|F_{i}| < 2$ to be already normalised. Pushing the `c' key will ignore this assumption and fit the continuum for the aperture. A full list of keyboard shortcuts available is available at any time by pushing the `h' key. Once the continuum has been fit to each aperture, the `Stitch Apertures' button in the normalisation tab will become enabled.

\subsection{Stitching}
Stitching involves the summation of apertures -- usually overlapping -- into a single continuous spectrum. \smh{} follows the normalisation and stacking technique for overlapping apertures as described by Wako Aoki\footnote{http://www.naoj.org/Observing/Instruments/HDS/specana200810e.pdf}. Prior to stitching the apertures together, a linear dispersion map covering the spectral extent of all apertures is created using the smallest pixel size in any aperture. The flux from each aperture is interpolated on to the new dispersion map\footnote{The astute reader will recognise that this step will inevitably over-sample the spectrum in some region, yielding more pixels than what physically exists on the CCD. This is important for the error determination in global $\chi^2$-fitting approaches as the flux in each pixel becomes more correlated with its adjacent pixels. However, this does not affect the accuracy of results obtained with \smh{} in our EW-fitting algorithm (Section \ref{sec:smh-eqw-measurements}).}, and summed on a per-pixel scale. Similarly, the continuum fit to each aperture is evaluated and summed for each pixel on the new dispersion map. This provides two stacked spectra with continuous coverage: a total observed flux per pixel, and the total continuum per pixel. The total flux is divided by the total continuum, yielding a normalised continuous spectrum. When the aperture stitching is complete, the normalised spectrum becomes visible in the `Doppler Correct' tab of the \smh{} GUI.


\subsection{Doppler Measurement and Correction}
\begin{figure}
	\includegraphics[width=\columnwidth]{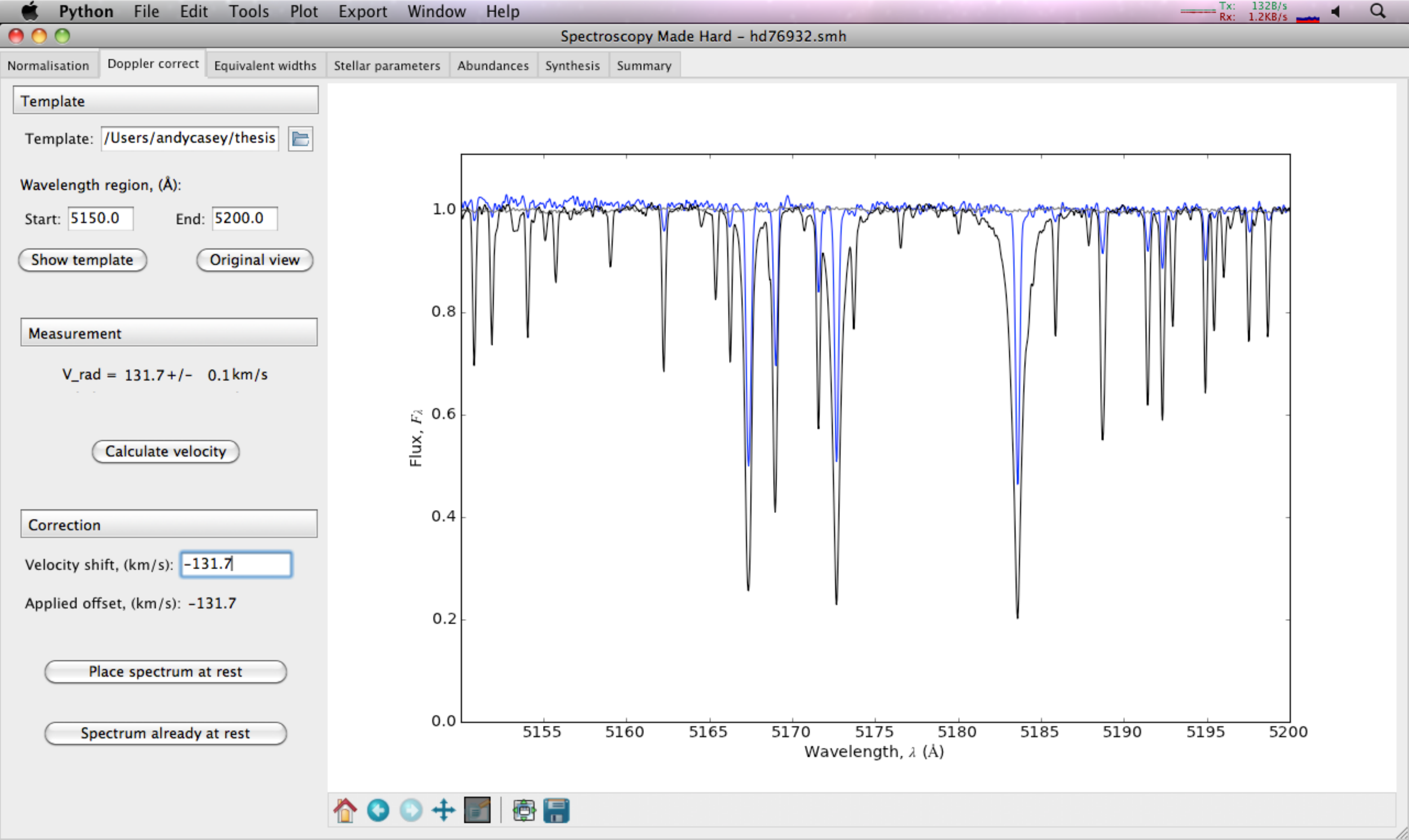}
	\caption[The `Doppler Correct' tab in Spectroscopy Made Hard for radial velocity measurement and correction]{The `Doppler Correct' tab in SMH after measuring and correcting \starname{} for radial velocity. The spectral region surrounding the Mg I triplet (515-520\,nm) is shown for the rest-frame spectrum (black), a template spectrum (blue), and a telluric spectrum (grey).}
\label{fig:doppler-correct}
\end{figure}

Radial velocities are calculated following by cross-correlation of the observed spectrum against a pre-calculated synthetic spectrum \citep{tonry;davis_1978}. The radial velocity is saved to the current session, and can be used to correct for the doppler shift with the `Set Spectrum at Rest' button (Figure \ref{fig:doppler-correct}). To minimise interpolation errors, no flux is interpolated; only the dispersion map is scaled to rest. The rest spectrum is shown in addition to the template spectrum, allowing the user to confirm the accuracy of the radial velocity measurement. If the correction is unsuitable, the user can specify the velocity offset to apply in order to place the spectrum at rest. Both the measured and applied velocity offset are stored in the \smh{} session file. After correcting for doppler shift, the rest spectrum immediately becomes present in the `Equivalent Widths' tab. Heliocentric and galactocentric rest-frame corrections are calculated from information stored within the input FITS headers, where available.

\subsection{Measuring Equivalent Widths}
\label{sec:smh-eqw-measurements}

The EWs for all absorption lines can be measured automatically or interactively in \smh{} by fitting Voigt, Gaussian or Lorentzian profiles. The initial rest wavelength of the line is necessary for automatic fitting, requiring the spectra to be at rest to within $\sim{}\pm5$\,km s$^{-1}$ and an atomic line list to be provided as input. The selected function (e.g., Gaussian) is iteratively fitted to every atomic absorption line while the local continuum (default within 10\,\AA{}) either side of the rest wavelength is determined by fitting a low order polynomial or spline. Measuring the local continuum ensures any errors in our initial normalisation or aperture stitching do not propagate through to our EW measurements\footnote{Theoretically it also permits the precise measurement of EWs without the need for any blaze removal or continuum determination. However, this is neither tested or recommended.}.

When determining the local continuum, pixels that are deemed to be part of an absorption or emission line are rejected. Any group of pixels next to each other that all deviate significantly ($>2\sigma$ default) from the local continua either belong to a residual cosmic ray, or they are peak/trough points of a nearby emission/absorption profile respectively. We attempt to fit an absorption profile (Gaussian by default) to each group of deviating pixels, using the current best estimate for the local continuum. If an emission or absorption profile is successfully fitted to the deviating group and its surrounding pixels, we assume those outlier points -- and all pixels within $5\sigma$ of the peak of that profile -- belong to a spectral feature, and are therefore excluded from the iterative continuum determination. 

When fitting the requisite absorption profile, the $\chi^2$ difference between the observed spectra and the fitted function is iteratively minimised using the Nelder-Mead Simplex algorithm \citep{nelder-mead}. In order to account for crowded spectral regions where the $\chi^2$ value might be influenced by a blended or nearby strong profile, our $\chi^2$ function is weighted by the distance to the rest wavelength ($\lambda_{r}$) with the function,

\begin{equation}
	W_{\lambda_{i}} = 1 + \exp{\left(\frac{-(\lambda_{i} - \lambda_{r})^2}{4\sigma^2}\right)}
\label{eq:chi-weight}
\end{equation}

\noindent{}where the rest wavelength and profile sigma\footnote{Recall $\rm{FWHM} = 2\sigma\sqrt{2\log_e{2}}$} ($\sigma$) are free parameters in the iterative fitting process. The free parameters are updated in each step of the continuum fitting. As a consequence of Equation \ref{eq:chi-weight}, pixels near the rest wavelength are weighted higher than those on the wings, forcing the fitting scheme to disregard blended or crowded surroundings. Although this approach relies on a reasonably accurate initial radial velocity correction of $<|5|$\,km s$^{-1}$ -- which is easily achievable -- it greatly improves the accuracy and robustness of the EW measurements.


Interestingly, the results of our iterative fitting approach are \textit{completely insensitive} to the initial FWHM guess. Increasing the initial FWHM estimate from 0.1\,\AA{} to 1 or 2\,\AA{} -- which are unphysically large values for high-resolution spectra -- does not alter either the measured FWHM or the EW for any of our absorption lines. Only a small increase in computational cost is observed. This reduces the complexity of the input parameters for the user.

\begin{figure}
	\includegraphics[width=\textwidth]{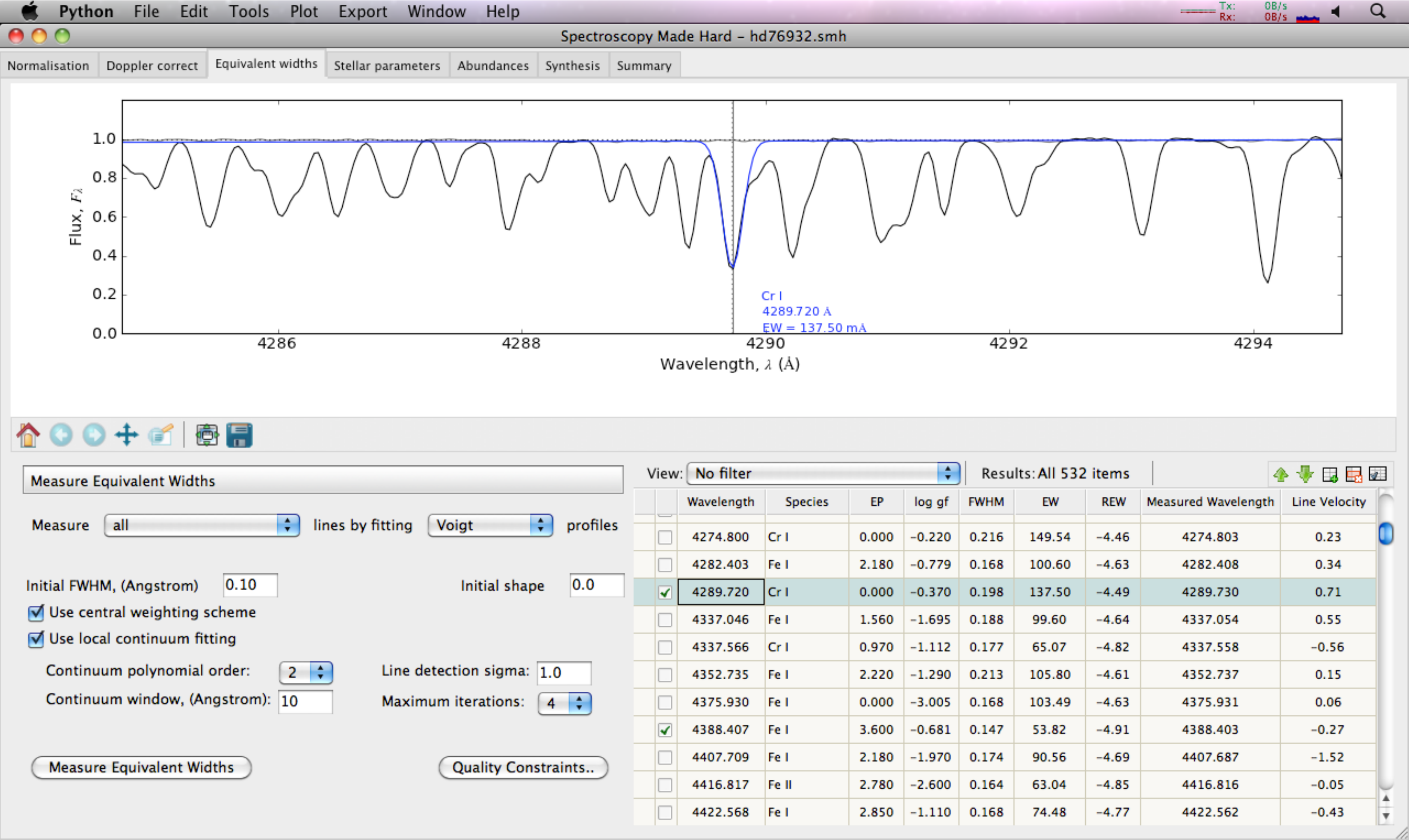}
	\caption[`Equivalent Width' tab in Spectroscopy Made Hard]{A screenshot highlighting the `Equivalent Width' tab for the analysis of \starname{} (black) in \smh{}. An atomic Cr\,I transition is highlighted, and the automatically fitted profile is shown (blue).}
\label{fig:eqw}
\end{figure}

The EW measurements provided by \smh{} are extremely reliable and accurate. Comparisons have already been published between EWs measured with \smh{} against trustworthy literature sources\footnote{In fact \citet{frebel;et-al_2013} use \smh{} and present the best agreement with manual measurements of EWs that we could find anywhere in the literature, by an order of magnitude.} \citep[e.g., see][]{casey;et-al_2013c, casey;et-al_2013b, frebel;et-al_2013}. These measurements agree excellently with manual measurements presented in the literature: to fractions of a m{\AA} in both the mean offset and standard deviation. No systematic effects appear present, even for saturated lines which would normally be discarded. The agreement is limited either by the $S/N$ of the data, or by precision of EWs listed in the literature. Nevertheless, we stress that every measured profile ought to be inspected by the user for sanity, and comparisons should be made between the software and a trusted source of measurement. Absorption lines that are deemed unsatisfactory can be interactively measured in \smh{} by using the `shift' key and selecting a point to define the continuum on either side of the line. The fitted profile and measured EW is immediately displayed.

\subsection{Quality Constraints}
\label{sec:smh-quality-constraints}
The quality of the EW measurements will significantly affect the accuracy of stellar parameters and chemical abundances. False-positive measurements of absorption lines (e.g., where only noise was fitted) can drive the stellar parameter determination in either unrealistic, or circular paths. Only EW measurements marked as `acceptable' (e.g., ticked in the GUI in Figure \ref{fig:eqw}) are utilised for stellar parameter determination and chemical abundances. Similarly, only data from lines marked as `acceptable' are outputted to publishable \LaTeX{} or machine-readable ASCII tables. It is critical to identify any spurious EW measurements and mark them as unacceptable before continuing the analysis.

\smh{} has several in-built filters to easily identify measurements that may be unrealistic. Lower and upper acceptable limits can be placed on the rest wavelength, residual line velocity, measured FWHM and EWs, as well as reduced equivalent widths (REWs) and the detection significance (in standard deviations, $\sigma_d$). The detection significance for a given absorption line is calculated as a multiple of the minimum detectable EW, as defined by a revised version of the \citet{cayrel} formula:

\begin{equation}
	{\rm EW}_{\rm min} = \left(\frac{S}{N}\right)^{-1}\sqrt{1.5\times FWHM \times \delta\lambda}
\end{equation}

\noindent{}where the FWHM is the smallest measurable FWHM as limited by the instrumental broadening, and $\delta\lambda$ is the pixel size in {\AA}ngstroms.

Bounded limits allow for physical boundaries that one can provide {\it a priori} without knowing anything about the data. These quality constraints can be applied through the interactive GUI from the `Equivalent Widths' tab. Filters and search capabilities are also included to only display measurements that match particular heuristics, or quickly identify certain absorption lines.


\subsection{Model Atmospheres}
\label{sec:smh-model-atmospheres}
Grids of existing model atmospheres come included with the \smh{} software. These include the 1D plane-parallel models of \citet{castelli;kurucz_2003}, the spherical 1D MARCs models \citep{marcs_models}, as well as the spatially and temporarily averaged 3D Stagger-Grid model atmospheres \citep{stagger_grid}, denoted as $\langle$3D$\rangle$. The assumptions, statistical input physics, and thermal stratification vary significantly between different model atmosphere grids, and some models will be more- or less-applicable depending on the stage of stellar evolution being examined. The user is encouraged to investigate the most appropriate models for their own analyses.

\smh{} employs logarithmic or linear (whichever is appropriate for the particular quantity) interpolation of temperatures, opacities, gas pressures and relevant quantities at all photospheric depths. This is performed using the Quickhill algorithm \citep{barber;et-al_1996}, which is reliant on Delaunay triangulation. This approach allows for the simultaneous interpolation of multivariate data with limitless dimensions. Linear, cubic and quadratic interpolation is available using this scheme, although linear interpolation is employed by default. A limitation to this approach is that Delaunay triangulation suffers from extremely skewed tetrahedral cells when the data axes vary in size by orders of magnitude by one another, as $T_{\rm eff}$ values do compared to $\log{g}$ or [(Fe,$\alpha$)/H]. If unaccounted for, the errors accumulated in the numerical approximations of sines and cosines of minute angles propagate through as significant errors in atmospheric properties across all photospheric depths. We scale all stellar parameters between zero and unity prior to interpolation in order to minimise these effects.

All atmospheres are interpolated in an identical way to facilitate a fair comparison between different grids of model atmospheres. From the user's perspective only the input stellar parameters and preferred atmosphere type are required. \smh{} identifies neighbouring models from the file names of model atmospheres and employs the relevant models for interpolation. No assumptions or restrictions are made on grid spacing or node frequency. This allows the user to generate a finer grid of model atmospheres and copy them into the appropriate directory. \smh{} will immediately utilize the correct model atmospheres.

\subsection{Stellar Parameters}
\subsubsection{Effective Surface Temperature, $T_{\rm eff}$}


\label{sec:smh-effective-temperature}
When a sufficient number of neutral transitions are present with an acceptable range in excitation potentials, the effective surface temperature can be accurately inferred from the Saha-Boltzmann equation. The number of atoms of an element in a given species is proportional to the excitation potential ($\chi_i$) and the effective temperature ($T_{\rm eff}$),

\begin{equation}
\label{eq:excitation_potential}
	\log{n({\rm Fe\,I})} \propto e^{-\chi_{i}/kT_{\rm eff}}
\end{equation}

\noindent{}All neutral iron lines should yield the same abundance for a representative effective temperature, if accurate atomic data are provided\footnote{Excitation equilibria can be performed for other elements (e.g., Ti, Cr) in \smh{}, but for this discussion we focus on neutral iron.}. If the spread in excitation potential of neutral iron lines is too small, the solution is unconstrained and an alternative method should be employed (e.g., photometric colour-$T_{\rm eff}$ relationships).

\begin{figure*}[t!]
	\includegraphics[width=\textwidth]{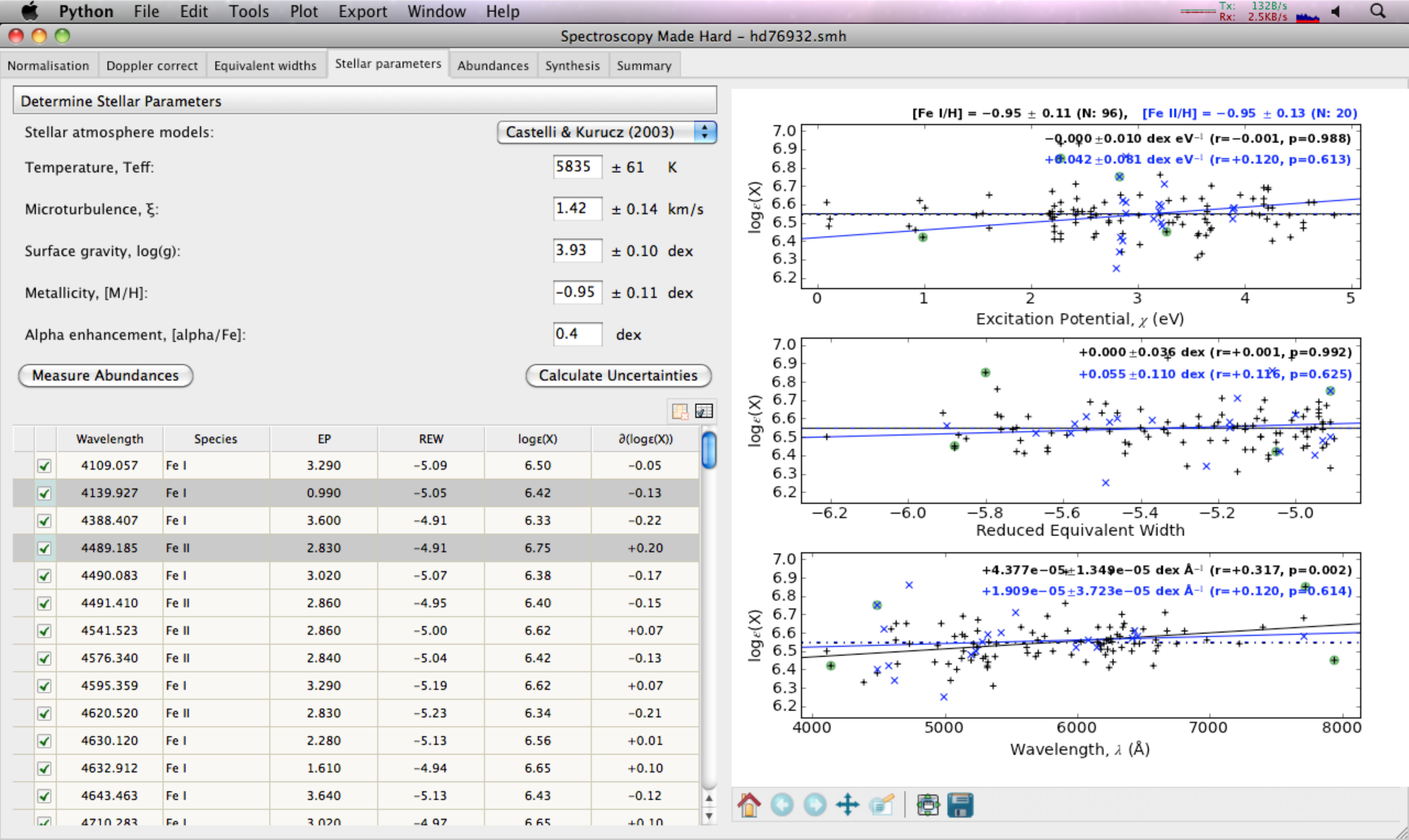}
	\caption[`Stellar Parameters' tab in Spectroscopy Made Hard, illustrating excitation and ionization equilibria]{Excitation and ionization balance in \smh{} for \starname{}. Four transitions have been selected from the figures, and the corresponding rows are highlighted in the table. The trend lines in Fe\,I and Fe\,II abundance with excitation potential $\chi_i$, REW, and wavelength $\lambda$, are shown. The standard error, correlation coefficient, and two-sided $p$-values are also quoted (see text).}
\label{fig:stellar-parameters}
\end{figure*}

Beginning with an initial guess of stellar parameters ($T_{\rm eff}$, $\log{g}$, [Fe/H], $\xi_t$), clicking the `Measure Abundances' button in the Stellar Parameters will interpolate a model atmosphere to the given input parameters. The atomic data for Fe lines, measured EWs and the interpolated model atmosphere are provided to the \textsc{abfind} routine in MOOG\footnote{All calls to MOOG are actually background calls to the non-interactive version, MOOGSILENT.} \citep{sneden_1973}. MOOG assumes local thermal equilibrium (LTE) and employs 1D model atmospheres to calculate the abundance required to produce the observed EW under the plasma conditions from the curve-of-growth for each transition. These abundances are read by \smh{} and immediately displayed in the Stellar Parameters tab (Figure \ref{fig:stellar-parameters}).

A linear fit is made to the $\chi_i$ and $\log\epsilon({\rm Fe\,I})$ data by least-squares regression. Following Equation \ref{eq:excitation_potential}, the effective temperature is ascertained by minimising the slope of this fit. The gradient in $\chi_i$ and $\log\epsilon({\rm Fe\,I})$, $m$, is quoted in \smh{} (Figure \ref{fig:stellar-parameters}), along with the standard error of the slope $\sigma_m$, the correlation coefficient $r$, and the two-sided $p$-value for a hypothesis test whose null hypothesis is that the slope is zero. No quality heuristics are employed by \smh{} with these statistics: the user may chose to adopt stellar parameters from the literature, even if they produce large $m$, $r$, or $p$-values. 

\subsubsection{Microturbulence, $\xi_t$}
Microturbulence is a necessary additive term in 1D model atmospheres to account for 3D turbulent motions larger than the mixing length. Strong atomic transitions that form deep in the photosphere will be most affected by microturbulence, whereas weak transitions forming at low optical depths will be largely unchanged. Weak and strong transitions of an element ought to yield the same abundance with the correct level of microturbulence. \smh{} solves for microturbulence by demanding a zero-trend in the neutral Fe (or similar) abundance with respect to REW, a measure of line strength. This is illustrated in the middle panel of Figure \ref{fig:stellar-parameters}, where the calculated slope and relevant statistics are shown.

Saturated lines near the flat part of the curve-of-growth should be excluded from analysis, as changes in stellar atmospheres will produce non-linear effects in abundance. This can be accomplished by employing a quality constraint (Section \ref{sec:smh-quality-constraints}) on REW values prior to stellar parameter determination (e.g., exclude transitions with $REW > -4.5$). Alternatively, lines can be marked as unacceptable directly from the `Stellar Parameters' tab (Figure \ref{fig:stellar-parameters}. Selecting a row in this table will highlight the point in all the figures in the `Stellar Parameters' tab. Similarly, points can be selected from these figures and the relevant table row will be highlighted. Multiple points can be selected using the `shift' key, and all selected transitions can be marked unacceptable or acceptable with the `u' or `a' keys. This allows for a quick interactive elimination of outlier measurements.

\subsubsection{Surface Gravity, $\log{g}$}
Following the Saha equation, the surface gravity for a star can be determined in \smh{} by the ionization balance of neutral and singly ionized iron abundances. A representative surface gravity will ensure that the mean of neutral and singly ionized iron abundances are the same\footnote{Ionization balance can be achieved with other transitions, but for the purpose of this discussion we focus on Fe.}. Like the determination of $T_{\rm eff}$ and $\xi_t$, this process is iterative. Similarly, in lieu of a requisite number of suitable Fe II transitions, the surface gravity can be set by employing other analysis methods (e.g., spectrophotometry, isochrones).

\subsubsection{Metallicity, ${\rm [M/H]}$}
The overall metallicity is generally taken as the mean Fe\,I abundance. The metallicity of a star has the least dependence on stellar parameters (compared to $T_{\rm eff}$, $\log{g}$, and $\xi_t$), and is typically the last stellar parameter to be solved. Since all of the stellar parameters are correlated to some degree, iterations from Section \ref{sec:smh-effective-temperature} are usually required.

\subsection{Chemical Abundances}
\begin{figure}[h]
	\includegraphics[width=\columnwidth]{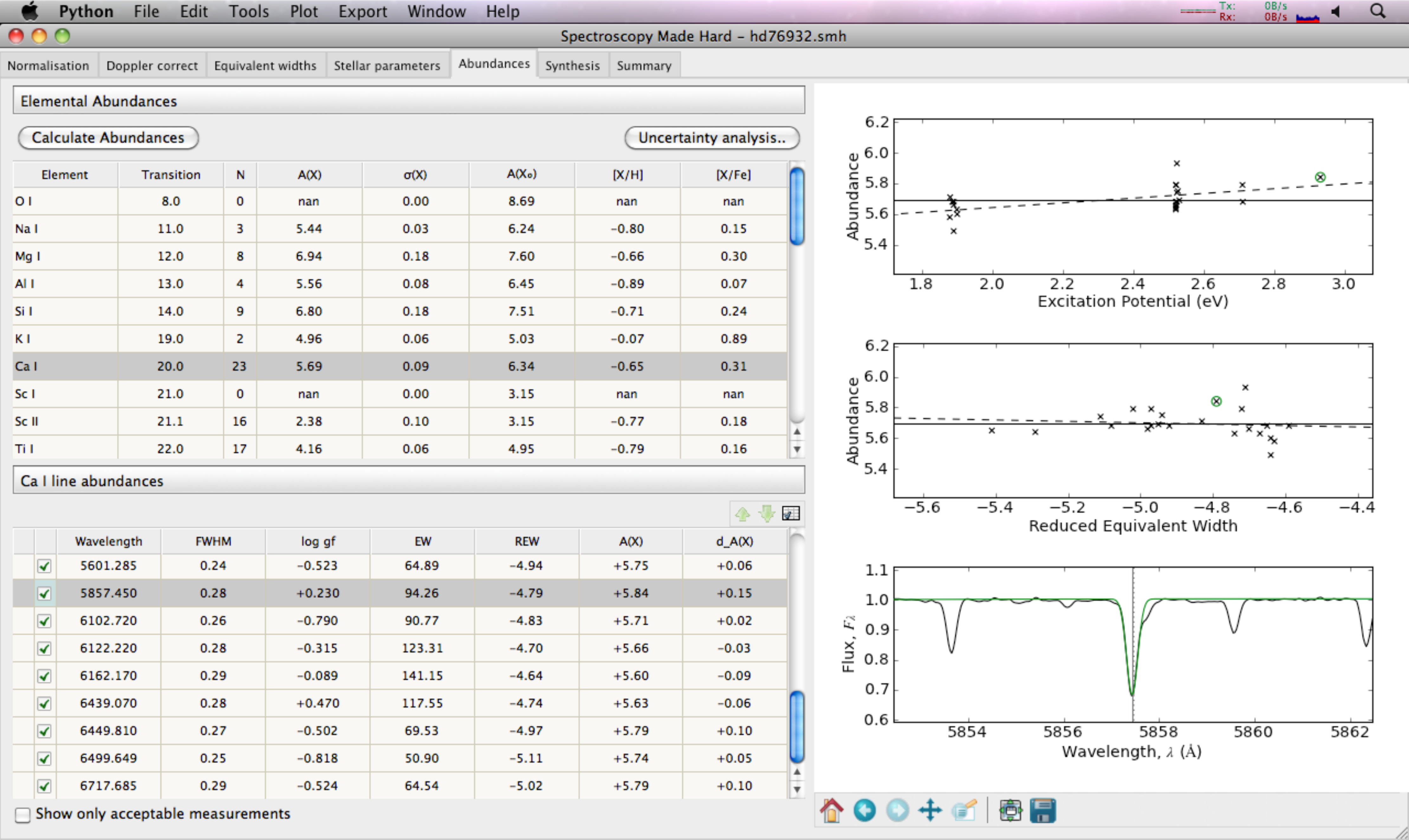}
	\caption[Screenshot of the `Chemical Abundances' tab in Spectroscopy Made Hard for \starname{}]{Screenshot of the `Chemical Abundances' tab in \smh{} for \starname{}. Ca\,I is highlighted in the top table, so all Ca\,I transitions are shown in the lower table and plotted in the figures. The 585.7\,nm transition is selected in the table and highlighted in the figures opposite. The fitted absorption profile is also shown, allowing for an interactive re-measurement.}
\label{fig:chemical-abundances}
\end{figure}

Once a suitable stellar atmosphere model has been generated, chemical abundances can be determined for relevant elements directly from measured EWs. \smh tab automatically calculates the mean abundance, standard deviation $\sigma$, [X/H] and [X/Fe] abundance ratios in the `Chemical Abundances' tab for all elements present in the atomic line list (Figure \ref{fig:chemical-abundances}). A tabulation of Solar abundances is required to calculate these abundance ratios, and by default the \citet{asplund;et-al_2009} composition is used.
 
Chemical abundances are only calculated for atomic lines that have EWs and have been marked as acceptable. The interactive mode is perfect for the careful inspection of EWs and abundances. Selecting a species\footnote{A species is defined by an element and ionisation state.} in the top table (Figure \ref{fig:chemical-abundances}) will immediately fill the lower table with all the line measurements for the selected element, and populate the relevant figures with the excitation potential $\chi$, REW, and $\log{\epsilon}(X)$ abundances for the selected species. Selecting a transition in the bottom table will highlight the point in the relevant figures, show the spectra surrounding that feature, including the fitted absorption profile. Similar to the `Stellar Parameters' tab, points can be selected directly from the figures and they will be highlighted in the table list. Any changes to the number of transitions marked as acceptable will automatically update the mean abundance and standard deviation.

Similar to the `Equivalent Widths' tab, the EW of the selected transition can be re-measured interactively from the lower plot in the `Chemical Abundances' tab by holding the `shift' key and selecting two continuum points. This will delete the calculated abundance for the selected line, as this needs to be re-calculated through MOOG. Pushing the `Calculate Abundances' button will update these abundances accordingly.

It is important to note the reference Fe abundance used to calculate [X/Fe] abundance ratios. These ratios appear in the chemical abundance analysis, during synthesis, in plots and when exporting data to databases or \LaTeX{} tables. By default the mean Fe I abundance is used as the reference $\log_{\epsilon}$Fe abundance -- even if only a single Fe line was measured. If no neutral Fe lines have been measured then the model atmosphere metallicity [M/H] will be employed, and the labels in \smh{} will switch from [X/Fe] to [X/M] to indicate this change. Similarly, if the model atmosphere metallicity is used as a reference and then a single Fe I line is measured, the labels and abundance ratios will automatically update to [X/Fe]. 

\subsection{Spectral Synthesis}
\label{sec:smh-synthesis}

Atomic transitions of some elements exhibit significant broadening due to hyperfine structure and/or isotopic splitting. Ignoring these effects leads to an underestimate of the line's EW and overestimated abundances. As such, these transitions -- as well as significantly blended or molecular features -- require a spectral synthesis approach.

\begin{figure}[h]
	\includegraphics[width=\columnwidth]{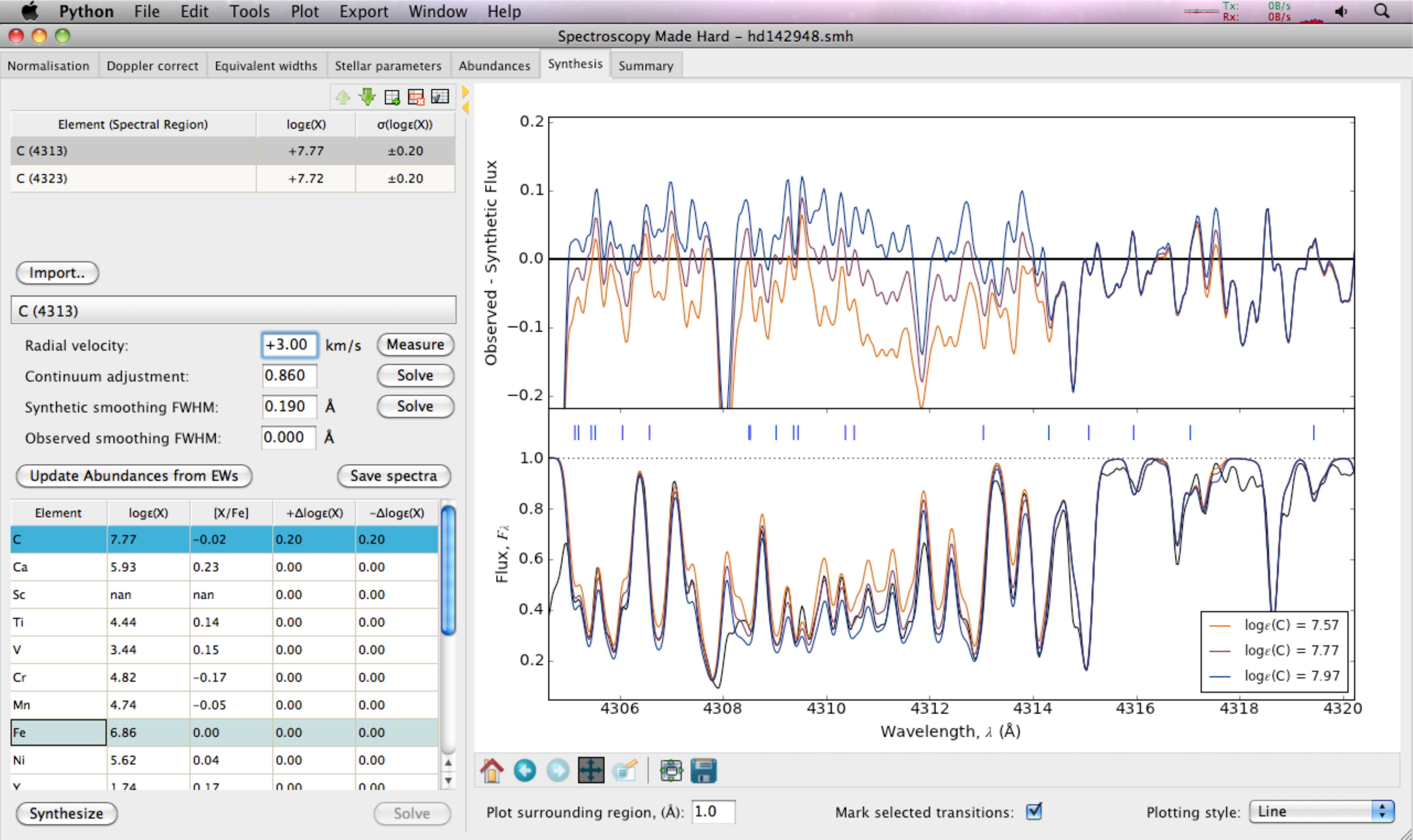}
	\caption[The `Synthesis' tab in Spectroscopy Made Hard]{A screenshot of \smh{} with a focus on the `Synthesis' tab. Smoothed synthesised spectra for the current setup (CH feature at 431.3\,nm) are overlaid upon the rest observed spectrum of \starname{} in this region.}
\label{fig:synthesis}
\end{figure}

Spectral synthesis can be performed in \smh{} once a model atmosphere has been created. Each synthesis of a spectral region requires a line list, the element of interest, and any relevant isotopic splitting as input. Multiple synthesis regions can be introduced to \smh{} at once from a properly formatted input file. \smh{} inspects the provided line list to determine the wavelength region of interest and identify relevant species. The observed spectrum is displayed and elemental abundances can be edited directly from the \smh{} GUI (Figure \ref{fig:synthesis}). Initially, abundances for all elements present in the line list are filled with mean abundances measured from EWs or other synthesised regions. These values can be adjusted by either varying the $\log\epsilon({\rm X})$ value or [X/Fe] abundance ratio. Adjusting one value will update its counterpart. Highlighting an element in this table will also display markers above the spectrum which indicate the rest wavelengths and hyperfine structure contributions for the selected element.

Upon generating a synthetic spectrum, the residual radial velocity, synthetic smoothing kernel and continuum scaling factor can be adjusted to fit the observed spectra. Any changes in these values are updated immediately in the figure. These values can also be measured automatically by minimising the $\chi^2$ difference between the observed and synthetic spectrum. This allows the user to perform interactive syntheses extremely quickly. The line list, and all edited abundances for each synthesis are saved to the current session, such that they allow any other spectroscopist to reproduce the results with the saved \smh{} session file.

\subsection{Uncertainties}
\subsubsection{Uncertainties in Stellar Parameters}
Uncertainties in stellar parameters can be calculated easily in \smh{}. Uncertainties in $T_{\rm eff}$ and $\xi_t$ are found by independently adjusting the effective temperature and microturbulence until the magnitude of the slope $m$ matches the formal uncertainty $\sigma_m$. Similarly, the uncertainty in $\log{g}$ is ascertained by adjusting the surface gravity until the mean difference in neutral and singly ionized iron abundances matches the quadrature sum of the standard error in Fe I and Fe II abundances. The uncertainty in overall metallicity is taken as the standard error about the mean for Fe I abundances. These uncertainties are calculated automatically by clicking the `Calculate Uncertainties' button in the `Stellar Parameters' tab (Figure \ref{fig:stellar-parameters}).

\subsubsection{Uncertainties in Chemical Abundances}
\begin{figure}[h]
	\includegraphics[width=\textwidth]{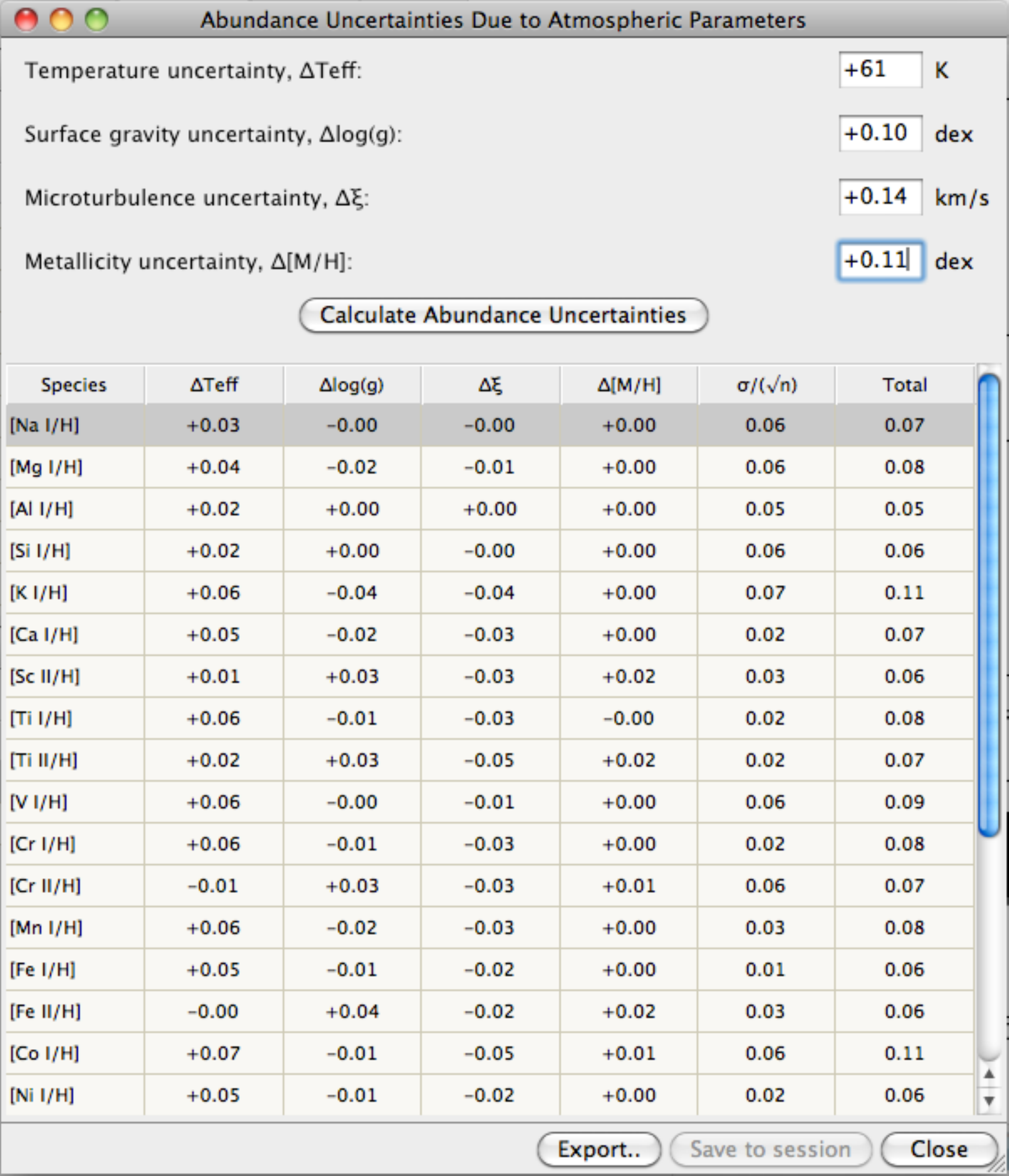}
	\caption[Dialog window for calculating elemental abundance uncertainties in Spectroscopy Made Hard]{\smh{} dialog window for calculating elemental abundances uncertainties, including the effects due to uncertainties in stellar parameters.}
\label{fig:uncertainties-due-to-stellar-parameters}
\end{figure}

Uncertainties in chemical abundances are due to a number of effects, which include uncertainties due to stellar parameters (i.e., systematic effects $\sigma_{sys}$), as well as the the random scatter in abundance measurements ($\sigma_{rand}$). To estimate these effects \smh{} independently varies the stellar parameters by the $1\sigma$ uncertainties in each parameter, and calculates the mean abundances from EWs for all elements. An example of this is demonstrated in Figure \ref{fig:uncertainties-due-to-stellar-parameters}. When calculating the standard error about the mean for abundance measurements, a minimum standard deviation of 0.10\,dex is adopted for all elements. This conservative approach accounts for scenarios when only a few atomic transitions were available for a given element. The total uncertainty in [X/H] for any element is calculated by adding the systematic errors in quadrature to provide a total systematic error $\sigma_{sys}$. The total systematic error is added in quadrature with the random error to provide a total uncertainty in $\log\epsilon({\rm X})$ or [X/H].

\section{Usage}
This software has already been used in a number of accepted and submitted publications. While the software is not currently open source, it is regularly used by almost two dozen stellar spectroscopists and has been used to analyse over a thousand high-resolution spectra. This software has also facilitated several extremely productive undergraduate student research projects. \smh{} allows students to perform a rigorous high-resolution spectroscopic analysis, and have their work interactively inspected by established spectroscopists. Naturally, the proper examination and scrutiny expected for scientific research should be made to their analysis and results. \smh{} allows for the students' results to be easily verified, reproduced, and compared with the literature. Therefore whilst the scrutiny afforded by the scientific method is maintained, the scientific output is significantly increased.

\section{Conclusions}
\label{sec:smh-conclusions}

We have presented intuitive, well-tested software to analyse high-resolution stellar spectra, and to ease the transition from small to seemingly insurmountable sample sizes. \smh{} differs from other approaches to solve this problem, by allowing a full manual adjustment of any part of the analysis via a GUI. This ensures accuracy, scientific reproducibility, and data provenance. It also allows spectroscopists to publish their \smh{} session files with their paper -- and this is encouraged -- as well as allowing for any reader with the \smh{} code to inspect and reproduce every aspect of their analyses.

\chapter{Hunting the Parent of the Orphan Stream I}
\label{chap:orphan_I}

\previouslypublished{Parts of this chapter have been previously published as \href{http://adsabs.harvard.edu/abs/2013ApJ...764...39C}{`Hunting the Parent of the Orphan Stream: Identifying Stream Members from Low-Resolution Spectroscopy', {Casey}, A.~R., {Da Costa}, G., {Keller}, S.~C., {Maunder}, E., 2013, ApJ, 764, 39C}. The work is presented here in expanded and updated form.} \\

We present candidate K-giant members in the Orphan stream which have been identified from low-resolution data taken with the AAOmega spectrograph on the Anglo-Australian Telescope. From modest $S/N$ spectra and independent cuts in photometry, kinematics, gravity and metallicity we yield self-consistent, highly probable stream members. We find a revised stream distance of $22.5\,\pm\,2.0$\,kpc near the celestial equator, and our kinematic signature peaks at $V_{GSR} = 82.1\,\pm\,1.4$\,km s$^{-1}$. The observed velocity dispersion of our most probable members is consistent with arising from the velocity uncertainties alone. This indicates that at least along this line-of-sight, the Orphan stream is kinematically cold. Our data indicates an overall stream metallicity of [Fe/H] $= -1.63\,\pm\,0.19$\,dex which is more metal-rich than previously found and unbiased by spectral type. Furthermore, the significant metallicity dispersion displayed by our most probable members, $\sigma(\mbox{[Fe/H]}) = 0.56$\,dex, suggests that the unidentified Orphan stream parent is a dSph satellite. We highlight likely members for high-resolution spectroscopic follow-up.

\section{Introduction}
\label{sec:ch2-introduction}



The Orphan stream was independently detected by both \citet{grillmair_2006} and \citet{belokurov;et-al_2006}, and is distinct from other substructures in the halo. The stream stretches over $60^\circ$ of arc on the sky, has a low surface brightness, and a narrow stream width of only $\sim$2$^\circ$. As the name suggests, the parent object largely remains a mystery. The stream extends past the celestial equator \--- outside the SDSS footprint \--- but has not been detected in existing southern surveys \citep{newberg;et-al_2010}. Whilst the parent system remains elusive, significant effort has been placed on associating the stream with known Milky Way satellites \citep{zucker;et-al_2006, fellhaur;et-al_2007,jin;lynden-bell_2007,sales;et-al_2008}. In contrast, there has been relatively limited observational work on the Orphan stream itself other than the original discovery papers \citep{grillmair_2006, belokurov;et-al_2006, belokurov;et-al_2007} and the work of \citet{newberg;et-al_2010}. This is largely to be expected given the absence of deep multi-band photometry in the southern sky, and the low total luminosity of the stream. This makes it difficult to reliably separate Orphan stream members from halo stars. Understanding the full extent of the stream awaits the SkyMapper and Pan-STARRS photometric surveys \citep{keller;et-al_2007, hodapp;et-al_2004}.

As \citet{sales;et-al_2008} point out, there is a natural observational bias towards more massive and recent mergers like Sagittarius. Consequently, the fainter end of this substructure distribution has yet to be fully recovered, or thoroughly examined. Interestingly, there are indications that  some fainter substructures like the Orphan stream and the Palomar 5 tidal tails \citep{odenkirchen;et-al_2009} have orbits which seem to be best-fit by Milky Way models with nearly 60\% less mass \citep{newberg;et-al_2010} than generally reported by \citet{xue;et-al_2008} and \citet{koposov;et-al_2010}. Such a discrepancy in the mass of the Milky Way is troublesome. More complete photometric and kinematic maps of these low total luminosity streams may provide the best test as to whether this mass discrepancy is real, or an artefact of incomplete observations. Whilst the full spatial extent of the Orphan stream remains unknown, we can examine the detailed chemistry of its members, investigate the stream history, and make predictions about the nature of the progenitor.

In this chapter we present a detailed, self-consistent analysis to identify K-giant members of the Orphan stream. Using our selection method we have catalogued the locations of nine highly probable Orphan stream candidates, all worthy of high-resolution spectroscopic follow up. In the following section we outline our photometric target selection. In Section \ref{sec:ch2-observations} we describe the low-resolution spectroscopic observations. The data analysis, including stream identification, is discussed in Section \ref{sec:ch2-analysis} and in Section \ref{sec:ch2-conclusions} the conclusions, predictions and future work are presented.

\section{Target Selection}
\label{sec:ch2-target-selection}

We have targeted K-giant members of the Orphan stream in order to investigate their detailed chemistry. Because K-giants are difficult to unambiguously detect from photometry alone, low-resolution spectroscopy is required to estimate stellar parameters and determine radial velocities. The Orphan stream has an extremely low spatial over-density, which makes it difficult to separate stream members from halo stars. However, there is a well described distance gradient along the stream \citep{belokurov;et-al_2007, newberg;et-al_2010} which provides an indication on where we should focus our spectroscopic efforts.

The Orphan stream is closest to us in two locations on the edge of the SDSS footprint: at the celestial equator \citep{belokurov;et-al_2007}, and along outrigger SEGUE Stripe 1540 \citep{newberg;et-al_2010}. These two locations are unequivocally the best place to recover bright stream members. Two fields centered on $(\alpha, \delta) =$ (10:48:15, 00:00:00) and (10:48:15, $-$02:30:00) have been targeted, and a combination of colour cuts with the SDSS DR 7 \citep{abazajian;et-al_2009} have been employed to identify likely K-giants:
\begin{eqnarray}
0.6 <& (g-i)_0 &< 1.7 \\
-15(g-i)_0 + 27 <& g_0 &< -3.75(g-i)_0 + 22.5 \\
15  <& i_0  &< 18 
\end{eqnarray}

Given our colour selection we expect to recover giants and contaminating dwarfs. Although the 2MASS $JHK$ colours can help to separate dwarfs and giants, our target K-giants stars are too faint to be detected in the 2MASS catalogue.

\section{Observations}
\label{sec:ch2-observations}

Observations took place on the Anglo-Australian Telescope using the AAOmega spectrograph in April 2009. AAOmega is a fibre-fed, dual beam multi-object spectrograph which is capable of simultaneously observing spectra of 392 (science and sky) targets across a $2\,^\circ$ field of view. We used the 570\,nm dichroic in combination with the 1000I grating in the red arm, and the 580V grating in the blue arm. This provides a spectral coverage between $800 \leq \lambda \leq 950$\,nm in the red at $\mathcal{R} \approx 4400$, and between $370 \leq \lambda \leq 580$\,nm with a lower spectral resolution of $\mathcal{R} \approx 1300$ in the blue.

The data were reduced using the standard 2DFDR reduction pipeline\footnote{http://www.aao.gov.au/2df/aaomega/aaomega\_2dfdr.html}. After flat-fielding, throughput calibration for each fibre was achieved using the intensity of skylines in each fibre. The median flux of dedicated sky fibres was used for sky subtraction, and wavelength calibration was performed using Th-Ar arc lamp exposures taken between science frames. Three thirty minute science exposures were median-combined to assist with cosmic ray removal. The median $S/N$ obtained in the red arm for our fields is modest at 35 per pixel, although this deteriorates quickly for our fainter targets. With the presence of strong Ca\,II triplet lines in the red arm, we are able to ascertain reliable radial velocities and reasonable estimates on overall metallicity \citep[][and references therein]{starkenburg;et-al_2010}. Our spectral region also includes gravity-sensitive magnesium lines: Mg\,I at 880.7\,nm, and the Mg\,I\,b 3$p$-4$s$ triplet lines at $\sim$517.8\,nm. These lines are sufficient to discriminate dwarfs from giants even with weak signal.

The blue and red arm spectra were normalised using a third order cubic spline after multiple iterations of outlier clipping. We used defined knot spacings of 20\,nm in the red arm, and 5\,nm in the blue arm in order to accommodate often poor $S/N$, and varying strengths of molecular band-heads.

\section{Analysis \& Discussion}
\label{sec:ch2-analysis}

We have employed a combination of separate criteria to identify likely Orphan stream members: kinematics, a giant/dwarf indication from Mg\,I lines, and selecting stars with consistent metallicities derived from both isochrone fitting and the strength of the Ca\,II triplet lines. Each criteria is discussed here separately.

\subsection{Kinematics}
Radial velocities were measured by cross-correlating our normalised spectra against a synthetic spectrum of a metal-poor giant across the spectral range $845 \leq \lambda \leq 870$\,nm. Heliocentric velocities were translated to the galactic rest frame by adopting the LSR velocity as 220\,km s$^{-1}$ towards $(l, b) = (53\,^\circ, 25\,^\circ)$ \citep{kerr;lynden-bell_1986, mihalas;binney_1981}\footnote{Where $V_{GSR} = V_{HELIO} + 220\sin{l}\cos{b} + 16.5\times[\sin{b}\sin{25^\circ} + \cos{b}\cos{25^\circ}\cos{(l - 53^\circ)}]$}.

\begin{figure}[h]
	\includegraphics[width=\columnwidth]{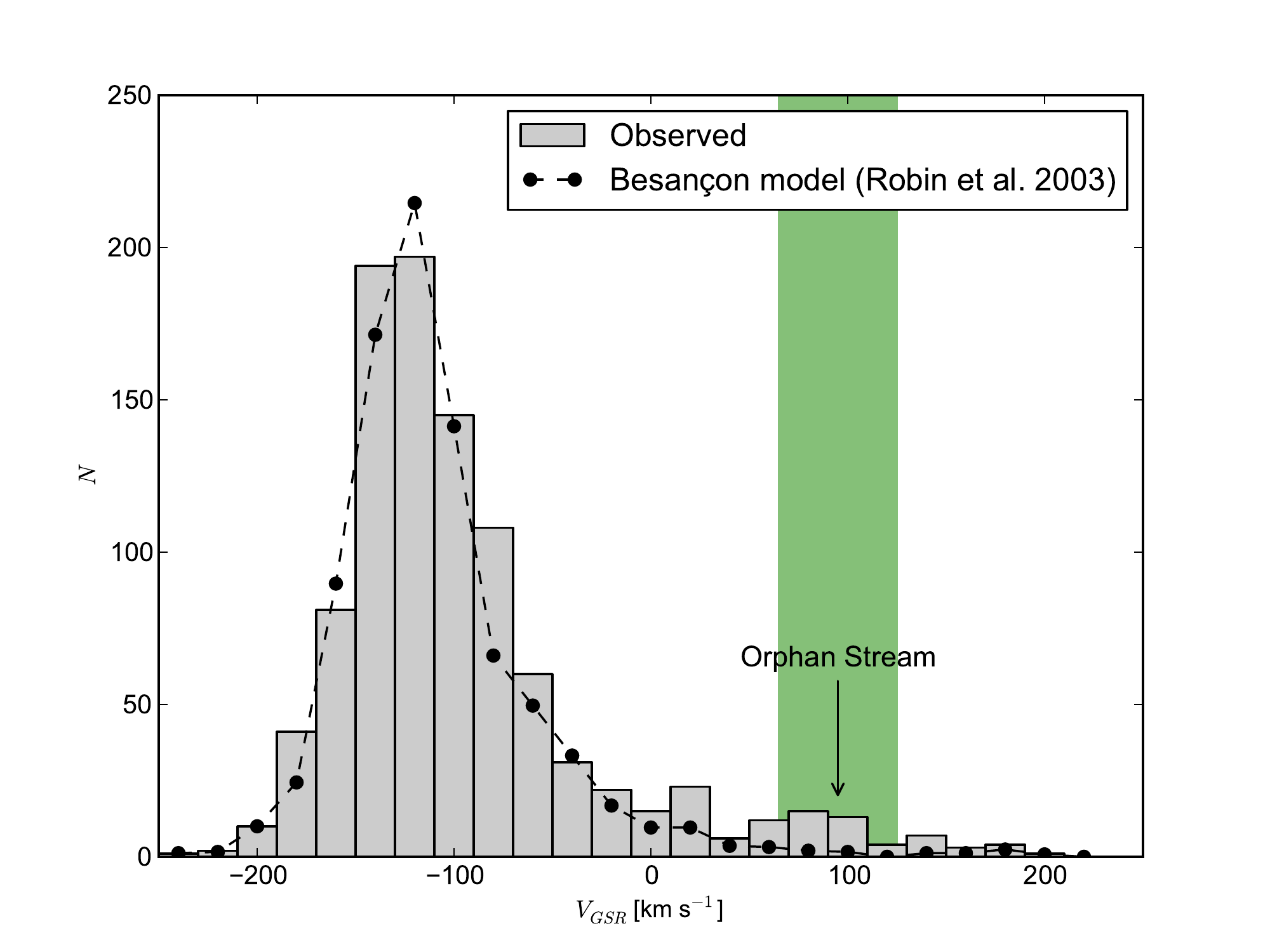}
	\caption[Predicted and observed galactocentric rest frame velocities]{Galactocentric rest frame velocities for stars in both our observed fields (grey), and predicted Besan\c{c}on velocities which have been scaled to match our observed sample size. The expected kinematic signature from \citet{newberg;et-al_2010} for the Orphan stream is highlighted, as is our kinematic selection window (green).}
	\label{fig:ch2-velocities}
\end{figure}

Figure \ref{fig:ch2-velocities} shows a histogram of our galactocentric velocities, compared to the predicted smooth line-of-sight velocity distribution for this region from the Besan\c{c}on model \citep{robin;et-al_2003}. We have selected particles from the Besan\c{c}on model using the same criteria outlined in Section \ref{sec:ch2-target-selection} after employing the \citet{jordi;et-al_2006} colour transformations. It is clear that our target selection has yielded mostly nearby disk dwarf stars with $V_{GSR} \approx -120$\,km\,s$^{-1}$.

In a nearby ($\Delta\Lambda_{Orphan} \sim 4^\circ$) region of the stream, \citet{newberg;et-al_2010} detected the Orphan stream with a $V_{GSR} = 101.4$\,km s$^{-1}$ from BHB stars. Differences in accounting for the local standard of rest between this work and \citet{newberg;et-al_2010} means that this corresponds to approximately 95\,km s$^{-1}$ on our $V_{GSR}$ scale.  This is discussed further in Section \ref{sec:ch2-newberg}. The expected Orphan stream kinematic peak is labelled in Figure \ref{fig:ch2-velocities}. There is no obvious sharp kinematic peak representative of the Orphan stream in our sample. From kinematics alone, our targets appears largely indistinguishable from a smooth halo distribution. To isolate potential Orphan stream members we have nominated a relatively wide selection criteria between $65 \leq V_{GSR} \leq 125$\,km s$^{-1}$ (shown in Figure \ref{fig:ch2-velocities}), which yields 28 Orphan stream candidates. The typical uncertainty in our velocities is $\pm{}5.0$\,km s$^{-1}$.

\subsection{Dwarf/Giant Discrimination}
\label{sec:ch2-dwarf-giant}

We have measured the EW of the gravity-sensitive Mg\,I line at 880.7\,nm to distinguish dwarfs from giants \citep{battaglia;starkenburg_2012}. At a given temperature (or $g - r$) and metallicity, giant stars present narrower Mg\,I absorption lines than their dwarf counterparts. Given the target selection, our sample is likely to contain many more dwarfs than giants. In some cases no Mg\,I 880.7\,nm line was apparent, so an upper limit was estimated based on the $S/N$ of the spectra. In these cases the candidate was considered a ``non-dwarf" because we cannot exclusively rule out a metal-poor sub-giant with this criteria alone. For these purposes we are only looking for a simple indication as to whether a star is likely a dwarf or not. 

\begin{figure}[h]
	\includegraphics[width=\columnwidth]{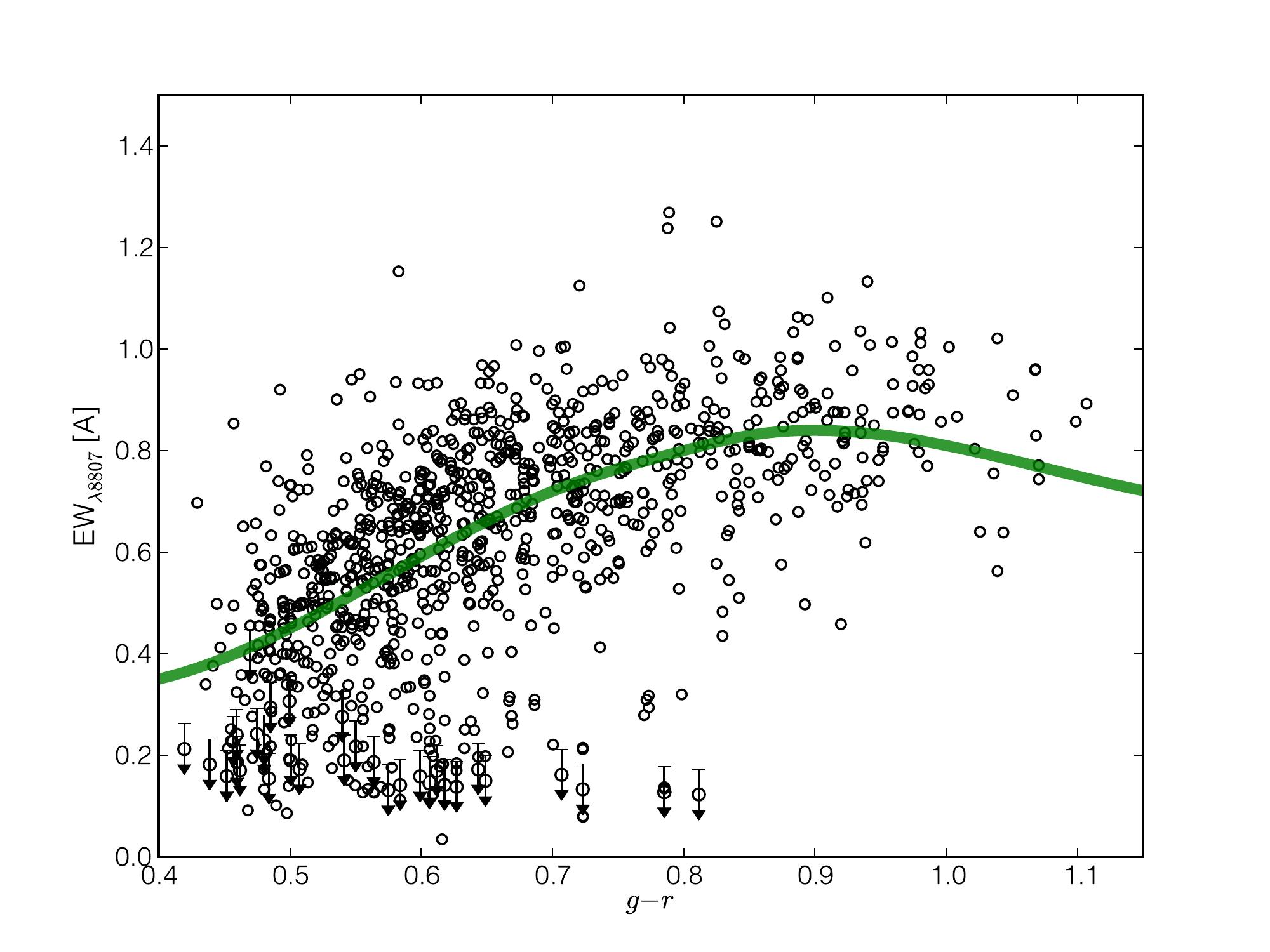}
	\caption[Dwarf/giant discriminant based on EW of the $\lambda$8807 Mg\,I transition]{SDSS $g - r$ against the measured EW of the Mg\,I transition at 880.7\,nm. Dwarf contaminants occupy the more populous upper branch. Our separation line between dwarfs and giants is shown in green.}
	\label{fig:ch2-ew-mg}
\end{figure}

Figure \ref{fig:ch2-ew-mg} illustrates the trend with $EW_{\lambda880.7}$ against SDSS de-reddened\footnote{All magnitudes presented in this chapter are de-reddened using the \citet{schlegel;et-al_1998} dust maps.} $g - r$, illustrating the dominant upper dwarf branch we wish to exclude. Giant stars populate the lower, sparser branch. A separation line has been adopted to distinguish dwarfs from giants, and is shown in Figure \ref{fig:ch2-ew-mg}. If we were to place this line higher, the total number of true giant stars may increase, but the dwarf contamination rate will rise dramatically. A compromise must be made between the rate of giant recoverability and the dwarf contamination. Our dwarf/giant separation line lies just below the main dwarf population. On its own, this dwarf/giant separation line would typically result in far too many dwarf contaminants. However, we are  employing selections on multiple observables (kinematics, metallicity, proper motions) in order to refine our Orphan stream giant sample. 

This dwarf/giant separation method was also employed using the total EW of the Mg\,I\,b triplet lines. Both analyses were entirely consistent with each other: essentially the same candidate list was found using both techniques. However, given slightly poorer signal at the Mg\,I\,b triplet, we were forced to adopt many more upper limits than when the 880.7\,nm line was employed. Because we classify all upper limits as being ``non-dwarfs" (i.e., potential giants), we deduced a slightly larger candidate sample for the Mg\,I\,b analysis, which was primarily populated by upper limits. In conclusion, we found the Mg\,I line at 880.7\,nm appeared to be a more consistent dwarf discriminant given our weak $S/N$ \--- particularly for our fainter stars. Thus, we have used the 880.7\,nm Mg\,I selection throughout the rest of our analysis.

The adopted dwarf/giant separation line shown in Figure \ref{fig:ch2-ew-mg} yields 425 potential giants. Upon taking the intersection of our kinematic and gravity selections, we find 20 stars that appear to be likely Orphan stream giants.

\subsection{Metallicities}
\label{sec:ch2-metallicities}

We have measured the metallicities for the stars that meet our kinematic and surface gravity criteria in two ways: with the strength of their Ca\,II triplet lines, and by isochrone-fitting. After correcting for luminosity, the EW of the Ca\,II triplet lines provide a good indication of the overall metallicity of a RGB star \citep{armandroff;da_costa_1991}. We have employed the \citet{starkenburg;et-al_2010} relationship and corrected for luminosity in $g$ against the horizontal branch magnitude at $g_{HB}$ = 17.1 \citep{newberg;et-al_2010}. Strictly speaking, the Ca\,II\---[Fe/H] calibration is only valid for stars brighter than the horizontal branch, although the relationship only becomes significantly inappropriate near $g \-- g_{HB} \sim +1$ \citep{saviane;et-al_2012}. Many of our candidates are fainter than this valid luminosity range, and therefore they should not be excluded solely because of their derived metallicities, as these could be uncertain. Stars fainter than $g_{HB}$ will have slightly lower metallicities than predicted by our Ca\,II\---[Fe/H] relationship, and for these stars we will only use metallicities to assign a relative qualitative likelihood for stream membership.

\begin{figure}[t!]
	\includegraphics[width=\columnwidth]{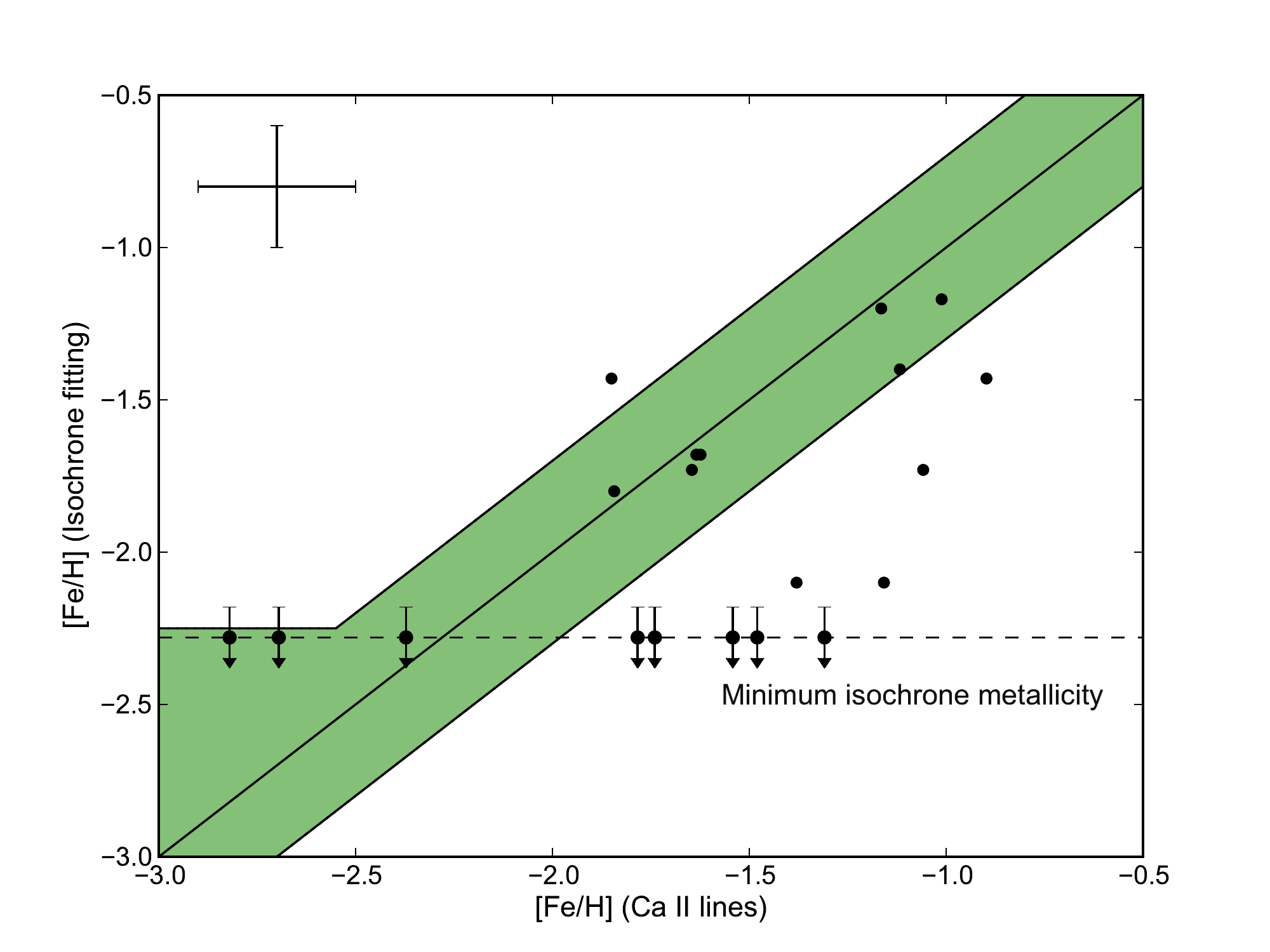}
	\caption[Metallicities of Orphan stream candidates from isochrone fitting and Ca\,II triplet lines]{Metallicities from the Ca\,II triplet lines versus those found from fitting isochrones to the 20 stars that meet our kinematic and surface gravity criteria. Both abundance determinations imply these stars are RGB members of the Orphan stream at a distance of $\sim$21.4\,kpc \citep{newberg;et-al_2010}. Consistency between these methods indicates highly likely stream membership (shaded region). The minimum isochrone [Fe/H], and a representative uncertainty of 0.2\,dex for abundance measurements is shown.}
	\label{fig:ch2-feh}
\end{figure}

Given a distance estimate to the Orphan stream, we can also deduce a star's metallicity through isochrone fitting. We have used a 10\,Gyr \citet{girardi;et-al_2004} isochrone at 21.4\,kpc \citep{newberg;et-al_2010} and found metallicities for all 20 likely stream members from their best-fitting isochrone. Derived metallicities from Ca\,II line strengths and isochrone fitting that are consistent (within $\pm0.3$\,dex) indicates these measurements are reliable, and that these stars are indeed at a distance of $\sim$21.4\,kpc. We find 10 highly likely stream members with consistently derived metallicities. They fall within the shaded region illustrated in Figure \ref{fig:ch2-feh}. 

\begin{figure}[t!]
	\includegraphics[width=\columnwidth]{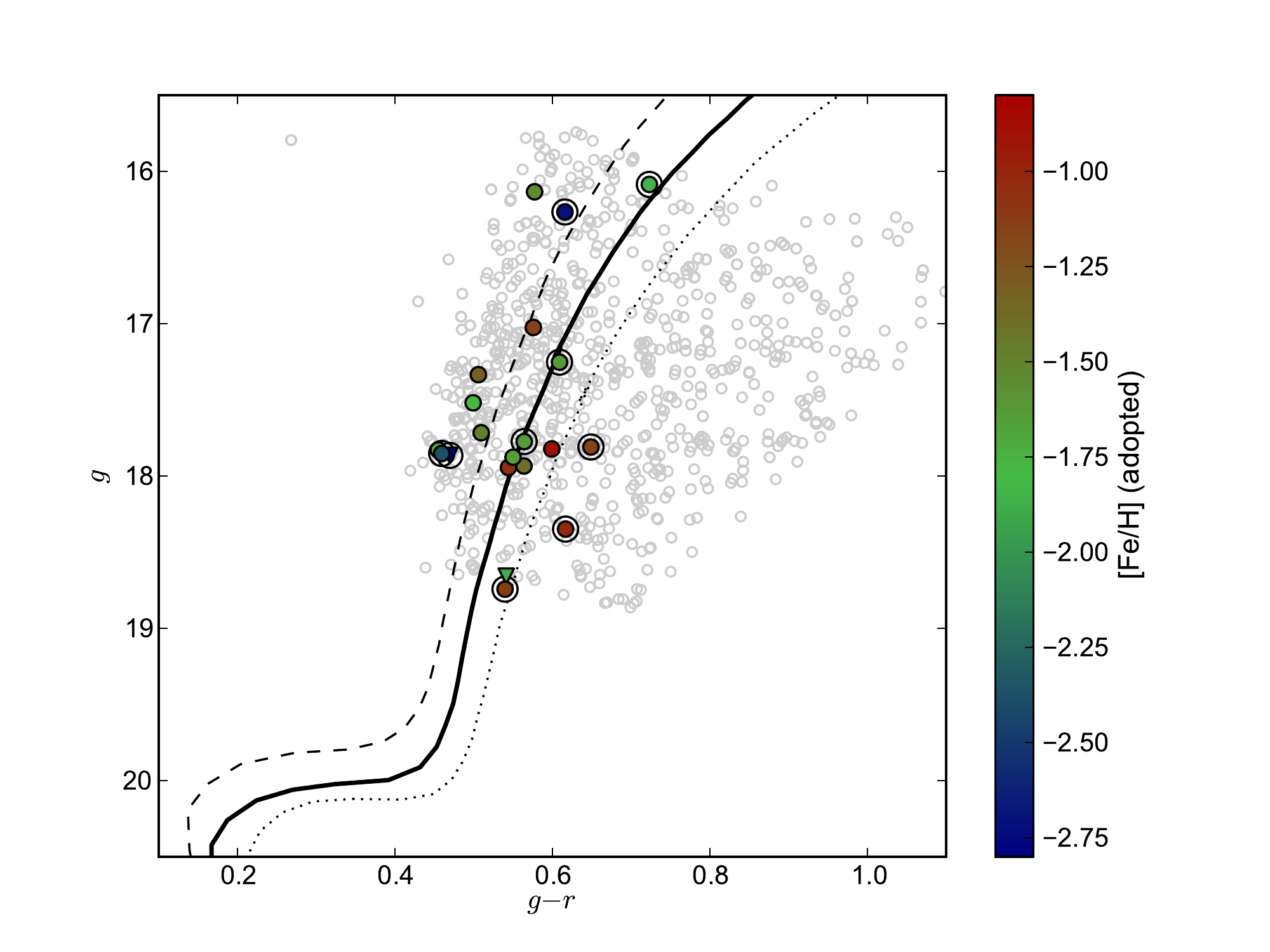}
	\caption[Color magnitude diagram of observed stars, where Orphan stream candidates are highlighted]{Color magnitude diagram showing our observed candidates (grey). Observations fulfilling kinematic and gravity cuts are colored by their metallicity, and those with upper limits for surface gravity are marked as triangles ($\triangledown$). Highly probable stream members (see text) are circled. Relevant 10\,Gyr \citet{girardi;et-al_2004} isochrones at $[\mbox{Fe/H}] = -1.5$ (dotted), $-2.0$ (dashed) at 21.4\,kpc \citep{newberg;et-al_2010} are shown, as well as our best-fitting 10\,Gyr isochrone of $[\mbox{Fe/H}] = -1.63$ at 22.5\,kpc (solid).}
	\label{fig:ch2-cmd}
\end{figure}

A final metallicity value for each star has been adopted based on the quality of our [Fe/H] measurements. These values are tabulated in Table \ref{tab:oss-members}. From our highly likely stream members we find an overall stream metallicity of $[\mbox{Fe/H}] = -1.63$ with a dispersion of $\sigma({\rm [Fe/H]}) = 0.56$\,dex. This abundance spread is larger than typically seen in globular cluster stars and is more representative of the chemical spread seen in dSph satellites.

\begin{table}[h!]
\centering
\caption{Identified Orphan stream candidates from low-resolution spectra\label{tab:oss-members}}
\resizebox{\textwidth}{!}{%
\begin{tabular}{lccccrcrcrrccccc}
\hline
\hline
Star & $\alpha$ & $\delta$ & $g$ & $g - r$ & \multicolumn{2}{c}{$\mu_\alpha$} & \multicolumn{2}{c}{$\mu_\delta$} & \multicolumn{2}{c}{$V_{GSR}$} & $EW_{\lambda8807}$ & [Fe/H]$_{Ca}$ & [Fe/H]$_{iso}$ & [Fe/H]\footnote{Final adopted [Fe/H] value based on quality of two metallicity measurements.} & Stream \\
Name & (J2000) & (J2000) & & & \multicolumn{2}{c}{(mas yr$^{-1}$)} & \multicolumn{2}{c}{(mas yr$^{-1}$)} & \multicolumn{2}{c}{(km s$^{-1}$)} & (m\AA{}) & (dex) & (dex) & (dex) & Prob. \\
\hline
OSS 1  & 10:46:21.9 & $+$00:43:21.8 & 17.52 & 0.50 & 4.1&$\pm$ 4.5 & $-$34.0 &$\pm$ 4.5 & 73.3 &$\pm$ \phn9.3 & 0.273 & --1.78 &$<$--2.28\phn\,& --1.78 & Low \\
OSS 2  & 10:46:29.3 & $-$00:19:38.5 & 17.77 & 0.56 & $-$1.7 &$\pm$ 4.3 & $-$2.2 &$\pm$ 4.3 & 78.4 &$\pm$ \phn5.2 & 0.126 & --1.63 & --1.68  & --1.63 & High \\
OSS 3  & 10:46:50.4 & $-$00:13:15.6 & 17.33 & 0.51 & 1.8 &$\pm$ 4.3 & $-$4.6 &$\pm$ 4.3 & 77.0 &$\pm$ \phn4.0 & 0.416 & --1.31 &$<$--2.28\phn\,& --1.31 & Low \\
OSS 4  & 10:47:06.1 & $-$01:56:03.9 & 18.74 & 0.54 & $-$6.3 &$\pm$ 4.9 & 4.6 &$\pm$ 4.9 & 74.9 &$\pm$    17.6 & 0.452 &:--1.12\footnote{Sufficiently fainter than $g_{HB}$ to qualify this measurement as uncertain.}& --1.40  & --1.40 & High \\
OSS 5  & 10:47:15.0 & $-$03:15:03.9 & 18.66 & 0.54 & $-$8.2 &$\pm$ 5.2 & 1.7 &$\pm$ 5.2 & 109.5 &$\pm$ \phn9.0 &$<$0.19&:--1.85\textsuperscript{b}& --1.43  & --1.43 & Medium \\
OSS 6  & 10:47:17.6 & $+$00:25:07.7 & 16.09 & 0.72 & $-$0.8 &$\pm$ 4.0 & $-$5.2 &$\pm$ 4.2 & 79.2 &$\pm$ \phn3.3 & 0.212 & --1.84 & --1.80  & --1.84 & High \\
OSS 7  & 10:47:29.1 & $-$02:02:22.6 & 17.86 & 0.47 & \multicolumn{2}{c}{\nodata} & \multicolumn{2}{c}{\nodata}  & 93.2 &$\pm$    29.8 &$<$0.40& --2.82 &$<$--2.28\phn\,& --2.82 & High \\
OSS 8  & 10:47:30.1 & $-$00:01:24.5 & 17.25 & 0.61 & $-$4.0 &$\pm$ 4.2 & $-$5.2 &$\pm$ 4.2 & 83.6 &$\pm$ \phn3.5 & 0.123 & --1.62 & --1.68  & --1.62 & High \\
OSS 9  & 10:48:20.9 & $+$00:26:34.4 & 17.88 & 0.55 & $-$8.1 &$\pm$ 4.3 & $-$5.4 &$\pm$ 4.3 & 118.9 &$\pm$    11.7 & 0.467 & --1.65 & --1.73  & --1.65 & High \\
OSS 10 & 10:48:27.8 & $+$00:55:24.0 & 17.72 & 0.51 & $-$14.4 &$\pm$ 4.6 & $-$5.3 &$\pm$ 4.6 & 124.5 &$\pm$ \phn6.7 & 0.182 & --1.48 &$<$--2.28\phn\,& --1.48 & Low \\
OSS 11 & 10:48:31.9 & $+$00:03:35.7 & 17.02 & 0.58 & $-$3.6 &$\pm$ 4.1 & $-$7.7 &$\pm$ 4.1  & 105.1 &$\pm$ \phn5.1 & 0.234 & --1.12 & --2.10  & --1.12 & Low \\
OSS 12 & 10:48:44.4 & $-$02:53:08.8 & 18.35 & 0.62 & $-$3.4 &$\pm$ 4.7 & $-$1.8 &$\pm$ 4.7 & 108.2 &$\pm$ \phn9.0 & 0.183 &:--1.01\textsuperscript{b}& --1.17  & --1.17 & High \\
OSS 13 & 10:48:46.9 & $-$00:32:27.8 & 17.85 & 0.46 & $-$28.4 &$\pm$ 4.8 & $-$12.3 &$\pm$ 4.8 & 109.3 &$\pm$ \phn8.1 & 0.324 & --2.37 &$<$--2.28\phn\,& --2.37 & Medium \\
OSS 14 & 10:49:08.3 & $+$00:02:00.2 & 16.27 & 0.62 & 4.9 &$\pm$ 4.0 & $-$6.0 &$\pm$ 4.0& 81.5 &$\pm$ \phn4.6 & 0.034 & --2.70 &$<$--2.28\phn\,& --2.70 & High \\
OSS 15 & 10:49:13.4 & $+$00:04:03.8 & 17.83 & 0.46 & 3.4 &$\pm$ 4.7 & $-$6.8 &$\pm$ 4.7 & 65.3 &$\pm$ \phn5.4 & 0.252 & --1.74 & $<$--2.28\phn\,& --1.74 & Medium \\
OSS 16 & 10:50:13.1 & $+$00:33:52.7 & 16.13 & 0.58 & $-$3.4 &$\pm$ 4.0 & $-$7.7 &$\pm$ 4.0 & 94.7 &$\pm$ \phn5.1 & 0.391 & --1.54 &$<$--2.28\phn\,& --1.54 & Low \\
OSS 17 & 10:50:24.2 & $-$01:49:05.4 & 17.94 & 0.54 & 3.4 &$\pm$ 4.6 & $-$2.2 &$\pm$ 4.6 & 109.9 &$\pm$    25.4 & 0.151 & --1.06 & --1.73  & --1.06 & Low \\
OSS 18 & 10:50:33.8 & $+$00:12:19.1 & 17.82 & 0.60 & $-$7.4 &$\pm$ 5.0 & $-$3.3 &$\pm$ 5.0 & 97.5 &$\pm$ \phn5.9 & 0.596 & --0.90 & --1.43  & --0.90 & Medium \\
OSS 19 & 10:51:19.7 & $+$00:05:15.5 & 17.81 & 0.65 & 4.2 &$\pm$ 4.7 & $-$11.8 &$\pm$ 4.7 & 82.7 &$\pm$ \phn5.0 & 0.198 & --1.16 & --1.20  & --1.16 & High \\
OSS 20 & 10:51:35.4 & $+$00:00:46.4 & 17.93 & 0.56 & --0.5 &$\pm$ 4.5 & $-$3.0 &$\pm$ 4.5 & 66.7 &$\pm$ \phn8.7 & 0.128 & :--1.38 & --2.10 & --2.10 & Medium \\
\hline
\end{tabular}}
\end{table}

\break
\subsection{Proper Motions \& Distances}

We have found proper motions for 19 of our candidates in the PPMXL proper motion catalogue \citep{roeser;et-al_2010}. One highly probably candidate (OSS 13 in Table \ref{tab:oss-members}) has a listed proper motion that is different from that of the other nine highly likely members at the $6\sigma$ level. Consequently, we have reduced the membership likelihood of this star from ``High" to ``Medium". Given the uncertainties in proper motions, we cannot reliably alter the membership probability for any other candidate.

Since our Orphan stream giants cover a wide evolutionary range along the giant branch (Figure \ref{fig:ch2-cmd}), we are in a good position to revise the distance estimate to the stream. Given a 10\,Gyr \citet{girardi;et-al_2004} isochrone at $[\mbox{Fe/H}] = -1.63$, we find a best-fitting distance to the stream of $22.5 \pm 2.0$\,kpc at $(l, b) = (250^\circ,\,50^\circ)$. This isochrone is shown in Figure \ref{fig:ch2-cmd}. Our derived distance is in reasonably good agreement with the measurement of $21.4 \pm 1.0$\,kpc independently deduced by \citet{grillmair_2006} and \citet{newberg;et-al_2010}.

\subsection{Comparison with \citet{newberg;et-al_2010}}
\label{sec:ch2-newberg}
\begin{figure}[h!]
	\includegraphics[width=\columnwidth]{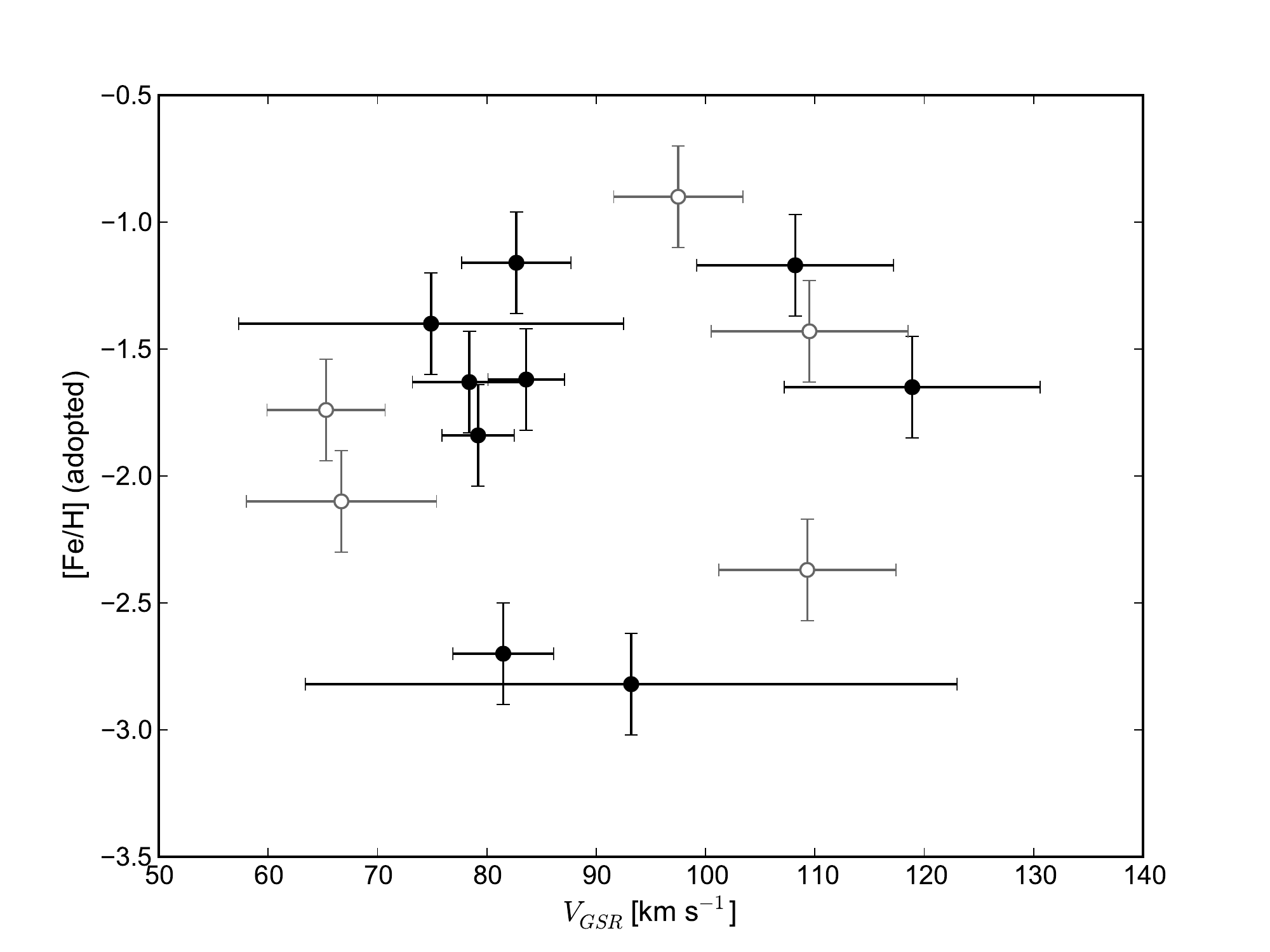}
	\caption[Galactocentric velocities and adopted metallicities for Orphan stream candidates]{Galactocentric velocities and adopted metallicities for the highest likely Orphan stream members (black $\bullet$), and those with probabilities assigned as ``Medium'' (grey $\circ$; see text).}
	\label{fig:ch2-vgsr-feh}
\end{figure}

\citet{newberg;et-al_2010} traced the Orphan stream using BHB stars selected from the SEGUE survey, allowing them to derive an orbit for the stream and make a strong prediction for the location of the undiscovered progenitor. Their closest stream detection to this study is at $\Lambda_{Orphan} = 18.4^\circ$, approximately $\Delta\Lambda_{Orphan} \sim 4^\circ$ away from our fields. At this location, \citet{newberg;et-al_2010} found the velocity of the stream to be $V_{GSR} = 101.4 \pm 8.9$\,km s$^{-1}$ based on 12 BHB stars. We note that this is $\sim95$\,km s$^{-1}$ on our scale, given the differences in accounting for the local standard of rest. The velocities and metallicities of our `High' and `Medium' probability candidates are illustrated in Figure \ref{fig:ch2-vgsr-feh}. Although we recover some candidates with velocities up to $V_{GSR} \sim 110$\,km s$^{-1}$, our kinematic distribution peaks near $V_{GSR} \sim 85$\,km s$^{-1}$, roughly 10\,km s$^{-1}$ lower than that of \citet{newberg;et-al_2010}.

There is a known velocity gradient along the Orphan stream which can account for this discrepancy. As $\Lambda_{Orphan}$ increases towards the edge of the SDSS boundary, galactocentric velocity quickly decreases. For the Orphan stream detection in the outrigger SEGUE Strip 1540 at $\Lambda_{Orphan} = 36^\circ$, \citet{newberg;et-al_2010} find $V_{GSR} = 38$\,km s$^{-1}$. This work presents likely Orphan stream K-giant candidates at $\Lambda_{Orphan} \sim 23^\circ$. Given the velocity gradient reported by \citet{newberg;et-al_2010}, a galactocentric velocity of $80-85$\,km s$^{-1}$ (on our scale) is perfectly reasonable. We note that since the velocities of BHB stars can have significant uncertainties, it was practical for us to assume a wide initial selection in kinematics to identify potential members. 

\begin{figure}[h!]
	\includegraphics[width=\columnwidth]{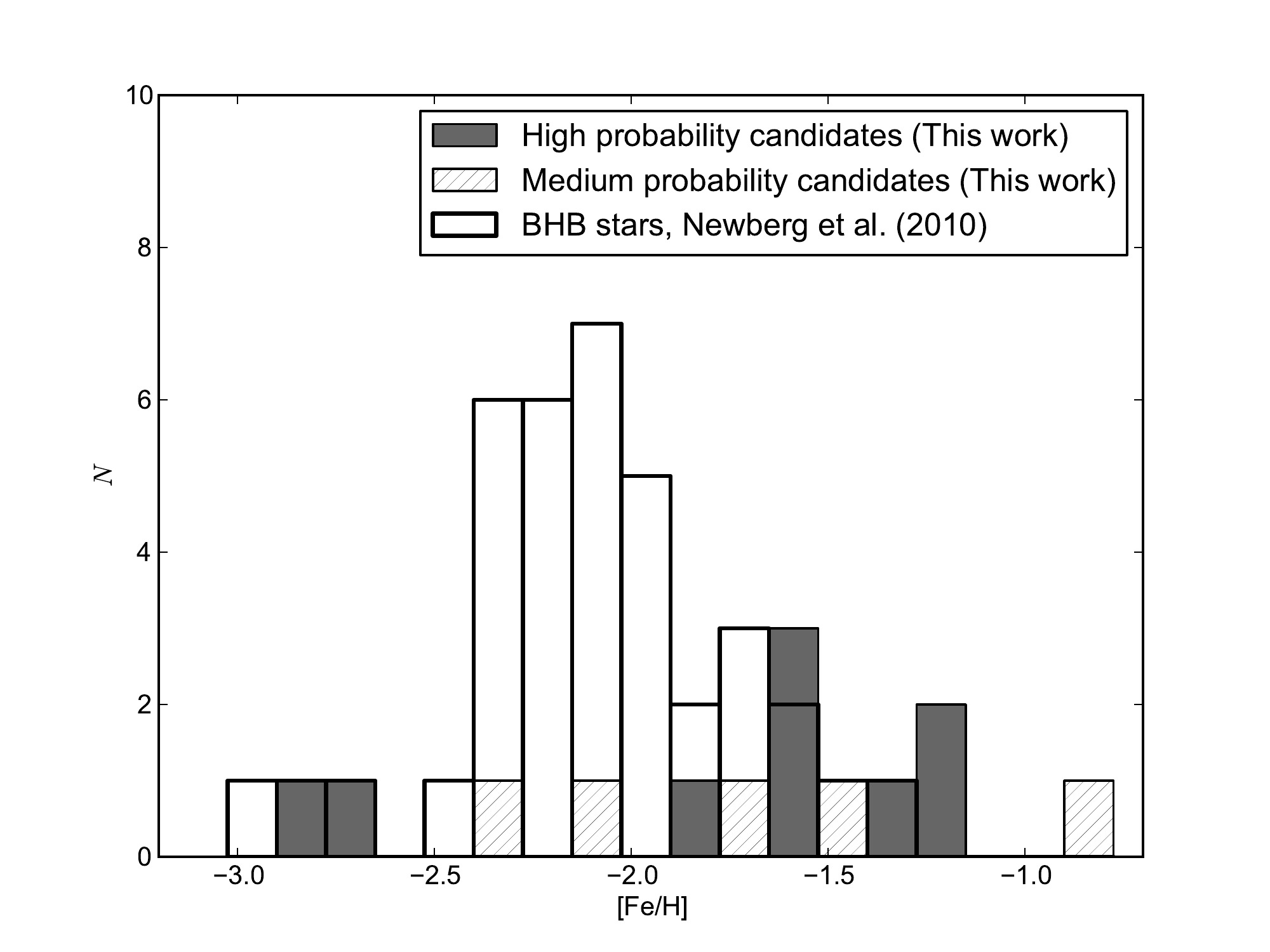}
	\caption[Observed metallicity distribution for Orphan stream candidates]{Observed metallicity distribution function for the Orphan stream candidates identified here, with comparisons to the distribution found by \citet{newberg;et-al_2010} from BHB stars.}
	\label{fig:ch2-newberg-feh}
\end{figure}

The adopted metallicities of our Orphan stream candidates are generally higher than those found by \citet{newberg;et-al_2010}. Our highly likely stream members have a mean metallicity of [Fe/H] = $-1.63$, with a dispersion of $\sigma({\rm [Fe/H]}) = 0.56$\,dex. As illustrated in Figures \ref{fig:ch2-vgsr-feh} and \ref{fig:ch2-newberg-feh}, there are two very metal-poor candidates which largely drive this dispersion, but we have no reason to suspect they are non-members. The \citet{newberg;et-al_2010} sample contains 37 BHB stars identified over a 60$^\circ$ arc on the sky, and has a peak metallicity at [Fe/H] = $-2.10 \pm 0.10$. The closest detection bin in the \citet{newberg;et-al_2010} sample was the most populous, yielding 7 BHB stars. For comparison, we identify 9 giant stars across $\sim{}4^\circ$. Given BHB stars are known to trace a somewhat more metal-poor population, and we are calculating statistics with marginal sample sizes, we conclude that the accuracy of these two metallicity distributions are not mutually exclusive. It is entirely possible that we are sampling the same distribution, but a larger sample size is required.

\section{Conclusions}
\label{sec:ch2-conclusions}

We have presented a detailed analysis to isolate individual Orphan stream K-giants from low-resolution spectroscopy using a combination of photometric, kinematic, gravity, metallicity, and proper motion information. Although each individual criterion is likely to induce some level of contamination, their intersection reveals nine highly probable, self-consistent, Orphan stream K-giants.  We deduce a median stream metallicity of $[\mbox{Fe/H}] = -1.63 \pm 0.19$ and find an intrinsically wide metallicity spread of $\sigma({\rm [Fe/H]}) = 0.56$\,dex, indicative of a dSph origin. Unlike other stellar tracers, K-type giants can exist at all metallicities, hence our derived metallicity spread is likely representative of the true stream MDF. Recall that the metallicity determination was performed after kinematic and gravity cuts, and three of our most probable members lay perfectly on a 10\,Gyr isochrone of $[\mbox{Fe/H}] = -1.63$. However, it is clear that more data are required to fully characterize the stream metallicity distribution function. Our data indicate a distance to the stream of $22.5 \pm 2.0$\,kpc at $(l, b) = (250^\circ,\,50^\circ)$, in agreement with that deduced by \citet{grillmair_2006} and \citet{newberg;et-al_2010}.

Given the stream orbit derived by \citet{newberg;et-al_2010}, they excluded all possible known halo objects except for the dissolved star cluster, Segue 1. \citet{simon;et-al_2011} obtained spectroscopy for six members in Segue 1 and found an extremely wide metallicity dispersion: from [Fe/H] $<-3.4$ to $-1.63$ dex. On the basis of the extremely low metallicity in the cluster and the wide chemical dispersion, they conclude that Segue 1 is a dwarf spheroidal galaxy. Although the data presented here indicates the Orphan stream progenitor is a disrupted dwarf spheroidal galaxy, we cannot reliably associate Segue 1 as the parent without additional observational data. 

If the Orphan stream continues through SEGUE Stripe 1540 at $(l, b) = (271^\circ,\,38^\circ)$ as \citet{newberg;et-al_2010} found, then the stream is even closer there than in the region analysed here. Thus, if our observations and analyses are repeated at $(271^\circ,\,38^\circ)$, we predict K-giant stream members of brighter apparent magnitude will be recovered. 

Using a maximum-likelihood estimation we find the stream velocity at $(l, b) = (250^\circ, 50^\circ)$ from nine stars to be $V_{GSR} = 85.3 \pm 4.4$\,km s$^{-1}$ and the dispersion to be $6.5 \pm 7.0$\,km s$^{-1}$. If we exclude three stars with low $S/N$ ratios \--- and hence large ($> 10$\,km s$^{-1}$) velocity uncertainties \--- the peak occurs at $82.1 \pm 1.4$\,km s$^{-1}$ and the intrinsic dispersion is found to be {$0.2\,\pm\,3.1$\,km s$^{-1}$}. Hence, the observed stream dispersion is dominated by the velocity uncertainties, indicating that the intrinsic dispersion is small. 

The K-giants presented here can provide great insight into the chemistry and history of the Orphan stream. High-resolution spectra have been obtained for some targets, and a detailed chemical analysis is presented in Chapter \ref{chap:orphan_II}. Individual chemical abundances can help determine both the nature of the progenitor before it is discovered, and allows us to compare peculiar chemical signatures with those of the known Milky Way satellites in order to associate likely parents. However, at least for the moment, the Orphan stream remains appropriately named.

\chapter{Hunting the Parent of the Orphan Stream II}
\label{chap:orphan_II}


\previouslypublished{Parts of this chapter have been previously submitted to the Astrophysical Journal as \href{http://arxiv.org/abs/1309.3563}{`Hunting the Parent of the Orphan Stream II: The First High-Resolution Spectroscopic Study'}, {Casey}, A.~R., {Keller}, S.~C., {Da Costa}, G., {Frebel}, A., {Maunder}, E., 2013. The work is presented here in expanded and updated form.} \\

We present the first high-resolution spectroscopic study on the Orphan Stream for five stream candidates, observed with the Magellan Inamori Kyocera Echelle (MIKE) spectrograph on the Magellan Clay telescope. The targets were selected from the low-resolution catalog of \citet{casey;et-al_2013a}: 3 high-probability members, 1 medium and 1 low-probability stream candidate were observed. Our analysis indicates the low and medium-probability target are metal-rich field stars. The remaining three high-probability targets range $\sim$1\,dex in metallicity, and are chemically distinct compared to the other 2 targets and all standard stars: low [$\alpha$/Fe] abundances are observed, and lower limits are ascertained for [Ba/Y], which sit well above the Milky Way trend. These chemical signatures demonstrate that the undiscovered parent system is unequivocally a dwarf spheroidal galaxy, consistent with dynamical constraints inferred from the stream width and arc. As such, we firmly exclude the proposed association between the globular cluster NGC 2419 and the Orphan stream. A wide range in metallicities adds to the similarities between the Orphan stream and Segue 1, however many open questions must be answered before Segue 1 could possibly be claimed as the `parent' of the Orphan stream. The parent system could well remain undiscovered in the southern sky.

\section{Introduction}

Amongst the known substructures in the halo, the Orphan stream is particularly interesting. \citet{grillmair_2006} and \citet{belokurov;et-al_2006} independently discovered the stream, spanning over $60^\circ$ in the sky from Ursa Major in the north to near Sextans in the south, where the SDSS coverage ends. The stream has properties that are unique from other halo substructures. It has an extremely low surface brightness, ranging from 32-40\,mag arcsec$^{-2}$, and a full width half-maximum of $\sim$2$^\circ$ on sky. At a distance of $\sim$20\,kpc \citep{belokurov;et-al_2007,grillmair_2006}, this corresponds to $\sim$700\,pc. The cross-section and luminosity of the stream are directly related to the mass and velocity dispersion of the parent satellite \citep{johnston_1998}. The Orphan stream width is significantly broader than every known globular cluster tidal tail \citep{odenkirchen;et-al_2003,grillmair_johnson_2006,grillmair_dionatos_2006a,grillmair_dionatos_2006b} and larger than the tidal diameter of all known globular clusters \citep[][2010 edition]{harris_1996}. As \citet{grillmair_2006} notes, if the cross-section of the stream is circular, then in a simple logarithmic potential with $v_c = 220$\,km s$^{-1}$ -- a reasonable first-order approximation for the Milky Way --  the expected random velocities of stars would be required to be ${\sqrt{\langle\sigma_{v}^{2}\rangle}} > 20$\,km\,s$^{-1}$ in order to produce the stream width. Such a velocity dispersion is significantly larger than the expected random motions of stars that have been weakly stripped from a globular cluster, implying that the Orphan stream's progenitor mass must be much larger than a classical globular cluster. Photometry indicates the stream is metal-poor, implying that negligible star formation has occurred since infall began several billion years ago. For a stream of this length to remain structurally coherent over such a long timescale, the progenitor is also likely to be dark matter-dominated. In their discovery papers, \citet{grillmair_2006} and \citet{belokurov;et-al_2007} concluded that the likely parent of the Orphan stream is a low-luminosity dSph galaxy.

 
\citet{newberg;et-al_2010} were able to map the distance and velocity of the stream across the length of the SDSS catalogue using BHB and F-turnoff stars. They found the stream distance to vary between 19-47\,kpc, extending the 20-32\,kpc distance measurements made by \citet{belokurov;et-al_2007}, and has been extended to 55\,kpc by \citet{sesar;et-al_2013}. \citet{newberg;et-al_2010} note an increase in the density of the Orphan stream near the celestial equator $(l, b) = (253^\circ, 49^\circ)$, proposing the progenitor may be close to this position. The authors attempted to extend their trace of the stream using southern survey data \citep[e.g., SuperCOSMOS,][]{newberg;et-al_2010}, but to no avail; the stream's surface brightness is lower than the survey faint limit. It is still unclear whether the stream extends deep into the southern sky. \citet{newberg;et-al_2010} observe an increase in surface brightness near the celestial equator, but given the stream is closest near the celestial equator \citep{belokurov;et-al_2007}, an increase in stream density may be expected given a constant absolute magnitude. Nevertheless, with SDSS photometry and radial velocities from the SEGUE catalog, \citet{newberg;et-al_2010} were able to derive a prograde orbit with an eccentricity, apogalacticon and perigalacticon of $e = 0.7$, 90\,kpc and 16.4\,kpc respectively. The ability to accurately trace the Orphan stream to such extreme distances would make a powerful probe for measuring the galactic potential \citep[e.g., see][]{adrn_2013}. From their simulations, \citet{newberg;et-al_2010} find a halo and disk mass of $M(R < 60\,{\rm kpc}) = 2.6 \times 10^{11} M_\odot$, and similarly \citet{sesar;et-al_2013} find $2.7\times 10^{11} M_\odot$, which is $\sim$60\% lower than that found by \citet{xue;et-al_2008} and \citet{koposov;et-al_2008}, and slightly lower than the virial mass of $7 \times 10^{11} M_\odot$ derived by \citet{sales;et-al_2008}.
 
Metallicities derived from SEGUE medium-resolution spectra confirm photometric estimates indicating that the stream is metal-poor. A mean metallicity of [Fe/H]$ = -2.1 \pm 0.1$\,dex is found from BHB stars, with a range extending from $-$1.3 to $\sim-$3\,dex \citep{newberg;et-al_2010}. \citet{sesar;et-al_2013} also find a wide range in metallicities by tracing RR Lyrae stars along the stream: [Fe/H] = $-1.5$ to $-2.7$\,dex. However if F-turnoff stars from the \citet{newberg;et-al_2010} sample are included, the metallicity distribution function extends more metal-rich from $\sim-$3 to $-$0.5\,dex. 

The situation is complicated by interlopers and small number statistics, so the full shape of the MDF is unknown. To this end, \citet{casey;et-al_2013a} observed low-resolution spectroscopy for hundreds of stars towards the stream at the celestial equator. The authors targeted the less numerous K-giants \citep[where a mere 1.3 red giant branch stars are expected per square degree]{sales;et-al_2008,morrison_1993} and found a very weak detection of the stream from kinematics alone. Using wide selections in velocity, distance, proper motions, metallicities, and surface gravity, they identified nine highly likely Orphan stream giants. The velocity dispersion of their candidates is within the observational errors ($\sigma_{v} < 4$\,km s$^{-1}$), suggesting the stream is kinematically cold along the line-of-sight. Like \citet{newberg;et-al_2010} (and independently confirmed by \citet{sesar;et-al_2013}), they also found an extended range in metallicities: two stars below [Fe/H] $\leq-2.70$ and two stars near [Fe/H] $\sim -1.17$\,dex, all of which are consistent with stream membership. The mean metallicity of their sample was [Fe/H]$ = -1.63 \pm 0.19$\,dex, with a wide dispersion of $\sigma({\rm [Fe/H]}) = 0.56$\,dex. It appears the Orphan stream may have an extremely wide range in metallicity, consistent with the internal chemical enrichment typically observed in dSph galaxies \citep{mateo_1998,tolstoy;et-al_2009,kirby;et-al_2011,frebel;et-al_2010}.


As the name suggests, the Orphan stream's parent satellite has yet to be found. In an effort to identify a progenitor, a number of systems have been identified as being plausibly associated with the Orphan stream. These include the linear Complex A H\,I clouds, as well as the globular clusters Palomar 1, Arp 2, Terzan 7 and Ruprecht 106. These systems were all identified along the great circle path of the stream. The low-luminosity dwarf satellites Segue 1 and Ursa Major II also lie along the great circle, although Segue 1 was considered an extended globular cluster until recently \citep{norris;et-al_2010,simon;et-al_2011}. 

\citet{belokurov;et-al_2007} first noted a possible association between the Orphan stream, Ursa Major II and Complex A. \citet{jin;lynden-bell_2007} and \citet{fellhaur;et-al_2007} explored the possible association between the Orphan stream with Complex A and Ursa Major II, respectively. In the best-fitting Complex A model, the predicted heliocentric velocities did not match those found by \citet{belokurov;et-al_2007}, or later observations by \citet{newberg;et-al_2010}. The expected distances were also under-estimated by a factor of $\sim$3 when compared with observations, making the association with Complex A tenuous at best. In the Ursa Major II scenario, the stream's on-sky position was required to exactly overlap with a previous wrap, a somewhat contrived and unlikely scenario. \citet{newberg;et-al_2010} found that the Ursa Major II--Orphan stream connection was also not compelling, as the observed stream kinematics were not consistent with the Ursa Major II model.

Simulations involving the Complex A and Ursa Major II associations introduced an a priori assumption that the object (e.g., Ursa Major II or Complex A) was related to the Orphan stream, and consequently found an orbit to match. In contrast, \citet{sales;et-al_2008} approached the problem by fitting an orbit to a single wrap of the data, without assuming a parent satellite a priori. Their $N$-body simulations were inconsistent with either a Complex A or Ursa Major II association. Instead, the authors favor a progenitor with a luminosity $L \sim 2.3 \times 10^4 L_\odot$ or an absolute magnitude $M_r \sim -6.4$, consistent with the observation by \citet{belokurov;et-al_2007} of $M_r \sim -6.7$. Simulations by \citet{sales;et-al_2008} suggest the progenitor may be similar to the present-day `classical' Milky Way dwarfs like Carina, Draco, Leo II or Sculptor, but would be very close to being fully disrupted, which they suggest has occurred over the last 5.3 gigayears. Time of infall is a critical inference. Longer timescales produce streams that are too wide and diffuse, whereas shorter timescales do not reproduce the $\sim{}$60$^\circ$ stream length. The degeneracies between these simulation parameters are important to note, but in any case, there are robust conclusions that can be drawn irrespective of those degeneracies. For example, \citet{sales;et-al_2008} note that a globular cluster has a central density too high ($\sim10^{12} M_\odot$\,kpc$^{-3}$) to be fully disrupted along their Orphan stream orbit within a Hubble time. Given this constraint, and the lower limit on luminosity ($L > 2\times10^5 L_\odot$), a globular cluster progenitor seems unlikely from their models.

In addition to the work by \citet{sales;et-al_2008}, $N$-body simulations by \citet{newberg;et-al_2010} exclude all known halo globular clusters as possible progenitors. They conclude with the postulation of two possible scenarios: the progenitor is either an undiscovered satellite located between $(l, b) = (250^\circ, 50^\circ)$ and $(270^\circ, 40^\circ)$, or Segue 1 is the parent system. Segue 1, an ultrafaint dwarf galaxy \citep{simon;et-al_2011,norris;et-al_2010}, resides at a similar distance ($\sim$23\,kpc) along the great circle path of the Orphan stream. Segue 1 also shares velocities that are consistent with the Orphan stream at its nearest point. Moreover, both Segue 1 and the Orphan stream exhibit low velocity dispersions: on the order of $\sim$4\,km s$^{-1}$. The similarities in position, distance, and velocities between the two systems are indeed striking. However, it is not the only system alleged to be associated with the Orphan stream.


Given the extended apogalaction of 90\,kpc in the stream orbit found by \citet{newberg;et-al_2010}, \citet{bruns;kroupa_2011} reasoned the stream may be the tidal tail of the massive globular cluster NGC 2419. This system is the most distant and luminous outer halo globular cluster known ($\sim$85\,kpc, $M_V = -9.6$). It is unlike any other globular cluster in the Milky Way. Spectroscopic studies have confirmed photometric observations by \citet{photometry_for_ngc2419} that the system is metal-poor ([Fe/H] $= -2.15$), and identified a remarkable anti-correlation between Mg and K abundances \citep{cohen;et-al_2011,mucciarelli;et-al_2012}. The level of magnesium depletion ([Mg/Fe] $> -1.40$) is not seen anywhere else in the Galaxy or its satellite systems, neither is the enhanced potassium enrichment (up to [K/Fe] = 2) at the opposite end of this peculiar Mg-K anti-correlation. If NGC 2419 is the parent of the Orphan stream, then this unprecedented chemical signature ought to exist in disrupted stream members as an example of chemical tagging \citep[e.g., see][]{freeman;bland-hawthorn_2002,de_silva;et-al_2007,wylie-de-boer;et-al_2010,majewski;et-al_2012}.

\begin{table}[h!]
\centering
\caption{Observation details for standard stars and Orphan stream candidates\label{tab:observations}}
\resizebox{\textwidth}{!}{%
\begin{tabular}{lccccccccccc}
\hline
\hline
Object\tablenotemark{a}	& $\alpha$ & $\delta$ & $V$\tablenotemark{b} & UT Date & UT Time & Airmass & Exp. Time & S/N\tablenotemark{c} & $V_{\mbox{hel}}$ & $V_{\mbox{err}}$ \\
 & (J2000) & (J2000) & (mags) & & & & (secs) & (px$^{-1}$) & (km s$^{-1}$) & (km s$^{-1}$) \\
\hline
HD 41667		& 06:05:03.7	& $-$32:59:36.8	& 8.52	& 2011-03-13	& 23:40:52	& 1.005	& 90			& 272	& 297.8 	& 1.0 \\
HD 44007		& 06:18:48.6	& $-$14:50:44.2	& 8.06	& 2011-03-13	& 23:52:18	& 1.033	& 30 		& 239	& 163.4 	& 1.3 \\
HD 76932		& 08:58:44.2	& $-$16:07:54.2	& 5.86	& 2011-03-14	& 00:16:47	& 1.158	& 30 		& 289	& 119.2 	& 1.2 \\
HD 122563		& 14:02:31.8	& $+$09:41:09.9	& 6.02 	& 2011-03-13	& 07:15:04	& 1.283	& 30			& 230	& --23.4	& 1.0 \\
HD 136316		& 15:22:17.2	& $-$53:14:13.9	& 7.65	& 2011-03-14	& 09:37:26	& 1.118	& 90	 		& 335	& --38.2	& 1.1 \\
HD 141531		& 15:49:16.9	& $+$09:36:42.5	& 9.08	& 2011-03-14	& 09:52:00	& 1.309	& 90	 		& 280	& 2.6 		& 1.0 \\
HD 142948		& 16:00:01.6	& $-$53:51:04.1	& 8.03	& 2011-03-14	& 09:45:12	& 1.107	& 90	 		& 271	& 30.3 	& 0.9 \\
OSS 3 (L)		& 10:46:50.6	& $-$00:13:17.9	& 17.33 	& 2011-03-14	& 01:51:07	& 1.363	&4$\times$2500	& 48	& 217.9	& 1.0 \\ 
OSS 6 (H)		& 10:47:17.8	& $+$00:25:06.9	& 16.09 	& 2011-03-14	& 00:25:37	& 1.995	&3$\times$1600	& 59	& 221.2	& 1.0 \\ 
OSS 8 (H)		& 10:47:30.3	& $-$00:01:22.6	& 17.25	& 2011-03-14	& 04:44:04	& 1.160	&5$\times$1900	& 49	& 225.9	& 1.0 \\ 
OSS 14 (H)	& 10:49:08.3	& $+$00:01:59.3	& 16.27 	& 2011-03-15	& 00:32:17	& 1.881	&4$\times$1400	& 48 	& 225.1	& 1.0 \\ 
OSS 18 (M)	& 10:50:33.7	& $+$00:12:18.3	& 17.82 	& 2011-03-15	& 02:12:46	& 1.295	&4$\times$2100	& 31 	& 247.8	& 1.2 \\
\hline
\multicolumn{11}{c}{\tablenotetext{a}{Probability of membership (Low, Medium, High) listed for Orphan stream candidates as defined by \citet{casey;et-al_2013a}.}
\tablenotetext{b}{$V$--band magnitudes for Orphan stream targets are estimated as $g$--band.}
\tablenotetext{c}{S/N measured at 600\,nm for each target.}}
\end{tabular}}
\end{table}

In this study, we present an analysis of high-resolution spectroscopic observations for five Orphan stream candidates. The observations and data reduction are outlined in Section \ref{sec:orphan_II-observations}. In Section \ref{sec:orphan_II-analysis} we describe the details of our analysis to infer stellar parameters and chemical abundances. We discuss our results in Section \ref{sec:orphan_II-discussion}, including the implications for association between the Orphan stream and the two currently alleged stream progenitors: Segue 1 and NGC 2419. Finally, we conclude in Section \ref{sec:orphan_II-conclusions} with a summary of our findings.

\section{Observations \& Data Reduction}
\label{sec:orphan_II-observations}


High-resolution spectra for five Orphan stream candidates and seven well-studied standard stars have been obtained with the MIKE spectrograph \citep{bernstein;et-al_2003} on the Magellan Clay telescope. These objects were observed in March 2011 using a 1\arcsec{} wide slit in mean seeing of 0.9\arcsec. This slit configuration provides a continuous spectral coverage from 333\,nm to 915\,nm, with a spectral resolution of $\mathcal{R} = 25,000$ in the blue arm and $\mathcal{R} = 28,000$ in the red arm. A minimum of 10 exposures of each calibration type (biases, flat fields, and diffuse flats) were observed in the afternoon of each day, with additional flat-field and Th-Ar arc lamp exposures performed throughout the night to ensure an accurate wavelength calibration. The details of our observations are tabulated in Table \ref{tab:observations}. The $S/N$ ratio for the standard stars exceeds 200 per pixel, and varies between 30-60 for the Orphan stream candidates.

The candidates were chosen from the low-resolution spectroscopic study of \citet{casey;et-al_2013a}. From their classification, three of the selected stream candidates have high probability of membership to the Orphan stream. One target was classified with medium probability, and another with a low probability of membership. 

Initially we planned to observe many more high-priority targets. However, after our last exposure of target star OSS 18, inclement weather forced us to relinquish the remainder of our observing time. The data were reduced using the CarPy pipeline written by D. Kelson\footnote{http://obs.carnegiescience.edu/Code/mike}. Every reduced echelle aperture was carefully normalised using cubic splines with defined knot spacing. Extracted apertures were stacked together and weighted by their inverse variance to provide a continuous normalised spectrum for each object.

\section{Analysis}
\label{sec:orphan_II-analysis}
Each normalised, stitched spectrum was cross-correlated against a synthetic template to measure the radial velocity of each star. This was performed using a Python\footnote{http://www.python.org} implementation of the \citet{tonry;davis_1978} method. The wavelength region employed was from $845 \leq \lambda \leq 870$\,nm, and a synthetic spectrum of a metal-poor giant was used as the rest template. Heliocentric corrections have been applied to our radial velocity measurements, and the resultant heliocentric velocities are shown in Table \ref{tab:observations}.


Atomic data for absorption lines has been taken from \citet{yong;et-al_2005}, and these data are listed with their measured EWs in Appendix \ref{app:orphan_ews}. EWs for all atomic transitions were measured using the automatic profile fitting algorithm described in \citet{casey;et-al_2013b}. While this technique is accurate, and robust against blended lines as well as strong changes in the local continuum, every fitted profile was visually inspected for quality. Spurious or false-positive measurements were removed and of order $\sim$5 EW measurements (of 528 atomic transitions in our line list\footnote{Although our final line list includes atomic data for 528 transitions, we have omitted data for 17 transitions from Appendix \ref{app:orphan_ews} as they were not significantly detected in any stars. See Section \ref{sec:orphan_II-chemical-abundances} for details.}) were manually re-measured for each object. We excluded strong transitions with $REW > -4.5$ in order to avoid using lines near the flat region of the curve-of-growth.

Given the range in metallicities for the Orphan stream candidates (see Section \ref{sec:orphan_II-analysis-metallicity}), these restrictions resulted in only 14 Fe\,I and 4 Fe\,II acceptable transitions for our most metal-poor candidate, OSS 4. With so few lines available, minute changes in stellar parameters caused large trends and variations in the Fe line abundances. This resulted in a poor solution while performing the excitation and ionization equilibria. At this point, we opted to supplement our line list with transitions from \citet{roederer;et-al_2010}. Each additional transition was inspected in our most metal-rich candidate (OSS 18) to ensure that it was not blended with other features. Blended transitions were not added. As a result, the minimum number of acceptable Fe\,I and Fe\,II transitions for any star increased to 48 and 12, respectively.

\subsection{Stellar Parameters}
We  employed the 1D plane-parallel model atmospheres of \citet{castelli;kurucz_2004} to infer stellar parameters. These $\alpha$-enhanced models assume that absorption lines form under local thermal equilibrium (LTE), ignore convective overshoot and any center-to-limb variations. We have interpolated within a grid of these model atmospheres following the prescription in Chapter \ref{chap:smh}.

\subsubsection{Effective Temperature, $T_{\rm eff}$}
Effective temperatures for all stars have been found by excitation balance of neutral iron lines.  Since all EW measurements were visually inspected, we generally identified no outlier measurements during this stage. The most number of Fe\,I outliers ($>3\sigma$) removed while determining the effective temperature for any star was three. For each iteration in temperature, a linear fit was made to the data ($\chi$, $\log\epsilon$(Fe\,I)). This slope was minimized with successive iterations of effective temperature. Final slopes less than $|10^{-3}|$\,dex eV$^{-1}$ were considered to be converged. 

The effective temperatures of our standard stars are in excellent agreement with the literature. On average, our effective temperatures are 13\,K cooler than the references listed in Table \ref{tab:stellar-parameters}. The largest discrepancy exists for HD 122563, a cool metal-poor giant. For extremely metal-poor stars, effective temperatures found through excitation balance are known to produce systematically cooler temperatures than those deduced by other methods \citep[e.g.,][]{frebel;et-al_2013}. We have chosen to remain consistent with the excitation balance approach, and accept the systematically cooler temperature of 4358\,K. The other noteworthy temperature deviant is HD 142948, where we find a temperature 337\,K hotter than that found by \citet{gratton;et-al_2000}. The reason for this discrepancy is not obvious.

\subsubsection{Microturbulence, $\xi$}
Microturbulence is necessary in 1D model atmospheres to represent large scale, 3D turbulent motions. The correct microturbulence will ensure that lines of the same species that are formed in deep and shallow photospheric depths will yield the same abundance. We have solved for the microturbulence by demanding a zero-trend in REW and abundance for all neutral iron lines. The resultant gradient between REW and abundance is typically $<|0.001|$\,dex, and the largest slope in any star is --0.004\,dex. 


\subsubsection{Surface Gravity, $\log{g}$}
The surface gravity for all stars has been inferred through the ionization balance of neutral and single ionized Fe lines. We iterated on surface gravity until the mean Fe\,I abundance matched the mean Fe\,II abundance to within 0.01\,dex. This process was performed in concert while solving for all other stellar parameters. As a consequence of a systematically cooler temperature in HD 122563, we have obtained a somewhat lower surface gravity for this star, such that it sits above a metal-poor isochrone on a Hertzsprung-Russell diagram. Modulo HD 122563, the surface gravities for our standard stars are in good agreement with the literature: a mean offset of --0.22\,dex is observed (our $\log{g}$ values are lower). This difference remains within the mutual 1$\sigma$ uncertainties.

\begin{table}[h!]
\centering
\caption{Stellar parameters for standard stars and Orphan stream candidates\label{tab:stellar-parameters}}
\resizebox{\textwidth}{!}{%
\begin{tabular}{lcccccccccl}
\hline
\hline
& \multicolumn{4}{c}{\textbf{This Study}} & & \multicolumn{5}{c}{\textbf{Literature}} \\
 \cline{2-5} \cline{7-11}
Object & $T_{\rm eff}$ & $\log{g}$ & $\xi$ & [Fe/H] && $T_{\rm eff}$ & $\log{g}$ & $\xi$ & [Fe/H] & Reference \\
& (K) & (dex) & (km\,s$^{-1}$) & (dex) && (K) & (dex) & (km\,s$^{-1}$) & (dex) \\
\hline
\\
\multicolumn{11}{c}{\textbf{Standard Stars}} \\
\hline
HD 41667		& 4643	& 1.54	& 1.81	&  --1.18 && 4605 & 1.88 & 1.44 		& --1.16 & \citet{gratton;et-al_2000} 		\\ 
HD 44007		& 4820	& 1.66	& 1.70	&  --1.69 && 4850 & 2.00 & 2.20 		& --1.71 & \citet{fulbright_2000} 		\\ 
HD 76932		& 5835	& 3.93	& 1.42	&  --0.95 && 5849 & 4.11 & \nodata	& --0.88 & \citet{nissen;et-al_2000} 		\\ 
HD 122563		& 4358	& 0.14	& 2.77	&  --2.90 && 4843 & 1.62 & 1.80 		& --2.54 & \citet{yong;et-al_2013} 		\\ 
HD 136316		& 4347	& 0.44	& 2.15	&  --1.93 && 4414 & 0.94 & 1.70 		& --1.90 & \citet{gratton_sneden_1991} 	\\ 
HD 141531		& 4373	& 0.52	& 2.05	&  --1.65 && 4280 & 0.70 & 1.60 		& --1.68 & \citet{shetrone_1996} 		\\ 
HD 142948		& 5050	& 2.39	& 1.83	&  --0.64 && 4713 & 2.17 & 1.38 		& --0.77 & \citet{gratton;et-al_2000} 		\\ 
\hline
\\
\multicolumn{11}{c}{\textbf{Orphan Stream Candidates}} \\
\hline
OSS 3 (L)		& 5225	& 3.16	& 1.10	&  --0.86 && \nodata	& \nodata	& \nodata	&  --1.31	& \citet{casey;et-al_2013a}\\
OSS 6 (H)		& 4554	& 0.70	& 2.00	&  --1.75 && \nodata	& \nodata	& \nodata	&  --1.84	& \citet{casey;et-al_2013a}\\
OSS 8 (H)		& 4880	& 1.71	& 1.86	&  --1.62 && \nodata	& \nodata	& \nodata	&  --1.62	& \citet{casey;et-al_2013a}\\
OSS 14 (H)		& 4675	& 1.00	& 2.53	&  --2.66 && \nodata	& \nodata	& \nodata	&  --2.70	& \citet{casey;et-al_2013a}\\
OSS 18 (M)		& 5205	& 2.91	& 1.88	&  --0.62 && \nodata	& \nodata	& \nodata	&  --0.90	& \citet{casey;et-al_2013a}\\
\hline
\end{tabular}}
\end{table}






\subsubsection{Metallicity, ${\rm [M/H]}$}
\label{sec:orphan_II-analysis-metallicity}

The final stage of iterative stellar parameter analysis is to derive the total metallicity. For these analyses we adopt the mean [Fe\,I/H] abundance as the overall metallicity [M/H]. A difference of ${|\textrm{[M/H]} - \langle\textrm{[Fe\,I/H]}\rangle| \leq 0.01}$\,dex was considered an acceptable model for the data. Excluding HD 122563, the mean metallicity difference for all other standard stars is 0.03\,dex less than values in the literature, with a standard deviation of 0.05\,dex. When the entire standard sample is included, these values change to --0.07$\pm$0.13\,dex. Acknowledging that our systematically low temperature for HD 122563 has resulted in differing stellar parameters, the rest of our standard stars exhibit excellent agreement with previous studies. We note that this effect is likely significantly smaller for the metal-poor star OSS 14, as this star is further down the giant branch with an effective temperature $\sim$300\,K hotter than HD\,122563. Thus, we can be confident in the metallicity determination for OSS 14.

The offsets in metallicities between the values we derive from high-resolution spectroscopy and those found by \citet{casey;et-al_2013a} from low-resolution spectroscopy are noticeable. For the high probability targets (OSS 6, 8 and 14) the agreement is excellent: $\Delta$[Fe/H] = +0.04, 0.0, +0.09\,dex, respectively. We note that the uncertainties adopted in \citet{casey;et-al_2013a} are of the order $\pm0.3$\,dex; the discrepancies are well within the 1$\sigma$ uncertainties of {\it either} study. The largest difference between this study and that of \citet{casey;et-al_2013a} is in the lowest probability target (OSS 3), where we find a metallicity that is $+$0.45\,dex higher. 

It is reassuring that the metallicities of our high probability stars show the best agreement with the low-resolution measurements. Candidates were classified by \citet{casey;et-al_2013a} to have a low, medium, or high probability of membership with the Orphan stream. This classification was dependent on a number of observables, including metallicity. However, the metallicity determinations by \citet{casey;et-al_2013a} were calculated with the implied assumption that these stars were at a distance of $\sim$20\,kpc. The fact that our metallicities from high-resolution spectra are in excellent agreement with these initial metallicities indicates that our initial assumption was correct, and these high probability targets are at the approximate distance to the Orphan stream.\footnote{We note that to avoid bias from the low-resolution work of \citet{casey;et-al_2013a}, these high-resolution analyses took place with the information headers removed from each spectrum, and filenames were re-named to random strings. Original filenames were cross-matched only after the analysis was complete.}

\subsection{Uncertainties in Stellar Parameters}
\label{sec:orphan_II-stellar-parameter-uncertainties}
During excitation and ionization equilibria assessments, each fitted slope has an associated uncertainty due to the scatter in iron abundances. We have varied the effective temperature and microturbulence to match the formal uncertainty in each Fe\,I slope. Although these parameters are correlated, we have independently varied each parameter to reproduce the slope uncertainty. 

In order to estimate the uncertainty in $\log{g}$, the surface gravity has been adjusted until the mean difference in Fe\,I and Fe\,II matches the quadrature sum of the variance in Fe\,I and Fe\,II abundances. These uncertainties are listed in Table \ref{tab:stellar-parameter-uncertainties} for all standard and program stars. As these uncertainties are calculated with the assumption that they are uncorrelated, they are therefore possibly underestimated. As such we have assumed a minimum uncertainty of 125\,K in $T_{\rm eff}$, 0.3\,dex in $\log{g}$ and 0.2\,km s$^{-1}$ in $\xi$. Thus, the adopted uncertainty for each star is taken as the maximum of these values and those listed in Table \ref{tab:stellar-parameter-uncertainties}.\\

\begin{table}[h!]
\centering
\caption{Uncorrelated uncertainties in stellar parameters of Orphan stream candidates\label{tab:stellar-parameter-uncertainties}}
\begin{tabular}{lccc}
\hline
\hline
Object 	& $\sigma(T_{\rm eff})$	& $\sigma(\log{g})$	& $\sigma(\xi)$ \\
		& (K) 					& (dex) 			& (km\,s$^{-1}$) \\
\hline
HD 41667	&	47 &	0.08 &	0.05 \\
HD 44007	&	51 &	0.08 &	0.17 \\
HD 76932	&	61 &	0.10 &	0.14 \\
HD 122563	&	15 &	0.01 &	0.07 \\
HD 136316	&	34 &	0.02 &	0.05 \\
HD 141531	&	40 &	0.03 &	0.05 \\
HD 142948	&	62 &	0.12 &	0.09 \\
OSS 3 (L)	&	73 &	0.08 &	0.11 \\
OSS 6 (H)	&	39 & 0.04 &	0.15 \\
OSS 8 (H)	&	74 & 0.17 &	0.23 \\
OSS 14 (H)&	51 & 0.05 &	0.13 \\
OSS 18 (M)&	135& 0.15 &	0.28 \\
\hline
\end{tabular}
\end{table}

\subsection{Chemical Abundances}
\label{sec:orphan_II-chemical-abundances}
Chemical abundances are described following the standard nomenclature\footnote{$\log\epsilon(X) = \log{\left(\frac{N_X}{N_H}\right)} + 12$}, and comparisons made with reference to the Sun\footnote{$\left[\frac{X}{H}\right] = \log\epsilon{\left(\frac{N_X}{N_H}\right)}_{\rm Star} - \log\epsilon{\left(\frac{N_X}{N_H}\right)}_{\rm Sun}$} have been calculated using the solar composition described in \citet{asplund;et-al_2009}. We note that some transitions have been removed from our line list (Appendix \ref{app:orphan_ews}) when they were not detected in any star (e.g., Sc\,I, Zr\,I, Zr\,II, Rb\,I, and Ce\,II transitions).

\subsubsection{Carbon}

Carbon has been measured from synthesizing the 431.3\,nm and 432.3\,nm CH features. The best fitting abundances inferred from these two syntheses were within 0.05\,dex for every star. Nevertheless, have adopted a conservative total uncertainty of $\pm$0.15\,dex for these measurements. The mid-point of these two measurements is listed in Table \ref{tab:chemical-abundances}. Only the 432.3\,nm region was synthesized for OSS 18, as significant absorption of atomic lines was present in the bluer band head, and we deemed the 432.3\,nm region to yield a more precise determination of carbon abundance.

\subsubsection{Light Odd-Z Elements (Na, Al, K)}
Sodium, aluminium and potassium are primarily produced through carbon, neon and oxygen burning in massive stars, before they are ejected into the interstellar medium \citep{woosley;weaver_1995}. 

Although our line list includes three clean, unblended Na lines, not all were detectable in the Orphan stream candidates. Generally only one Na and Al line was available for the Orphan stream stars. In these cases, a minimum standard deviation of 0.10\,dex has been assumed when calculating abundance uncertainties (see Section \ref{sec:orphan_II-chemical-abundance-uncertainties}). Wherever there were no detectable lines for a given element, an upper limit was determined by synthesis of the strongest transition in our line list. At least one K line was detected in every standard and program star. These lines at 766\,nm and 769\,nm are quite strong, but fall directly between a strong telluric band head. Usually both lines were detected, but one was dominated by the Earth's atmospheric absorption. In these cases we rejected the contaminated line and adopted the single, unaffected K line.

\subsubsection{$\alpha$-elements (O, Mg, Ca, Si, Ti)}

The $\alpha$-elements (O, Mg, Ca, Si and Ti) are produced during hydrostatic burning of carbon, neon and silicon by $\alpha$-particle capture. Following Type II supernovae, the $\alpha$-enriched material is dispersed into the interstellar medium and contributes to the next generation of star formation. Although Ti is not formally an $\alpha$-element ($Z = 22$), it generally tracks with $\alpha$-elements, and has been included here to facilitate a comparison with the Local Group study of \citet{venn;et-al_2004}.

Generally, the Orphan stream candidates have lower $\alpha$-element abundances compared to iron than their halo counterparts. The situation is a little ambiguous for [O/Fe] compared to other $\alpha$-elements. We have employed the forbidden lines at 630\,nm and 636\,nm to measure oxygen abundances in our stars. Given the weakness of these lines, oxygen was immeasurable in all of the Orphan stream candidates. In place of abundances, upper limits have been determined from spectral synthesis of the region. The synthesis line list includes the Ni\,I feature hidden within the 630\,nm absorption profile \citep{allende-prieto;et-al_2001}. In the low $S/N$ ratio case of OSS 18, the forbidden oxygen region was sufficiently dominated by telluric absorption such that we deemed even a robust upper limit to be indeterminable.

\begin{figure}[h]
	\includegraphics[width=\textwidth]{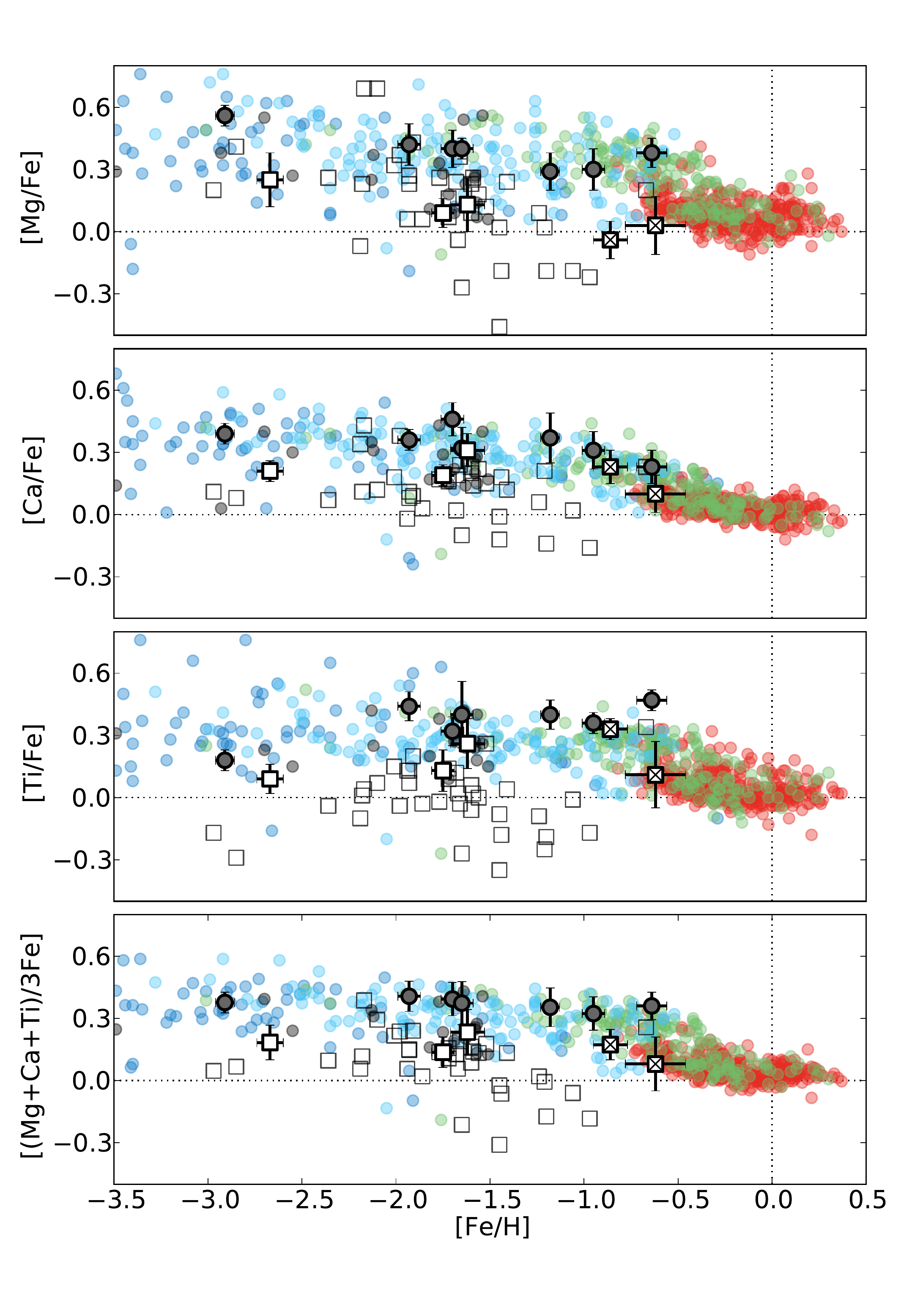}
	\caption[$\alpha$-element abundance ratios for standard and program stars]{$\alpha$-element abundance ratios (Mg\,I, Ca\,I, Ti\,II shown) for field standards (grey, filled), high-probability Orphan stream candidates (white squares), and lower-probability Orphan stream candidates (white squares with crosses) that are deemed to be interlopers. Data compiled by \citet{venn;et-al_2004,venn;et-al_2006} for the present-day dSph galaxies (grey squares) and the Milky Way is shown, using the same colour scheme adopted in \citet{venn;et-al_2004}: thin disk  (red), thick disk (green), halo (cyan), high velocity halo (dark blue), and retrograde halo stars (black).
	}
	\label{fig:alpha-fe}
\end{figure}

The [Mg/Fe] abundance ratios for the Orphan stream targets are noticeably lower than any of the other $\alpha$-elements. In fact, OSS 18 and 3 exhibit solar or sub-solar [Mg/Fe] abundances, and the mean [$\alpha$/Fe] abundances in the lower panel of Figure \ref{fig:alpha-fe} are most affected by [Mg/Fe]. Nevertheless, [Ca/Fe] and [Si/Fe] abundances for Orphan stream targets are systematically lower than their Milky Way counterparts. In Figure \ref{fig:alpha-fe}, Ti\,II has been adopted for [Ti/Fe] as few reliable Ti\,I transitions were available for analysis.

Lower [$\alpha$/Fe] abundance ratios are less apparent for our lower probability Orphan stream candidates, OSS 18 and OSS 3. Although their mean [$\alpha$/Fe] $\sim$ +0.20\,dex, with [Fe/H] $\sim$ --0.70\,dex they are significantly more metal-rich than expected for the Orphan stream. Therefore their [$\alpha$/Fe] abundance ratios are consistent with the [$\alpha$/Fe]-[Fe/H] trend of the Milky Way. As discussed further in Section \ref{sec:orphan_II-stream-membership}, these low probability targets are unlikely to be true Orphan stream members, and are marked appropriately in Figure \ref{fig:alpha-fe}.

\subsubsection{Fe-peak elements (Sc, V, Cr, Mn, Co, Ni, Cu, Zn)}
The Fe-peak elements ($Z = 23$ to 30) are primarily produced by explosive nucleosynthesis during oxygen, neon, and silicon burning. The ignition of these burning phases occurs both from Type II SN of massive stars, and once an accreting white dwarf exceeds the Chandrasekhar limit, causing spontaneous ignition of carbon and an eventual Type Ia supernova. The abundance of the Fe-peak elements with respect to iron in the Milky Way are typically either flat (i.e., [X/Fe] $\sim$ 0; Cr\,II, Ni) or trend positively (Sc, Cr\,I, Mn, Cu) with overall [Fe/H] \citep[e.g.,][]{ishigaki;et-al_2013,yong;et-al_2013}.

Scandium absorption profiles have appreciable broadening due to hyperfine splitting. As such, we have determined Sc abundances for all stars from spectral synthesis with hyperfine splitting taken into account. However, for OSS\,3 we found the broadening due to hyperfine structure to be negligible, and the variance in our synthesis measurements was larger than our abundance measurements. The mean Sc\,II abundance to be $\log\epsilon({\rm Sc}) = 2.22 \pm 0.23$ from synthesis of 5 lines, where the EWs of 11 Sc\,II lines yields $\log\epsilon({\rm Sc}) = 2.24 \pm 0.09$. Therefore in the case of OSS\,3, we have adopted the Sc\,II abundances from EWs.

Other Fe-peak elements with hyperfine structure (e.g., V, Mn, Co, Cu) have been synthesised with the relevant isotopic and/or hyperfine splitting included. The random measurement scatter in V, Cr, and Mn abundances is typically low ($<$0.05\,dex) for all the standard and program stars. However for Co, there was a noticeable increase in line-to-line scatter for the Orphan stream stars compared to the standard stars. There is a factor of $\sim$6 difference in $S/N$ ratio between the two samples which can explain this variance. In the Orphan stream stars, no synthesised Co profiles were sufficiently `worse' than each other to qualify exclusion. The case was quite different for Ni, where a plethora of clean lines (without appreciable hyperfine structure) are available, and the uncertainties in $\log\epsilon({\rm Ni})$ abundances due to the uncertainties in stellar parameters generally cancel with $\log\epsilon({\rm Fe})$, yielding excellent measurements of [Ni/Fe]. Cu abundances and upper limits are derived from the synthesis of a single neutral Cu line at 510.5\,nm. The inclusion of hyperfine structure was most important for Cu, as a Cu abundance measured directly from an EW produced a systematically higher $\log\epsilon({\rm Cu})$, on the order of +0.4\,dex. Zn abundances were calculated directly from the EWs of the 472.2\,nm and 481.0\,nm transitions where available. All of our Fe-peak abundances with respect to iron are consistent with the observed chemical trends in the Milky Way \citep{ishigaki;et-al_2013,yong;et-al_2013}.

\subsubsection{Neutron-capture elements (Sr, Y, Ba, La, Nd, Eu)}

The atomic absorption lines of these heavy elements have appreciable broadening due to hyperfine structure and isotopic splitting. Where applicable, the relevant hyperfine and/or isotopic splitting has been employed, and solar isotopic compositions have been adopted.

Sr and Y are elements of the first $n$-capture peak. While only one line (421.5\,nm) was synthesised for Sr\,II, generally three unblended Y lines were available. When no Y lines were detected above $3\sigma$, an upper limit was ascertained from the 520\,nm line -- the strongest in our line list. Sr and Y abundances generally agree with each other in our stars, with the exception of high probability stream candidates OSS 6 and OSS 8. In these cases only upper limits could be determined for Y\,II, and those limits are $\sim$0.6\,dex lower than our measured Sr\,II abundances. The remaining high probability candidate, OSS 14, also has an upper limit for Y\,II ([Y/Fe] $< -0.40$) but this limit is much closer to our Sr\,II measurement ([Sr/Fe] = $-0.42$). The Y upper limits for our high probability Orphan stream stars results in lower limits for [Ba/Y], which for OSS 6 and OSS 8, are well in excess of the Milky Way trend (Figure \ref{fig:ba-y}).

Lanthanum and neodymium abundances for our stars are consistent with the chemical evolution of the Milky Way. Europium, a $r$-process dominated element, has been measured by synthesising the 664.5\,nm Eu\,II transition. Although the spread in [Eu/Fe] abundances for our stars is wide, no significant $r$-process enhancement is observed. 

\subsection{Chemical Abundance Uncertainties Due to Stellar Parameters}
\label{sec:orphan_II-chemical-abundance-uncertainties}
Although the standard error about the mean ($\sigma_{\bar{x}}$) abundance in Table \ref{tab:chemical-abundances} can be quite low, the uncertainties in stellar parameters will significantly contribute to the total error budget for any given abundance. Furthermore, the total uncertainty in abundance ratios (e.g., [A/B]) depends on how the uncertainties in elements A and B are correlated with stellar parameters. \\

\begin{longtable}{lcrcrrc}
\tiny
\label{tab:chemical-abundances}\\
\hline
\hline
Species & N & $\log\epsilon(X)$ & $\sigma$ & [X/H] & [X/Fe] & $\sigma_{\bar{x}}$ \\
\hline
\endfirsthead
\hline
Species & N & $\log\epsilon(X)$ & $\sigma$ & [X/H] & [X/Fe] & $\sigma_{\bar{x}}$ \\
\hline
\endhead
\hline
\multicolumn{7}{r}{Continued..}
\endfoot
\endlastfoot

\\
\multicolumn{7}{c}{\textbf{HD 41667}} \\
\hline


C (CH) &   2 &    6.95 &    0.20 &  --1.48 &  --0.24 &    0.15 \\
   O I &   1 &    7.85 & \nodata &  --0.84 &    0.34 & \nodata \\
  Na I &   3 &    4.95 &    0.10 &  --1.29 &  --0.10 &    0.06 \\
  Mg I &   7 &    6.71 &    0.13 &  --0.89 &    0.29 &    0.05 \\
  Al I &   4 &    5.27 &    0.11 &  --1.18 &  --0.00 &    0.05 \\
  Si I &   8 &    6.58 &    0.14 &  --0.93 &    0.26 &    0.05 \\
\end{longtable}

\setlength\tabcolsep{1pt}
\begin{longtable}{lrrrccc}
\label{tab:chemical-abundance-uncertainties} \\
\hline
\hline
 & &  & & & \multicolumn{2}{c}{\textbf{Total Uncertainty}} \\
\cline{6-7}
Species & $T_{\rm eff}+\sigma(T_{\rm eff})$ & $\log{g}+\sigma(\log{g})$ & ${v_t}+\sigma(v_t)$ & $Max(0.10, S.D.)/\sqrt(N)$ & [X/H] & [X/Fe] \\
        & $\Delta$abundance & $\Delta$abundance & $\Delta$abundance & (dex) & (dex) & (dex) \\
\hline
\endfirsthead
\hline
 & &  & & & \multicolumn{2}{c}{\textbf{Total Uncertainty}} \\
\cline{6-7}
Species & $T_{\rm eff}+\sigma(T_{\rm eff})$ & $\log{g}+\sigma(\log{g})$ & ${v_t}+\sigma(v_t)$ & $Max(0.10, S.D.)/\sqrt(N)$ & [X/H] & [X/Fe] \\
        & $\Delta$abundance & $\Delta$abundance & $\Delta$abundance & (dex) & (dex) & (dex) \\
\hline

\endhead
\hline
\multicolumn{7}{r}{Continued..}
\endfoot
\endlastfoot
\hline
\\
\multicolumn{7}{c}{\textbf{HD 41667}} \\
\hline
 O \textsc{I}   	&  +0.07	&  +0.10	&  +0.06	& 0.10	& 0.17	& 0.10	\\
 Na \textsc{I}   	&  +0.02	& --0.02	& --0.01	& 0.06	& 0.07	& 0.10	\\
 Mg \textsc{I}   	&  +0.02	& --0.02	& --0.03	& 0.05	& 0.06	& 0.09	\\
 Al \textsc{I}   	&  +0.02	& --0.01	& --0.01	& 0.05	& 0.06	& 0.08	\\
 Si \textsc{I}   	&  +0.01	&  +0.01	&  +0.00	& 0.05	& 0.05	& 0.05	\\
 K \textsc{I}   	&  +0.04	& --0.02	& --0.05	& 0.07	& 0.10	& 0.12	\\
\end{longtable}

In order to investigate the abundance uncertainties due to stellar parameters, we have generated model atmospheres for a $\pm1\sigma$ offset in each stellar parameter (Section \ref{sec:orphan_II-stellar-parameter-uncertainties}; $T_{\rm eff}$, $\log{g}$, $\xi$) independently, and calculated the resultant mean abundance offset from our EWs. The resultant abundance changes were added in quadrature with $\sigma_{\bar{x}}$ to provide a total uncertainty in [X/H]. These total uncertainties are tabulated in Table \ref{tab:chemical-abundance-uncertainties} for all standard and program stars. While this yields an uncertainty for our abundance ratios in [X/H], we are often interested in how elemental abundances vary with respect to iron. We have calculated these uncertainties following \citet{johnson_2002}:
\begin{equation}
	\sigma^{2}\left({\rm A/B}\right) = \sigma^{2}\left({\rm A}\right) + \sigma^{2}\left({\rm B}\right) - 2\sigma_{{\rm A,B}}
\end{equation}

\noindent{}where the covariance between elements A and B ($\sigma_{{\rm A,B}}$) is given by
\small
\begin{multline}
	\sigma_{\rm A,B} = \left(\frac{\partial\log{\epsilon_{\rm A}}}{\partial T_{\rm eff}}\right)\left(\frac{\partial\log\epsilon_{\rm B}}{\partial T_{\rm eff}}\right)\sigma_{T_{\rm eff}}^2 
	+ \left(\frac{\partial\log{\epsilon_{\rm A}}}{\partial\log{g}}\right)\left(\frac{\partial\log\epsilon_{\rm B}}{\partial\log{g}}\right)\sigma_{\log{g}}^2	+ \left(\frac{\partial\log{\epsilon_{\rm A}}}{\partial\xi}\right)\left(\frac{\partial\log\epsilon_{\rm B}}{\partial\xi}\right)\sigma_{\xi}^2 \\
	+ \left[\left(\frac{\partial\log{\epsilon_{\rm A}}}{\partial T_{\rm eff}}\right)\left(\frac{\partial\log\epsilon_{\rm B}}{\partial\log{g}}\right) + \left(\frac{\partial\log{\epsilon_{\rm A}}}{\partial\log{g}}\right)\left(\frac{\partial\log\epsilon_{\rm B}}{\partial T_{\rm eff}}\right)\right]\sigma_{T_{\rm eff},\log{g}}
	+ \left[\left(\frac{\partial\log{\epsilon_{\rm A}}}{\partial\xi}\right)\left(\frac{\partial\log\epsilon_{\rm B}}{\partial\log{g}}\right) + \left(\frac{\partial\log{\epsilon_{\rm A}}}{\partial\log{g}}\right)\left(\frac{\partial\log\epsilon_{\rm B}}{\partial\xi}\right)\right]\sigma_{\xi,\log{g}}
\end{multline}
\normalsize
The covariance between effective temperature and surface gravity has been calculated by sampling about the effective temperature and performing an ionization balance for the adjusted temperature. The resultant covariance is given as:

\begin{equation}
	\sigma_{T_{\rm eff},\log{g}} = \frac{1}{N}\sum_{i=1}^{N}\left(T_{{\rm eff},i} - \overline{T_{\rm eff}}\right)\left(\log{g_i} - \overline{\log{g}}\right)
\end{equation}

\noindent{}Covariance between $\xi$ and $\log{g}$ is calculated in the same manner. The total abundance uncertainty with respect to H and Fe (e.g., $\sigma($[X/H]), $\sigma$([X/Fe])) for all species is given in Table \ref{tab:chemical-abundance-uncertainties}. These total uncertainties have been adopted in all figures throughout this text. Due to the cancellation of systematic effects, some abundance ratios with respect to iron have lower uncertainties than their quoted absolute abundance uncertainties.

\section{Discussion}
\label{sec:orphan_II-discussion}

\subsection{Stream Membership}
\label{sec:orphan_II-stream-membership}

Given the low surface brightness of the Orphan stream, separating true members from interlopers can be particularly challenging. Before inferring any properties of the undiscovered parent satellite from our sample, we must examine whether our targets are stars truly from the disruption of the Orphan stream progenitor.

When compared against the mean of our high priority targets, the velocities of the low and medium probability candidates are +6.2 and $-23.7$\,km s$^{-1}$ different, respectively. Given this region of the stream has a low intrinsic velocity dispersion \citep{casey;et-al_2013a}, the significantly lower velocity of the medium-probability target, OSS 18, is intriguing. Perhaps more concerning is that the low and medium probability candidates are markedly more metal-rich than the high-priority targets: [Fe/H] = $-0.86$ and $-0.62$\,dex for OSS 3 and OSS 18, a difference of $+0.45$ and $+$0.28\,dex from the low-resolution measurements, respectively. Spectroscopic studies suggest the stream is significantly more metal-poor than [Fe/H] $\sim -0.8$\,dex \citep{belokurov;et-al_2007,newberg;et-al_2010,casey;et-al_2013a,sesar;et-al_2013}, making the association between the Orphan stream and OSS 3 or OSS 18 tenuous. It is also worth noting that these targets are the farthest candidates ($|B_{\rm Orphan}| \sim 0.5$\,$^\circ$) from the best-fit Orphan stream orbital plane deduced by \citet{newberg;et-al_2010}. On the basis of the observables we deduce that the lower probability members, OSS 3 and OSS 18, are unlikely to be disrupted Orphan stream members.

The three high-probability targets (OSS 6, 8 and 14) have velocities within 2.4\,km s$^{-1}$ of each other, consistent with the Orphan stream velocity. These velocities from high-resolution spectra confirm the very low line-of-sight velocity dispersion in this part of the stream. The metallicities of these targets are also in reasonable agreement with those found from low-resolution spectroscopy, implying the candidates are $\sim$20\,kpc away -- at approximately the same distance as the Orphan stream. For the remainder of this discussion, we consider OSS 6, 8 and 14, to be true disrupted members of the Orphan stream. 

\subsection{Metallicity Distribution Function}
Although the sample size is small, the three Orphan stream members have a wide range in metallicity, ranging from [Fe/H] = $-1.58$ to as metal-poor [Fe/H] = $-2.66$. From three members not much can be said about the metallicity distribution, other than to say it is wide, and inconsistent with a mono-metallic population (e.g., a globular cluster). \citet{newberg;et-al_2010} and \citet{sesar;et-al_2013} found the stream to have a mean metallicity of [Fe/H] $= -2.1$ from BHB and RR Lyrae stars. According to the metallicity gradient along the stream reported by \citet{sesar;et-al_2013}, we might expect a higher mean metallicity than [Fe/H] $> -2.1$ in our sample closer to the celestial equator. We note that our metallicity spread of [Fe/H] = $-1.58$ to $-2.66$ is consistent with that of \citet{sesar;et-al_2013}: [Fe/H] = $-1.5$ to $-2.7$, and the somewhat wider spread found by \citet{newberg;et-al_2010}: [Fe/H] = $-1.3$ to $-3$.

A total metallicity spread as wide as $\sim$1\,dex, or a standard deviation of $\sigma({\rm [Fe/H]}) = 0.56$ \citep{casey;et-al_2013a}, is consistent with the stochastic chemical enrichment of the present-day dSph galaxies \citep{mateo_1998,kirby;et-al_2011}. This is also compatible with the constraints from the dynamical constraints placed inferred by the arc and width of the stream, which suggest the progenitor is a dark matter-dominated system.

\subsubsection{[$\alpha$/Fe] abundance ratios}
The abundance trends of individual elements track the star formation history of a cluster or galaxy. In the Milky Way, this is most evident by the evolution of $\alpha$-element abundances with increasing metallicity. The $\alpha$-elements are produced in massive stars before being ejected to the interstellar medium by Type II supernovae. At later times, Type Ia supernovae begin to contribute to the Galaxy's chemical enrichment. Type Ia supernovae expel iron (like Type II), but do not produce significant amounts of $\alpha$-elements. As such, a net decrease in [$\alpha$/Fe] is observed upon the onset of Type Ia supernovae. This inflexion occurs in the Milky Way near [Fe/H] $\sim -0.7$, decreasing towards Solar-like values.

\citet{tolstoy;et-al_2003} first noted that stars in the present-day dSph galaxies were separated in [$\alpha$/Fe] from the majority of the Milky Way stars. In the dSph galaxies, a significant contribution of Type Ia supernovae is observable in [$\alpha$/Fe] abundance ratios near [Fe/H] $> -1.7$\,dex, with the exception of Draco, which shows low [$\alpha$/Fe] at all metallicities. Some dwarf spheroidal stars also have a large range in [$\alpha$/Fe] values, increasing up to values seen in field stars.

The Orphan stream stars have low [$\alpha$/Fe] abundance ratios ($\sim$0.22) with respect to the Milky Way sample ($\sim$0.40\,dex). The low- and medium-probability members -- which we deem to be field interlopers -- also have low [$\alpha$/Fe] abundance ratios, but given their overall metallicity these $\alpha$-element ratios are consistent with the thick disk enrichment of the Milky Way. Since our Orphan stream members (i.e., the high-probability members OSS 6, 8 and 14) are metal-poor, we can be confident we are not confusing their low-$\alpha$ signature for the standard chemical enrichment of the Milky Way. 

We stress that [$\alpha$/Fe], on its own, should never be used as a litmus test for accretion on to the Milky Way. One must carefully select candidates that appear to be stream members based on all the observables, and then examine their detailed abundances in order to infer the chemical evolution of the disrupted host. We are already quite confident these stars are truly disrupted Orphan stream members, and as such, we can say that the low [$\alpha$/Fe] abundance ratios observed in the Orphan stream are consistent with the low $\alpha$-element enhancement of the present-day dSph galaxies.

\subsubsection{[Ba/Y]}
Differences in [Ba/Y] ratios between the present-day dSph stars and the Milky Way have been observed by a number of groups \citep{shetrone;et-al_2003,venn;et-al_2004}. A number of possible explanations have been proposed to explain this offset, including changes in SNe II yields -- or by changing the frequency of SNe II explosions via adjustments to the initial mass function -- or altering the influence of $\alpha$-rich freeze-out \citep[e.g., see][for a discussion]{venn;et-al_2004}.

\begin{figure}[h]
	\includegraphics[width=\textwidth]{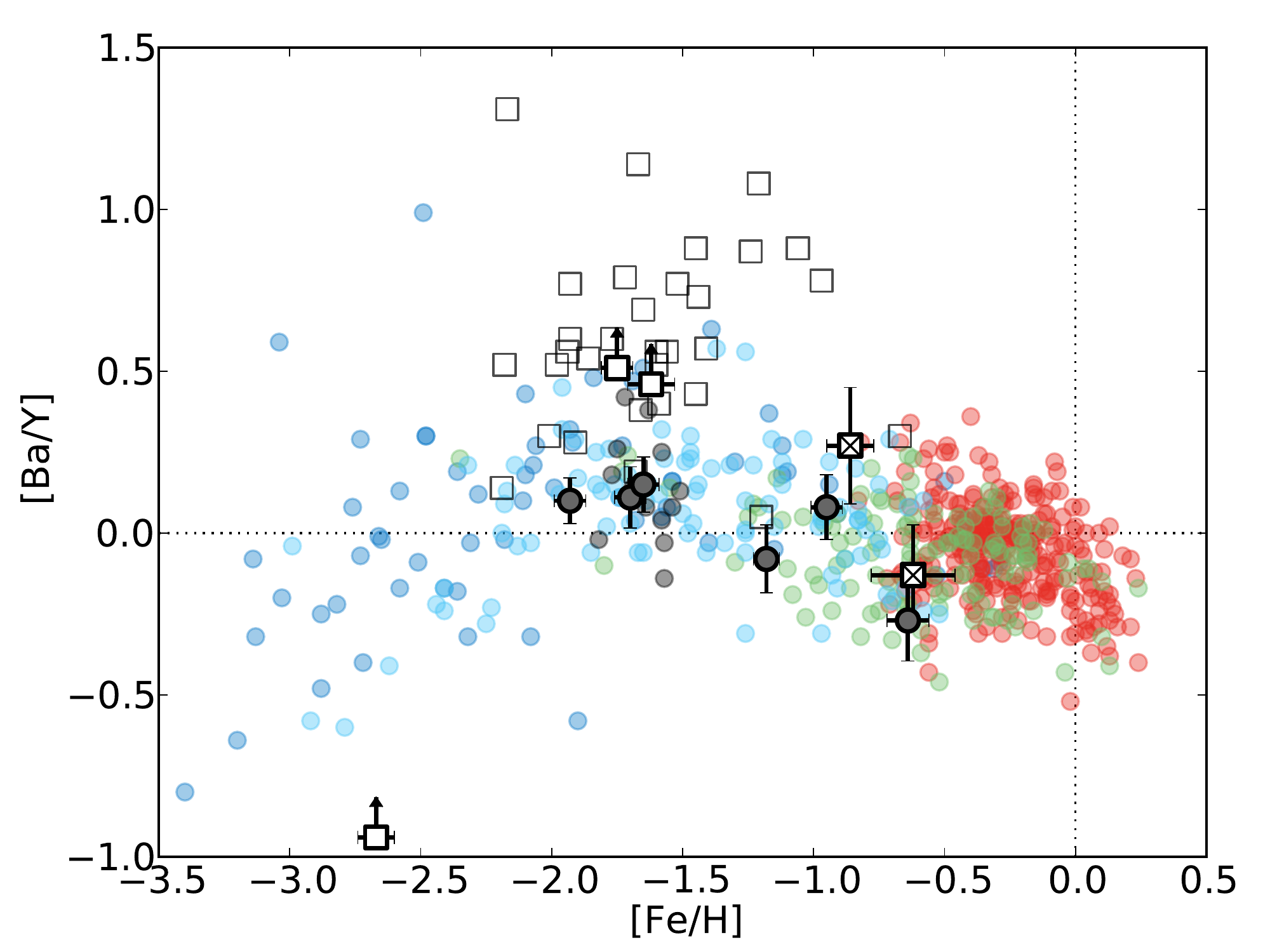}
	\caption[${\rm [Ba/Y]}$ abundance ratios for the standard and program stars]{[Ba/Y] abundance ratios for the Orphan stream stars and field standards. Milky Way and dSph data compiled by \citet{venn;et-al_2004} is also shown. Markers and colours are the same described in Figure \ref{fig:alpha-fe}.}
	\label{fig:ba-y}
\end{figure}

Arbitrarily adjusting SNe II yields between two different nucleosynthesis sites appears a somewhat unlikely scenario. By employing SN frequency corrections in addition to using adjusted yields, \citet{qian;wasserburg} can reproduce the low [Ba/Y] trends in the Galaxy, but not the high [Ba/Y] ratios observed in the dSph galaxies. Leaving SNe II yields unchanged, these offsets could alternatively be reproduced by simply adjusting the frequency of SNe II events through the truncation of the upper initial mass function (IMF). In effect, massive stars would be less numerous, which are thought to be the primary production sites for the first $n$-capture peak elements (e.g., yttrium in this case). Because higher $\alpha$-element yields are also expected for massive stars, this may affect both the [Ba/Y] abundance and the overall [$\alpha$/Fe] abundance. However, whether the upper IMF differs between the dSph and the Milky Way remains an open question.


We have lower limits in [Ba/Y] for all Orphan stream stars. This is because Y was not detected in any of the stream stars. The standard stars and lower probability stream targets all have [Ba/Y] ratios that are consistent with the Milky Way for their given overall metallicity. This is shown in Figure \ref{fig:ba-y}. Although we have only lower limits for [Ba/Y], in OSS 6 and OSS 8 these are $\sim$+0.5\,dex offset from the main component of the Milky Way. This is a similar offset as that observed in the present-day dSphs, as compiled by \citet{venn;et-al_2004}. For the most metal-poor stream star, OSS 14, our limit on Y abundance yields a weak -- but consistent -- lower limit of [Ba/Y] $> -0.97$\,dex. We note that robust lower limits for [Ba/Y] in the most metal-poor dSph stars ([Fe/H]$ < -2$) are difficult to ascertain.

\subsection{Possible Parent Systems}
A number of associations have been proposed between the Orphan stream and known Milky Way satellites. However, most have been shown to be either unlikely or implausible. Here we discuss the possibility of association between the Orphan stream and NGC 2419, as well as Segue 1.

\subsubsection{NGC 2419}
The width and length of a tidal tail are a clear indication to the nature of the parent satellite. For the case of the Orphan stream, these characteristics favor a dark-matter dominated system (e.g., a dwarf galaxy). However, given the large apogalacticon of 90\,kpc for the Orphan stream, a relationship has been proposed between the Orphan stream and NGC 2419 \citep{bruns;kroupa_2011}. NGC 2419 is the most distant ($D \sim 85$\,kpc) -- and most luminous ($M_V \sim -9.6$) -- globular cluster in the Milky Way halo ($D > 20$\,kpc). Just like the Orphan stream, the system is also quite metal poor: [Fe/H] = --2.15.

Stars in globular clusters are known to exhibit peculiar chemical patterns, most notably in an anti-correlation between sodium and oxygen abundances \citep[e.g.,][]{carretta;et-al_2009a}. Following this, NGC 2419 is particularly special because it shows an unusual anti-correlation between magnesium and potassium, even stronger than the classical Na-O anti-correlation. This chemical signature is so far unique to NGC 2419, and illustrates the most extreme cases of Mg-depletion and K-enhancement seen anywhere in the Milky Way, or its satellite systems \citep{cohen;et-al_2011,mucciarelli;et-al_2012}.

\begin{figure}[h]
	\includegraphics[width=\textwidth]{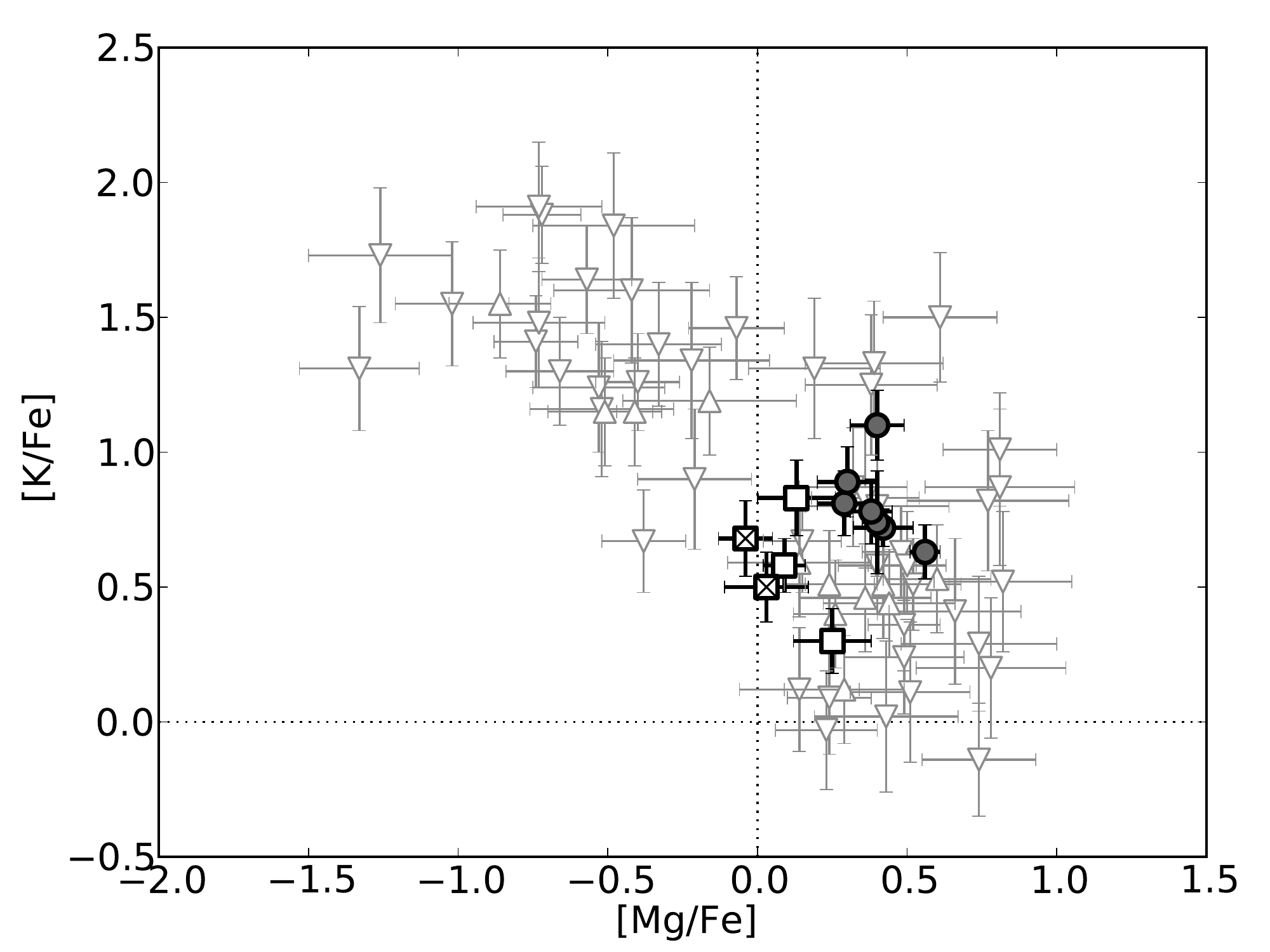}
	\caption[Magnesium and potassium abundances for stars in the globular cluster NGC 2419 compared with the standard and program stars observed]{Magnesium and potassium abundance ratios for stars in the globular cluster NGC 2419 from \citet{cohen;et-al_2011} ($\vartriangle$) and \citet{mucciarelli;et-al_2012} ($\triangledown$). Standard stars and Orphan stream candidates are shown with the same markers used in Figure \ref{fig:alpha-fe}.}
	\label{fig:mg-k}
\end{figure}

We do not find any Orphan stream stars to be extremely depleted in [Mg/Fe] or enhanced in [K/Fe] (Figure \ref{fig:mg-k}), as found in NGC 2419 \citep{cohen;et-al_2010,cohen;et-al_2011,mucciarelli;et-al_2012}. If we consider two sub-populations in NGC 2419: a Mg-poor, K-rich and a Mg-normal, K-poor sample, then a two sample Kolmogorov-Smirnoff test demonstrates that we can be at least 99.7\% confident that our Orphan stream members are inconsistent with being drawn from the Mg-poor NGC 2419 population. We cannot make such statements about the Mg-normal population, since abundances for that sample are typical for the Milky Way. The Mg-poor and Mg-normal sub-populations each comprise about half of the observed NGC 2419 population \citep{mucciarelli;et-al_2012}. Given the uniqueness of the Mg-K anti-correlation, a single Orphan stream star with significant Mg-depletion or K-enhancement would have provided an exciting hint for the Orphan stream parent. However, given the stream's large metallicity spread \citep{casey;et-al_2013a,sesar;et-al_2013}, this indicates the stream is unlikely to be associated with NGC 2419 given the cluster's small dispersion in overall metallicity \citep[0.17\,dex;][]{mucciarelli;et-al_2012}. We therefore firmly rule out any association between NGC 2419 and the Orphan stream.



\subsubsection{Segue 1}


The only known satellite system with an ambiguous association with the Orphan stream is Segue 1. The system resides in a particularly crowded region of the sky: it lies extremely close to the perigalacton of the Orphan stream, and just south of the bifurcated Sagittarius stream. The original discovery of Segue 1 by \citet{belokurov;et-al_2007} suggested the system was an extended globular cluster, until recent studies showed the system was an ultrafaint dwarf galaxy \citep{geha;et-al_2009,norris;et-al_2010,simon;et-al_2011}. The similarities between Segue 1 and the Orphan stream are striking. In addition to being nearby on the sky, Segue 1's distance of 23\,kpc is consistent with the closest portion of the Orphan stream. The velocities between the two systems are also coincident: +114\,km s$^{-1}$ for Segue 1 and $\sim$+120\,km s$^{-1}$ for the nearest portion of the stream. Additionally, both systems exhibit extremely low velocity dispersions, on the order of 3-4\,km s$^{-1}$. However, stars linking the two systems have yet to be found. The association between the two is not obvious. As \citet{gilmore} summarises, `{\it are they just ships passing in the dark night?}'

The characteristics of the Orphan stream are clues to the nature of its progenitor: the arc length and intrinsic width suggests the parent system must be dark matter-dominated. Segue 1 is the most dark matter-dominated system known, such that it has become a prime focus for indirect dark matter detection experiments \citep{essig,baushev}. However, there are some doubts as to whether Segue 1 has experienced any tidal effects \citep{ostholt,simon;et-al_2011}. Segue 1 is well within the galactic tidal field, and initial indications of east-west tidal effects \citep{ostholt} seem to have been complicated by the crowded region bordering the system \citep{simon;et-al_2011}. Presently, no stars or tails clearly connecting the two systems have been found.

The chemistry of Segue 1 is particularly relevant here. Segue 1 hosts an extremely wide range in stellar metallicities: the mean is found to be [Fe/H] $= -2.5$ and the spread extends from [Fe/H] = $-3.4$ to $-1.6$\,dex \citep{simon;et-al_2011}. There is significant overlap between the metallicity distributions reported for Segue 1 and the Orphan stream. The high-resolution spectra in this study confirm the wide metallicity range of Orphan stream members found by \citet{casey;et-al_2013a}, and independently reported by \citet{sesar;et-al_2013} ([Fe/H] $= -1.5$ to $-2.7$). Thus, in addition to on-sky position, distances, velocity, and velocity dispersion, Segue 1 and the Orphan stream share a wide and consistent range in metallicities. In contrast to these observables, \citet{vargas;et-al_2013} find Segue 1 members to be extremely $\alpha$-enhanced ([$\alpha$/Fe] $> 0.4$), unlike what we observe in disrupted Orphan stream members. 

We note that although the two systems appear to be related in some respected, if Segue 1 were the `parent' of the Orphan stream, the differing [$\alpha$/Fe] abundances and lack of tidal features surrounding Segue 1 -- after extensive examination -- is somewhat puzzling. Furthermore if Segue 1 is the disrupted host, and if the stream metallicity gradient presented by \citet{sesar;et-al_2013} is correct, that implies the stream becomes more metal-rich at greater distances from the parent system, contrary to what is typically observed. 

\section{Conclusions}
\label{sec:orphan_II-conclusions}

We present a chemical analysis of five Orphan stream candidates, three of which we confirm are true members from the disrupted Orphan stream parent satellite. The two non-members were both originally identified to have `low' or `medium' probability of stream membership from the low-resolution spectroscopy by \citet{casey;et-al_2013a}. We encourage high-resolution spectroscopic follow-up of the remaining high-probability Orphan stream members presented in \citet{casey;et-al_2013a}.

This work demonstrates the first detailed chemical study of the Orphan stream from high-resolution spectra. A large metallicity spread is present in the Orphan stream members, confirming the work by \citet{casey;et-al_2013a} and \citet{sesar;et-al_2013} that the Orphan stream is not mono-metallic. The spread in overall metallicity is consistent with the internal chemical evolution of the present-day dwarf galaxies. Detailed chemical abundances confirm this scenario. Low [$\alpha$/Fe] element ratios are observed in the stream stars, and lower limits of [Ba/Y] are ascertained, which sit well above the bulk component of the Milky Way for two of the three stream members. Thus, we present the first detailed chemical evidence that the parent of the Orphan stream is a dwarf galaxy.

On the basis of chemistry, we exclude the extended globular cluster NGC 2419 as a plausible parent to the Orphan stream. No firm link between Segue 1 and the Orphan stream has been identified. While the wide range in metallicities adds to the phase-space similarities between the two systems, the substantial difference in [$\alpha$/Fe] abundance ratios places doubt on any association. It appears the disrupting dwarf galaxy parent may reside in the southern sky, just waiting to be discovered.


\chapter{The Aquarius Moving Group is Not a Disrupted Classical Globular Cluster}
\label{chap:aquarius}


\previouslypublished{This chapter has been submitted to the Monthly Notices of the Royal Astronomical Society as \href{http://arxiv.org/abs/1309.3562}{`The Aquarius Co-Moving Group is Not a Disrupted Classical Globular Cluster'}, {Casey}, A.~R., {Keller}, S.~C., {Alves-Brito}, A., {Frebel}, A., {Da Costa}, G., {Karakas}, A., {Yong}, D., {Schlaufman}, K., {Jacobson}, H.~R., {Yu}, Q., {Fishlock}, C. 2013. The work is presented here in expanded form.} \\

We present a detailed analysis of high-resolution, high $S/N$ spectra for 5 Aquarius stream stars observed with the MIKE spectrograph on the Magellan Clay telescope. Our sample represents one third of the 15 known members in the stream. We find the stream is not mono-metallic: the metallicity ranges from [Fe/H] = $-$0.63 to $-$1.58. No anti-correlation in Na--O abundances is present, and we find a strong positive Mg--Al relationship, similar to that observed in the thick disk. We find no evidence that the stream is a result of a disrupted classical globular cluster, contrary to a previously published claim. High [(Na, Ni, $\alpha$)/Fe] and low [Ba/Y] abundance ratios in the stream suggests it is not a tidal tail from a disrupted dwarf galaxy, either. The stream is chemically indistinguishable from Milky Way field stars with the exception of one candidate, C222531-145437. From its position, velocity, and detailed chemical abundances, C222531-145437 is likely a star that was tidally disrupted from $\omega$-Centauri. We propose the Aquarius stream is Galactic in origin, and could be the result from a disk-satellite perturbation in the Milky Way thick disk on the order of a few Gyr ago: derived orbits, $UVW$ velocities, and angular momenta of the Aquarius members offer qualitative support for our hypothesis. Assuming C222531-145437 is a tidally disrupted member of $\omega$-Centauri, this system is the most likely disk perturber. In the absence of compelling chemical and/or dynamical evidence that the Aquarius stream is the tidal tail of a disrupted satellite, we advocate the ``Aquarius group'' as a more appropriate description. Like the Canis Major over-density, as well as the Hercules and Monoceros groups, the Aquarius group joins the list of kinematically-identified substructures that are not actually accreted material: they are simply part of the rich complexity of the Milky Way structure.

\section{Introduction}
Wide-field deep imaging surveys have proved excellent sources for finding stellar streams \citep[e.g., ][]{belokurov;et-al_2007}. Dozens of streams have been identified through careful photometric selections and matched-filtering techniques, with some to a Galactocentric distance of 100\,kpc \citep[e.g., see][]{drake;et-al_2013}. This suggests that a large fraction of the stellar halo has been built up by accretion. However, as \citet{Helmi;White_1999} point out, these detection strategies are most successful for identifying streams that are sufficiently distant from the solar neighborhood. A nearby stream, within $\sim$10\,kpc, will not appear as a photometric over-density because the stars would be sparsely positioned across the sky. Such substructures would only be detectable by their kinematics, or perhaps with precise elemental abundances through a ``chemical tagging'' approach (e.g., see \citealt[][]{freeman;bland-hawthorn_2002}). The confirmation of such substructures would serve to substantially increase the fraction of the known accreted material in the Galaxy.

It is therefore necessary to spectroscopically survey stars in the solar neighbourhood to reveal any nearby substructures. The RAVE team began such a survey in 2003 and has taken spectra of over 500,000 stars across 17,000\,deg$^{2}$. The primary goal of RAVE is to obtain radial velocities for stars in the solar neighbourhood and beyond. In an attempt to remain kinematically unbiased, RAVE candidates were selected solely by their apparent magnitude ($9 < I < 13$). Almost all have radial velocities published in the RAVE data releases \citep{steinmetz;et-al_2006}, and for a subset of stars with a sufficient $S/N$ ratio, stellar parameters have been derived by a $\chi^2$-minimisation technique \citep{zwitter;et-al_2008, siebert;et-al_2011}. 

Using these data, \citet{williams;et-al_2011} identified a co-moving group of nearby ($0.5\,\mbox{kpc} \lesssim D \lesssim 10\,\mbox{kpc}$) stars near ${(l, b) = (60^\circ, -55^\circ)}$, in the vicinity of the Aquarius constellation. Thus, the co-moving group was named the Aquarius stream. The stream is most apparent when examining heliocentric velocities against Galactic latitude for stars within $-70^\circ < b < -50^\circ$. \citet{williams;et-al_2011} employed a selection criteria of $-250\,\mbox{km s}^{-1} < V_{\rm hel} < -150\,\mbox{km s}^{-1}$, $30^\circ < l < 75^\circ$ and $J > 10.3$ to maximize the contrast between the stream and stellar background, identifying 15 stars in the process. The average heliocentric velocity of these members was found to be $V_{\rm hel} = -199\,\mbox{km s}^{-1}$, with a dispersion of 27\,km s$^{-1}$. The radial velocity uncertainties provided by the RAVE catalog are described to be $\sim$2\,km s$^{-1}$, so the stream's wide velocity distribution appears to be real.

Through a statistical comparison with predictions of stellar positions and kinematics from the Galaxia \citep{sharma;et-al_2011} and Besan\c{c}on \citep{robin;et-al_2003} models of the Milky Way, \citet{williams;et-al_2011} found the stream \ to be statistically significant ($>$4$\sigma$). The choice of model, cell dimension, or extinction rate made no substantial difference to the detection significance. The authors concluded the over-density was genuine, and inferred that the co-moving group is a stellar stream. Based on the phase space information available, \citet{williams;et-al_2011} concluded that the newly discovered stream could not be positively associated with the Sagittarius or Monoceros stream, the Hercules-Aquila cloud, or either the Canis Major or Virgo over-densities.

RAVE data suggest the Aquarius stream has a metallicity of [Fe/H] $= -1.0 \pm 0.4$\,dex\footnote{\citet{williams;et-al_2011} formally quote [M/H], but for the sake of a consistent discussion we assume ${{\rm [M/H]} \equiv {\rm [Fe/H]}}$ throughout this study.}, whereas field stars at the same distance show ${\mbox{[Fe/H]} = -1.1 \pm 0.6}$\,dex after the same selection cuts had been employed. Of the 15 Aquarius stream stars in the \citet{williams;et-al_2011} discovery sample, the metallicity range determined from medium-resolution spectroscopy is wide: from [Fe/H] = --2.02 to --0.33. High-resolution spectra with high $S/N$ are necessary to accurately characterise the stream's MDF.

To this end, \citet{wylie-de-boer;et-al_2012} obtained high-resolution ($\mathcal{R} = 25,000$) spectra with a modest $S/N$ ratio of $\sim$30 for six Aquarius stream stars using the echelle spectrograph on the Australian National University's 2.3m telescope. Their data indicate a surprisingly narrow spread in metallicity compared to previous work: ${\mbox{[Fe/H]} = -1.09 \pm 0.10}$\,dex, with a range extending only from $-1.25$ to {--0.98\,dex}. Samples with such small dispersions in metallicity are typically observed in mono-metallic environments (e.g., globular or open clusters). 

In addition to ascertaining stellar parameters, \citet{wylie-de-boer;et-al_2012} measured elemental abundances for the Aquarius stream stars -- the only study to date to do so. The authors primarily focussed on Na, O, Mg, Al, and Ni. These elements have been extensively studied in globular cluster stars, where unique abundance patterns are observed. Specifically, an anti-correlation between sodium and oxygen content appears ubiquitous to stars in globular clusters \citep{carretta;et-al_2009a}. \citet{wylie-de-boer;et-al_2012} identified two stream stars with slightly higher [Na/Fe] abundance ratios than halo stars of the same metallicity. No strong oxygen depletion was evident in the data, and no overall {Na-O} anti-correlation was present. \citet{wylie-de-boer;et-al_2012} also found [Ni/Fe] abundance ratios similar to thick disk/globular cluster stars, markedly higher than those reported for the Fornax dSph galaxy, which has a comparable mean metallicity to the Aquarius stream.

Combined with the low level of [Fe/H] scatter present in their sample, these chemical abundances led \citet{wylie-de-boer;et-al_2012} to conclude that the Aquarius stream is the result of a tidally disrupted globular cluster. We note, though, that \citet{williams;et-al_2011} previously excluded this scenario after modelling an Aquarius-like progenitor falling towards the Milky Way. The predicted positions and velocities from their simulations could not be reconciled with any known globular cluster, except for $\omega$-Centauri, although no explicit link was argued. Alternatively, any parent cluster may have been totally disrupted, leaving no identifiable remnant for discovery. 

We seek to investigate the nature of the Aquarius stream, specifically the  globular cluster origin claimed by \citet{wylie-de-boer;et-al_2012}. Details of the observations and data reduction  are outlined in the following section. The bulk of our analysis is presented in Section \ref{sec:aquarius-analysis} and our chemical abundance analysis is chronicled separately in Section \ref{sec:aquarius-chemical-abundances}. A detailed discussion of our results is made in Section \ref{sec:aquarius-discussion}, and we conclude in Section \ref{sec:aquarius-conclusions} with a summary of our findings and critical interpretations.

\begin{table}
\caption{Observations and radial velocities\label{tab:observations}}
\resizebox{\textwidth}{!}{%
\begin{tabular}{lcccccccccc}
\hline
\hline
Designation & $\alpha$ & $\delta$ & UT & UT & Airmass & Seeing & $t_{\rm exp}$ & $S/N$\tablenotemark{a} & $V_{\rm hel}$ \\
 & (J2000) & (J2000) & Date & Time &  & (\arcsec) & (secs) & (px$^{-1}$) & (km s$^{-1}$) \\
\hline
\\
\multicolumn{10}{c}{\textbf{Standard Stars}} \\
\hline
HD 41667	& 06:05:03.7 & $-$32:59:36.8	& 2011-03-13	& 23:40 & 1.01	& 1.0 & 90 	& 340 & 297.1 		\\
HD 44007	& 06:18:48.6 & $-$14:50:44.2 	& 2011-03-13	& 23:52 & 1.03	& 1.0 & 120	& 280 & 161.8 		\\
HD 76932	& 08:58:44.2 & $-$16:07:54.2	& 2011-03-14	& 00:16 & 1.16	& 1.0 & 25	& 330 & 117.8		\\
HD 136316	& 15:22:17.2 & $-$53:14:13.9	& 2011-03-14	& 09:37 & 1.12	& 0.9 & 120 	& 400 & $-$38.8 	\\
HD 141531	& 15:49:16.9 & $+$09:36:42.5	& 2011-03-14	& 09:52 & 1.31	& 0.9 & 120	& 350 & 2.8		\\
HD 142948	& 16:00:01.6 & $-$53:51:04.1	& 2011-03-14	& 09:45 & 1.11	& 1.0 & 90 	& 320 & 29.9 		\\
\hline
\\
\multicolumn{10}{c}{\textbf{Program Stars}} \\
\hline
C222531-145437	& 22:25:31.7 & $-$14:54:39.6	& 2011-07-30	& 06:52 & 1.03	& 0.8	& 650 & 135	& $-$156.4 \\
C230626-085103	& 23:06:26.6 & $-$08:51:04.8	& 2011-07-30	& 08:15 & 1.10	& 1.0	& 650 & 100	& $-$221.1 \\
J221821-183424	& 22:18:21.2 & $-$18:34:28.3	& 2011-07-30	& 05:58 & 1.03	& 0.9 	& 650 & 115	& $-$159.5 \\
J223504-152834	& 22:35:04.5 & $-$15:28:34.9	& 2011-07-30 & 07:34 & 1.05	& 1.0 	& 650 & 130	& $-$169.7 \\
J223811-104126	& 22:38:11.6 & $-$10:41:29.4	& 2011-07-30	& 08:57 & 1.22	& 1.2 	& 650 & 115	& $-$235.7  \\
\hline
\tablenotetext{a}{$S/N$ measured per pixel ($\sim$0.09\,{\AA} px$^{-1}$) at 600\,nm for each target.}
\end{tabular}}
\end{table}

\section{Observations \& Data Analysis}

The most complete sample of Aquarius stream stars is presented in the discovery paper of \citet{williams;et-al_2011}. We have obtained high-resolution, high $S/N$ spectra for 5 Aquarius stream candidates using the MIKE spectrograph \citep{bernstein;et-al_2003} on the Magellan Clay telescope. Although these observations were carried out independently of the \citet{wylie-de-boer;et-al_2012} study, by chance there are four stars common to both samples. The additional star in this sample, C2306265-085103, was observed by the RAVE survey but had a $S/N$ ratio too low for stellar parameters to be accurately determined. All program stars were observed in July 2011 in $\sim$1\arcsec{} seeing at low airmass (Table \ref{tab:observations}), and six standard stars were observed in March 2011. All observations were taken using a 1.0\arcsec{} slit without  spectral or spatial binning, providing a spectral resolution in excess of $\mathcal{R} = 28,000$ in the blue arm and $\mathcal{R} = 25,000$ in the red arm. The exposure time for our program stars was 650\,seconds per star in order to ensure a $S/N$ ratio in excess of 100\,pixel$^{-1}$ at 600\,nm. 

Calibration frames were taken at the start of each night, including 20 flat-field frames (10 quartz lamp, 10 diffuse flats) and 10 Th-Ar arc lamp exposures. The data were reduced using the CarPy pipeline\footnote{http://code.obs.carnegiescience.edu/mike}. For comparison purposes one of the standard stars, HD 41667, was also reduced using standard extraction and calibration methods in IRAF. The resultant spectra from both approaches were compared for residual fringing, $S/N$, and wavelength calibration. No noteworthy differences were present, and the CarPy pipeline was utilized for the remainder of the data reduction. Each reduced echelle order was carefully normalized using a cubic spline with defined knot spacing. Normalized orders were stitched together to provide a single one-dimensional spectrum from 333 to 916\,nm. A portion of normalised spectra for the program stars is shown in Figure \ref{fig:spectra}.
 
The white dwarf HR 6141 was observed in March 2011 as a telluric standard. The $S/N$ ratio for HR 6141 exceeds that of any of our standard or program stars. Although the atmospheric conditions at Las Campanas Observatory are certain to change throughout the night and between observing runs, we are primarily using this spectrum to identify stellar absorption lines that are potentially affected by telluric absorption. 

\section{Analysis}
\label{sec:aquarius-analysis}

\subsection{Radial Velocities}
\label{sec:aquarius-radial-velocities}

The radial velocity for each star was determined in a two step process. An initial estimate of the radial velocity was ascertained by cross-correlation with a synthetic spectrum of a giant star with $T_{\rm eff} = 4500$\,K, $\log{g} = 1.5$, and ${\mbox{[Fe/H]} = -1.0}$ across the wavelength range 845 to 870\,nm. The observed spectrum was shifted to the pseudo-rest frame using this initial velocity estimate. Equivalent widths (EWs) were measured for $\sim$160 atomic transitions by integrating fitted Gaussian profiles (see Section \ref{sec:aquarius-line-measurements}). In each case a residual line velocity was calculated from the expected rest wavelength and the measured wavelength. The mean residual velocity offset correction is small in all cases ($<$1\,km s$^{-1}$), and this residual correction is applied to the initial velocity measurement from cross-correlation. The final heliocentric velocities are listed in Table \ref{tab:observations}, where the typical uncertainty is $\pm0.1$\,km s$^{-1}$. These velocities agree quite well with those compiled by \citet{williams;et-al_2011} as part of the RAVE survey: the mean offset is ${\langle\Delta{V}\rangle = 2.5 \pm 2.7}$\,km s$^{-1}$, demonstrating the accuracy of the velocities published in the RAVE catalog.

\subsection{Line Measurements}
\label{sec:aquarius-line-measurements}


For the measurement of atomic absorption lines, we employed the line list of \citet{yong;et-al_2005} with additional transitions of Cr, Sc, Zn, and Sr from \citet{roederer;et-al_2010}. The list has been augmented with molecular CH data from \citet{plez;et-al_2008}. Isotopic and hyperfine splitting data was taken from \citet{sneden}. For molecular features (e.g., CH), or lines with hyperfine and/or isotopic splitting (Sc, V, Mn, Co, Cu, Ba, La, Eu), we determined the abundance using spectral synthesis with the relevant data included. For all other transitions, abundances were obtained using the measured EWs.

\begin{figure}[t]
	\includegraphics[width=\textwidth]{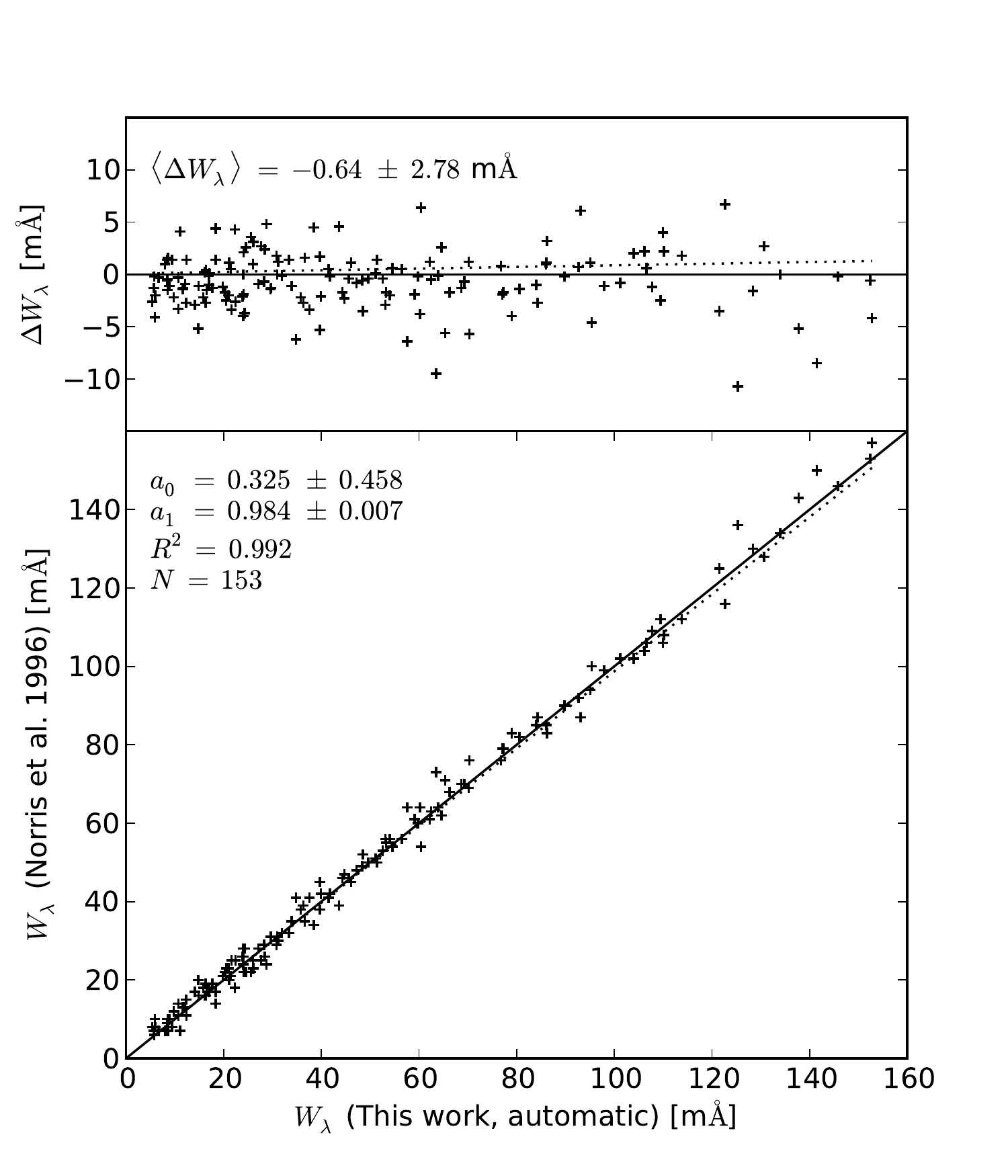}
	\caption[Comparison of equivalent widths measured automatically for HD 140283 against manual measurements by \citet{norris;et-al_1996}]{Comparison showing equivalent widths measured for HD\,140283 using our automatic routine (see Section \ref{sec:aquarius-line-measurements}), and manual measurements by \citet{norris;et-al_1996}. No systematic trend is present, and the mean difference between these studies is ${\langle\Delta{}W_\lambda\rangle = -0.64 \pm 2.78}$\,m\AA{}. The offset ($a_0$) and the slope ($a_1$) of the fit are shown.}
	\label{fig:ew-compare}
\end{figure}

\begin{figure*}[t!]
	\includegraphics[width=\textwidth]{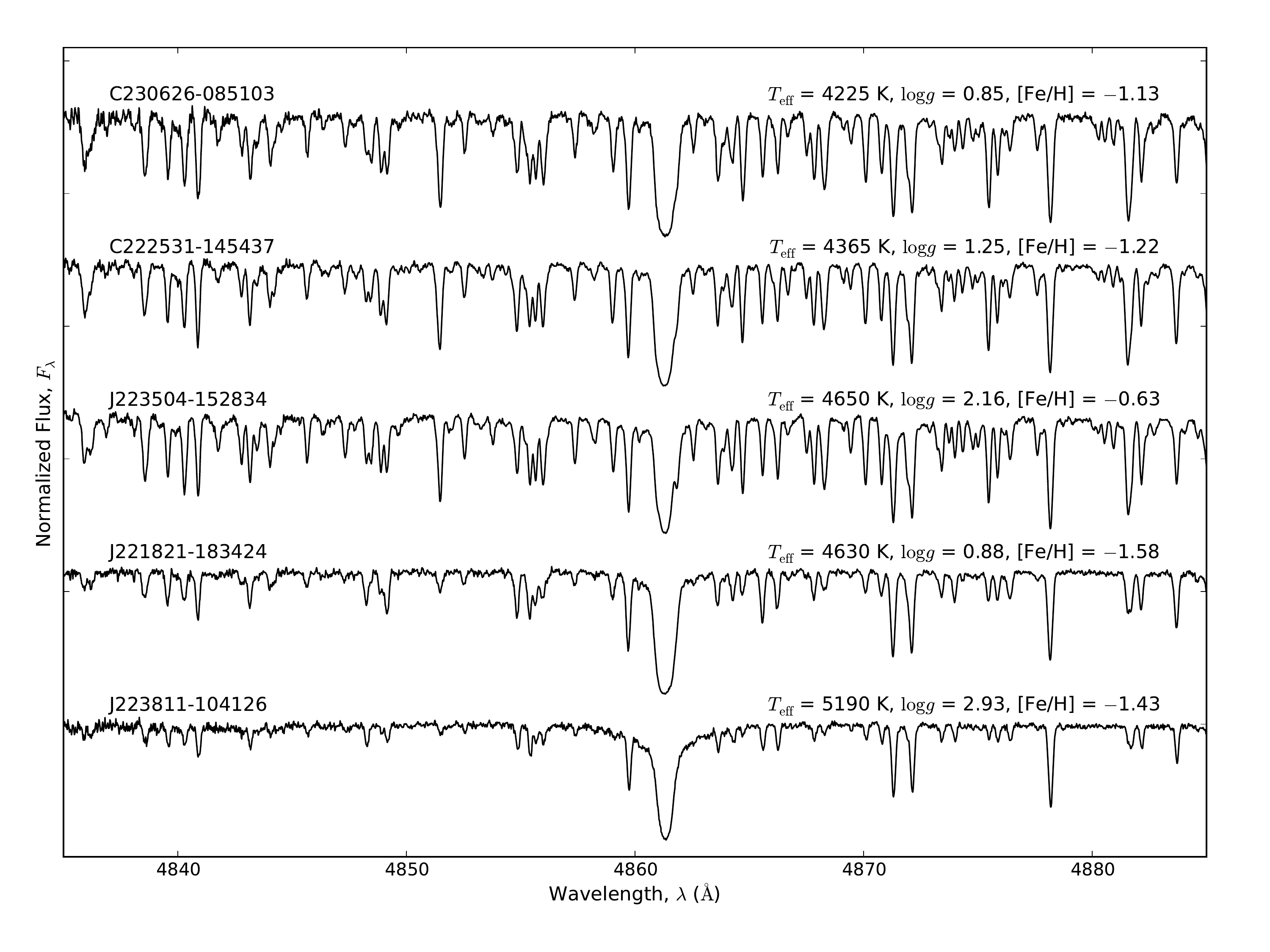}
	\caption[Normalized rest-frame spectra surrounding the H-$\beta$ absorption line for all Aquarius stars]{Normalized rest-frame spectra surrounding the H-$\beta$ absorption line for all Aquarius stream candidates with offset fluxes. The effective temperature, surface gravity and metallicity is shown for all stars.}
	\label{fig:spectra}
\end{figure*}

The EWs for all absorption lines were measured automatically using software written during this study. The local continuum surrounding every atomic transition is determined, and a Gaussian profile is iteratively fit to the absorption feature of interest. Our algorithm accounts for crowded or blended regions by weighting pixels as a function of difference to the rest wavelength. For this study we have verified our approach by comparing EWs of 156 lines measured by hand and tabulated in \citet{norris;et-al_1996}. We only included measurements in the \citet{norris;et-al_1996} study that were not marked with questionable line quality parameters. Excellent agreement is found between the two studies, which is illustrated in Figure \ref{fig:ew-compare}. The mean difference is a negligible ${-0.64 \pm 2.78}$\,m\AA{}, and no systematic trend is present. The scatter can be attributed to the lack of significant digits in the \citet{norris;et-al_1996} study, as well as the $S/N$ of the data. Other studies \citep[e.g.][]{frebel;et-al_2013} using the same algorithm used here find better agreement for spectra with higher $S/N$ ratios: $0.20 \pm 0.16$\,m{\AA} when we compare our results with manual measurements by \citet{aoki;et-al_2007}, and a difference of $0.25 \pm 0.28$\,m{\AA} is found between manual measurements by \citet{cayrel;et-al_2004} and our automatic results. Although we are extremely confident in our EW measurements, \textit{every} absorption profile was repeatedly examined by eye for quality, and spurious measurements were removed.

We list the atomic data and measured EWs in Appendix \ref{app:aquarius_ews}. Transitions near the flat portion of the curve-of-growth have been excluded by removing measurements with ${REW > -4.5}$. A minimum detectable EW was calculated as a function of wavelength, $S/N$ and spectral resolution by following \citet{norris;et-al_2001}, and only lines that exceeded a 3$\sigma$ detection significance were included for this analysis. 

\subsection{Model Atmospheres}
We have employed the ATLAS9 plane-parallel stellar atmospheres of \citet{castelli;kurucz_2003}. These one-dimensional models ignore any center-to-limb spatial variations, assume hydrostatic equilibrium and no convective overshoot from the photosphere. The stellar parameter spacing between models is 250\,K in temperature, 0.5\,dex in surface gravity, 0.5\,dex in [M/H] and 0.4\,dex in [$\alpha$/Fe]. We interpolated the temperature, gas and radiative pressure, electron density and opacities between atmosphere models using the Quickhull algorithm \citep{barber;et-al_1996} (see Section \ref{sec:smh-model-atmospheres}).

\subsection{Stellar Parameters}
\label{sec:aquarius-stellar-parameter-derivation}
The May 2011 version of the MOOG \citep{sneden_1973} spectral synthesis code has been used to derive individual line abundances and stellar parameters. This version employs Rayleigh scattering \citep{sobeck;et-al_2011} instead of treating scattering as true absorption, which is particularly important for transitions blue-ward of 450\,nm. This is noteworthy, but is less relevant for these analyses as most of the atomic transitions utilized here are red-ward of 450\,nm.

\subsubsection{Effective Temperature}
\label{sec:aquarius-effective-teffs}
The effective temperature, $T_{\rm eff}$, for each star was found by demanding a zero-trend in excitation potential and line abundance for measurable Fe\,I transitions. The data were fitted with a linear slope, and gradients less than $|10^{-3}|$\,dex eV$^{-1}$ were considered to be converged. For comparison, photometric temperatures were calculated after our spectroscopic temperatures had been derived, and these are discussed in Section \ref{sec:aquarius-photometric-temperatures}.

\subsubsection{Microturbulence and Surface Gravity}
The microturbulence for each star was found by forcing a zero-trend in the REW and abundance for Fe\,I lines. Similar to the effective temperature, linear slopes in REW and abundance of less than $|10^{-3}|$\,dex were considered converged. The surface gravity for all stars was found by forcing the mean Fe\,I and Fe\,II abundances to be equal. A tolerance of $|\langle$Fe\,I$\rangle - \langle$Fe\,II$\rangle| \leq 0.05$ was deemed acceptable. The process is iterative: a zero trend with the excitation potential, REW and abundances must be maintained. A solution was only adopted when the all criteria were simultaneously satisfied.

\subsubsection{Metallicity}
The model atmosphere metallicity was exactly matched to that of our mean Fe\,I abundance. Individual Fe line abundances that were unusually deviant (e.g., $>$3$\sigma$) from the mean abundance were removed. The largest number of outlier measurements removed for any observation was nine for C222531-145437. These were transitions near the flat part of the curve-of-growth with REWs $\sim -4.5$, leaving 60 Fe\,I and 10 Fe\,II lines for the analysis of C222531-145437. Usually only one outlier measurement was removed for the other candidates. The minimum number of Fe transitions employed for stellar parameter determination was 42 lines (33 Fe\,I and 9 Fe\,II), which occurred for our hottest star, J223811-104126.

\begin{table*}[t!]
\caption{Stellar parameters for standard and program stars\label{tab:stellar-parameters}}
\resizebox{\textwidth}{!}{%
\begin{tabular}{lccccccccccccccc}
\hline
\hline
& \multicolumn{4}{c}{{\bf This Study}} && \multicolumn{5}{c}{{\bf Literature}} \\
\cline{2-5} \cline{7-11}
Designation & $T_{\rm eff}$ & $\log{g}$ & $\xi_t$ & [Fe/H] & & $T_{\rm eff}$ & $\log{g}$ & $\xi_t$ & [Fe/H] & Reference \\
& (K) & (dex) & (km\,s$^{-1}$) & (dex) && (K) & (dex) & (km s$^{-1}$) & (dex) \\
\hline
\\
\multicolumn{11}{c}{\textbf{Standard Stars}} \\
\hline
HD\,41667		& 4660	& 1.71	& 1.84 	& $-$1.20&
				& 4605	& 1.88	& 1.44	& $-$1.16
				& \citet{gratton;et-al_2000} \\
HD\,44007		& 4835	& 1.78	& 1.95	& $-$1.77&
				& 4850	& 2.00 	& 2.20	& $-$1.71
				& \citet{fulbright_2000} \\
HD\,76932		& 5800 & 3.88 & 1.65 & $-$1.05&
				& 5849 & 4.11 & \nodata & $-$0.88 
				& \citet{nissen;et-al_2000} \\
HD\,136316		& 4355 & 0.58 & 2.06 & $-$1.93&
				& 4414 & 0.94 & 1.70 & $-$1.90 
				& \citet{gratton_sneden_1991} \\
HD\,141531		& 4345 & 0.63 & 2.07 & $-$1.69 &
				& 4280 & 0.70 & 1.60 & $-$1.68
				& \citet{shetrone_1996} \\
HD\,142948		& 5025	& 2.25	& 2.05	& $-$0.74&
				& 4713 	& 2.17 	& 1.38	& $-$0.77
				& \citet{gratton;et-al_2000} \\
\hline \\

\multicolumn{11}{c}{\textbf{Program Stars}} \\
\hline 

C222531-145437	& 4365				& 1.25				& 1.94				& $-$1.22	&
				& 4235\,$\pm$\,118 	& 1.45\,$\pm$\,0.21	& 1.96\,$\pm$\,0.11 	& $-$1.20\,$\pm$\,0.14 
				& \citet{wylie-de-boer;et-al_2012} \\
C230626-085103	& 4225	& 0.85	& 1.92 	& $-$1.13&
				& \dots	& \dots	& \dots	& \dots
				& \dots \\	
J221821-183424	& 4630				& 0.88				& 2.16				& $-$1.58&
				& 4395\,$\pm$\,205 	& 1.45\,$\pm$\,0.35 	& 1.96\,$\pm$\,0.18 	& $-$1.15\,$\pm$\,0.21
				& \citet{wylie-de-boer;et-al_2012} \\
J223504-152834	& 4650				& 2.16 				& 1.55				& $-$0.63&
				& 4597\,$\pm$\,158 	& 2.40\,$\pm$\,0.14 	& 1.47\,$\pm$\,0.07 	& $-$0.98\,$\pm$\,0.17
				& \citet{wylie-de-boer;et-al_2012} \\
J223811-104126	& 5190				& 2.93				& 1.62				& $-$1.43&
				& 5646\,$\pm$\,147 	& 4.60\,$\pm$\,0.15 	& 1.09\,$\pm$\,0.11 	& $-$1.20\,$\pm$\,0.20
				& \citet{wylie-de-boer;et-al_2012} \\
\hline
\end{tabular}}
\end{table*}

\subsubsection{Photometric Effective Temperatures}
\label{sec:aquarius-photometric-temperatures}
 
As a consistency check for our spectroscopic temperatures, we have estimated effective temperatures using the colour-$T_{\rm eff}$ empirical relationship for giant stars from \citet{ramirez;melendez_2005}. The $V-K$ colour has been employed as its calibration has the lowest residual fit. This relationship has a slight dependence on metallicity, and as such we have adopted the spectroscopic [Fe/H] values in Table \ref{tab:stellar-parameters} for these calculations. Optical $V$-band photometry has been sourced from the APASS catalogue \citep{henden;et-al_2012}, and $K$-band magnitudes have been sourced from the 2MASS catalog \citep{skrutskie;et-al_2006}. The reddening maps of \citet{schlegel;et-al_1998} estimate that the extinction for our stars varies between ${E(B-V) = 0.03}$ to {0.07\,mags}, and these values have been used to de-redden our $V-K$ colour.

\begin{table*}[h]
\centering
\caption{Reddening \& photometric temperatures for program stars\label{tab:photometric-temperatures}}
\begin{tabular}{lccccccc}
\hline
\hline
Designation & $E(B-V)$ & $(V-K)_0$ & $T_{\rm phot}$ & $T_{\rm spec}$ & $\Delta{}T$ \\
	& (mag) & (mag) & (K) & (K) & (K) \\
\hline
C222531-145437 	& 0.03\phn	& 	2.86 & 4285	& 4365 & $-$80 \\	
C230626-085103		& 0.05\phn	&	3.00 & 4196	& 4225 & $-$29 \\	
J221821-183424 	& 0.03\phn	&	2.36 & 4685	& 4630 &   +55  \\	
J223504-152834 	& 0.04\phn	&	2.50 & 4557	& 4650 & $-$93 \\	
J223811-104126 	& 0.07\phn	&	1.84 & 5240	& 5190 &   +50  \\
\hline
\end{tabular}
\end{table*}

Calculated photometric temperatures are listed in Table \ref{tab:photometric-temperatures}. The mean difference between the photometric temperatures and those found by excitation balance is $-$19\,K, where the largest variation is $-$93\,K for J223504-152834. While these photometric temperatures serve as a confirmation for our spectroscopically-derived values, for the remainder of this analysis we have employed effective temperatures determined by excitation balance.

\subsection{Uncertainties in Stellar Parameters}
\label{sec:aquarius-uncertainties-in-stellar-parameters}
Due to scatter in neutral iron lines measurements, there is a formal uncertainty in our calculated trend line between excitation potential and abundance, as well as between the reduced equivalent width and abundance. We have calculated 1$\sigma$ uncertainties in effective temperature and microturbulence by independently varying each stellar parameter until the relevant slope matches that formal uncertainty. This process is repeated for positive and negative offsets in temperature and microturbulence to allow for asymmetric uncertainties. The largest absolute offset is taken as the $1\sigma$ uncertainty. For surface gravity, the uncertainty has been calculated by varying $\log{g}$ until the difference in mean Fe\,I -- Fe\,II abundance matches the standard error about the mean for Fe\,I and Fe\,II in quadrature. The calculated uncertainties are tabulated in Table \ref{tab:stellar-parameter-uncertainties}. 

These uncertainties ignore any correlations between stellar parameters, and therefore are likely to be under-estimated. As such, we have assumed the total uncertainty in stellar parameters to be $\sigma(T_{\rm eff}) = \pm125$\,K, $\sigma(\log{g}) = \pm0.30$\,dex, and $\sigma(\xi_t) = \pm0.20$ km s$^{-1}$. These adopted uncertainties are higher than those listed in Table \ref{tab:stellar-parameter-uncertainties}, and can be regarded as extremely conservative.

\begin{table*}[h!]
\centering
\caption{Uncorrelated uncertainties in stellar parameters for standard and program stars\label{tab:stellar-parameter-uncertainties}}
\begin{tabular}{lccc}
\hline
\hline
Designation & $\sigma(T_{\rm eff})$ & $\sigma(\xi_t)$ & $\sigma(\log{g})$ \\
 & (K) & (km s$^{-1}$) & (dex) \\
\hline
HD\,41667  	&    53   	&   0.09   &   0.13   \\
HD\,44007 		&    81   	&   0.29   &   0.09   \\
HD\,76932 		&   107  	&   0.08   &   0.19   \\
HD\,136316		&    33   	&   0.15   &   0.12   \\
HD\,141531		&    25   	&   0.05   &   0.13   \\
HD\,142948		&    47   	&   0.10   &   0.11   \\
J221821-183424 	&    42   	&   0.11   &   0.09   \\
C222531-145437  	&    46   	&   0.05   &   0.12   \\
J223504-152834  	&    61   	&   0.08   &   0.05   \\
J223811-104126  	&    49   	&   0.16   &   0.08   \\
C230626-085103  	&    52   	&   0.05   &   0.04   \\
\hline
\end{tabular}
\end{table*}

\subsection{Distances}
\label{sec:aquarius-distances}

\begin{table*}[t!]
\caption{Parameters and uncertainties for Monte-Carlo realisations\label{tab:monte-carlo-data}}
\resizebox{\textwidth}{!}{%
\begin{tabular}{lccccccccc}
\hline
\hline
& \multicolumn{7}{c}{{\bf Input Parameters for Monte-Carlo Simulation}} && {\bf Output} \\
\cline{2-8} \cline{10-10}
Designation & $T_{\rm eff}$ & $\log{g}$ & $J$ & $E(B-V)$ & $V_{\rm hel}$ & $\mu_\alpha$ & $\mu_\delta$ && $D$ \\
& (K) & (dex) & (mag) & (mag) & (km s$^{-1}$) & (mas yr$^{-1}$) & (mas yr$^{-1}$) & &(kpc) \\
\hline
C222531-145437 & $4365 \pm 125$ & $1.25 \pm 0.20$ & $10.341 \pm 0.022$ & $0.03 \pm 0.01$ & $-156.4 \pm 0.1$ & \phn\phn\,$3.5 \pm 2.1$ & $-14.7 \pm 2.2$ &\,& $5.1^{+1.1}_{-0.8}$\\
C230626-085103 & $4225 \pm 125$ & $0.85 \pm 0.20$ & $10.312 \pm 0.025$ & $0.05 \pm 0.01$ & $-221.1 \pm 0.1$ & \phn$-2.5 \pm 2.8$ & $-15.4 \pm 2.7$ &\,& $6.5^{+1.4}_{-1.1}$\\
J221821-183424 & $4630 \pm 125$ & $0.88 \pm 0.20$ & $10.340 \pm 0.021$ & $0.03 \pm 0.01$ & $-159.5 \pm 0.1$ & $-10.6 \pm 2.5$ & $-19.3 \pm 2.5$ &\,& $5.6^{+1.3}_{-0.9}$ \\
J223504-152834 & $4650 \pm 125$ & $2.16 \pm 0.20$ & $10.363 \pm 0.025$ & $0.04 \pm 0.01$ & $-169.7 \pm 0.1$ & \phn\,$15.9 \pm 2.2$ & $-12.8 \pm 2.2$ &\,& $1.9^{+0.5}_{-0.4}$ \\
J223811-104126 & $5190 \pm 125$ & $2.93 \pm 0.20$ & $10.420 \pm 0.018$ & $0.07 \pm 0.01$ & $-235.7 \pm 0.1$ & $-25.3 \pm 2.1$ & $-99.5 \pm 2.1$ &\,& $1.1^{+0.3}_{-0.2}$ \\
\hline
\end{tabular}}
\end{table*}

Many groups have determined distances for stars in the RAVE survey catalog, which includes all Aquarius stream members. \citet{williams;et-al_2011} tabulated a range of distances inferred by different techniques. Not every measurement technique was applicable to all Aquarius stars. The reduced proper motion distance technique was the only method to estimate distances for all Aquarius candidates. The variations between distance measurements are large. In particular, the distance for {C222531-146537} ranged from $1.4 \pm 0.1$\,kpc \citep{burnett_binney_2010} to $10.3 \pm 2.4$\,kpc \citep{breddels;et-al_2010}, where both groups claim to have the ``most likely'' distances.

\begin{figure}[h]
	\includegraphics[width=\textwidth]{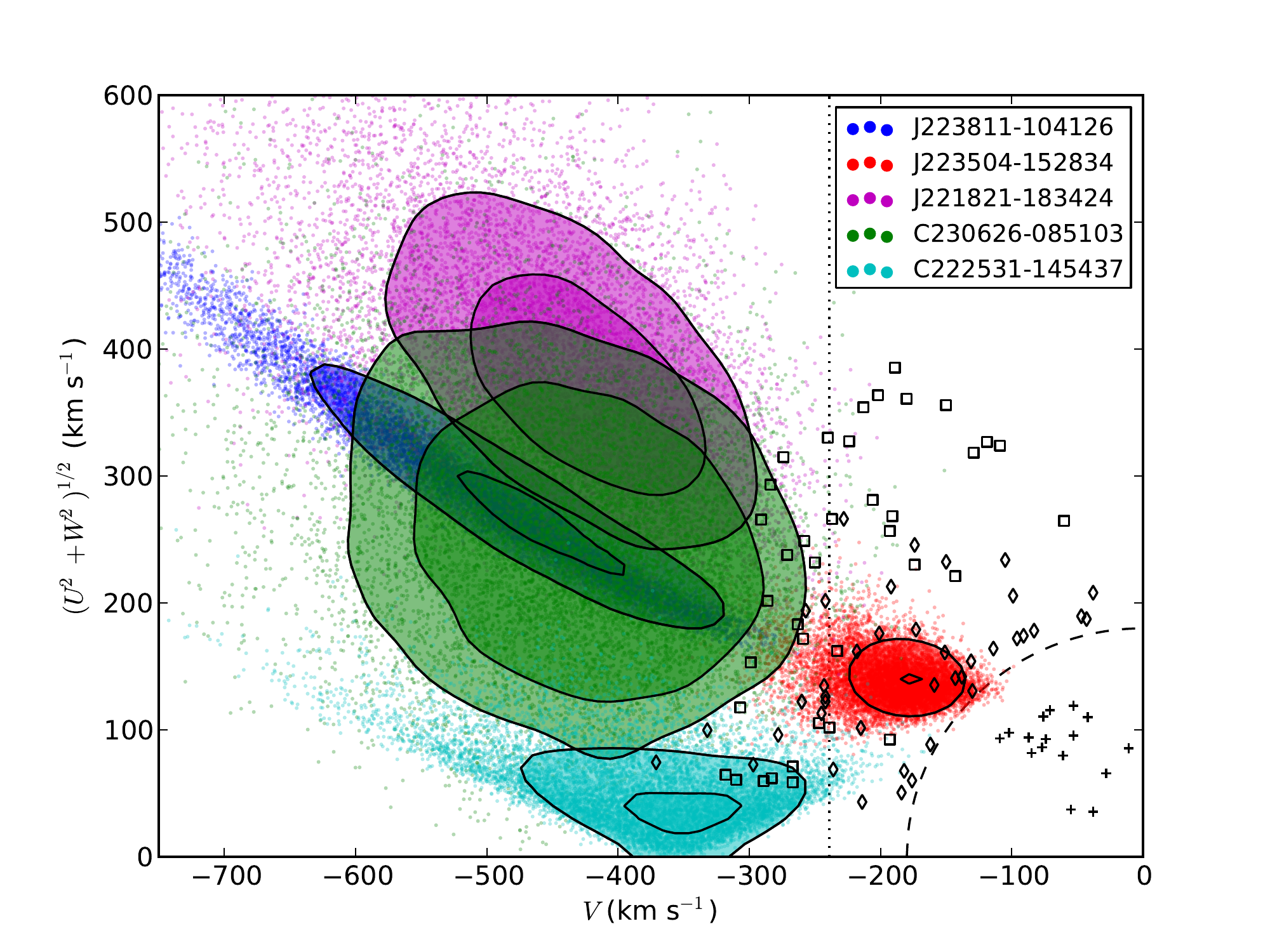}
	\caption[Galactic plane rotational planar and out-of-plane velocities for Aquarius stars]{Galactic plane rotational velocities versus out-of-plane total velocities. The contours of each star represent the 68\% and 95\% confidence intervals from 10,000 Monte-Carlo realisations of the parameter distributions shown in Table \ref{tab:monte-carlo-data}. A sample of thick disk data from \citet{nissen;schuster_2010} is shown ($+$), as well as their high- and low-alpha halo populations ($\diamond$ and $\circ$ respectively).}
	\label{fig:toombre}
\end{figure}

Using the stellar parameters tabulated in Table \ref{tab:stellar-parameters}, we have calculated distances by isochrone fitting. The \citet{Dotter;et-al_2008} $\alpha$-enhanced isochrones were used for these calculations, and an age of 10\,Gyr was assumed for all stars \citep{williams;et-al_2011,wylie-de-boer;et-al_2012}. The closest point to the isochrone was found by taking the uncertainties in $T_{\rm eff}$ and $\log{g}$ (see Section \ref{sec:aquarius-uncertainties-in-stellar-parameters}) into account and measuring the distance modulus in the $J$ band. Given the (i) number of uncertain measurements involved in calculating distances ($T_{\rm eff}$, $\log{g}$, $E(B-V)$, $J$) and (ii) the resultant asymmetric uncertainties, distances were determined from 10,000 Monte-Carlo realisations. Table \ref{tab:monte-carlo-data} lists the input parameters and uncertainties adopted for the Monte-Carlo realisations, as well as the emergent distances and uncertainties. Uncertainties in input parameters were assumed to be normally distributed. Of the distance scales collated in \citet{williams;et-al_2011}, our distances are in most agreement with the \citet{zwitter;et-al_2010} system. In fact, we find the best agreement with the mean of all the distance scales tabulated in \citet{williams;et-al_2011}. The uncertainties in our distance determinations are on the order of twenty per cent.

\subsection{Dynamics}

Velocity vectors and Galactic orbits have been determined in the same Monte-Carlo realisations outlined in Section \ref{sec:aquarius-distances}, which includes uncertainties in distances, proper motions\footnote{The proper motions in Table 1 of \citet{williams;et-al_2011} are erroneous in that they are associated with the wrong stars. The error was typographical and did not affect the transverse velocity calculations (M.E.K. Williams, private communication). The proper motions listed in our Table \ref{tab:monte-carlo-data} are correct.} and heliocentric velocities. We assume no uncertainty in on-sky position $(\alpha, \delta)$. Orbital energy calculations have assumed a three-component (bulge, disk, halo) model of the Galactic potential that reasonably reproduces the Galactic rotation curve. The bulge is represented by a Hernquist potential:

\begin{equation}
	\Phi_{\rm bulge}(x, y, z) = \frac{GM_b}{r + a}
\end{equation}

\noindent where $a = 0.6$\,kpc. The disk is modelled as a Miyamoto-Nagai potential \citep{miyamoto;nagai_1975} where:

\begin{equation}
	\Phi_{\rm disk}(x, y, z) = \frac{GM_{\rm disk}}{\sqrt{x^2 + y^2 + (b + \sqrt{z^2 + c^2})^2}}
\end{equation}

\noindent with $b = 4.5$\,kpc and $c = 0.25$\,kpc and the Galactic halo is represented by a Navarro-Frenk-White dark matter halo \citep{navarro;et-al_1997}:

\begin{equation}
	\Phi_{\rm halo} = -\frac{GM_{\rm vir}}{r\left[\log{(1 + c)} - c/(1 + c)\right]}\log{\left(1 + \frac{r}{r_s}\right)}
\end{equation}

\noindent{}with the three components scaled such that the disk provides 85\% of the radial force at $R_{\rm GC}$, in order to yield a flat circular-speed curve at $R_{\rm GC}$. The solar motion of \citet{schonrich;et-al_2012} has been adopted, where $R_{\rm GC}$ = 8.27\,kpc and a circular velocity speed $V_c = 238$\,km s$^{-1}$.

\begin{figure}[h!]
	\includegraphics[width=\textwidth]{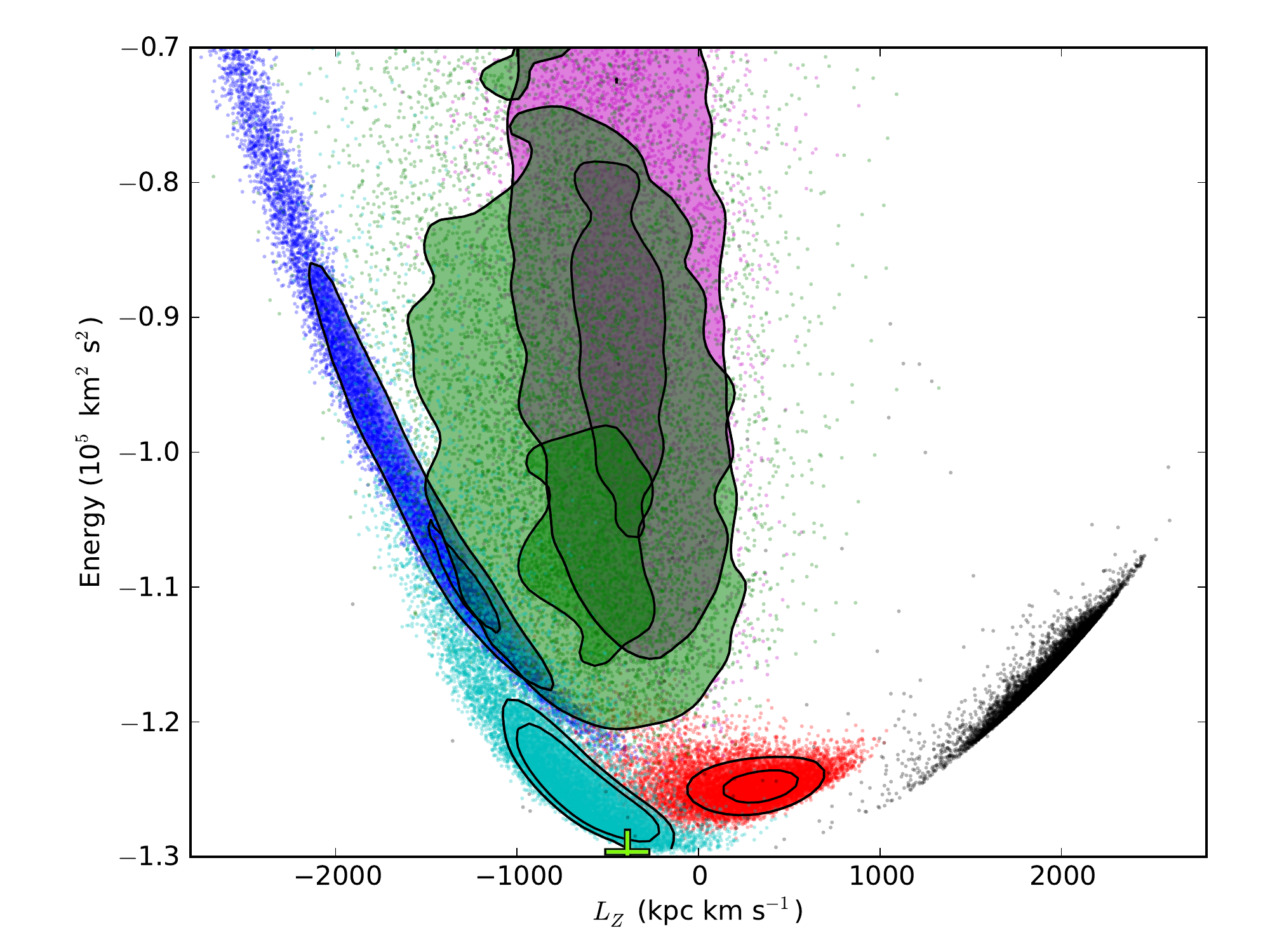}
	\caption[Linblad diagram showing angular momenta and orbital energies for Aquarius stars]{A Linblad ($L_Z-E$) diagram showing angular momenta and orbital energies after 10,000 Monte-Carlo realisations for each Aquarius stream star. Iso-contours represent the 68\% and 95\% confidence intervals. $\omega$-Centauri is shown as a lime green marker \citep{wylie-de-boer;et-al_2010}. The black points without contours are from the Geneva-Cophenagen Survey sample \citep{nordstrom;et-al_2004}, which primarily consists of nearby disk stars and serves as a validation of our orbital energy calculations. Colors are the same as in Figure \ref{fig:toombre}.}
	\label{fig:orbits}
\end{figure}

The Aquarius stream members have bound orbits, all of which are probably retrograde except for {J223504-152834} (Figure \ref{fig:toombre}). Orbital energies and angular momenta from Monte-Carlo simulations are illustrated in Figure \ref{fig:orbits}. The 16,686 stars from the Geneva-Cophenhagen Survey sample \citep{nordstrom;et-al_2004} are also shown as a reference, which primarily consists of nearby disk stars.

\section{Chemical Abundances}
\label{sec:aquarius-chemical-abundances}
We have scaled our chemical abundances to Solar values using the chemical composition described in \citet{asplund;et-al_2009}. The abundances for the standard and program stars are shown in Tables \ref{tab:standard-star-abundances} and \ref{tab:program-star-abundances}, respectively. The discussion of comparable elements are grouped accordingly. 

\subsection{Carbon}
Carbon is produced by the triple-$\alpha$ process and ejected through supernovae events, or by mass-loss from asymptotic giant branch (AGB) stars \citep{kobayashi;et-al_2011}. 

\begin{figure}[h!]
	\includegraphics[width=\textwidth]{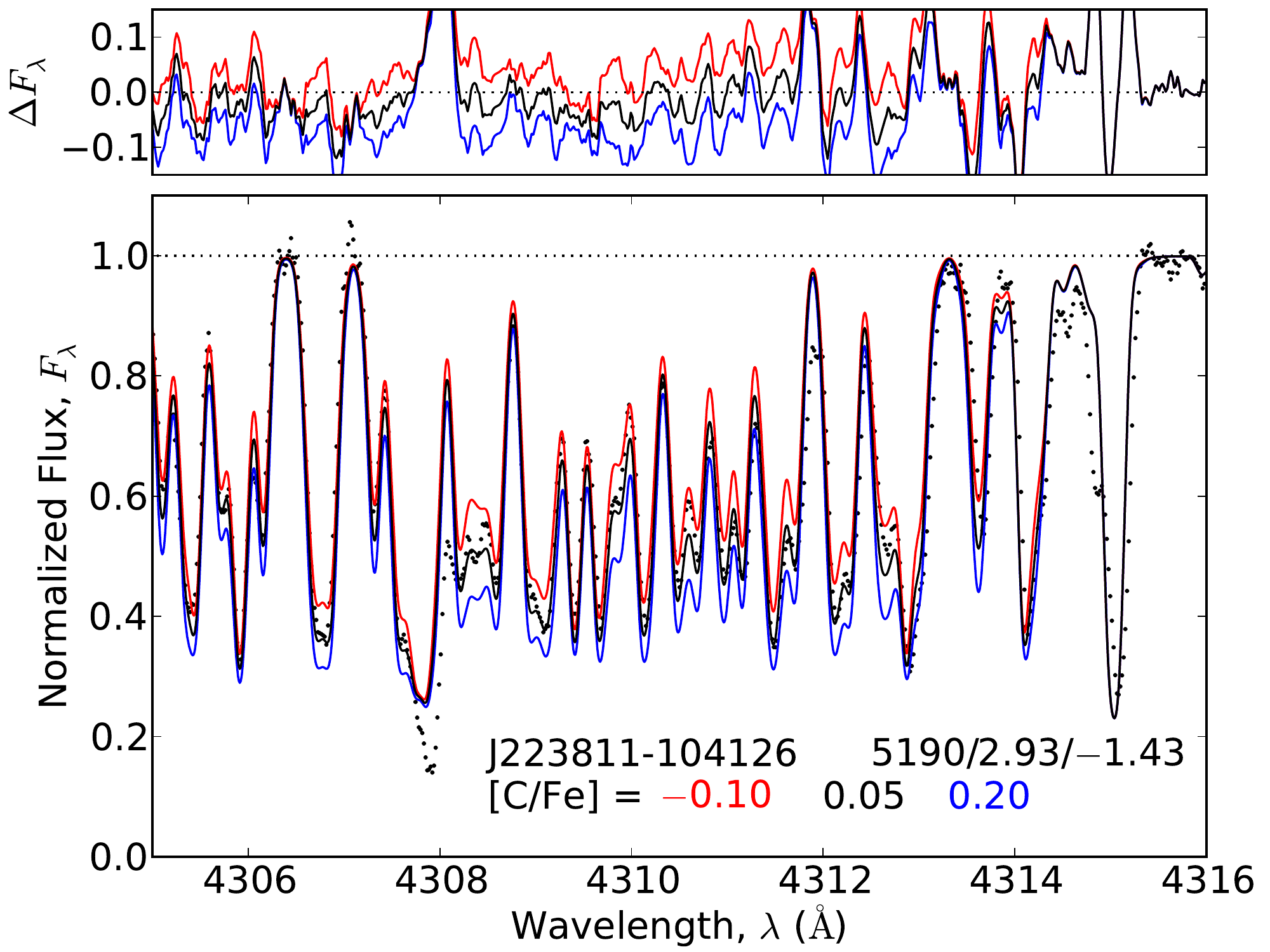}
	\caption[Best-fitting synthesised and observed spectra surrounding the CH feature near 431.3\,nm for J223811-104126]{The carbon CH feature near 431.3\,nm in program star J223811-104126. The best-fit synthetic spectra is shown, with synthetic spectra for $\pm0.15$\,dex about the best-fitting abundance.}
	\label{fig:carbon}
\end{figure}

We have measured carbon abundances for all stars from the G-band head near 431.3\,nm and the CH molecular feature at 432.3\,nm, by comparing observed spectra with synthetic spectra for different carbon abundances. The synthetic spectra were convolved with a Gaussian kernel where the width was determined from nearby atomic lines with known abundances. Carbon was measured separately for both features, and in all stars the two measurements agree within 0.10\,dex. An example fit to this spectral region for J223811-104126 is shown in Figure \ref{fig:carbon}.

Carbon abundances in our standard stars agree well with the literature. For HD\,136316 we find {${\rm [C/Fe]} = -0.50 \pm 0.15$}, where \citet{gratton;et-al_2000} find {[C/Fe] $= -0.66$}. Our ${{\rm [C/Fe]} = -0.48}$ measurement for HD\,141531 agrees with \citet{gratton;et-al_2000} to within 0.06\,dex. Most program stars have near-solar carbon abundances, ranging from ${{\rm [C/Fe]} = -0.30}$ for {J221821-183424}, and +0.05 for {J223811-104126}.

\subsection{Sodium and Aluminium}
\label{sec:aquarius-sodium-abundances}
Our line list includes three clean, unblended sodium lines at $\lambda$5688, $\lambda$6154 and $\lambda$6161. Not all three of these lines were detectable in each star. In the hottest and most metal-poor stars, J223811-104126 and J221821-183424 respectively, only the $\lambda$5688 line was measurable. For stars where multiple sodium lines were available, the line-to-line scatter is usually around 0.04\,dex with a maximum of 0.09\,dex in HD\,41667. However, in calculating total abundance uncertainties (see Section \ref{sec:aquarius-chemical-abundance-uncertainties}) we have conservatively assumed a minimum random scatter of $\pm0.10$\,dex for all stars.

Our [Na/Fe] abundances appear systematically higher than values found in the literature by $\sim$0.10\,dex. For HD\,142948 we find {[Na/Fe] = 0.22}, which is +0.10\,dex higher than that found by \citet{gratton;et-al_2000}, and similarly we find HD\,76932 to be +0.10\,dex higher than reported by \citet{fulbright_2000}. \citet{gratton;et-al_2000} also found HD\,136316 to have {[Na/Fe] = $-$0.29}, where we find {[Na/Fe] = $-0.14$}, yet excellent agreement is found in the stellar parameters in \citet{gratton;et-al_2000} and this study. Different solar compositions employed between this study and earlier work can account for $\sim$0.08\,dex of this effect, leaving the residual difference well within the observational uncertainties. However, it is important to note that the [Na/Fe] abundance ratios presented in this study may be slightly higher compared to previous studies. While a systematic offset may be present, no intrinsic abundance dispersion in [Na/Fe] is present in the Aquarius sample.

There are six aluminium transitions in our optical spectra. The strongest of these lines occur at $\lambda$3944 and $\lambda$3961 and are visible in all of our stars. However this is a particularly crowded spectral region: the lines fall between the strong Ca H and K lines, with the $\lambda$3961 transition clearly located in the wing of the Ca H line. Additionally, the $\lambda$3944 and $\lambda$3961 lines have appreciable departures from the assumption of LTE, resulting in under-estimated abundances by up to $\sim$0.6\,dex \citep{baumueller;gehren_1997}. Instead, we have measured Al abundances from other available transitions: the {Al\,I} lines at $\lambda$6696, $\lambda$6698, $\lambda$7835 and $\lambda$7836. Generally the four {Al\,I} lines are in reasonable agreement with one another, yielding random scatter of less than 0.05\,dex. 

\subsection{$\alpha$-elements (O, Mg, Si, Ca and Ti)}
\label{sec:aquarius-alpha-elements}

The $\alpha$-elements (O, Mg, Si, Ca and Ti) are forged through $\alpha$-particle capture during hydrostatic burning of carbon, neon and silicon. Material enriched in $\alpha$-elements is eventually dispersed into the interstellar medium following Type II core-collapse SN. 

Oxygen can be a particularly difficult element to measure. There are only a handful of lines available in an optical spectrum: the forbidden {[O\,I]} lines at $\lambda$6300 and $\lambda$6363 and the O\,I triplet lines at $\sim$777\,nm. The forbidden lines are very weak and become difficult to measure in hot and/or metal-poor stars (${\mbox{[Fe/H]} \lesssim -1.5}$\,dex). When they are present, depending on the radial velocity of the star, the {[O\,I]} lines can be significantly affected by telluric absorption. Moreover, the $\lambda$6300 line is blended with a Ni\,I absorption line \citep{allende-prieto;et-al_2001}. Hence the region requires careful consideration. Although the O\,I triplet lines at $\sim$777\,nm are stronger than the forbidden lines, they are extremely susceptible to non-LTE effects, surface granulation \citep{asplund;perez_2001}, and are sensitive to changes in microturbulence. Our forbidden {[O\,I]} abundances for HD\,136316 agree well with those from \citet{gratton;et-al_2000} -- the difference is only 0.07\,dex.

The {[O\,I]} lines were measurable in four of our Aquarius stream candidates. The $\lambda$6300 line in one of our candidates, {C2306265-085103}, was sufficiently affected by telluric absorption such that we deemed the line unrecoverable. Thus, only the $\lambda$6363 transition was used to derive an oxygen abundance for {C2306265-085103}. In our hottest star, {J223811-104126}, the forbidden oxygen lines were not detected above a 3$\sigma$ significance. After synthesising the region, we deduce a very conservative upper limit of ${\mbox{[O/Fe]} < 0.50}$ from the {[O\,I]} lines. This is consistent with the rest of our candidates, with [O/Fe] abundances varying between 0.43 to 0.49\,dex.

\begin{figure}[h!]
	\includegraphics[width=\textwidth]{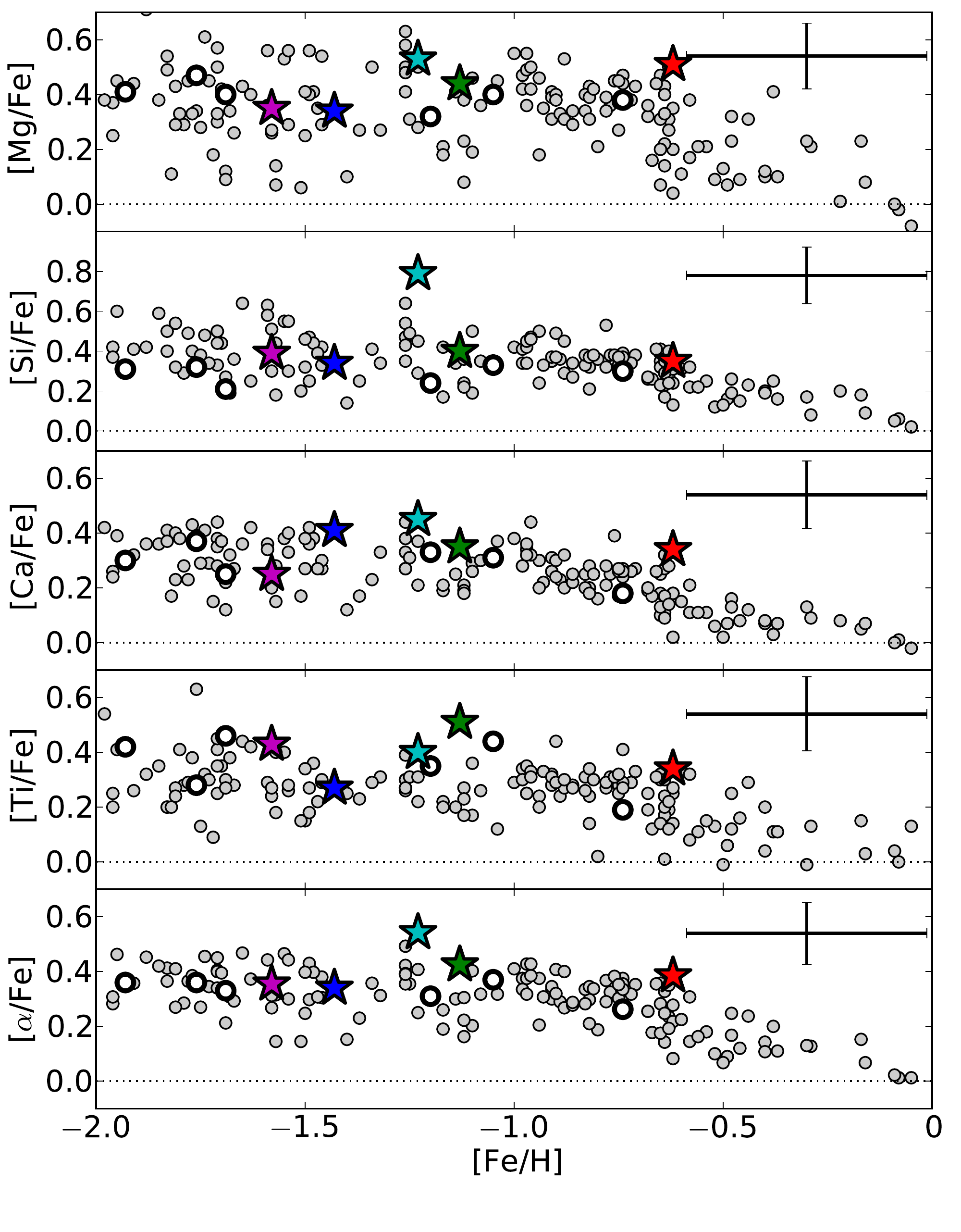}
	\caption[$\alpha$-element abundances for Aquarius and standard stars compared to the Milky Way trend]{$\alpha$-element abundances with respect to iron content. The mean [$\alpha$/Fe] abundance from these is shown in the bottom panel. The Solar element-to-iron ratio is marked as a dotted line in each panel, and a linear trend line between the Aquarius stars is shown in grey. Colors are as per Figure \ref{fig:toombre}. Standard stars are shown as open circles. Mean conservative \emph{total uncertainties} (random and systematic) for stars in this study are shown in each panel. Filled circles represent Milky Way field stars from \citet{fulbright_2000}. Oxygen abundances are shown separately in Figure \ref{fig:o-na}.}
	\label{fig:alpha-elements}
\end{figure}

In order to derive an oxygen measurement for {J223811-104126}, we were forced to use the triplet lines at $\sim$777\,nm. We extended these measurements for all Aquarius stars, and a mean abundance for each candidate was found from the synthesis of the permitted triplet lines. Oxygen abundances inferred from the triplet lines in all other stars were systematically $+0.27$\,dex higher than abundances calculated from the {[O\,I]} forbidden lines. \citet{perez;et-al_2006} found the same result from stars with similar stellar parameters: [O/Fe] values based on the {O\,I} permitted triplet lines are on average ${+0.19 \pm 0.07}$\,dex higher than those found from the forbidden lines, which did not include non-LTE corrections of $+$0.08\,dex. Thus, we attribute our $+$0.27\,dex offset between measurements of the {[O\,I]} and {O I} triplet lines to non-LTE and 3D effects. \citet{perez;et-al_2006} concluded that the forbidden lines, when not too weak, probably give the most reliable estimate of oxygen abundance. From the permitted O\,I triplet in {J223811-104126}, we derive an oxygen abundance of ${\mbox{[O/Fe]} = 0.42 \pm 0.01}$\,dex (random scatter). This measurement will be systematically higher than the `true' abundance if it were discernible from the [O\,I] lines, on the order of ${\sim}$+0.27\,dex. When we apply this crude offset derived from the rest of our sample, we arrive at a corrected abundance of ${\mbox{[O/Fe]} = 0.15 \pm 0.13}$ (total uncertainty) for {J223811-104126}. This is the most oxygen-deficient star in our sample by a factor of two.

Depending on the radial velocity of the star, some magnesium lines were affected by telluric absorption, particularly the $\lambda$6318 and $\lambda$6965 transitions. Atmospheric absorption was most notable for C222531-145437, where three of the four Mg transitions in our line list suffered some degree of telluric absorption, requiring an attentive correction. Every amended absorption profile was carefully examined, and lines with suspicious profiles were excluded from the final magnesium abundance. All [$\alpha$/Fe] abundance ratios in the standard stars are in excellent agreement with the literature. Typically the difference is 0.01\,dex, with the largest discrepancy of {$\Delta$[Ti/Fe] = +0.13\,dex} for HD\,76932 when compared with \citet{fulbright_2000}.

While \citet{wylie-de-boer;et-al_2012} find almost no scatter {($\pm0.02$\,dex)} in [Mg/Fe] for stars common to both studies, we find {C222531-145437} and {J223504-152834} to be almost {$+0.20$\,dex} higher than the rest of the sample. Of the {Mg\,I} line profiles measured, only two transitions are common to both line lists: $\lambda$6318 and $\lambda$6319. The oscillator strengths differ between studies; in these two lines the $\log{gf}$ differs by $-0.24$ and {$-0.27$\,dex} respectively (our oscillator strengths are lower). This indicates that the difference in oscillator strengths may explain the $\sim{}$0.2\,dex offset in [Mg/Fe] between this study and \citet{wylie-de-boer;et-al_2012}.

Of all the $\alpha$-elements, calcium has the smallest measurement scatter in our stars. The mean was formed from four line measurements in each star, with a typical random scatter of 0.01\,dex. Nevertheless, the aforementioned conservative minimum of 0.10\,dex for random scatter applies, and uncertainties in stellar parameters will contribute to the total error budget. As shown in Figure \ref{fig:alpha-elements}, all Aquarius stream candidates show super-solar [Ca/Fe], ranging between +0.23 to +0.43\,dex, consistent with [(Mg,Si,Ti)/Fe] measurements. 

{C222531-145437} has an unusually high silicon abundance (${\mbox{[Si/Fe]} = 0.79}$), well outside the uncertainties of the rest of our sample. The 5 silicon line abundances in this star are in relatively good agreement with each another. If we exclude the highest measurement, then the mean abundance drops only slightly to ${\mbox{[Si/Fe]} = 0.73 \pm 0.04}$ (random scatter). The lowest silicon line abundance for {C222531-145437} is ${\mbox{[Si/Fe]} = 0.61}$, which is still significantly higher than the mean abundance for any other star.

Titanium abundance ratios for the stream show typical levels of $\alpha$-enhancement. Our mean titanium abundances are derived from four to seven clean, unblended Ti\,II lines. In our hottest and most metal-poor stars, no suitable Ti\,I transitions were available. 

\subsection{Iron-peak Elements}
\label{sec:aquarius-fe-peak-abundances}

The Fe-peak elements (Sc to Zn) are primarily synthesized by the explosive nucleosynthesis of oxygen, neon, and silicon burning. Ignition can occur from Type II SN explosions of massive stars, or once a white dwarf accretes enough material to exceed the Chandrasekhar mass limit and spontaneously ignite carbon, leading to a Type Ia SN. 

Although not all {Fe-peak} elements are created equally, many Fe-peak elements generally exhibit similar trends with overall metallicity. All exhibit a positive trend with increasing iron abundance, with varying gradients.

\begin{figure}[t!]
	\includegraphics[width=\textwidth]{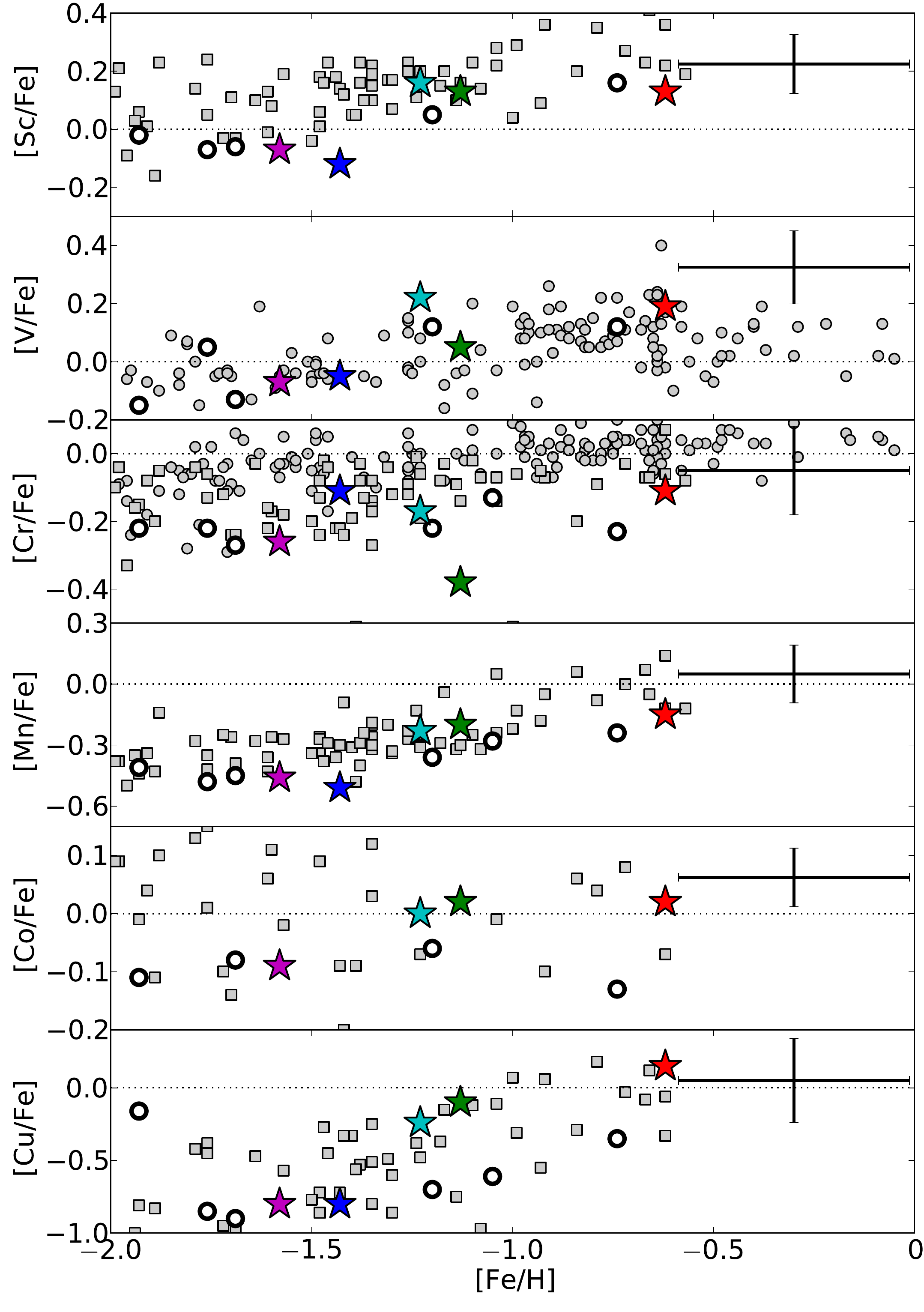}
	\caption[Iron-peak element abundances with respect to iron for all Aquarius stream stars]{Iron-peak element abundances (Sc, V, Cr, Mn, Co and Cu) with respect to iron for all Aquarius stream stars. Ni, an additional Fe-peak element, is discussed in Section \ref{sec:aquarius-na-ni-relationship} and shown in Figure \ref{fig:na-ni}. Colors are as per Figure \ref{fig:toombre}. Standard stars are shown as open circles. Mean conservative \emph{total uncertainties} (random and systematic) for this study are shown in each panel. Filled circles and squares represent Milky Way field stars from \citet{fulbright_2000} and \citet{ishigaki;et-al_2013}, respectively. Unlike Figure \ref{fig:alpha-elements}, panels have different $y$-axis ranges to accommodate the data.}
	\label{fig:fe-peak-elements}
\end{figure}

The [Sc/Fe] measurements presented in Figure \ref{fig:fe-peak-elements} are averaged from six clean {Sc\,II} lines, and there is very little line-to-line scatter, the largest of which is {0.06\,dex}. The number of clean, suitable {Cr\,I} lines available between members fluctuated from three to twelve. Very little line-to-line scatter is present in both {Cr\,I} and Cr\,II: the random scatter is below {0.04\,dex} for most stars. Chromium abundances are only available for one of the standard stars, where our {[Cr/Fe] = 0.03} is in excellent agreement with \citet{fulbright_2000}, where they find {${\rm [Cr/Fe]} = 0.04$}.

Manganese demonstrates a strong trend with increasing iron abundance (Figure \ref{fig:fe-peak-elements}). A significant source of Mn comes from Type Ia SN, and the strong Mn--Fe correlation is consistent with chemodynamical simulations \citep{kobayashi;nakasato_2011}, as well as thick disk observations by \citet{reddy;et-al_2006}.  Although Mn is known to demonstrate significant departures from LTE, we have not applied any non-LTE corrections to our abundances. 

Abundances of Co\,I lines were calculated by synthesis, as they demonstrate appreciable broadening due to hyperfine structure. Although they are known to suffer significant departures from LTE \citep{bergemann;et-al_2010}, no corrections have been made for these data. In general, [Co/H] follows [Fe/H] in our candidates.

Most Aquarius stream stars have seven clean {Ni\,I} transitions available. These lines are in excellent agreement, with a typical scatter of {0.03\,dex}. Nickel abundances have been published for two of our standard stars: HD\,76932 \citep{fulbright_2000} and HD\,141531 \citep{shetrone_1996}. In both cases, our [Ni/Fe] abundance ratios are slightly higher by +0.08 and +0.10\,dex respectively. The different solar compositions employed by these studies can only account for 0.01\,dex of this discrepancy, and the differences in oscillator strengths for common Ni\,I lines are negligible. Overall, the [Ni/Fe] abundance ratios in the Aquarius stream stars do not deviate greatly from the Solar ratio.

Hyperfine structure data has been included for the synthesis of Cu abundances. Only the Cu\,I line at The offsets between EW and synthesis abundances for Cu were significant: $\sim$0.4\,dex higher for some stars without the inclusion of hyperfine structure information. Cu abundances have also been determined by synthesis, and are consistent with the Milky Way trend.

\subsection{Neutron-capture Elements}
Neutron-capture elements (Sr to Eu; ${38 \leqslant Z \leqslant 63}$) can be forged through multiple nucleosynthetic processes. The two primary processes that produce these elements are the rapid ($r$-) process and the slow ($s$-) process. While the $r$-process is theorised to occur in SN explosions, the $s$-process takes place foremost in AGB stars with a significant contribution from massive stars at higher metallicities \citep[e.g.,][]{meyer_1994}, although models of rotating massive stars may change this picture at the very lowest metallicities \citep{frischknecht;et-al_2012}. 

\begin{figure}[t!]
	\centering
	\includegraphics[height=0.9\textheight]{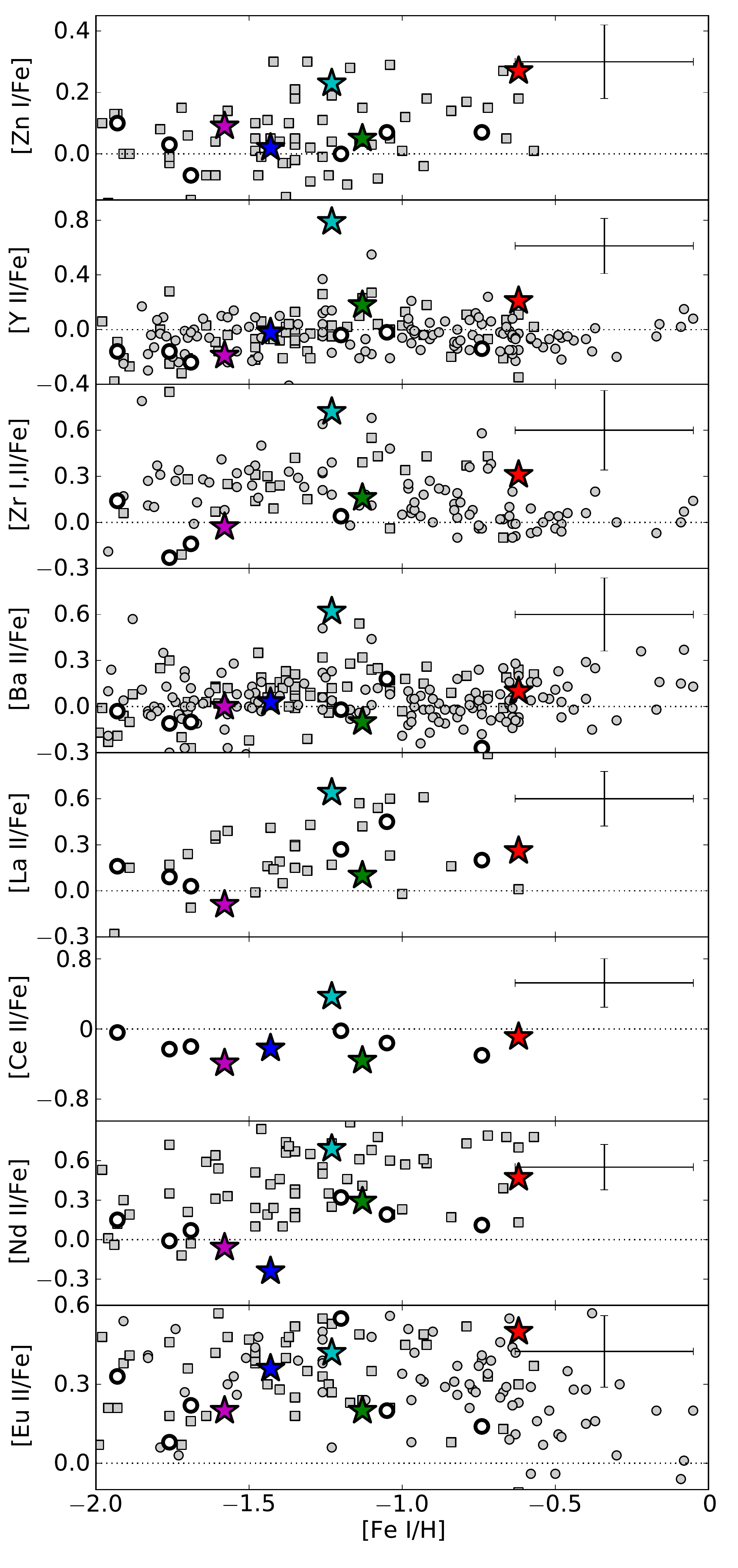}
	\caption[Heavy element abundance ratios for Aquarius stream stars]{Element ratios for Aquarius stream stars. In the case of [Zr/Fe], {[Zr\,I/Fe]} is taken where available and {[Zr\,II/Fe]} if no measurement was available for {Zr\,I}. See Table \ref{tab:program-star-abundances} for details. Standard stars are shown as open circles. Mean conservative \emph{total uncertainties} (random and systematic) for this study are shown in each panel. Filled circles and squares represent Milky Way field stars from \citet{fulbright_2000} and \citet{ishigaki;et-al_2013}, respectively. Panels have varying $y$-axis ranges to accommodate the data.}
	\label{fig:x-on-fe}
\end{figure}

\subsubsection{Strontium, Yttrium and Zirconium}
\label{sec:aquarius-weak-s-process-elements}

These neutron-capture elements belong to the first $s$-process peak, and generally increase in lock-step with each another. {[Y\,II/Fe]} and {[Zr\,I, Zr II/Fe]} are in good agreement among all candidates. The {Sr\,II} lines at $\lambda$4077 and $\lambda$4215 were not detected in any program or standard stars.

The Aquarius stream candidates have Y abundances consistent with halo field stars, with the exception of {C222531-145437}. With {[Y/Fe] = 0.79}, {C222531-145437} is significantly over-abundant in Y for its metallicity \citep[see Figure 4 of][]{travaglio;et-al_2004}. {C222531-145437} is consistently over-abundant in Zr, too. All other program and standard stars have first $n$-capture peak abundances and trends that are consistent with the chemical evolution of the Milky Way.

\subsubsection{Barium and Lanthanum}

Barium and Lanthanum belong to the second $s$-process peak. Ba has appreciable hyperfine and isotopic splitting, and its measurement requires some careful consideration. Solar Ba isotopic ratios have been adopted. Our standard stars have [Ba/Fe] abundances typical of the Milky Way halo. Two standard stars have existing [Ba/Fe] measurements from high-resolution spectra: HD\,44007 and HD\,76932. We find HD\,44007 to have ${\mbox{[Ba/Fe]} = 0.03}$, which is in good agreement with \citet{burris;et-al_2000}, who find {0.05\,dex}. For HD\,76932 our measurement of {$\mbox{[Ba/Fe]} = 0.18$} is in reasonable agreement with the \citet{fulbright_2000} value of $-0.02$\,dex, especially when differences in adopted solar composition are considered.

With one exception, the Aquarius stream candidates have [Ba/Fe] abundance ratios that are indistinguishable from field stars, ranging between ${\mbox{[Ba/Fe]} = -0.10}$ to {0.10\,dex}. The exception is {C222531-145437}, which has an anomalously high barium abundance of ${\mbox{[Ba/Fe]} = 0.62}$. This is $\sim$0.60\,dex higher than the Milky Way trend at its given metallicity of ${\mbox{[Fe/H]} = -1.26}$ \citep{ishigaki;et-al_2013}. Our two {Ba\,II} lines in {C222531-145437} are in excellent agreement with each other: [Ba/Fe] = 0.63, and 0.61. 

Lanthanum abundances have been determined by synthesis of the $\lambda$4558 and $\lambda$5805 lines with hyperfine splitting data included. All stars have La abundances that are consistent with the chemical enrichment of the Galaxy \citep{ishigaki;et-al_2013}, with the exception of the Ba-rich star C222531-145437, where [La/Fe] = 0.64 is observed.

\subsubsection{Cerium, Neodymium and Europium}

Europium is primarily produced by the $r$-process, whereas the production of Ce and Nd is split between $s$- and $r$-process. Europium abundances have been determined by synthesising the $\lambda$6645 Eu\,II transition with hyperfine splitting data from \citet{sneden}.

We chose not to use the $\lambda$6437 Eu\,II line as it is appreciably blended by a nearby Si\,I line \citep{Lawler;et-al_2001}, and our measurements were consistent with a hidden blend: the $\lambda$6437 Eu\,II abundance was systematically higher than the $\lambda$6645 counterpart. One Aquarius stream candidate, {C222531-145437}, appears enhanced in all [$s$-process/Fe] abundance ratios compared to the program and standard sample. However no noteworthy difference in Eu, which is generally considered to be a $r$-process dominated element, was observed.\\

\setlength\tabcolsep{3pt}
\begin{longtable}[h!]{lcrcrrclcrcrrcc}
\label{tab:standard-star-abundances}\\
\hline
\hline
Species & $N$ & $\log\epsilon(X)$ & $\sigma$ & [X/H] & [X/Fe] && 
Species & $N$ & $\log\epsilon(X)$ & $\sigma$ & [X/H] & [X/Fe] \\ 
\hline
\endfirsthead
\hline
Species & $N$ & $\log\epsilon(X)$ & $\sigma$ & [X/H] & [X/Fe] && 
Species & $N$ & $\log\epsilon(X)$ & $\sigma$ & [X/H] & [X/Fe] \\ 
\hline\endhead
\hline
\multicolumn{13}{r}{Continued..}
\endfoot
\endlastfoot
\\
\multicolumn{6}{c}{\textbf{HD\,41667}} &  & \multicolumn{6}{c}{\textbf{HD\,44007}} \\
\cline{1-6} \cline{8-13}
   C (CH)       &   2 &    6.95 &    0.20 &  --1.48 &  --0.28 && 
   C (CH)       &   2 &    6.66 &    0.20 &  --1.77 &  --0.01 \\ 
   O I &   2 &    7.95 &    0.06 &  --0.74 &    0.46 && 
   O I &   1 &    7.41 &    0.00 &  --1.28 &    0.48 \\ 
  Na I &   3 &    4.90 &    0.18 &  --1.34 &  --0.14 &&
  Na I &   2 &    4.44 &    0.09 &  --1.80 &  --0.04 \\
  Mg I &   4 &    6.72 &    0.10 &  --0.88 &    0.32 &&
  Mg I &   2 &    6.30 &    0.06 &  --1.29 &    0.47 \\
  Al I &   4 &    5.18 &    0.11 &  --1.27 &  --0.07 &&
  Al I &   1 &   :4.80 & \nodata & :--1.65 &   :0.11 \\
  Si I &   5 &    6.55 &    0.06 &  --0.96 &    0.24 &&
  Si I &   5 &    6.07 &    0.07 &  --1.44 &    0.32 \\
   K I &   1 &    4.64 & \nodata &  --0.39 &    0.81 &&
   K I &   1 &    4.31 & \nodata &  --0.72 &    1.04 \\
  Ca I &   4 &    5.47 &    0.06 &  --0.87 &    0.33 &&
  Ca I &   4 &    4.95 &    0.02 &  --1.39 &    0.37 \\
 Sc II &   5 &    2.00 &    0.12 &  --1.15 &    0.05 &&
 Sc II &   5 &    1.32 &    0.12 &  --1.85 &  --0.07 \\
  Ti I &   4 &    3.96 &    0.04 &  --0.99 &    0.21 &&
  Ti I &   1 &    3.48 & \nodata &  --1.47 &    0.29 \\
 Ti II &   3 &    4.09 &    0.25 &  --0.86 &    0.35 &&
 Ti II &   4 &    3.47 &    0.15 &  --1.48 &    0.28 \\
   V I &   4 &    2.85 &    0.11 &  --1.08 &    0.12 &&
   V I &   1 &    2.22 & \nodata &  --1.72 &    0.05 \\
  Cr I &  10 &    4.22 &    0.08 &  --1.42 &  --0.22 &&
  Cr I &  15 &    3.65 &    0.07 &  --1.99 &  --0.22 \\
 Cr II &   2 &    4.54 &    0.05 &  --1.09 &    0.11 &&
 Cr II &   3 &    4.00 &    0.01 &  --1.64 &    0.12 \\
 Mn I  &   3 &    3.87 &    0.04 &  --1.56 &  --0.36 &&
 Mn I  &   2 &    3.21 &    0.06 &  --2.22 &  --0.48 \\
  Fe I &  61 &    6.30 &    0.12 &  --1.20 &    0.00 &&
  Fe I &  51 &    5.74 &    0.13 &  --1.76 &    0.00 \\
 Fe II &  13 &    6.35 &    0.05 &  --1.15 &    0.05 &&
 Fe II &  15 &    5.74 &    0.10 &  --1.76 &  --0.00 \\
  Co I &   3 &    3.73 &    0.06 &  --1.26 &  --0.06 &&
  Co I &   0 & \nodata & \nodata & \nodata & \nodata \\
  Ni I &   7 &    4.94 &    0.12 &  --1.28 &  --0.08 &&
  Ni I &   4 &    4.47 &    0.05 &  --1.75 &    0.01 \\
  Cu I &   1 &    2.29 & \nodata &  --1.90 &  --0.70 &&
  Cu I &   1 &    1.58 & \nodata &  --2.61 &  --0.85 \\
  Zn I &   2 &    3.36 &    0.08 &  --1.20 &    0.00 &&
  Zn I &   2 &    2.83 &    0.05 &  --1.73 &    0.03 \\
  Y II &   5 &    0.97 &    0.19 &  --1.24 &  --0.04 &&
  Y II &   6 &    0.28 &    0.11 &  --1.93 &  --0.16 \\
  Zr I &   2 &    1.42 &    0.05 &  --1.17 &    0.04 &&
  Zr I &   0 & \nodata & \nodata & \nodata & \nodata \\
 Zr II &   1 &    1.28 & \nodata &  --1.30 &  --0.10 &&
 Zr II &   1 &    0.59 & \nodata &  --1.99 &  --0.23 \\
 Ba II &   2 &    0.95 &    0.07 &  --1.23 &  --0.02 &&
 Ba II &   2 &    0.31 &    0.06 &  --1.87 &  --0.11 \\
 La II &   1 &    0.17 & \nodata &  --0.93 &     0.27 &&
 La II &   2 &  --0.57 &    0.05 &  --1.67 &     0.09 \\
 Ce II &   4 &    0.35 &    0.18 &  --1.23 &  --0.02 &&
 Ce II &   3 &  --0.41 &    0.12 &  --1.99 &  --0.23 \\
 Nd II &   9 &    0.54 &    0.10 &  --0.88 &    0.32 &&
 Nd II &   9 &  --0.36 &    0.11 &  --1.78 &  --0.01 \\
 Eu II &   1 &  --0.13 & \nodata &  --0.65 &    0.55 &&
 Eu II &   1 &  --1.16 & \nodata &  --1.68 &    0.08 \\

\cline{1-6} \cline{8-13} \\ \\
\multicolumn{6}{c}{\textbf{HD\,76932}} && \multicolumn{6}{c}{\textbf{HD\,136316}} \\
\cline{1-6} \cline{8-13}
   C (CH)       &   2 &    7.52 &    0.20 &  --0.91 &    0.14 &&
   C (CH)       &   2 &    5.95 &    0.20 &  --2.48 &  --0.50 \\
   O I &   1 &    8.05 & \nodata &  --0.64 &    0.41 &&
   O I &   1 &    7.17 & \nodata &  --1.52 &    0.41 \\
  Na I &   3 &    5.37 &    0.04 &  --0.87 &    0.18 &&
  Na I &   2 &    4.17 &    0.04 &  --2.08 &  --0.14 \\
  Mg I &   3 &    6.95 &    0.20 &  --0.65 &    0.40 &&
  Mg I &   2 &    6.08 &    0.24 &  --1.52 &    0.41 \\
  Al I &   4 &    5.45 &    0.07 &  --1.00 &    0.05 &&
  Al I &   0 & \nodata & \nodata & \nodata & \nodata \\
  Si I &   5 &    6.79 &    0.06 &  --0.72 &    0.33 &&
  Si I &   4 &    5.89 &    0.05 &  --1.62 &    0.31 \\
   K I &   1 &    4.94 & \nodata &  --0.09 &    0.96 &&
   K I &   1 &    3.91 & \nodata &  --1.12 &    0.81 \\
  Ca I &   4 &    5.60 &    0.02 &  --0.74 &    0.31 &&
  Ca I &   4 &    4.71 &    0.02 &  --1.63 &    0.30 \\
 Sc II &   4 &    2.10 &    0.05 &  --1.05 &    0.01 &&
 Sc II &   4 &    1.20 &    0.08 &  --1.95 &  --0.02 \\
  Ti I &   1 &    4.36 & \nodata &  --0.59 &    0.46 &&
  Ti I &   3 &    3.19 &    0.03 &  --1.76 &    0.17 \\
 Ti II &   3 &    4.33 &    0.04 &  --0.62 &    0.44 &&
 Ti II &   3 &    3.44 &    0.10 &  --1.51 &    0.42 \\
   V I &   1 &   :3.33 & \nodata & :--0.60 &   :0.45 &&
   V I &   3 &    1.85 &    0.01 &  --2.08 &  --0.15 \\
  Cr I &  15 &    4.46 &    0.05 &  --1.18 &  --0.13 &&
  Cr I &  12 &    3.49 &    0.05 &  --2.15 &  --0.22 \\
 Cr II &   3 &    4.76 &    0.02 &  --0.88 &    0.17 &&
 Cr II &   2 &    3.90 &    0.02 &  --1.74 &    0.19 \\
  Mn I &   3 &    4.09 &    0.06 &  --1.34 &  --0.28 &&
  Mn I &   3 &    3.09 &    0.03 &  --2.34 &  --0.41 \\
  Fe I &  51 &    6.45 &    0.10 &  --1.05 &    0.00 &&
  Fe I &  62 &    5.57 &    0.11 &  --1.93 &    0.00 \\
 Fe II &  13 &    6.50 &    0.07 &  --1.00 &    0.05 &&
 Fe II &  14 &    5.61 &    0.12 &  --1.89 &    0.04 \\
  Co I &   1 &    3.94 & \nodata &  --1.05 &    0.00 &&
  Co I &   2 &    2.95 &    0.11 &  --1.09 &  --0.11 \\
  Ni I &   5 &    5.29 &    0.02 &  --0.93 &    0.13 &&
  Ni I &   5 &    4.22 &    0.11 &  --2.00 &  --0.07 \\
  Cu I &   1 &    2.53 & \nodata &  --1.66 &  --0.61 &&
  Cu I &   1 &    1.36 & \nodata &  --2.09 &  --0.16 \\
  Zn I &   2 &    3.58 &    0.03 &  --0.98 &    0.07 &&
  Zn I &   2 &    2.72 &    0.03 &  --1.83 &    0.10 \\
  Y II &   5 &    1.14 &    0.05 &  --1.07 &  --0.02 &&
  Y II &   7 &    0.12 &    0.11 &  --2.09 &  --0.16 \\
  Zr I &   0 & \nodata & \nodata & \nodata & \nodata &&
  Zr I &   1 &    0.79 & \nodata &  --1.79 &    0.14 \\
 Zr II &   0 & \nodata & \nodata & \nodata & \nodata &&
 Zr II &   1 &    0.68 & \nodata &  --1.90 &    0.03 \\
 Ba II &   2 &    1.31 &    0.07 &  --0.87 &    0.18 &&
 Ba II &   2 &    0.22 &    0.02 &  --1.96 &  --0.03 \\
 La II &   1 &    0.50 & \nodata &  --0.60 &    0.45 &&
 La II &   1 &  --0.68 & \nodata &  --1.78 &    0.16 \\
 Ce II &   2 &    0.37 &    0.03 &  --1.21 &  --0.16 &&
 Ce II &   5 &  --0.39 &    0.18 &  --1.97 &  --0.04 \\
 Nd II &   3 &    0.56 &    0.06 &  --0.86 &    0.19 &&
 Nd II &  10 &  --0.36 &    0.04 &  --1.78 &    0.15 \\
 Eu II &   1 &  --0.33 & \nodata &  --0.85 &    0.20 &&
 Eu II &   1 &  --1.06 & \nodata &  --1.58 &    0.33 \\

\cline{1-6} \cline{8-13} \\ \\
\multicolumn{6}{c}{\textbf{HD\,141531}} && \multicolumn{6}{c}{\textbf{HD\,142948}} \\
\cline{1-6} \cline{8-13}
   C (CH)       &   2 &    6.33 &    0.20 &  --2.10 &  --0.48 &&
   C (CH)       &   2 &    7.72 &    0.20 &  --0.71 &    0.03 \\
   O I &   2 &    7.33 &    0.01 &  --1.35 &    0.34 &&
   O I &   2 &    8.43 &    0.02 &  --0.26 &    0.47 \\
  Na I &   2 &    4.28 &    0.05 &  --1.96 &  --0.27 &&
  Na I &   3 &    5.73 &    0.13 &  --0.51 &    0.22 \\
  Mg I &   2 &    6.30 &    0.15 &  --1.29 &    0.40 &&
  Mg I &   3 &    7.24 &    0.12 &  --0.36 &    0.38 \\
  Al I &   2 &    4.74 &    0.10 &  --1.71 &  --0.02 &&
  Al I &   4 &    5.94 &    0.08 &  --0.51 &    0.23 \\
  Si I &   5 &    6.03 &    0.10 &  --1.48 &    0.21 &&
  Si I &   5 &    7.07 &    0.05 &  --0.44 &    0.30 \\
   K I &   1 &    3.99 & \nodata &  --1.04 &    0.65 &&
   K I &   1 &    5.04 & \nodata &    0.01 &    0.75 \\
  Ca I &   4 &    4.90 &    0.03 &  --1.44 &    0.25 &&
  Ca I &   4 &    5.78 &    0.01 &  --0.56 &    0.18 \\
 Sc II &   5 &    1.40 &    0.11 &  --1.75 &  --0.06 &&
 Sc II &   5 &    2.57 &    0.12 &  --0.58 &    0.16 \\
  Ti I &   4 &    3.33 &    0.07 &  --1.62 &    0.07 &&
  Ti I &   4 &    4.44 &    0.09 &  --0.51 &    0.23 \\
 Ti II &   4 &    3.71 &    0.08 &  --1.24 &    0.46 &&
 Ti II &   3 &    4.40 &    0.21 &  --0.55 &    0.19 \\
   V I &   4 &    2.10 &    0.07 &  --1.83 &  --0.13 &&
   V I &   5 &    3.31 &    0.04 &  --0.62 &    0.12 \\
  Cr I &  12 &    3.68 &    0.06 &  --1.96 &  --0.27 &&
  Cr I &  13 &    4.67 &    0.15 &  --0.97 &  --0.23 \\
 Cr II &   2 &    4.11 &    0.02 &  --1.53 &    0.16 &&
 Cr II &   3 &    4.88 &    0.03 &  --0.76 &  --0.02 \\
  Mn I &   3 &    3.29 &    0.04 &  --2.14 &  --0.45 &&
  Mn I &   3 &    4.45 &    0.06 &  --0.98 &  --0.24 \\
  Fe I &  54 &    5.81 &    0.06 &  --1.69 &    0.00 &&
  Fe I &  61 &    6.76 &    0.10 &  --0.74 &    0.00 \\
 Fe II &  13 &    5.86 &    0.03 &  --1.64 &    0.05 &&
 Fe II &  13 &    6.75 &    0.06 &  --0.75 &  --0.02 \\
  Co I &   3 &    3.22 &    0.12 &  --1.77 &  --0.08 &&
  Co I &   3 &    4.36 &    0.11 &  --0.63 &  --0.13 \\
  Ni I &   7 &    4.42 &    0.12 &  --1.80 &  --0.11 &&
  Ni I &   5 &    5.62 &    0.04 &  --0.60 &    0.13 \\
  Cu I &   1 &    1.60 & \nodata &  --2.59 &  --0.90 &&
  Cu I &   1 &    3.10 & \nodata &  --1.09 &  --0.35 \\
  Zn I &   2 &    2.80 &    0.04 &  --1.76 &  --0.07 &&
  Zn I &   2 &    3.89 &    0.06 &  --0.67 &    0.07 \\
  Y II &   6 &    0.27 &    0.13 &  --1.94 &  --0.24 &&
  Y II &   6 &    1.33 &    0.32 &  --0.88 &  --0.14 \\
  Zr I &   0 & \nodata & \nodata & \nodata & \nodata &&
  Zr I &   0 & \nodata & \nodata & \nodata & \nodata \\
 Zr II &   1 &    0.75 & \nodata &  --1.83 &  --0.14 &&
 Zr II &   0 & \nodata & \nodata & \nodata & \nodata \\
 Ba II &   2 &    0.39 &    0.05 &  --1.79 &  --0.10 &&
 Ba II &   2 &    1.17 &    0.01 &  --1.01 &  --0.27 \\
 La II &   1 &  --0.56 & \nodata &  --1.67 &    0.03 &&
 La II &   1 &    0.56 & \nodata &  --0.54 &    0.20 \\
 Ce II &   4 &  --0.31 &    0.12 &  --1.89 &  --0.20 &&
 Ce II &   3 &    0.54 &    0.20 &  --1.04 &  --0.30 \\
 Nd II &  10 &  --0.20 &    0.08 &  --1.62 &    0.07 &&
 Nd II &   6 &    0.79 &    0.10 &  --0.63 &    0.11 \\
 Eu II &   1 &  --0.95 & \nodata &  --1.47 &    0.22 &&
 Eu II &   1 &    0.08 & \nodata &  --1.55 &    0.14 \\
\hline
\caption{Elemental abundances for standard stars}
\end{longtable}

\begin{longtable}[h!]{lcrcrrclcrcrrcc}
\label{tab:program-star-abundances}\\
\hline
\hline
Species & $N$ & $\log\epsilon(X)$ & $\sigma$ & [X/H] & [X/Fe] && 
Species & $N$ & $\log\epsilon(X)$ & $\sigma$ & [X/H] & [X/Fe] \\ 
\hline
\endfirsthead
\hline
Species & $N$ & $\log\epsilon(X)$ & $\sigma$ & [X/H] & [X/Fe] && 
Species & $N$ & $\log\epsilon(X)$ & $\sigma$ & [X/H] & [X/Fe] \\ 
\hline\endhead
\hline
\multicolumn{13}{r}{Continued..}
\endfoot
\endlastfoot
\\
\multicolumn{6}{c}{\textbf{J221821-183424}} &  & \multicolumn{6}{c}{\textbf{C222531-145437}} \\
\cline{1-6} \cline{8-13}
   C (CH)       &   2 &    6.55 &    0.20 &  --1.88 &  --0.30 &&
   C (CH)       &   2 &    7.15 &    0.20 &  --1.28 &  --0.05 \\
   O I &   2 &    7.55 &    0.04 &  --1.13 &    0.45 &&
   O I &   2 &    7.96 & \nodata &  --0.73 &    0.49 \\
  Na I &   1 &    4.75 & \nodata &  --1.49 &    0.09 &&
  Na I &   2 &    5.12 &    0.02 &  --1.12 &    0.10 \\
  Mg I &   3 &    6.37 &    0.09 &  --1.23 &    0.35 &&
  Mg I &   2 &    6.90 &    0.08 &  --0.70 &    0.53 \\
  Al I &   1 &    5.08 & \nodata &  --1.37 &    0.21 &&
  Al I &   4 &    5.94 &    0.10 &  --0.51 &    0.71 \\
  Si I &   5 &    6.32 &    0.08 &  --1.19 &    0.39 &&
  Si I &   5 &    7.07 &    0.15 &  --0.44 &    0.79 \\
   K I &   1 &    4.34 & \nodata &  --0.69 &    0.89 &&
   K I &   1 &    4.42 & \nodata &  --0.61 &    0.62 \\
  Ca I &   4 &    5.01 &    0.04 &  --1.33 &    0.25 &&
  Ca I &   4 &    5.57 &    0.04 &  --0.77 &    0.45 \\
 Sc II &   4 &    1.50 &    0.12 &  --1.65 &  --0.07 &&
 Sc II &   4 &    2.08 &    0.13 &  --1.07 &    0.16 \\
  Ti I &   0 & \nodata & \nodata & \nodata & \nodata &&
  Ti I &   4 &    4.10 &    0.03 &  --0.85 &    0.37 \\
 Ti II &   4 &    3.80 &    0.13 &  --1.15 &    0.43 &&
 Ti II &   2 &    4.12 &    0.13 &  --0.83 &    0.40 \\
   V I &   3 &    2.28 &    0.01 &  --1.65 &  --0.07 &&
   V I &   5 &    2.91 &    0.10 &  --1.01 &    0.22 \\
  Cr I &  11 &    3.80 &    0.06 &  --1.84 &  --0.26 &&
  Cr I &   8 &    4.24 &    0.17 &  --1.40 &  --0.17 \\
 Cr II &   2 &    4.07 &    0.03 &  --1.57 &    0.01 &&
 Cr II &   1 &    4.38 & \nodata &  --1.26 &  --0.03 \\
  Mn I &   2 &    3.38 &    0.03 &  --2.05 &  --0.46 &&
  Mn I &   3 &    3.98 &    0.05 &  --1.45 &  --0.23 \\
  Fe I &  52 &    5.92 &    0.09 &  --1.58 &    0.00 &&
  Fe I &  60 &    6.27 &    0.10 &  --1.23 &    0.00 \\
 Fe II &  13 &    5.94 &    0.05 &  --1.56 &    0.02 &&
 Fe II &  10 &    6.30 &    0.06 &  --1.20 &    0.03 \\
  Co I &   1 &    3.32 & \nodata &  --1.67 &  --0.09 &&
  Co I &   4 &    3.77 &    0.09 &  --1.22 &    0.00 \\
  Ni I &   5 &    4.61 &    0.14 &  --1.61 &  --0.03 &&
  Ni I &   7 &    5.07 &    0.09 &  --1.15 &    0.08 \\
  Cu I &   1 &    1.81 & \nodata &  --2.38 &  --0.80 &&
  Cu I &   1 &    2.72 & \nodata &  --1.47 &  --0.24 \\
  Zn I &   1 &    3.07 & \nodata &  --1.49 &    0.09 &&
  Zn I &   2 &    3.56 &    0.24 &  --1.00 &    0.23 \\
  Y II &   3 &    0.44 &    0.02 &  --1.77 &  --0.19 &&
  Y II &   5 &    1.78 &    0.16 &  --0.43 &    0.79 \\
  Zr I &   0 & \nodata & \nodata & \nodata & \nodata &&
  Zr I &   3 &    2.07 &    0.05 &  --0.51 &    0.72 \\
 Zr II &   1 &    0.97 & \nodata &  --1.61 &  --0.03 &&
 Zr II &   0 & \nodata & \nodata & \nodata & \nodata \\
 Ba II &   1 &    0.60 & \nodata &  --1.58 &    0.00 &&
 Ba II &   2 &    1.58 &    0.01 &  --0.60 &    0.62 \\
 La II &   1 &  --0.58 & \nodata &  --1.67 &  --0.09 &&
 La II &   2 &    0.51 &    0.02 &  --0.59 &    0.64 \\
 Ce II &   3 &  --0.39 &    0.06 &  --1.97 &  --0.39 &&
 Ce II &   5 &    0.73 &    0.15 &  --0.85 &    0.37 \\
 Nd II &  10 &  --0.22 &    0.07 &  --1.64 &  --0.06 &&
 Nd II &   8 &    0.88 &    0.13 &  --0.54 &    0.69 \\
 Eu II &   1 &  --0.86 &    0.11 &  --1.38 &    0.20 &&
 Eu II &   1 &  --0.29 & \nodata &  --0.81 &    0.42 \\

\cline{1-6} \cline{8-13} \\ \\
\multicolumn{6}{c}{\textbf{J223504-152834}} && \multicolumn{6}{c}{\textbf{J223811-104126}} \\
\cline{1-6} \cline{8-13}
   C (CH)       &   2 &    7.71 &    0.30 &  --0.72 &  --0.10 &&
   C (CH)       &   2 &    7.05 &    0.25 &  --1.38 &    0.05 \\
   O I &   2 &    8.50 &    0.10 &  --0.19 &    0.43 &&
   O I\footnote{Abundance derived from the permitted O I triplet instead of the forbidden [O I] lines, see Section \ref{sec:aquarius-alpha-elements}} &   3 &    7.41 &    0.13 &  --1.28 &    0.15 \\
  Na I &   3 &    5.87 &    0.12 &  --0.37 &    0.26 &&
  Na I &   1 &    4.89 & \nodata &  --1.35 &    0.08 \\
  Mg I &   3 &    7.48 &    0.15 &  --0.12 &    0.51 &&
  Mg I &   2 &    6.51 &    0.03 &  --1.09 &    0.34 \\
  Al I &   3 &    6.12 &    0.09 &  --0.33 &    0.29 &&
  Al I &   2 &    5.13 &    0.13 &  --1.32 &    0.11 \\
  Si I &   5 &    7.24 &    0.10 &  --0.27 &    0.35 &&
  Si I &   3 &    6.42 &    0.04 &  --1.09 &    0.34 \\
   K I &   1 &    5.05 & \nodata &    0.02 &    0.64 &&
   K I &   1 &    4.50 & \nodata &  --0.53 &    0.90 \\
  Ca I &   4 &    6.06 &    0.03 &  --0.28 &    0.34 &&
  Ca I &   4 &    5.32 &    0.03 &  --1.02 &    0.41 \\
 Sc II &   5 &    2.65 &    0.10 &  --0.50 &    0.13 &&
 Sc II &   2 &    1.60 &    0.03 &  --1.55 &  --0.12 \\
  Ti I &   4 &    4.65 &    0.02 &  --0.30 &    0.32 &&
  Ti I &   0 & \nodata & \nodata & \nodata & \nodata \\
 Ti II &   1 &    4.67 & \nodata &  --0.28 &    0.34 &&
 Ti II &   4 &    3.79 &    0.09 &  --1.16 &    0.27 \\
   V I &   4 &    3.50 &    0.11 &  --0.43 &    0.19 &&
   V I &   1 &    2.45 & \nodata &  --1.48 &  --0.05 \\
  Cr I &   7 &    4.90 &    0.11 &  --0.74 &  --0.11 &&
  Cr I &  12 &    4.10 &    0.06 &  --1.54 &  --0.11 \\
 Cr II &   2 &    4.84 &    0.04 &  --0.79 &  --0.17 &&
 Cr II &   3 &    4.34 &    0.07 &  --1.30 &    0.12 \\
  Mn I &   3 &    4.66 &    0.04 &  --0.77 &  --0.15 &&
  Mn I &   2 &    3.50 &    0.01 &  --1.93 &  --0.51 \\
  Fe I &  63 &    6.88 &    0.12 &  --0.62 &    0.00 &&
  Fe I &  33 &    6.07 &    0.06 &  --1.43 &    0.00 \\
 Fe II &  12 &    6.87 &    0.07 &  --0.63 &  --0.01 &&
 Fe II &   9 &    6.04 &    0.07 &  --1.46 &  --0.03 \\
  Co I &   3 &    4.39 &    0.09 &  --0.60 &    0.02 &&
  Co I &   0 & \nodata & \nodata & \nodata & \nodata \\
  Ni I &   7 &    5.64 &    0.09 &  --0.58 &    0.05 &&
  Ni I &   2 &    4.84 &    0.04 &  --1.38 &    0.05 \\
  Cu I &   1 &    3.72 & \nodata &  --0.47 &    0.15 &&
  Cu I &   1 &    1.96 & \nodata &  --2.23 &  --0.80 \\
  Zn I &   2 &    4.21 &    0.03 &  --0.35 &    0.27 &&
  Zn I &   2 &    3.15 &    0.05 &  --1.41 &    0.02 \\
  Y II &   3 &    1.80 &    0.03 &  --0.41 &    0.21 &&
  Y II &   6 &    0.76 &    0.06 &  --1.45 &  --0.02 \\
  Zr I &   3 &    2.26 &    0.05 &  --0.32 &    0.31 &&
  Zr I &   0 & \nodata & \nodata & \nodata & \nodata \\
 Zr II &   0 & \nodata & \nodata & \nodata & \nodata &&
 Zr II &   0 & \nodata & \nodata & \nodata & \nodata \\
 Ba II &   2 &    1.65 &    0.02 &  --0.53 &    0.10 &&
 Ba II &   2 &    0.78 &    0.07 &  --1.40 &    0.03 \\
 La II &   1 &    0.76 & \nodata &  --0.34 &    0.28 &&
 La II &   0 & \nodata & \nodata & \nodata & \nodata \\
 Ce II &   3 &    0.87 &    0.13 &  --0.71 &  --0.09 &&
 Ce II &   2 &  --0.07 &    0.02 &  --1.65 &  --0.22 \\
 Nd II &   6 &    1.27 &    0.13 &  --0.15 &    0.47 &&
 Nd II &   1 &  --0.25 & \nodata &  --1.67 &  --0.24 \\
 Eu II &   1 &    0.40 & \nodata &  --0.12 &    0.50 &&
 Eu II &   1 &  --0.55 & \nodata &  --1.07 &    0.36 \\

\cline{1-6} \cline{8-13} \\ \\
\multicolumn{7}{c}{\textbf{C2306265-085103}} \\
\cline{1-6}
   C (CH)       &   2 &    7.20 &    0.20 &  --1.23 &  --0.10 \\
   O I &   2 &    8.02 &    0.04 &  --0.67 &    0.46 \\
  Na I &   2 &    5.31 &    0.01 &  --0.93 &    0.21 \\
  Mg I &   2 &    6.90 &    0.06 &  --0.70 &    0.44 \\
  Al I &   4 &    5.65 &    0.08 &  --0.80 &    0.33 \\
  Si I &   5 &    6.78 &    0.08 &  --0.73 &    0.40 \\
   K I &   1 &    4.46 & \nodata &  --0.57 &    0.56 \\
  Ca I &   4 &    5.56 &    0.04 &  --0.78 &    0.35 \\
 Sc II &   3 &    2.15 &    0.09 &  --1.00 &    0.13 \\
  Ti I &   4 &    4.13 &    0.03 &  --0.82 &    0.32 \\
 Ti II &   3 &    4.32 &    0.35 &  --0.63 &    0.51 \\
   V I &   4 &    2.85 &    0.06 &  --1.09 &    0.05 \\
  Cr I &   3 &    4.13 &    0.12 &  --1.51 &  --0.38 \\
 Cr II &   1 &    4.50 & \nodata &  --1.14 &  --0.01 \\
  Mn I &   3 &    4.10 &    0.05 &  --1.33 &  --0.20 \\
  Fe I &  62 &    6.37 &    0.12 &  --1.13 &    0.00 \\
 Fe II &  11 &    6.39 &    0.10 &  --1.11 &    0.02 \\
  Co I &   3 &    3.88 &    0.06 &  --1.11 &    0.02 \\
  Ni I &   7 &    5.11 &    0.07 &  --1.11 &    0.02 \\
  Cu I &   1 &    2.96 & \nodata &  --1.23 &  --0.10 \\
  Zn I &   2 &    3.48 &    0.15 &  --1.08 &    0.05 \\
  Y II &   4 &    1.26 &    0.26 &  --0.95 &    0.18 \\
  Zr I &   3 &    1.60 &    0.05 &  --0.98 &    0.16 \\
 Zr II &   0 & \nodata & \nodata & \nodata & \nodata \\
 Ba II &   2 &    0.95 &    0.10 &  --1.23 &  --0.10 \\
 La II &   1 &    0.07 & \nodata &  --1.03 &    0.10 \\
 Ce II &   2 &    0.09 &    0.02 &  --1.49 &  --0.36 \\
 Nd II &   7 &    0.58 &    0.21 &  --0.84 &    0.29 \\
 Eu II &   1 &  --0.41 & \nodata &  --0.93 &    0.20 \\
 \hline
\caption{Elemental abundances for program stars}
\end{longtable}

\subsection{Uncertainties in Chemical Abundances}
\label{sec:aquarius-chemical-abundance-uncertainties}

The uncertainties in chemical abundances are primarily driven by systematic uncertainties in stellar parameters, with a small contribution of random measurement scatter from individual lines. In order to calculate the abundance uncertainties due to stellar parameters, we have independently varied the stellar parameters by the adopted uncertainties, and measured the resultant change in chemical abundances. For lines requiring synthesis due to hyperfine structure, the difference in chemical abundances has been calculated from EWs. However, the effect of wing broadening due to hyperfine or isotopic splitting was generally small.

The individual abundance errors from varying each of the stellar parameters were added in quadrature to obtain the systematic error. To obtain the total error, we added in quadrature the systematic error with the standard error of the mean (random error). In some cases, the standard error about the mean is unrealistically small. As discussed earlier in Section \ref{sec:aquarius-sodium-abundances}, we have conservatively adopted an abundance floor of 0.10\,dex for the standard deviation (i.e., $\max(0.10, \sigma(\log_{\epsilon}X)$). These resultant changes in abundances and total uncertainties are listed for all stars in Table \ref{tab:chemical-abundance-uncertainties}. These total uncertainties have been used in all figures. This provides us with an uncertainty for all abundances, in all stars, of [X/H]. Generally though, we are most interested in the uncertainty in [X/Fe]. In order to calculate this uncertainty, the correlations in uncertainties due to stellar parameters between (X, Fe) need to be considered. We have followed the description in \citet{johnson_2002} to calculate these correlations, and the overall uncertainties in [X/Fe], which are listed in Table \ref{tab:chemical-abundance-uncertainties} for all program and standard stars.


\setlength\tabcolsep{2pt}
\begin{longtable}[h!]{lrrrccc}
\label{tab:chemical-abundance-uncertainties} \\
\hline
\hline
 & & & & & \multicolumn{2}{c}{\textbf{Total Uncertainty}} \\
 \cline{6-7}
 Species &  $T_{\rm eff} +125\,{\rm K}$ & $\log{g} +0.20\,{\rm dex}$ & ${\xi_t} +0.30$\,km s$^{-1}$ & $\max(0.10, \sigma)/\sqrt(N)$ & [X/H] & [X/Fe] \\
 & $\Delta$abund. & $\Delta$abund. & $\Delta$abund. & (dex) & (dex) & (dex) \\
\hline
\endfirsthead
\hline
 & & & & & \multicolumn{2}{c}{\textbf{Total Uncertainty}} \\
 \cline{6-7}
 Species &  $T_{\rm eff} +125\,{\rm K}$ & $\log{g} +0.20\,{\rm dex}$ & ${\xi_t} +0.30$\,km s$^{-1}$ & $\max(0.10, \sigma)/\sqrt(N)$ & [X/H] & [X/Fe] \\
 & $\Delta$abund. & $\Delta$abund. & $\Delta$abund. & (dex) & (dex) & (dex) \\
\hline
\endhead
\hline
\multicolumn{7}{r}{Continued..}
\endfoot
\endlastfoot
\\
\multicolumn{7}{c}{\textbf{HD 41667}} \\
\hline
O I  		& 0.03		& 0.08		&$-$0.01		& 0.07		& 0.11		& 0.16 \\
Na I  		& 0.13		& 0.00		&$-$0.02		& 0.10		& 0.16		& 0.21 \\
Mg I  		& 0.08		& 0.01		&$-$0.01		& 0.05		& 0.10		& 0.18 \\
Al I  		& 0.10		& 0.01		&$-$0.01		& 0.06		& 0.11		& 0.18 \\
Si I  		& 0.02		& 0.03		&$-$0.02		& 0.04		& 0.06		& 0.16 \\
K I  		& 0.14		&$-$0.03	&$-$0.17		& 0.10		& 0.24		& 0.28 \\
\end{longtable}

\section{Discussion}
\label{sec:aquarius-discussion}

\subsection{Stellar Parameter Discrepancies with \citet{wylie-de-boer;et-al_2012}}
We now seek to investigate the nature of the Aquarius stream and in particular, the globular cluster origin suggested by \citet{wylie-de-boer;et-al_2012}. Before proceeding, we now compare our stellar parameters with those of \citet{wylie-de-boer;et-al_2012} for the four stars in common. \citet{wylie-de-boer;et-al_2012} deduce their stellar parameters ($T_{\rm eff}$, $\log{g}$, $\xi_t$, [M/H]) by minimizing the $\chi^2$ difference between the observed spectra and synthetic spectra from the \citet{munari;et-al_2005} spectral library. For the four stars common to both samples, the stellar parameters reported in \citet{wylie-de-boer;et-al_2012} differ from our values listed in Table \ref{tab:stellar-parameters}.  In general, effective temperatures between the two studies agree within the uncertainties. The only aberration is {J223811-104126}, where we find an effective temperature of {5190\,K}, $\sim$450\,K cooler than the {5646\,K} found by \citet{wylie-de-boer;et-al_2012}. Similarly, \citet{williams;et-al_2011} report a hotter effective temperature of {5502\,K} from low-resolution spectra. This is the largest discrepancy we find in any of our standard or program stars.

Photometric temperature relationships support our spectroscopic temperature for J223811-104126. The \citet{ramirez;melendez_2005} relationship for giants suggests an effective temperature of 5240\,K, which is 50\,K warmer than our spectroscopically-derived temperature. Furthermore, the metallicity-independent $J-K$ colour-$T_{\rm eff}$ relationship for giants by \citet{alonso;et-al_1999} yields an effective temperature of 5215\,K, 25\,K warmer than our spectroscopic temperature. As a test, we set the temperature for {J223811-104126} to be 5600\,K -- within the temperature regime reported by \citet{williams;et-al_2011} and \citet{wylie-de-boer;et-al_2012}. The slopes and offsets in abundance with excitation potential and REW were large: ${m_{\rm Fe\,I} = -0.099}$\,dex\,eV$^{-1}$, 0.162\,dex, ${m_{\rm Fe\,II} = -0.133}$\,dex\,eV$^{-1}$, $-0.033$\,dex respectively, and in doing so we could not find a representative solution for this temperature. 

\citet{williams;et-al_2011} and \citet{wylie-de-boer;et-al_2012} find {J223811-104126} to be a sub-giant/dwarf, with a surface gravity {$\log{g} = 4.16$} and {4.60} respectively. We note that the \citet{williams;et-al_2011} and \citet{wylie-de-boer;et-al_2012} effective temperatures for J223811-104126 are 150-300\,K hotter than the \citet{casagrande;et-al_2010} $J-K$ photometric temperature calibration for dwarfs and sub-giants. We find the surface gravity for {J223811-104126} to be {$\log{g} = 2.93 \pm 0.30$\,dex}, placing this star at the base of the red giant branch.

With the exception of {J223811-104126}, our surface gravities are largely in agreement with \citet{wylie-de-boer;et-al_2012}. The only other noteworthy difference is for {J221821-183424}, where we find a lower gravity of {$\log{g} = 0.88 \pm 0.30$} and \citet{wylie-de-boer;et-al_2012} find {$\log{g} = 1.45 \pm 0.35$}. Given the difference in the $S/N$  between these studies, this difference is not too concerning. \citet{wylie-de-boer;et-al_2012} calculate the microturbulence from empirical relationships derived by \citet{reddy;et-al_2003} for dwarfs and \citet{fulbright_2000} for giants. These relationships are based on the effective temperature and surface gravity. Our published microturbulent velocities agree excellently with the values presented in \citet{wylie-de-boer;et-al_2012}, again with the exception of {J223811-104126}, where the difference in $v_{t}$ is directly attributable to the offsets in other observables.

Of all the stellar parameters, metallicities exhibit the largest discrepancy between the two studies. In the \citet{wylie-de-boer;et-al_2012} study, after the stellar parameters ($T_{\rm eff}$, $\log{g}$, $v_{t}$, and an initial [M/H] estimate) were determined through a $\chi^2$ minimisation, the authors synthesised individual {Fe\,I} and Fe\,II lines using MOOG. \citet{castelli;kurucz_2003} stellar atmosphere models were employed (K. Freeman, private communication, 2013) -- the same ones used in this study -- albeit the interpolation schemes will have subtle differences. The median abundance of synthesised {Fe\,I} lines was adopted as the overall stellar metallicity, and scaled relative to the Sun using the \citet{grevesse;sauval_1998} Solar composition.

The study of \citet{wylie-de-boer;et-al_2012} is of slightly lower resolution ($\mathcal{R} = 25,000$ compared to $\mathcal{R} = 28,000$ presented here), but with a much lower $S/N$ ratio: ${\sim{}25\,{\rm pixel}^{-1}}$ compared to {$>$100\,pixel$^{-1}$} achieved here. This offset translates into a factor of at least 15 electrons per pixel. The line list employed in the \citet{wylie-de-boer;et-al_2012} study utilized astrophysical oscillator strengths derived from a reverse solar analysis on the Solar spectrum. However, there are very few transitions listed in their line list: a maximum of 14 {Fe\,I} lines and 3 {Fe\,II} lines were available. For contrast, our analysis is based on 63 {Fe\,I} and 13 {Fe\,II} clean, unblended lines. 

We suspect that the discrepancy in derived metallicities can be primarily attributed to the difference in line lists. In order to test this hypothesis, we re-analysed our data using the \citet{wylie-de-boer;et-al_2012} line list and stellar parameters. Excitation or ionization equilibria were unachievable using any stellar parameters within the quoted uncertainties by \citet{wylie-de-boer;et-al_2012}, and we observed metallicities different to \citet{wylie-de-boer;et-al_2012}. Alternatively, if the \citet{wylie-de-boer;et-al_2012} line list is employed and we solve for stellar parameters (see Section \ref{sec:aquarius-stellar-parameter-derivation}), we obtain stellar parameters closer to \emph{our} existing measurements tabulated in Table \ref{tab:stellar-parameters}, which are also distinct from the \citet{wylie-de-boer;et-al_2012} values. From the four stars common between these studies, using the \citet{wylie-de-boer;et-al_2012} line list and our stellar parameters\footnote{Here we are referring to the stellar parameters found by excitation and ionization equilibria using the \citet{wylie-de-boer;et-al_2012} line list, \emph{not} those listed in Table \ref{tab:stellar-parameters}.}, we observe a metallicity dispersion of $\sigma$([Fe/H]) = 0.32\,dex. This is contrast to the $\sigma$([Fe/H]) = 0.10\,dex reported by \citet{wylie-de-boer;et-al_2012} from six Aquarius stream stars. There is no obvious reason to explain this situation.

Furthermore given the small numbers of Fe lines used by \citet{wylie-de-boer;et-al_2012}, even subtle changes to the stellar parameters produced large variations to both the individual and mean Fe abundances. Additionally, we note that one {Fe\,I} transition at $\lambda$6420 in the \citet{wylie-de-boer;et-al_2012} line list was either not detected at the {3$\sigma$} level -- even though the $S/N$ at this point exceeds {115\,pixel$^{-1}$} in every observation -- or it was blended with a stronger neighbouring transition.

\subsection{The Aquarius Stream Metallicity Distribution}

Given the overall data quality and the limited Fe line list used for analysis, it appears the Aquarius stream stars conspired to present a tight metallicity distribution of {$\sigma(\mbox{[Fe/H]}) = 0.10$\,dex} in the \citet{wylie-de-boer;et-al_2012} analysis (in our analysis we find {$\sigma(\mbox{[Fe/H]}) = 0.33$\,dex}). When viewed in light of enhanced [Ni/Fe] and [Na/Fe] abundance ratios, \citet{wylie-de-boer;et-al_2012} interpreted this chemistry as a signature of a globular cluster origin for the Aquarius stream. Our study of high-resolution spectra with high $S/N$ reveals a much broader metallicity distribution for the stream than previously reported. With just 5 stars we find the metallicity varies from {$\mbox{[Fe/H]} = -0.63$} to {$-1.58$}. Although this is a small sample, we find the mean abundance and standard deviation to be {$\mbox{[Fe/H]} = -1.20 \pm 0.33$}. 

If the metallicity dispersion were smaller, as found by \citet{wylie-de-boer;et-al_2012}, a globular cluster scenario may be plausible. Classical globular clusters typically exhibit very little dispersion in metallicity. An intrinsic [Fe/H] dispersion of 0.33\,dex -- ignoring error contribution -- is substantially larger than that seen in any globular cluster, with the exception of the unusual system $\omega$-Centauri. In that cluster the total abundance range is about $\Delta$[Fe/H] $\sim$ 1.4\,dex: from $-$2 to $-$0.6 \citep[e.g.][]{marino;et-al_2011}, and many sub-populations have been identified \citep[e.g.,][]{johnson_pilachowski_2010}. 

Other clusters with established intrinsic [Fe/H] dispersions include M54 -- a nuclear star cluster of the Sagittarius dSph -- where $\sigma_{\rm int}$([Fe/H])$ = 0.19$ \citep{carretta;et-al_2010}, and M22, where the interquartile range in [Fe/H] is $\sim{}$0.24 dex \citep{da_costa;et-al_2009,marino;et-al_2009,marino;et-al_2011}. There are a few clusters where the intrinsic dispersion is $\sim$0.10, namely NGC 1851 \citep{carretta;et-al_2011}, NGC 5824 \citep{saviane;et-al_2012}, and NGC 3201 \citep{simmerer;et-al_2013}. These globular clusters are outliers, and even amongst these unusual systems they largely do not match the abundance spread observed in the Aquarius stream \citep[e.g., see Figure 4 in][]{simmerer;et-al_2013}. In fact, the Aquarius stream metallicity distribution -- on its own -- is large enough to be reconcilable with dSph galaxies like Fornax \citep[e.g., ][]{letarte;et-al_2010}. Similarly, the mean Aquarius stream metallicity and the $\log(L)$, $\langle$[Fe/H]$\rangle$ relation of \citet{kirby;et-al_2011} also suggest a relatively luminous system with $L_{\rm tot} \sim 10^{7.5}L_\odot$ \citep{kirby;et-al_2011}.  However, the Aquarius stream stars exhibit very different abundance ratios to Fornax. For example, [Ba/Y] (e.g., heavy/light $s$-process) abundance ratios in the Aquarius stream vary between $-$0.24 and +0.19, significantly lower than the [Ba/Y] $\geq$ 0.5 level generally observed in the present day dSphs \citep{venn;et-al_2004}.

\subsection{The Na-O Relationship}

Sodium is primarily produced through carbon burning in massive stars by the dominant $^{12}\mbox{C}(^{12}\mbox{C}, p)^{23}\mbox{Na}$ reaction. The final Na abundance is dependent on the neutron excess of the star, which slowly increases during carbon burning due to weak interactions \citep{arnett;truran_1969}. Massive stars ($>10 M_\odot$) deliver their synthesized sodium to the interstellar-medium through {SN\,II} explosions. Because the eventual {SN\,II} explosion is devoid of any significant $\beta$-decay processes, the neutron excess of the ejected material is representative of the pre-explosive abundance. The ejected material eventually condenses to form the next generation of stars, which will have a net increase in their neutron excess with respect to their predecessors. Since the sodium production rate is correlated with the neutron excess, an overall increase in the total sodium content \textit{and} Na-production rate between stellar generations can be expected. The sodium content also becomes important for production of nickel during the {SN\,II} event (see Section \ref{sec:aquarius-na-ni-relationship}) because $^{23}$Na is the only stable isotope produced in significant quantities during C-burning.

Extensive studies of stars in globular clusters have revealed variations in light element abundances, most notably an anti-correlation in sodium and oxygen content \citep[see][and references therein]{norris;da_costa_1995,carretta;et-al_2009a}. This chemical pattern has been identified in every well-studied globular cluster, although the magnitude and shape of the anti-correlation vary from cluster to cluster. The direct connection between Na and O abundances requires an additional synthesis mechanism for Na, at least for the Na content that exceeds the Na in the primordial population.

Oxygen depletion is likely the result of complete CNO burning within the stellar interior. The nucleosynthetic pathways that produce the {Na-O} anti-correlation are well understood to be proton-capture nucleosynthesis at high temperatures \citep{prantzos;et-al_2007}. However, the temperatures required to produce these patterns are not expected within the interiors of globular cluster stars. While the exact mechanism for which these conditions occur remains under investigation, we can describe the abundance variation as an external oxygen depletion (or dilution) model with time. Through comparisons with existing globular clusters, we can make inferences on the star-formation history of a system by measuring sodium and oxygen abundances in a sample of its stars. 

\begin{figure*}[t!]
	\includegraphics[width=\textwidth]{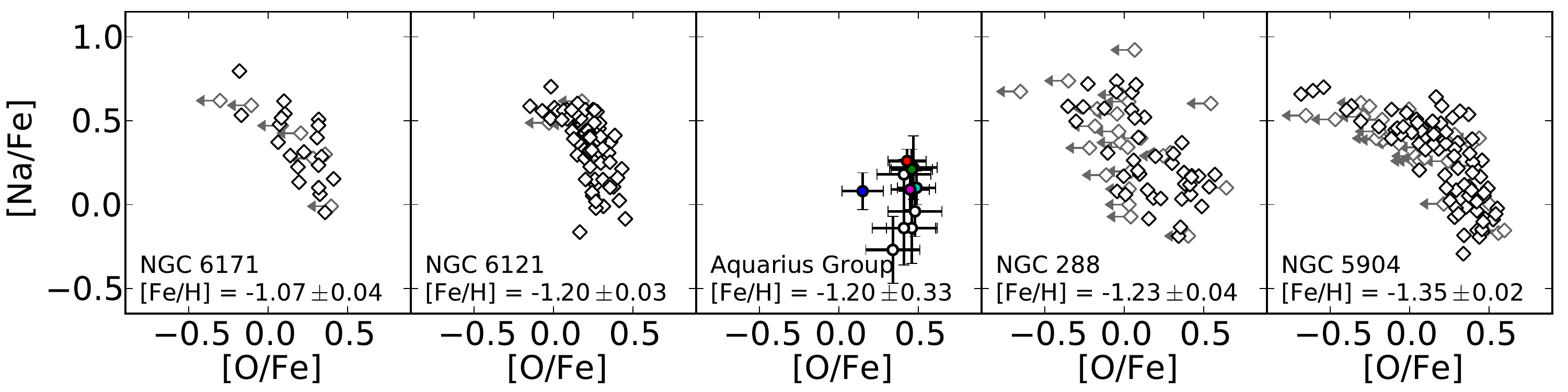}
	\caption[Oxygen and sodium abundances for the Aquarius stream stars, as well as classical globular clusters with similar mean metallicities to the Aquarius stream]{Oxygen and sodium abundances for 4 classical globular clusters with mean metallicities similar to the Aquarius stream \citep{carretta;et-al_2009a}. All clusters demonstrate a {Na-O} anti-correlation. [O/Fe] and [Na/Fe] abundances from this study for the 5 Aquarius stream stars and standard stars (open circles) are shown in the middle panel. Colors for Aquarius stars is as per Figure \ref{fig:toombre}.}
	\label{fig:o-na-clusters}
\end{figure*}

Such inferences must be made with careful consideration. In addition to the normal care afforded for measuring elemental abundances from high-resolution spectroscopic data, attention must be given to telluric absorption, contamination from Ni\,I, as well as non-LTE and 3D effects when determining oxygen abundances. Furthermore, when characterising the oxygen depletion rate -- the strength of the {Na-O} anti-correlation in a globular cluster -- it is vital to sample, where possible, stars belonging to all three components \citep[primordial, intermediate and extreme, see][]{carretta;et-al_2009a}. If only a primordial sample of stars is observed, their [Na/Fe] and [O/Fe] abundances will be, by definition, indistinguishable from field stars of a similar metallicity. In such a scenario any inferred anti-correlation could equally be explained by small abundance variations or observational uncertainties. 

\begin{figure}[t!]
	\includegraphics[width=\textwidth]{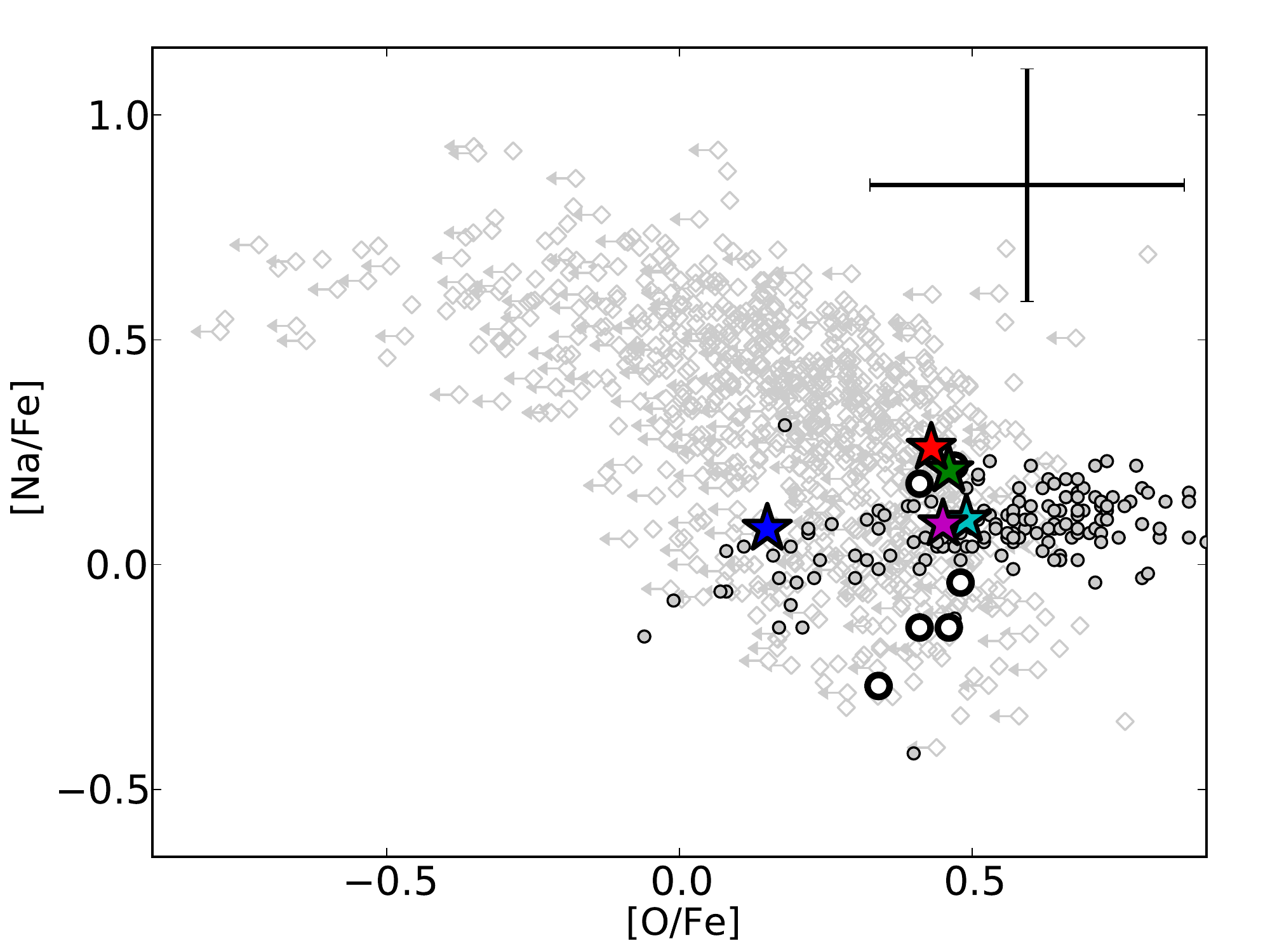}
	\caption[Oxygen and sodium abundances for disk/halo stars, globular clusters, and the Aquarius stream stars]{Oxygen and sodium abundances for disk/halo stars from \citet{reddy;et-al_2006} are shown as grey circles, and globular cluster stars from \citet{carretta;et-al_2009a} are shown as diamonds. The Aquarius stream stars are also shown -- following the same colors in Figure \ref{fig:toombre} -- illustrating how their [O/Fe], [Na/Fe] content is not dissimilar from Galactic stars. Although some standard stars (open circles) have lower [Na/Fe] abundances than the \citet{reddy;et-al_2006} sample, our values are consistent with \citet{ishigaki;et-al_2013}, who did not measure O abundances.}
	\label{fig:o-na}
\end{figure}

\citet{wylie-de-boer;et-al_2012} measured sodium and oxygen abundances for four of their six Aquarius stream members. These abundance measurements exist for only three stars common to this study and \citet{wylie-de-boer;et-al_2012}, as the data quality for {J223811-104126} in the \citet{wylie-de-boer;et-al_2012} was too low to permit oxygen measurements. We have measured sodium and oxygen abundances for all of our stars, which are plotted in Figures \ref{fig:o-na-clusters} and \ref{fig:o-na}\footnote{Although the \citet{reddy;et-al_2006} sample primarily consists of dwarfs/subgiants and we are observing primarily giants/subgiants, this does not affect our interpretation.}. These figures employ the corrected [O/Fe] value for {J223811-104126} rather than a conservative upper limit (see Section \ref{sec:aquarius-alpha-elements}).

The \citet{wylie-de-boer;et-al_2012} measurements show two stars with solar levels of [Na/Fe] -- identical to field star abundances for their metallicity -- and two stars with slightly enhanced sodium content: {J223504-152834} and {J232619-080808}. We also find {J223504-152834} to be slightly sodium-enhanced, whereas the second star in their study, {J232619-080808}, is not in our sample. We find the additional star not present in the \citet{wylie-de-boer;et-al_2012} sample, {C2306265-085103}, to be enhanced to almost the same level of {J223504-152834} with {$\mbox{[Na/Fe]} = 0.26$}. The sodium-enhanced stars are not enhanced significantly above the total uncertainties, and they do not exhibit depletion of oxygen: their chemistry is not representative of a {Na-O} anti-correlation. 

In the Aquarius sample we observe no intrinsic dispersion above the measurement uncertainty in [O/Fe] or [Na/Fe]. A dispersion of 0.14\,dex is observed for [O/Fe] (or 0.03\,dex when J223811-104126 is excluded), which is only marginally larger than the mean total uncertainty of $\sigma({\rm [O/Fe]}) = 0.12$\,dex. Similarly for [Na/Fe], a dispersion of $\sigma({\rm [Na/Fe]}) = 0.08$\,dex is observed, when taking the uncertainties into account, is consistent with zero dispersion. We also see no significant variation in [C/Fe] outside the uncertainties, or relationship between [C/Fe]-[Na/Fe] \citep[e.g., see][]{yong}.

Now we consider the possibility that the Aquarius stars did originate in a globular cluster. Given the negligible dispersions present in [(O, Na, C)/Fe] (among other elements), the stars would be considered as members the primordial component, which comprises $\sim$33\% of the total population for any globular cluster \citep{carretta;et-al_2009a}. The likelihood of randomly observing five globular cluster stars that all belong to the primordial component is 0.4\%. If the primordial component was a larger fraction (e.g., 40\% or 50\%), this probability raises marginally, to 1\% and 3\% respectively. Given the dispersion in overall metallicity though, such a globular cluster would be an unusual object.

If the Aquarius stream is the result of a disrupted globular cluster, a large part of the picture must still be missing. Almost all of the Aquarius stream stars studied to date (either in this sample or the \citet{wylie-de-boer;et-al_2012} study), would be unambiguously classified as belonging to a ``primordial'' component, with chemistry indistinguishable from field stars. Identifying more Aquarius stream members belonging to the intermediate component with strong oxygen depletion, or perhaps members of an extreme component, would be convincing evidence for a {Na-O} anti-correlation and a globular cluster origin. Three stream stars identified to date (including two from this sample) might tenuously be classified as members of an intermediate population, with only a slight enhancement in sodium and no oxygen-depletion. Recall our [Na/Fe] abundance ratios appear systematically higher in our standard stars when compared to the literature sources listed in Table \ref{tab:stellar-parameters}. Thus, if the strength of any {Na-O} relationship is to be used to vet potential disrupted hosts for the Aquarius stream, many more stream members will need to be identified and observed spectroscopically with high-resolution and high $S/N$. In the absence of such data, no evidence exists for a {Na-O} anti-correlation in the Aquarius stream.

\subsection{The Al-Mg Relationship}

Although not ubiquitous to every system, many globular clusters exhibit an anti-correlation between aluminium and magnesium. This is perhaps unsurprising, given the nucleosynthetic pathways for these elements. In addition to the CNO cycle operating during hydrogen burning, the {Mg-Al} chain can also operate under extreme temperatures \citep[$T \sim 8 \times 10^{6}\,$K;][]{arnould;et-al_1999}. Aluminium is produced by proton-capture onto magnesium, beginning with $^{25}$Mg to $^{26}$Al. The relative lifetime of $\beta$-decay to proton-capture allows for the production of unstable $^{27}$Si through proton-capture. Seconds later, the isotope decays to $^{27}$Al, completing the $^{27}$Si path of the Mg-Al chain. The alternative process from $^{26}$Al involves $\beta$-decay to $^{26}$Mg.

The Mg-Al cycle can explain the observed anti-correlation observed in some globular clusters, and some AGB models can predict such a pattern (e.g., \citet{ventura;et-al_2011} but see nucleosynthesis yields from \citealt{karakas;lattanzio_2007} and \citealt{karakas_2010}). However, at present the observed Mg isotope ratios are in tension with AGB model predictions \citep{yong;et-al_2006,da_costa;et-al_2013}. Thus, while the pathways for creating these chemical patterns are understood, the temperatures required to produce this chain are much higher than expected for low-mass stellar interiors, so the exact site and requisite conditions are lacking a full description.

\citet{wylie-de-boer;et-al_2012} published magnesium and aluminium abundances for five stars in their Aquarius sample. No inverse correlation is present in their data; their abundances are indistinguishable from field stars. The [(Mg, Al)/Fe] abundance ratios tabulated in Table \ref{tab:program-star-abundances} are generally in agreement with the \citet{wylie-de-boer;et-al_2012} sample, and we also find no {Mg-Al} anti-correlation. 

\begin{figure}[h!]
	\includegraphics[width=\textwidth]{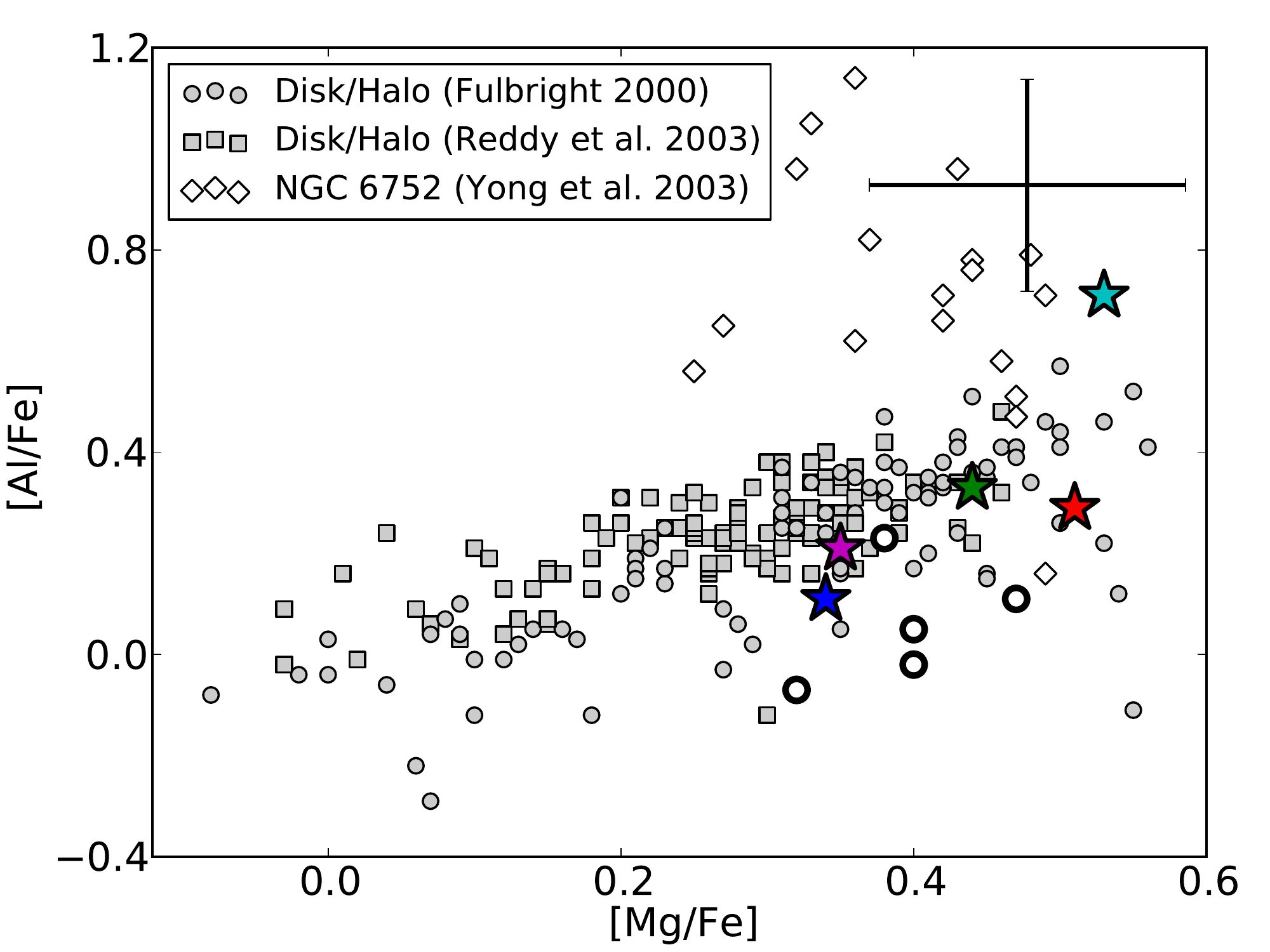}
	\caption[Magnesium and aluminium abundances for Aquarius stream stars compared with Milky Way field stars]{Magnesium and aluminium abundances for Aquarius stream stars, as well as Milky Way halo/disk stars from \citet{reddy;et-al_2003} and \citet{fulbright_2000}. Aquarius stars are colored as described in Figure \ref{fig:toombre}.}
	\label{fig:mg-al}
\end{figure}

However it {\it is} surprising that we find such a strong positive relationship in [Mg/Fe] and [Al/Fe], with a best-fitting slope of ${{\rm [Al/Fe]} = 2.08\times\mbox{[Mg/Fe]} - 0.57}$. If we exclude the chemically peculiar star {C222531-145437}, the slope decreases to {${\rm [Al/Fe]} = 0.96\times\mbox{[Mg/Fe]} - 0.16$}, a near 1:1 relationship. Even when a {Mg-Al} anti-correlation is not detected in globular clusters, there is generally more scatter in [Al/Fe] at near-constant [Mg/Fe] \citep[e.g., see Figure \ref{fig:mg-al} or ][]{carretta;et-al_2009a}. This is because Mg is much more abundant than Al, requiring only a small amount of Mg atoms to be synthesized to Al before the differences in Al abundance become appreciable, whilst the observed Mg abundance could remain within the uncertainties.

No classical globular clusters exhibit a positive correlation, and nor is such a pattern expected in globular clusters. However, a positive relationship between magnesium and aluminium can result from {SN\,II} contributions to the local interstellar medium. Intermediate-mass ($\gtrsim$4$M_\odot$) AGB models can also contribute towards a positive correlation between aluminium and magnesium. Under extreme temperatures ($T \gtrsim 300 \times 10^{6}\,$K), substantial $^{25}$Mg and $^{26}$Mg are produced by $\alpha$-capture onto $^{22}$Ne by the $^{22}$Ne($\alpha$,n)$^{25}$Mg and $^{22}$Ne($\alpha$,$\gamma$)$^{26}$Mg reactions respectively \citep[e.g.,][]{karakas;et-al_2006}. Depending on uncertain numerical details of stellar modelling , the third dredge-up can mix significant quantities of $^{25}$Mg and $^{26}$Mg into the photosphere, even more than the quantity of $^{26}$Al produced through the {Mg-Al} cycle. Therefore, a positive relationship between magnesium and aluminium can occur if there has been significant contributions from intermediate-mass AGB stars, however this should also produce a Na-O anti-correlation \citep[][but see results in \citealt{ventura;et-al_2011}]{karakas;lattanzio_2003}.
 
An unrealistically high fraction of intermediate-mass AGB stars would be required to produce the same chemical signature of a single {SN\,II} event. Given the efficiency of chemical mixing following supernovae, the observed positive {Mg-Al} correlation in the thick disk is likely the result of {SN\,II} events. Distinguishing between these processes observationally requires careful measurements of magnesium isotope abundances $^{24}$Mg (indicating supernovae mixing), $^{25}$Mg and $^{26}$Mg (suggesting significant AGB contribution), which is not possible given our $S/N$ or spectral resolution. In either scenario, we can summarise that the strong Mg-Al relationship provides additional chemical evidence against a globular cluster scenario for the Aquarius stream, and the chemistry is suggestive of Milky Way disk stars.

\subsection{The Na-Ni Relationship}
\label{sec:aquarius-na-ni-relationship}

Detailed chemical studies of nearby disk stars have noted a correlation with [Na/Fe] and [Ni/Fe] abundance ratios. This relationship was first hinted in \citet{nissen;schuster_1997}, where the authors found eight stars that were under-abundant in [$\alpha$/Fe], [Na/Fe] and [Ni/Fe]. Interestingly, the authors noted that stars at larger Galactocentric radii were most deficient in these elements. \citet{fulbright_2000} saw a similar signature: stars with low [Na/Fe] were only found at large {($R_{\rm GC} > 20$\,kpc)} distances. \citet{nissen;schuster_1997} proposed that since the outer halo is thought to have been largely built up by accretion, then the {Na-Ni} pattern may be a chemical indicator of merger history within the galaxy.

With additional data from \citet{nissen;schuster_2011}, the {Na-Ni} relationship was found to be slightly steeper than originally proposed. The pattern exists only for stars with {$-1.5 < \mbox{[Fe/H]} < -0.5$}, and is not seen in metal-poor dSph stars \citep{venn;et-al_2004}, providing a potentially useful indicator for investigating chemical evolution. However, it is crucial to note that although there are only a few dSph stars in the $-1.5 < \mbox{[Fe/H]} < -0.5$ metallicity regime with [Na/Fe] and [Ni/Fe] measurements, they agree reasonably well with the Galactic trend. 

The correlation between sodium and nickel content is the nucleosynthetic result of neutron-capture in massive stars. As previously discussed, the total Na abundance is controlled by the neutron excess, which limits the production of $^{58}$Ni during {SN\,II} events. When the inevitable supernova begins, the core photodissociates into neutrons and protons, allowing the temporary creation of $^{56}$Ni before it decays to $^{56}$Fe. A limited amount of $^{54}$Fe is also formed, which is the main source of production for the stable $^{58}$Ni isotope through $\alpha$-capture. The quantity of $^{54}$Fe (and hence $^{58}$Ni) produced is dependent on the abundance of neutron-rich elements during the explosion. As $^{23}$Na is a relatively plentiful neutron source with respect to other potential sources (like $^{13}$C), the post-supernova $^{58}$Ni abundance is driven by the pre-explosion $^{23}$Na content. Thus, through populations of massive stars undergoing C-burning, a positive correlation between sodium and nickel can be expected. 

\begin{figure}[h!]
	\includegraphics[width=\textwidth]{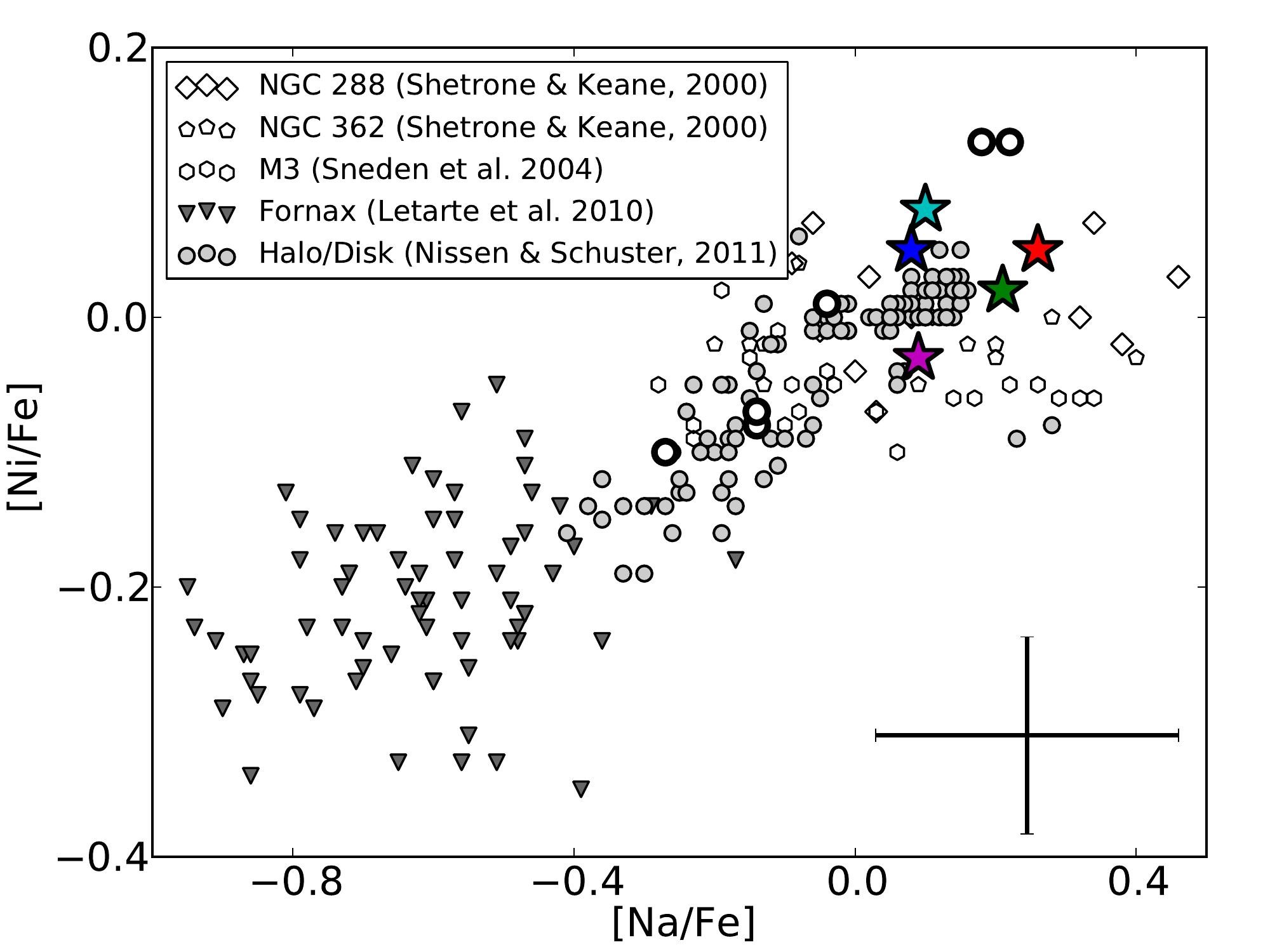}
	\caption[Sodium and nickel abundances for Aquarius stream stars, globular clusters, the Fornax dwarf gaalxy, and field stars]{[Na/Fe] and [Ni/Fe] for Aquarius stream stars and for globular cluster, dSph, and field (halo/disk) stars. Aquarius targets are colored as per Figure \ref{fig:toombre}. Stars from the most representative dSph galaxy, Fornax, are shown as downward triangles ($\blacktriangledown$). Fornax has been chosen as it lies closest to the luminosity that one would expect for an Aquarius host system, given its overall metallicity, metallicity dispersion, and the $\log(L)-\langle$[Fe/H]$\rangle$ relationship \citep{kirby;et-al_2011}. The \citet{nissen;schuster_2011} plotted sample included halo stars, as well as low- and high-$\alpha$ disk members. As noted by \citet{nissen;schuster_1997,nissen;schuster_2011}, a positive Na-Ni relationship exists for field stars, with dSph members exhibiting strong depletion in both elements and globular cluster stars consistently showing an enhancement. Sodium and nickel content for Aquarius members indicate a dSph accretion origin is unlikely.}
	\label{fig:na-ni}
\end{figure}

Stars originating in dSph galaxies and globular clusters have very different chemical enrichment environments. Consequently, both types of systems exhibit chemistry that reflects their nucleosynthetic antiquity. Stars in dSphs do not demonstrate enhanced sodium or nickel content with respect to iron, as there has been a relatively small lineage of massive stars undergoing supernova. In contrast, globular cluster stars do have elevated [Na/Fe] and [Ni/Fe] signatures. This sharp contrast between dSph and globular cluster star chemistry is illustrated in Figure \ref{fig:na-ni}. Given the extended star formation within the Milky Way disk, globular cluster stars and disk stars are indiscernible in the Na/Ni plane: they both show an extended contribution of massive stars. The most that can be inferred from the Na and Ni abundances of Aquarius stream stars is that their enrichment environment is less like a dSph galaxy, and more representative of either a globular cluster, or the Milky Way disk. 

\subsection{The Chemically Peculiar Star C222531-145437}

In almost every element with respect to iron, {C222531-145437} is distinct from the other Aquarius stream stars. It is over-abundant in light  and neutron-capture elements, with a high barium abundance of {$\mbox{[Ba/Fe]} = 0.68$}. This value is well in excess of the halo ({$\mbox{[Ba/Fe]} \sim 0.0$}) -- and our other Aquarius stream stars -- which vary between $-$0.02 to {0.15\,dex}.

Here we discuss the possibility that an unseen companion has contributed to the surface abundances of C222531-145437. Although no radial velocity variations were observed between exposures, we do not have a sufficient baseline to detect such variation. The abundances of heavy elements produced by AGB stars have a high dependence on the initial metallicity and mass. Low-mass ($\lesssim{}3M_\odot$) AGB stars of low metallicity produce high fractions of heavy $s$-process elements compared to their light $s$-process counterparts \citep{busso;et-al_2001}. As such, [Ba/Y] is a useful indicator for considering contributions from a low-mass AGB companion. For C222531-145437, ${{\rm [Ba/Y]} = -0.17}$, which is much lower than expected if a low-mass AGB star was responsible for the heavy element enhancements ([Ba/Y] $\sim$ 0.5 as shown in Figure \ref{fig:ba-y}; see also \citealt{cristallo;et-al_2009}). If mass transfer from an AGB companion has occurred very recently, non-negligible amounts of technetium, produced by the AGB star, remain visible in the companion's photosphere before $^{99}$Tc decays over $\sim$2\,Myr \citep{brown;et-al_1990,van_eck_jorrissen_1999, uttenthaler;et-al_2011}. We saw no technetium absorption at $\lambda$4049, $\lambda$4238 or $\lambda$4297 in the spectrum of C222531-145437. Intermediate-mass (3-5$M_\odot$) AGB stars also cannot explain the abundances for C222531-145437: using recently computed intermediate-mass AGB $s$-process yields for 3-5$M_\odot$ for a star of ${{\rm [Fe/H]} \approx -1.2}$ (Fishlock et al., in preparation) the resulting surface abundances do not match the observations (Figure \ref{fig:ba-y}). Therefore, we find no reason to suspect the heavy element enhancement in C222531-145437 is the result of mass transfer from an AGB companion.

\begin{figure}[h!]
	\includegraphics[width=\textwidth]{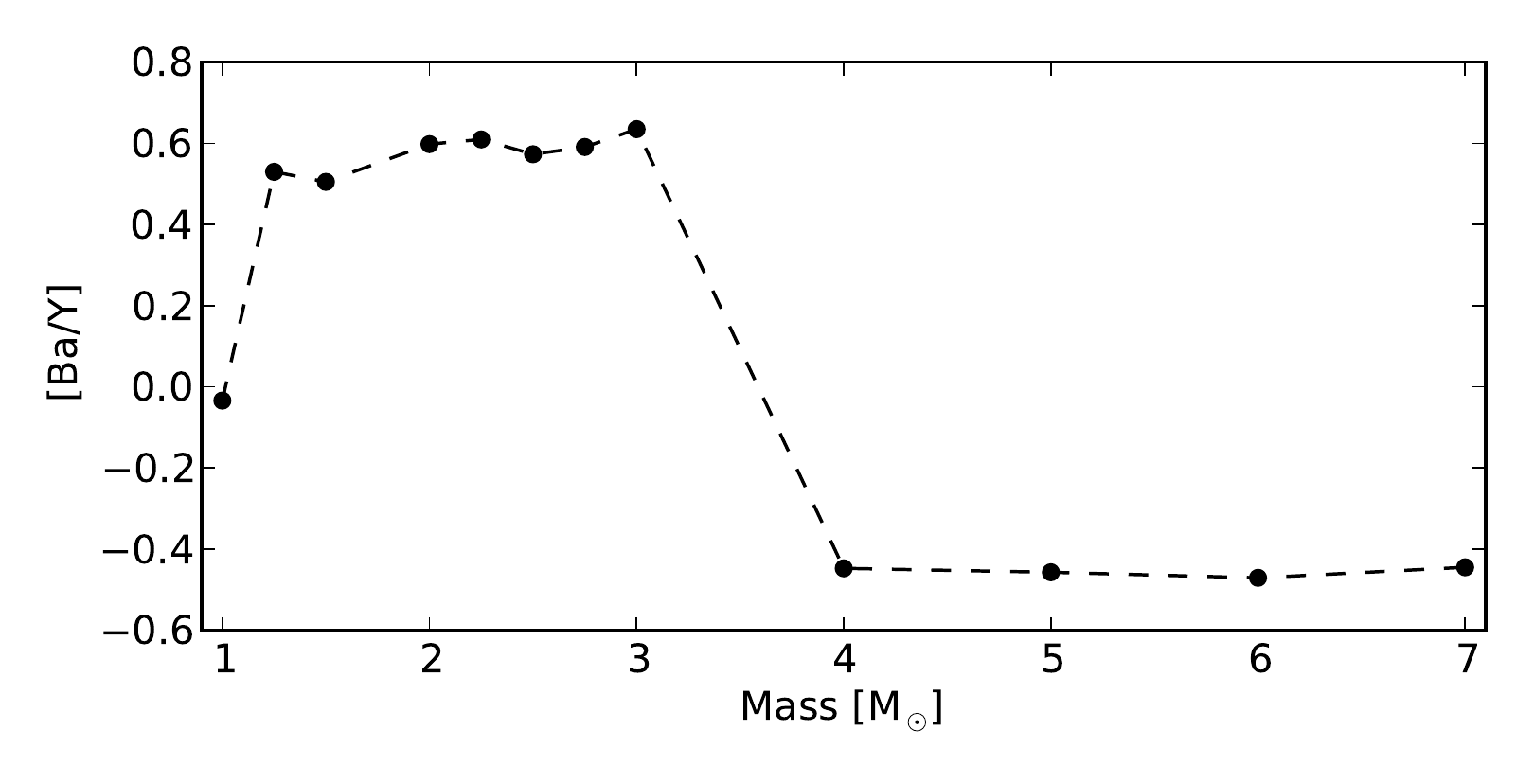}
	\caption[Distribution of final surface abundances of ${\rm [Ba/Y]}$ with initial AGB model mass]{Distribution of the final surface abundance of [Ba/Y] with initial mass for each of the AGB models calculated for $Z~=~0.001$ (Fishlock et al., in preparation). The ratio of [Ba/Y] can be used as an indicator of the initial mass of the AGB companion where the low-mass AGB models show a higher [Ba/Y] ratio compared to the intermediate-mass AGB models.}
	\label{fig:ba-y}
\end{figure}

Stars in $\omega$-Centauri show large over-abundances of $s$-process elements compared to the Galaxy \citep{norris;da_costa_1995,stanford;et-al_2010,johnson_pilachowski_2010}. M22 also hosts an $s$-process rich population \citep{marino;et-al_2011}. Like the Aquarius co-moving group, both clusters are relatively close to the Sun: {5.2\,kpc} and {3.2\,kpc}, respectively. M22 has a mean metallicity of {$\mbox{[Fe/H]} \sim -1.7$} and a range between {$-2.0 < \mbox{[Fe/H]} < -1.6$\,dex}, making an association between {C222531-145437} and M22 unlikely. Similarly, {C222531-145437} is unlikely to be associated with the metal-rich Argus association \citep[IC 2391;][]{de_silva;et-al_2013}, which also shows large enhancement in $s$-process abundances. Other groups have identified field stars enriched in $s$-process elements, which have generally been associated as tidal debris from $\omega$-Centuari \citep{wylie-de-boer;et-al_2010,majewski;et-al_2012}. The high $s$-process abundance ratios and overall metallicity of {C222531-145437} ({$\mbox{[Fe/H]} = -1.22$}) suggests this star may also be a remnant from the tidal disruption process.

\begin{figure}[h!]
	\includegraphics[width=\textwidth]{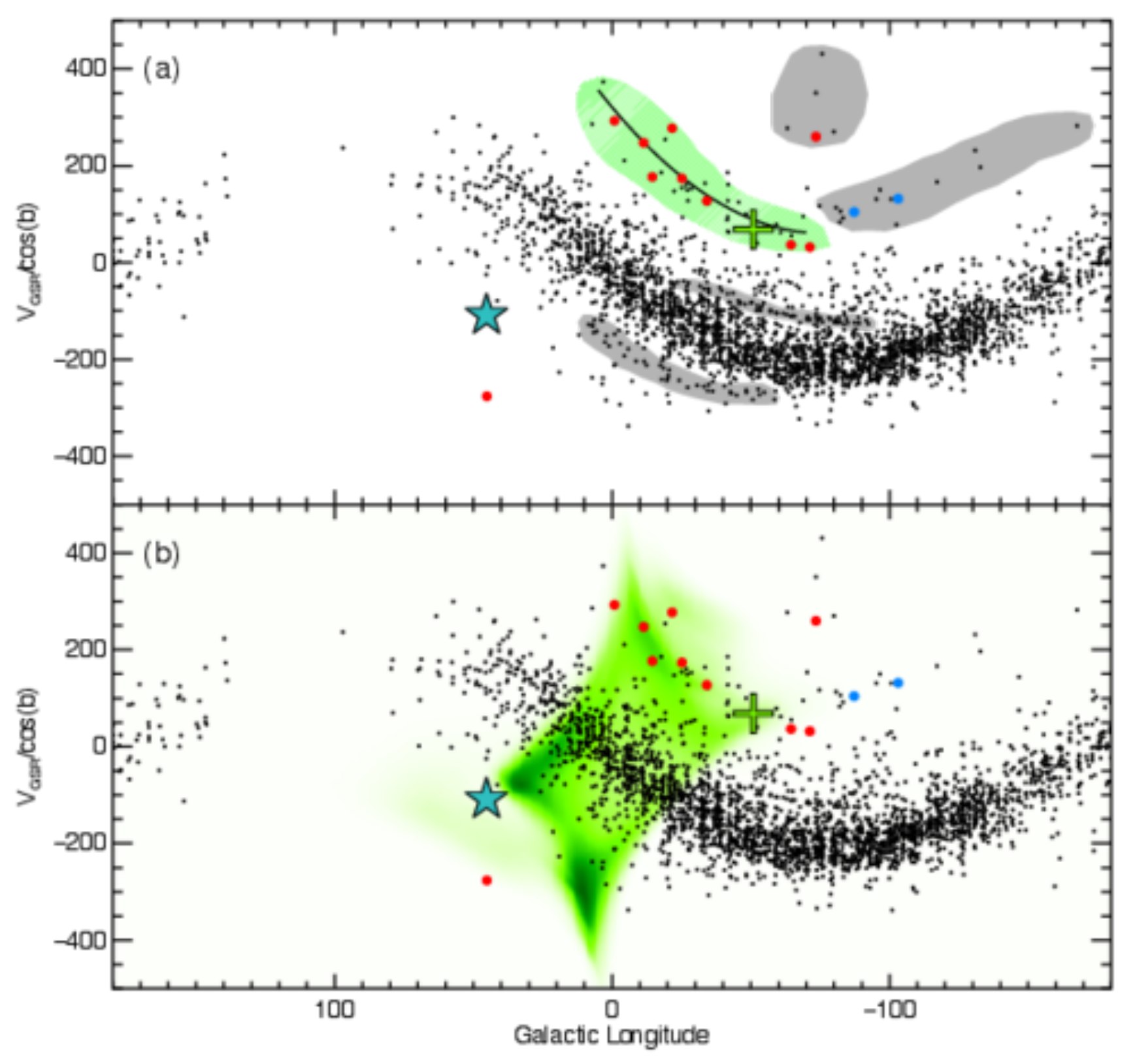}
	\caption[Position and velocity information for giant stars from the GGSS survey. C222531-145437 is shown in context of $\omega$-Centauri tidal debris.]{Panel (a) shows the distribution of giant stars in the GGSS \citep{majewski;et-al_2012} in Galactic longitude and $V_{\rm GSR}/cos(b)$ after excluding stars with $|b| > 60^\circ$. Stars from the GGSS sample believed to be $\omega$-Centauri tidal debris are shown in green shading. Red points are stars from the GGSS sample with abundances that follow the $\omega$-Centauri [Ba/Fe]--[Fe/H] pattern. Blue points are those with high-resolution spectra that do not follow this trend. Grey shading highlights other potential halo substructures from their study. Panel (b) shows the probability distribution of $\omega$-Centauri tidal debris from 4,050 simulations.  The $\omega$-Centauri core is shown as a green cross and the cyan star represents {C222531-145437}, falling almost precisely where a relatively high probability of $\omega$-Centauri tidal debris is expected.}
	\label{fig:omega-cen-tidal-debris}
\end{figure}

$\omega$-Centauri has a retrograde orbit with low inclination. Many groups simulating this orbit have predicted retrograde tidal debris to occur near the Solar circle \citep{dinescu_2002,tsuchiya_2003,tsuchiya_2004,bekki;freeman_2003}. Subsequent searches for $\omega$-Centauri debris in the Solar neighbourhood have led to tantalising signatures of debris. From over 4,000 stars targeted by \citet{da_costa;coleman_2008} in the vicinity of the cluster's tidal radius, only six candidate debris members were recovered, consistent with tidal stripping occurring long ago. Using data from the Grid Giant Star Survey (GGSS), an all-sky search looking for metal-poor giant stars, \citet{majewski;et-al_2012} identified 12 stream candidates. In addition, \citet{majewski;et-al_2012} performed 4,050 simulations in order to predict likely locations for $\omega$-Centauri tidal debris. The results of their simulation are replicated in Figure \ref{fig:omega-cen-tidal-debris}, where the location of {C222531-145437} is also shown. The velocity and position of {C222531-145437} align almost precisely where \citet{majewski;et-al_2012} predict a high probability of tidal debris. More interestingly, the angular momentum and orbital energy for {C222531-145437} (Figure \ref{fig:orbits}) matches excellently for $\omega$-Centurai cluster stars as well as its previously identified tidal remnants \citep{wylie-de-boer;et-al_2010}. The chemical and phase-space information strongly suggests that {C222531-145437} is associated with the remnants of tidal stripping that occurred as the proto-$\omega$-Centauri fell into the Galaxy \citep{bekki;freeman_2003}.

\begin{figure}[h!]
	\includegraphics[width=\textwidth]{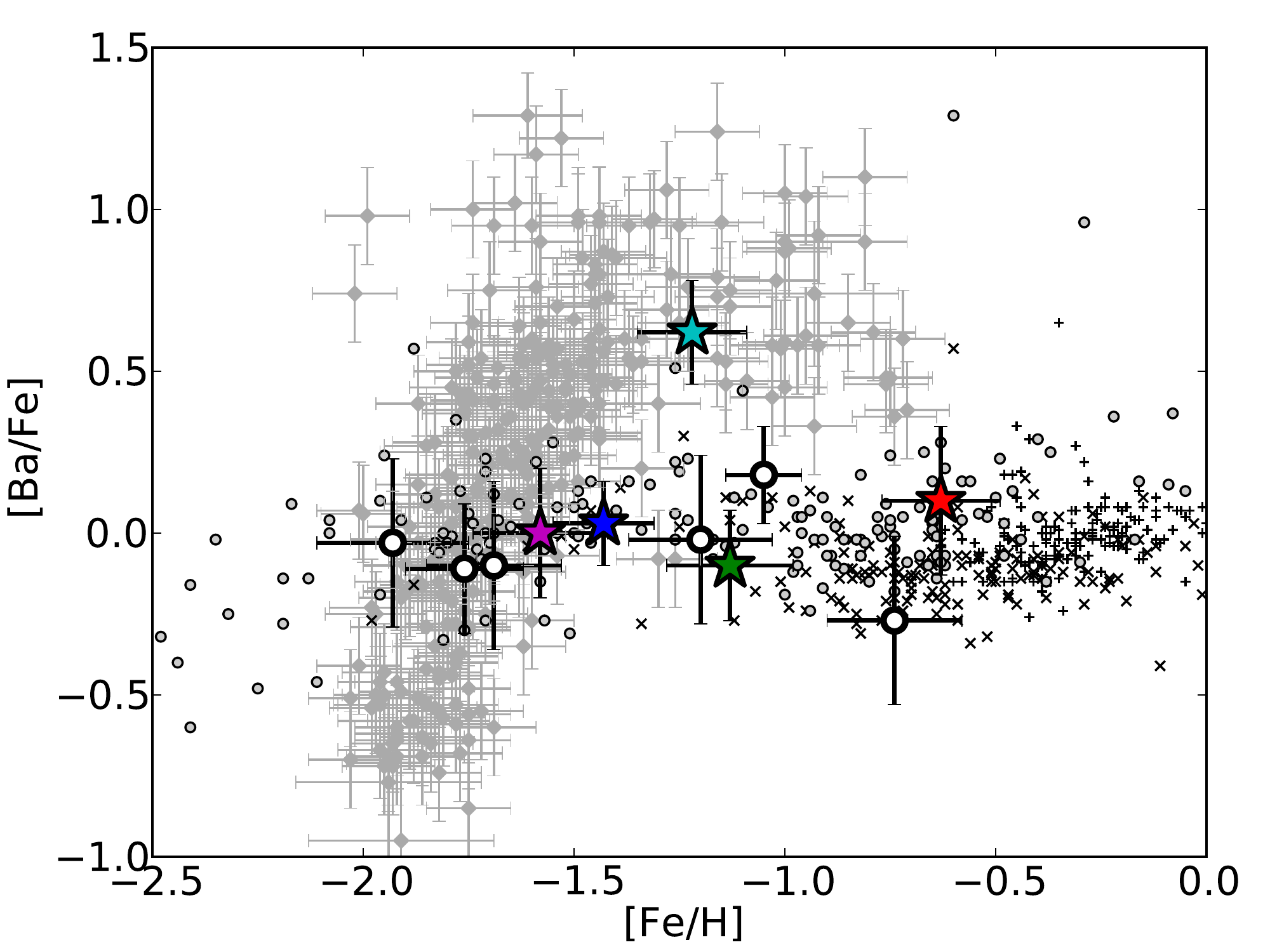}
	\caption[${\rm [Fe/H]}$ and ${\rm [Ba/Fe]}$ abundance ratios for field and $\omega$-Centauri red giant branch stars]{[Fe/H] and [Ba/Fe] for halo/disk stars (black) from \citet{fulbright_2000,reddy;et-al_2003,reddy;et-al_2006} and $\omega$-Centauri RGB stars (grey) from \citet{francois;et-al_1988,smith;et-al_2000,marino;et-al_2011}. Similar trends are observed for other heavy elements in $\omega$-Centauri members. Aquarius stars are colored as per Figure \ref{fig:toombre}.}
	\label{fig:omega-cen-barium}
\end{figure}

In the Aquarius stream discovery paper, \citet{williams;et-al_2011} attempted to exclude possible known progenitors for the Aquarius stream. On the basis of metallicity, distance, proper motions, transverse velocities and orbital energies, the authors were able to exclude all known Milky Way satellites with the notable exception of $\omega$-Centauri. Although the Aquarius stream metallicity distribution is not dissimilar from a known sub-population in $\omega$-Centauri, the individual chemical abundances are quite distinct. The strong $s$-process enhancement with overall metallicity is not observed in the rest of our sample. Thus, with the exception of {C222531-145437}, the Aquarius members do not have a chemistry that is synonymous with $\omega$-Centauri tidal debris. It will be most interesting to learn how many other members of the Aquarius stream are tidal remnants of $\omega$-Centauri, given the frequency of these objects is quite low \citep[e.g., see][]{da_costa;coleman_2008,majewski;et-al_2012}.

\subsection{Disrupted Disk/Halo Stars -- Signature of a Disk-Satellite Interaction?}

Since the Aquarius stream is kinematically coherent, it has been assumed that the moving group has been accreted onto the Milky Way from a tidally disrupted satellite.  The chemical abundances presented in this study do not favour an accretion scenario from a globular cluster or a dSph; there is conflicting evidence for either hypothesis. As it stands, the moving group  appears chemically indistinguishable from thick disk/halo stars. These results force us to consider other scenarios that may replicate the observations.

The Aquarius stream has an unusually wide intrinsic velocity distribution. Generally a stellar stream is considered kinematically `cold' when its velocity dispersion is $\lesssim 8$\,km s$^{-1}$. We find the velocity dispersion from five members to be $\sim$30\,km\,s$^{-1}$, consistent with \citet{williams;et-al_2011}. Hypotheses invoked to explain the Aquarius moving group must account for the high velocity dispersion.
 
There are other moving groups in the Milky Way that were initially considered as tidal tails from disrupted satellites but are no longer regarded as accretion events. We now list some examples. The Hercules moving group is significantly offset from the bulk of the velocity distribution observed in the field. Members of the Hercules group exhibit a wide range of metallicities and ages \citep{bensby;et-al_2007, bovy;hogg_2010}. Furthermore, Hercules group stars have [X/Fe] abundance ratios at a given [Fe/H] that are {\it not} substantially different from the thin or thick disk. The Hercules group kinematics are well replicated in simulations by stars in the outer disk resonating with the bar in the central region of the Milky Way \citep{dehnen_2000,fux_2001}, and strong predictions are made for disk velocity distributions that would lend further weight to this hypothesis \citep{bovy_2010}. Similarly, the Canis Major stellar over-density was also first considered to be an accretion feature from the postulated Canis Majoris dSph galaxy \citep{martin;et-al_2004}. However, \citet{momany;et-al_2004} demonstrate that the star counts, proper-motions, photometry and kinematics of the ``accreted feature'' can be easily explained by the warp and flare in the outer thick disk. The Monoceros ring \citep{Newberg;et-al_2002,Juric;et-al_2008} is perhaps another example of such an occurrence, as similar features naturally emerge as a consequence of galaxy-satellite interactions \citep{purcell;et-al_2011}, which has prompted considerable discussion \citep{lopez-corredoira;et-al_2012}. It is clear that not all kinematic groups are attributable to accretion events; in many scenarios a Galactic origin is more likely, and simpler.

We hypothesise that the Aquarius group is the result of displaced stars from a perturbation in the thick disk. That is, the stars are Galactic in origin but have been displaced by a disk-satellite interaction. Minor mergers can significantly disrupt the host galaxy \citep{villalobos;helmi_2008}, producing extended spatial and kinematic structure in the process. \citet{minchev;et-al_2009} proposed that such a perturbation would cause a Galactic ``ringing'' effect in the neighbourhood surrounding the merger site, analogous to the resulting compression wave propagating outwards from a stone falling in water. Stars move closer together in the wave peak, a signature which is observable in the velocities and orbital motions of nearby stars. This signature is most prominent in the $U$--$V$ velocity plane as concentric circles \citep{gomez;et-al_2012}, and dissolves over time (a few Gyr, depending on the mass of the perturber). After the $U$--$V$ velocity signature dissipates, a clear signature in angular momentum and orbital energy ($L_Z$, $E$) persists for long periods following the merger \citep[e.g., see][]{gomez;et-al_2012}.

Through Milky Way-Sagittarius simulations, \citet{purcell;et-al_2011} found that these disk-satellite interactions can explain ringing perturbations within the disk. Additionally, \citet{widrow;et-al_2012} and \citet{gomez;et-al_2012b} independently observed these phenomena -- a {\it``wavelike perturbation''}, as Widrow described -- in the SDSS and SEGUE catalogues. More recently, \citet{gomez;et-al_2013} proposed that these patterns were induced by the Sagittarius dSph interacting with the disk. Their simulations reproduce the observed north-south asymmetries and vertical wave-like structure, and show that the amplitude of these oscillations is strongly dependent on Galactocentric distance. Combined with the oscillating vertical motions with the $U$--$V$ velocity pattern, corrugated waves are observed as a result of the interaction.

The stars in these oscillations should exhibit a wide range of ages, metallicities and a large spread in velocity dispersion. Thus, resultant oscillations following a disk-satellite interaction can satisfactorily explain the existence of the Aquarius moving group. We do not observe a distinct coherence in the $U$--$V$ velocity plane in our data, but the angular momentum and orbital energies for Aquarius members qualitatively reproduces the theoretically predicted pattern by \citet{gomez;et-al_2012} in a retrograde direction. The extent and gradient of this $L_Z$--$E$ signature is dependent on the mass of the perturber and the time since infall. Although our sample size is minute -- and the sample size would still be small even if all Aquarius members had reliable orbits -- the fact that we see no $U$--$V$ velocity coherence (Figure \ref{fig:toombre}) is consistent with the observed $L_Z$--$E$ pattern: signatures in the $L_Z$--$E$ plane (Figure \ref{fig:orbits}) become more extended over time as the $U$--$V$ signature dissipates. This is consistent with a disk-satellite interaction occurring in the disk approximately a few Gyr ago.

The Aquarius moving group resides at a an intermediate latitude ($b \approx -55^\circ$) and with a radial distance of up to $\sim$5\,kpc for some stars, the stars are slightly out of the plane. This is not inconsistent with a disk-satellite interaction, as similar features in the Galactic field star population naturally emerge. \citet{gomez;et-al_2013} find that a significant fraction of the total energy goes into vertical perturbations. While the mean vertical distance $\langle{}Z\rangle$ in their simulations are near zero, this is an average of disk particles at all plane heights -- positive and negative -- and the dispersions around $\langle{}Z\rangle$ are very large (F. G{\'o}mez, private communication, 2013). Moreover, \citet{gomez;et-al_2013} were only able to reliably track particles up to $|Z| \approx 1.4$\,kpc due to a finite number of particles in each cell volume.

If the Aquarius group is a feature of a disk-satellite interaction, the perturber must have a mass on the order of a large globular cluster or a dSph satellite to produce the residual pattern in orbital energy and angular momenta. The Sagittarius dSph galaxy is an obvious candidate, but $\omega$-Centauri is also a possible perturber. On the basis on position, velocities, chemical abundances and orbit, we identify {C222531-145437} was highly likely stripped from $\omega$-Centauri in the past. Thus, it is plausible that $\omega$-Centauri has disrupted Galactic stars as it passed through the plane, adding to any other oscillating modes rippling through the disk, resulting in what we now observe as the Aquarius stream.

\section{Conclusions}
\label{sec:aquarius-conclusions}
We have presented a detailed chemical and dynamical analysis for 5 members of the recently discovered Aquarius stream from data taken with the MIKE spectrograph on the Magellan Clay telescope. Hereafter we solely refer to the discovery as a moving group instead of a stellar stream, as we find no evidence that the group is a tidal tail of a disrupted satellite. The main conclusions are as follows:

\begin{itemize}
\item The Aquarius stream is not mono-metallic. A wide spread in metallicities is observed, with [Fe/H] ranging from $-0.63$ to $-1.58$ in just 5 members. The mean of the sample is [Fe/H] $= -1.20$ and the dispersion is $\sigma({\rm [Fe/H]}) = 0.33$\,dex.

\item No Na-O anti-correlation is observed in the Aquarius group. Two members have slightly enhanced levels of sodium with respect to iron. If the candidates were \textit{known globular cluster members}, they would be classified as belonging to either the primordial component, or at most, tenuous membership could be argued for the lower envelope of the intermediate group.

\item We find no evidence that the Aquarius group is the result of a disrupted classical globular cluster. The large [Fe/H] variation severely limits the number of possible parent hosts, and both the extreme and intermediate component of the Na-O anti-correlation have not been observed. A strong positive Mg-Al relationship is observed, reminiscent of Milky Way field stars. In total, high-resolution spectra exists for more than half of the stream.

\item The moving group shows an $\alpha$-enhancement of ${[\alpha/{\rm Fe}] = +0.40}$\,dex, similar to the Milky Way, and distinct to that typically observed in stars in dSph galaxies with comparable metallicities.

\item Aquarius members are enhanced in [Na/Fe] and [Ni/Fe] to levels typically observed in either the thick disk or globular clusters. These levels of [Na/Fe] and [Ni/Fe] enhancement are not observed in stars from dSph galaxies. Low [Ba/Y] abundance ratios are also observed in the Aquarius group, in conflict with chemistry of present day dSph galaxies. Thus, on the basis of [(Na, Ni, $\alpha$)/Fe] and [Ba/Y] abundance ratios, it is unlikely the Aquarius moving group is the result of a tidally disrupted dSph galaxy.

\item One of the candidates, {C222531-145437}, has an abundance pattern that is clearly distinct from the other Aquarius members, most notably in barium where {[Ba/Fe] = 0.68}. We exclude the possibility that the abundance variations have resulted from an AGB companion.

\item The position and velocity of {C222531-145437} agrees excellently where simulations by \citet{majewski;et-al_2012} predict large amounts of $\omega$-Centauri tidal debris, and the orbital energy and angular momenta are consistent with the $\omega$-Centauri cluster. The chemical and phase-space information suggests that {C222531-145437} is a rare tidal debris remnant from the globular cluster $\omega$-Centauri. Removing C222531-145437 from the Aquarius sample does not extinguish or diminish any of the aforementioned conclusions.

\item While no evidence exists for an accreted origin, and the Aquarius group members are indistinguishable from thick disk/halo stars, we hypothesise the moving group is the result from a disk-satellite interaction. We see no coherent pattern in the $U$--$V$ plane from Monte-Carlo simulations, but the orbital energies and angular momenta for the Aquarius group qualitatively reproduces patterns predicted by \citet{gomez;et-al_2012}. This is consistent with a minor merger in the Milky Way thick disk occurring perhaps up to a few gigayear ago. Given the location and velocity of the Aquarius group, and the identification of {C222531-145437} as a star tidally stripped from $\omega$-Centauri, it is plausible that the Milky Way-$\omega$Cen interaction sufficiently perturbed outer disk/halo stars to produce what we now observe as the Aquarius group.
\end{itemize}

It is clear that not all moving groups are tidal tails of disrupted satellites, and that the structure of the Milky Way is indeed complex. While we find no chemical evidence that the Aquarius group is a tidal tail from a disrupted satellite, we propose the members are Galactic in origin, and the group is a result of a disk-satellite interaction. Thus, although the Aquarius group has not been accreted onto the Galaxy, it certainly adds to the rich level of kinematic substructure within the Milky Way.

\chapter{Conclusions}\label{chap:conclusions}

The Galaxy has a chaotic merger history, yet it provides an accessible laboratory to evaluate galaxy formation models and test cosmological structure theories. As \citet{eggen;et-al_1962} first demonstrated, studies of galactic archaeology is accessible through the analysis of stellar populations. While dynamical information does eventually dissipate, the chemical abundances of individual stars remain largely unchanged over a stellar lifetime. Moreover, the observed abundances in the upper photosphere of a star largely represents the integrated sum of chemical enrichment across all previous star formation environments. This allows us to explore the chemical evolution of the galaxy, and distinguish between stars that formed in different environments, which is particularly relevant when differentiating in-situ or accreted stars.

Consistent with the preferred paradigm of hierarchical formation, the Milky Way has accreted -- and disrupted -- a substantial number of satellite systems. This thesis has examined several known substructures with the aim to place these features in context with the rest of the Galaxy, and help reconstruct its formation history. These studies have primarily been observational. Each study included the utilisation of existing deep multi-band photometric survey data to identify stars that are likely members of a known substructure, preparing and undertaking low- and/or high-resolution spectroscopic observations, reducing those data, ascertaining stellar parameters and detailed chemical abundances from calibrated data, and inferring the most probabilistic scenario to explain the observations. Here we summarise critical points from these studies.

Hundreds of stars in the direction of the northern leading Sgr stream and Virgo Over-Density were observed with low-resolution spectroscopy. The Sgr stream stars are kinematically sensitive to the shape of the Galactic dark halo. These data are best represented by a tri-axial dark matter distribution, and we rule out a prolate dark halo as it predicts an insufficient amount of observed debris. The photometric selections employed for Chapter \ref{chap:sgr} were indadvertedly biased towards more metal-poor stars at the distances of the Sgr stream. Consequently, a significantly more metal-poor ([Fe/H] $= - 1.7$) sub-population of stars has been identified in the Sgr stream. These observations also demonstrate that the Sgr stream likely hosts a substantially larger abundance range than previously found.

Low-resolution spectra were also obtained in the direction of the Orphan stream. Although the Orphan stream has an extremely low surface brightness, we have reliably identified sparse red giant branch stars that were once bound to the Orphan stream parent system. From low-resolution spectra, these stars demonstrate a wide range in metallicities ($\sigma({\rm [Fe/H]}) = 0.56$\,dex) and a undetectably low dispersion in velocities ($<4$km s$^{-1}$). The intrinsic metallically dispersion is consistent with the undiscovered host having undergone independent chemical enrichment, and suggests a dwarf galaxy origin. This is consistent with dynamical constraints from the stream arc and width, which suggest the host system to be dark matter-dominated.

High-resolution spectra for a subset of Orphan stream candidates confirm the wide metallicity spread. Additionally, low [$\alpha$/Fe] abundance ratios and high lower limits for [Ba/Y] abundance ratios are observed. These detailed chemical abundances are consistent with stars observed in the present-day dwarf galaxies surrounding the Milky Way. Thus, we have presented the first detailed chemical evidence that the undiscovered Orphan stream parent is a dwarf galaxy, and not a globular cluster. On the basis of the observable chemical evidence, we have firmly excluded the globular cluster NGC 2419 as a plausible parent to the Orphan stream. At present, Segue 1 appears the only plausible \emph{known} satellite cannot be excluded as a potential host. The large spread in metallicities found in the Orphan stream is also consistent with that observed in Segue 1. However, we conclude that there are still significant inconsistencies with the Segue 1--Orphan stream connection, specifically the lack of tidal effects surrounding Segue 1 and the high [$\alpha$/Fe] abundance ratios found in Segue 1 stars. It appears the parent host is perhaps more likely to be an undiscovered dwarf satellite in the southern sky.

Our analysis of the Aquarius co-moving group has concluded that not all kinematically-identified substructures are actually accreted material. From high-resolution spectra for just five stars, a wide metallicity range is observed: [Fe/H] = $-$0.63 to $-$1.58. No anti-correlation in Na--O or Mg--Al abundances is present. High [($\alpha$, Ni, Na)/Fe] and low [Ba/Y] abundance ratios suggest the system is not a disrupted dwarf galaxy, either. Chemically, the group is largely indistinguishable from Milky Way field stars. We hypothesise the group is galactic in origin, having resulted from a disk-satellite interaction. Until compelling dynamical or chemical evidence is presented to the contrary, the `Aquarius Group' is suggested as a more appropriate characterisation. Like other kinematically-identified substructures in the Galaxy, the Aquarius group simply adds to the rich level of substructure present in the Milky Way.

This thesis has aimed to further our knowledge of the Galaxy's structure and merger history through the aforementioned photometric and spectroscopic studies of known substructures. The full extent of Milky Way substructure remains unknown, and these contributions examine only a small baryonic fraction of the Galaxy. However these studies help chronicle the merger history of the Galaxy into a coherent story. They help afford us a galactic perspective of the Milky Way, one which is easily obtainable for distant galaxies, yet difficult for our own: an understanding of the breadth, structural components, and evolution of our home, the Milky Way.

\section{Future prospects} 
\subsection{Spectroscopy Made Hard}
The software package described in Chapter \ref{chap:smh} is currently being used by dozens of spectroscopists for the analysis of high-resolution spectra. However, these analyses are all manually performed by the user. In order to continue a transition towards homogenous, automatic analysis, there are a number of improvements that are currently being tested or developed:

\begin{itemize}
\item Automatic stellar parameter determination by excitation and ionization equilibria. This approach will encompass a sophisticated optimization algorithm to quickly iterate to the correct solution and side-step local minima.

\item Effective temperature determination by employing the Hydrogen line-profile fitting approach developed by Paul Barklem. SMH will perform a background call to this routine, and provide the user with an interactive view of the fitted profile.

\item Improvements to the atmospheric interpolation scheme to allow for interpolating on a column-mass scale.

\item Non-LTE corrections for elements where grids of pre-computed corrections exist. These corrections can be automatically applied during excitation and ionization equilibria, as well as to specific chemical elements after the determination of stellar parameters. At present, only elements with `trustworthy' non-LTE corrections will be employed. Computed grids of corrections for additional elements will be added as they become available. 

\item Statistical quality assurances for survey data sets may also be added. The principal investigator for a spectroscopic survey using the automatic version of SMH may be able to add their co-investigators to verify the quality control of the results provided by SMH. Co-investigators will be randomly asked to verify that the results for a given star make sense based on diagnostic plots. Upon receiving three independent verifications from co-investigators, those results can be marked as reliable. Quality control metrics will be used to identify stars with questionable results, as well as randomly sampling stars that look perfectly normal. This will allow a quantitative measure of the per cent of acceptable results, with only a subset of the sample requiring quality control verification.
\end{itemize}

All automatic analyses will be saved in the SMH format, such that they can be inspected, adjusted, and reproduced by anyone with the SMH software. This allows for full data provenance of an entire survey data set, regardless of the sample size. After rigorous scientific testing by numerous groups, this code will eventually be opened to the public. \\

\subsection{Understanding the Milky Way}
A wealth of information is available from photospheric light in a single star. The evolutionary state, stellar parameters, photospheric turbulence, as well as atomic, molecular and isotopic abundances are all inferred from starlight. A significant portion of astrophysics has been inferred directly from starlight. Collectively, spectra from massive samples of stars allow us to undertake measurements on a larger, Galactic scale. These stellar fossils have kinematics and chemistry that reflect the signatures of Galaxy formation from early and recent times. To further understand the evolution of the Milky Way it is clear that high-resolution spectra for unprecedented numbers of stars are required.

With these data the full structure of the Milky Way can be unveiled, and the interplay between Galactic components can be accurately characterised. Perhaps dozens of new stellar streams and satellite systems will be discovered, easing the tension between observations and cosmological simulations. The transition point between in situ- and accretion-dominated formation can be inferred, as well as the distribution of substructure in the stellar halo. Robust constraints on the nature and Galactic distribution of dark matter can be deduced. When the results of these spectral data are coupled with accurate distances from Gaia, we will be able to measure the mass of the Milky Way more accurately than ever before. In fact, we will be able to start to adequately answer the simple question: compared to what we see surrounding us in the vast cosmos, is the Milky Way normal?

It is clear that high-resolution spectra of millions of stars will drastically clarify our understanding of cosmological structure, galaxy formation and evolution, stellar structure, and the nucleosynthesis of the elements. We are witnessing a rapid transition into a new era of stellar astronomy, facilitating an unprecedented understanding of astrophysics. Thus it is perhaps fitting to finish by following \emph{Edward R. Harrison}'s famous hydrogen quote: that with these data we will truly step closer to understanding where we came from, and where we are going.

%
%
\cleardoublepage

%
\phantomsection\addcontentsline{toc}{chapter}{Bibliography}
\bibliographystyle{astroads}

\begin{thebibliography}{338}
\expandafter\ifx\csname natexlab\endcsname\relax\def\natexlab#1{#1}\fi
\expandafter\ifx\csname href\endcsname\relax
  \def\href#1#2{}\fi
\expandafter\ifx\csname urllinklabel\endcsname\relax
  \def\urllinklabel{(Link)}\fi
\expandafter\ifx\csname adsurllinklabel\endcsname\relax
  \def\adsurllinklabel{(ADS~entry)}\fi

\bibitem[{{Abazajian} {et~al.}(2009){Abazajian}, {Adelman-McCarthy},
  {Ag{\"u}eros}, {Allam}, {Allende Prieto}, {An}, {Anderson}, {Anderson},
  {Annis}, {Bahcall}, \& et~al.}]{abazajian;et-al_2009}
{Abazajian}, K.~N., {et~al.} 2009, \apjs, 182, 543
 \href{http://adsabs.harvard.edu/abs/2009ApJS..182..543A}{\adsurllinklabel}
{}

\bibitem[{{Abel} {et~al.}(2002){Abel}, {Bryan}, \& {Norman}}]{abel;et-al_2002}
{Abel}, T., {Bryan}, G.~L., \& {Norman}, M.~L. 2002, Science, 295, 93
 \href{http://adsabs.harvard.edu/abs/2002Sci...295...93A}{\adsurllinklabel}
{}

\bibitem[{{Ahn} {et~al.}(2013){Ahn}, {Alexandroff}, {Allende Prieto}, {Anders},
  {Anderson}, {Anderton}, {Andrews}, {Aubourg}, {Bailey}, {Bastien}, \&
  et~al.}]{apogee_dr10}
{Ahn}, C.~P., {et~al.} 2013, ArXiv e-prints
 \href{http://adsabs.harvard.edu/abs/2013arXiv1307.7735A}{\adsurllinklabel}
{}

\bibitem[{{Allende Prieto} {et~al.}(2001){Allende Prieto}, {Lambert}, \&
  {Asplund}}]{allende-prieto;et-al_2001}
{Allende Prieto}, C., {Lambert}, D.~L., \& {Asplund}, M. 2001, \apjl, 556, 63
 \href{http://adsabs.harvard.edu/abs/2001ApJ...556L..63A}{\adsurllinklabel}
{}

\bibitem[{{Allende Prieto} {et~al.}(2008){Allende Prieto}, {Majewski},
  {Schiavon}, {Cunha}, {Frinchaboy}, {Holtzman}, {Johnston}, {Shetrone},
  {Skrutskie}, {Smith}, \& {Wilson}}]{apogee}
{Allende Prieto}, C., {et~al.} 2008, Astronomische Nachrichten, 329, 1018
 \href{http://adsabs.harvard.edu/abs/2008AN....329.1018A}{\adsurllinklabel}
{}

\bibitem[{{Alonso} {et~al.}(1999){Alonso}, {Arribas}, \&
  {Mart{\'{\i}}nez-Roger}}]{alonso;et-al_1999}
{Alonso}, A., {Arribas}, S., \& {Mart{\'{\i}}nez-Roger}, C. 1999, \aaps, 140,
  261
 \href{http://adsabs.harvard.edu/abs/1999A%26AS..140..261A}{\adsurllinklabel}
{}

\bibitem[{{Aoki} {et~al.}(2007){Aoki}, {Honda}, {Beers}, {Takada-Hidai},
  {Iwamoto}, {Tominaga}, {Umeda}, {Nomoto}, {Norris}, \&
  {Ryan}}]{aoki;et-al_2007}
{Aoki}, W., {et~al.} 2007, \apj, 660, 747
 \href{http://adsabs.harvard.edu/abs/2007ApJ...660..747A}{\adsurllinklabel}
{}

\bibitem[{{Armandroff} \& {Da Costa}(1991)}]{armandroff;da_costa_1991}
{Armandroff}, T.~E., \& {Da Costa}, G.~S. 1991, \aj, 101, 1329
 \href{http://adsabs.harvard.edu/abs/1991AJ....101.1329A}{\adsurllinklabel}
{}

\bibitem[{{Arnett} \& {Truran}(1969)}]{arnett;truran_1969}
{Arnett}, W.~D., \& {Truran}, J.~W. 1969, \apj, 157, 339
 \href{http://adsabs.harvard.edu/abs/1969ApJ...157..339A}{\adsurllinklabel}
{}

\bibitem[{{Arnould} {et~al.}(1999){Arnould}, {Goriely}, \&
  {Jorissen}}]{arnould;et-al_1999}
{Arnould}, M., {Goriely}, S., \& {Jorissen}, A. 1999, \aap, 347, 572
 \href{http://adsabs.harvard.edu/abs/1999A%26A...347..572A}{\adsurllinklabel}
{}

\bibitem[{{Asplund} \& {Garc{\'{\i}}a P{\'e}rez}(2001)}]{asplund;perez_2001}
{Asplund}, M., \& {Garc{\'{\i}}a P{\'e}rez}, A.~E. 2001, \aap, 372, 601
 \href{http://adsabs.harvard.edu/abs/2001A%26A...372..601A}{\adsurllinklabel}
{}

\bibitem[{{Asplund} {et~al.}(2009){Asplund}, {Grevesse}, {Sauval}, \&
  {Scott}}]{asplund;et-al_2009}
{Asplund}, M., {Grevesse}, N., {Sauval}, A.~J., \& {Scott}, P. 2009, \araa, 47,
  481
 \href{http://adsabs.harvard.edu/abs/2009ARA%26A..47..481A}{\adsurllinklabel}
{}

\bibitem[{Barber {et~al.}(1996)Barber, Dobkin, \&
  Huhdanpaa}]{barber;et-al_1996}
Barber, C., Dobkin, D., \& Huhdanpaa, H. 1996, ACM Transactions on Mathematical
  Software (TOMS), 22, 469
{}
{}

\bibitem[{{Barkana} \& {Loeb}(2007)}]{barkana;loeb_2007}
{Barkana}, R., \& {Loeb}, A. 2007, Reports on Progress in Physics, 70, 627
 \href{http://adsabs.harvard.edu/abs/2007RPPh...70..627B}{\adsurllinklabel}
{}

\bibitem[{{Battaglia} {et~al.}(2008){Battaglia}, {Irwin}, {Tolstoy}, {Hill},
  {Helmi}, {Letarte}, \& {Jablonka}}]{battaglia;et-al_2008}
{Battaglia}, G., {Irwin}, M., {Tolstoy}, E., {Hill}, V., {Helmi}, A.,
  {Letarte}, B., \& {Jablonka}, P. 2008, \mnras, 383, 183
 \href{http://adsabs.harvard.edu/abs/2008MNRAS.383..183B}{\adsurllinklabel}
{}

\bibitem[{{Battaglia} \& {Starkenburg}(2012)}]{battaglia;starkenburg_2012}
{Battaglia}, G., \& {Starkenburg}, E. 2012, \aap, 539, A123
 \href{http://adsabs.harvard.edu/abs/2012A%26A...539A.123B}{\adsurllinklabel}
{}

\bibitem[{{Battaglia} {et~al.}(2005){Battaglia}, {Helmi}, {Morrison},
  {Harding}, {Olszewski}, {Mateo}, {Freeman}, {Norris}, \&
  {Shectman}}]{battaglia;et-al_2005}
{Battaglia}, G., {et~al.} 2005, \mnras, 364, 433
 \href{http://adsabs.harvard.edu/abs/2005MNRAS.364..433B}{\adsurllinklabel}
{}

\bibitem[{{Baumueller} \& {Gehren}(1997)}]{baumueller;gehren_1997}
{Baumueller}, D., \& {Gehren}, T. 1997, \aap, 325, 1088
 \href{http://adsabs.harvard.edu/abs/1997A%26A...325.1088B}{\adsurllinklabel}
{}

\bibitem[{{Baushev} {et~al.}(2012){Baushev}, {Federici}, \& {Pohl}}]{baushev}
{Baushev}, A.~N., {Federici}, S., \& {Pohl}, M. 2012, \prd, 86, 063521
 \href{http://adsabs.harvard.edu/abs/2012PhRvD..86f3521B}{\adsurllinklabel}
{}

\bibitem[{{Bekki} \& {Freeman}(2003)}]{bekki;freeman_2003}
{Bekki}, K., \& {Freeman}, K.~C. 2003, \mnras, 346, L11
 \href{http://adsabs.harvard.edu/abs/2003MNRAS.346L..11B}{\adsurllinklabel}
{}

\bibitem[{{Bell} {et~al.}(2008){Bell}, {Zucker}, {Belokurov}, {Sharma},
  {Johnston}, {Bullock}, {Hogg}, {Jahnke}, {de Jong}, {Beers}, {Evans},
  {Grebel}, {Ivezi{\'c}}, {Koposov}, {Rix}, {Schneider}, {Steinmetz}, \&
  {Zolotov}}]{bell;et-al_2008}
{Bell}, E.~F., {et~al.} 2008, \apj, 680, 295
 \href{http://adsabs.harvard.edu/abs/2008ApJ...680..295B}{\adsurllinklabel}
{}

\bibitem[{{Bellazzini} {et~al.}(2006){Bellazzini}, {Correnti}, {Ferraro},
  {Monaco}, \& {Montegriffo}}]{Bellazzini;et-al_2006}
{Bellazzini}, M., {Correnti}, M., {Ferraro}, F.~R., {Monaco}, L., \&
  {Montegriffo}, P. 2006, \aap, 446, L1
 \href{http://adsabs.harvard.edu/abs/2006A%26A...446L...1B}{\adsurllinklabel}
{}

\bibitem[{{Belokurov} {et~al.}(2006{\natexlab{a}}){Belokurov}, {Evans},
  {Irwin}, {Hewett}, \& {Wilkinson}}]{belokurov;et-al_2006_5466}
{Belokurov}, V., {Evans}, N.~W., {Irwin}, M.~J., {Hewett}, P.~C., \&
  {Wilkinson}, M.~I. 2006{\natexlab{a}}, \apjl, 637, L29
 \href{http://adsabs.harvard.edu/abs/2006ApJ...637L..29B}{\adsurllinklabel}
{}

\bibitem[{{Belokurov} {et~al.}(2006{\natexlab{b}}){Belokurov}, {Zucker},
  {Evans}, {Gilmore}, {Vidrih}, {Bramich}, {Newberg}, {Wyse}, {Irwin},
  {Fellhauer}, {Hewett}, {Walton}, {Wilkinson}, {Cole}, {Yanny}, {Rockosi},
  {Beers}, {Bell}, {Brinkmann}, {Ivezi{\'c}}, \&
  {Lupton}}]{belokurov;et-al_2006}
{Belokurov}, V., {et~al.} 2006{\natexlab{b}}, \apjl, 642, L137
 \href{http://adsabs.harvard.edu/abs/2006ApJ...642L.137B}{\adsurllinklabel}
{}

\bibitem[{{Belokurov} {et~al.}(2007{\natexlab{a}}){Belokurov}, {Evans},
  {Irwin}, {Lynden-Bell}, {Yanny}, {Vidrih}, {Gilmore}, {Seabroke}, {Zucker},
  {Wilkinson}, {Hewett}, {Bramich}, {Fellhauer}, {Newberg}, {Wyse}, {Beers},
  {Bell}, {Barentine}, {Brinkmann}, {Cole}, {Pan}, \&
  {York}}]{belokurov;et-al_2007}
---. 2007{\natexlab{a}}, \apj, 658, 337
 \href{http://adsabs.harvard.edu/abs/2007ApJ...658..337B}{\adsurllinklabel}
{}

\bibitem[{{Belokurov} {et~al.}(2007{\natexlab{b}}){Belokurov}, {Zucker},
  {Evans}, {Kleyna}, {Koposov}, {Hodgkin}, {Irwin}, {Gilmore}, {Wilkinson},
  {Fellhauer}, {Bramich}, {Hewett}, {Vidrih}, {De Jong}, {Smith}, {Rix},
  {Bell}, {Wyse}, {Newberg}, {Mayeur}, {Yanny}, {Rockosi}, {Gnedin},
  {Schneider}, {Beers}, {Barentine}, {Brewington}, {Brinkmann}, {Harvanek},
  {Kleinman}, {Krzesinski}, {Long}, {Nitta}, \&
  {Snedden}}]{belokurov;et-al_2007_quintet}
---. 2007{\natexlab{b}}, \apj, 654, 897
 \href{http://cdsads.u-strasbg.fr/abs/2007ApJ...654..897B}{\adsurllinklabel}
{}

\bibitem[{{Bensby} {et~al.}(2003){Bensby}, {Feltzing}, \&
  {Lundstr{\"o}m}}]{bensby;et-al_2003}
{Bensby}, T., {Feltzing}, S., \& {Lundstr{\"o}m}, I. 2003, \aap, 410, 527
 \href{http://adsabs.harvard.edu/abs/2003A%26A...410..527B}{\adsurllinklabel}
{}

\bibitem[{{Bensby} {et~al.}(2007){Bensby}, {Oey}, {Feltzing}, \&
  {Gustafsson}}]{bensby;et-al_2007}
{Bensby}, T., {Oey}, M.~S., {Feltzing}, S., \& {Gustafsson}, B. 2007, \apjl,
  655, L89
 \href{http://adsabs.harvard.edu/abs/2007ApJ...655L..89B}{\adsurllinklabel}
{}

\bibitem[{{Bergemann} {et~al.}(2010){Bergemann}, {Pickering}, \&
  {Gehren}}]{bergemann;et-al_2010}
{Bergemann}, M., {Pickering}, J.~C., \& {Gehren}, T. 2010, \mnras, 401, 1334
 \href{http://adsabs.harvard.edu/abs/2010MNRAS.401.1334B}{\adsurllinklabel}
{}

\bibitem[{{Bernstein} {et~al.}(2003){Bernstein}, {Shectman}, {Gunnels},
  {Mochnacki}, \& {Athey}}]{bernstein;et-al_2003}
{Bernstein}, R., {Shectman}, S.~A., {Gunnels}, S.~M., {Mochnacki}, S., \&
  {Athey}, A.~E. 2003, in Society of Photo-Optical Instrumentation Engineers
  (SPIE) Conference Series, Vol. 4841, Society of Photo-Optical Instrumentation
  Engineers (SPIE) Conference Series, ed. M.~{Iye} \& A.~F.~M. {Moorwood},
  1694--1704
 \href{http://adsabs.harvard.edu/abs/2003SPIE.4841.1694B}{\adsurllinklabel}
{}

\bibitem[{{Blumenthal} {et~al.}(1984){Blumenthal}, {Faber}, {Primack}, \&
  {Rees}}]{blumenthal;et-al_1984}
{Blumenthal}, G.~R., {Faber}, S.~M., {Primack}, J.~R., \& {Rees}, M.~J. 1984,
  \nat, 311, 517
 \href{http://adsabs.harvard.edu/abs/1984Natur.311..517B}{\adsurllinklabel}
{}

\bibitem[{{Bonaca} {et~al.}(2012){Bonaca}, {Geha}, \&
  {Kallivayalil}}]{bonaca;et-al_2012}
{Bonaca}, A., {Geha}, M., \& {Kallivayalil}, N. 2012, \apjl, 760, L6
 \href{http://adsabs.harvard.edu/abs/2012ApJ...760L...6B}{\adsurllinklabel}
{}

\bibitem[{{Bonifacio} {et~al.}(2004){Bonifacio}, {Sbordone}, {Marconi},
  {Pasquini}, \& {Hill}}]{Bonifacio;et-al_2004}
{Bonifacio}, P., {Sbordone}, L., {Marconi}, G., {Pasquini}, L., \& {Hill}, V.
  2004, \aap, 414, 503
 \href{http://adsabs.harvard.edu/abs/2004A%26A...414..503B}{\adsurllinklabel}
{}

\bibitem[{{Bovy}(2010)}]{bovy_2010}
{Bovy}, J. 2010, \apj, 725, 1676
 \href{http://adsabs.harvard.edu/abs/2010ApJ...725.1676B}{\adsurllinklabel}
{}

\bibitem[{{Bovy} \& {Hogg}(2010)}]{bovy;hogg_2010}
{Bovy}, J., \& {Hogg}, D.~W. 2010, \apj, 717, 617
 \href{http://adsabs.harvard.edu/abs/2010ApJ...717..617B}{\adsurllinklabel}
{}

\bibitem[{{Bovy} {et~al.}(2012){Bovy}, {Rix}, \& {Hogg}}]{bovy_2011}
{Bovy}, J., {Rix}, H.-W., \& {Hogg}, D.~W. 2012, \apj, 751, 131
 \href{http://adsabs.harvard.edu/abs/2012ApJ...751..131B}{\adsurllinklabel}
{}

\bibitem[{{Breddels} {et~al.}(2010){Breddels}, {Smith}, {Helmi},
  {Bienaym{\'e}}, {Binney}, {Bland-Hawthorn}, {Boeche}, {Burnett}, {Campbell},
  {Freeman}, {Gibson}, {Gilmore}, {Grebel}, {Munari}, {Navarro}, {Parker},
  {Seabroke}, {Siebert}, {Siviero}, {Steinmetz}, {Watson}, {Williams}, {Wyse},
  \& {Zwitter}}]{breddels;et-al_2010}
{Breddels}, M.~A., {et~al.} 2010, \aap, 511, A90
 \href{http://adsabs.harvard.edu/abs/2010A%26A...511A..90B}{\adsurllinklabel}
{}

\bibitem[{{Bromm} {et~al.}(1999){Bromm}, {Coppi}, \&
  {Larson}}]{bromm;et-al_1999}
{Bromm}, V., {Coppi}, P.~S., \& {Larson}, R.~B. 1999, \apjl, 527, L5
 \href{http://adsabs.harvard.edu/abs/1999ApJ...527L...5B}{\adsurllinklabel}
{}

\bibitem[{{Bromm} \& {Yoshida}(2011)}]{bromm;yoshida_2011}
{Bromm}, V., \& {Yoshida}, N. 2011, \araa, 49, 373
 \href{http://adsabs.harvard.edu/abs/2011ARA%26A..49..373B}{\adsurllinklabel}
{}

\bibitem[{{Brooks} \& {Zolotov}(2012)}]{brooks;et-al_2012}
{Brooks}, A.~M., \& {Zolotov}, A. 2012, ArXiv e-prints
 \href{http://adsabs.harvard.edu/abs/2012arXiv1207.2468B}{\adsurllinklabel}
{}

\bibitem[{{Brown} {et~al.}(1990){Brown}, {Smith}, {Lambert}, {Dutchover},
  {Hinkle}, \& {Johnson}}]{brown;et-al_1990}
{Brown}, J.~A., {Smith}, V.~V., {Lambert}, D.~L., {Dutchover}, Jr., E.,
  {Hinkle}, K.~H., \& {Johnson}, H.~R. 1990, \aj, 99, 1930
 \href{http://adsabs.harvard.edu/abs/1990AJ.....99.1930B}{\adsurllinklabel}
{}

\bibitem[{{Br{\"u}ns} \& {Kroupa}(2011)}]{bruns;kroupa_2011}
{Br{\"u}ns}, R.~C., \& {Kroupa}, P. 2011, \apj, 729, 69
 \href{http://adsabs.harvard.edu/abs/2011ApJ...729...69B}{\adsurllinklabel}
{}

\bibitem[{{Burnett} \& {Binney}(2010)}]{burnett_binney_2010}
{Burnett}, B., \& {Binney}, J. 2010, \mnras, 407, 339
 \href{http://adsabs.harvard.edu/abs/2010MNRAS.407..339B}{\adsurllinklabel}
{}

\bibitem[{{Burris} {et~al.}(2000){Burris}, {Pilachowski}, {Armandroff},
  {Sneden}, {Cowan}, \& {Roe}}]{burris;et-al_2000}
{Burris}, D.~L., {Pilachowski}, C.~A., {Armandroff}, T.~E., {Sneden}, C.,
  {Cowan}, J.~J., \& {Roe}, H. 2000, \apj, 544, 302
 \href{http://adsabs.harvard.edu/abs/2000ApJ...544..302B}{\adsurllinklabel}
{}

\bibitem[{{Busso} {et~al.}(2001){Busso}, {Gallino}, {Lambert}, {Travaglio}, \&
  {Smith}}]{busso;et-al_2001}
{Busso}, M., {Gallino}, R., {Lambert}, D.~L., {Travaglio}, C., \& {Smith},
  V.~V. 2001, \apj, 557, 802
 \href{http://adsabs.harvard.edu/abs/2001ApJ...557..802B}{\adsurllinklabel}
{}

\bibitem[{{Cacciari} {et~al.}(2002){Cacciari}, {Bellazzini}, \&
  {Colucci}}]{Cacciari;et-al_2002}
{Cacciari}, C., {Bellazzini}, M., \& {Colucci}, S. 2002, in IAU Symposium, Vol.
  207, Extragalactic Star Clusters, ed. {D.~P.~Geisler, E.~K.~Grebel, \&
  D.~Minniti}, 168
 \href{http://adsabs.harvard.edu/abs/2002IAUS..207..168C}{\adsurllinklabel}
{}

\bibitem[{{Carlberg}(2013)}]{carlberg;et-al_2013}
{Carlberg}, R.~G. 2013, \apj, 775, 90
 \href{http://adsabs.harvard.edu/abs/2013ApJ...775...90C}{\adsurllinklabel}
{}

\bibitem[{{Carlberg} \& {Grillmair}(2013)}]{carlberg;et-al_2012}
{Carlberg}, R.~G., \& {Grillmair}, C.~J. 2013, \apj, 768, 171
 \href{http://adsabs.harvard.edu/abs/2013ApJ...768..171C}{\adsurllinklabel}
{}

\bibitem[{{Carretta} {et~al.}(2011){Carretta}, {Lucatello}, {Gratton},
  {Bragaglia}, \& {D'Orazi}}]{carretta;et-al_2011}
{Carretta}, E., {Lucatello}, S., {Gratton}, R.~G., {Bragaglia}, A., \&
  {D'Orazi}, V. 2011, \aap, 533, A69
 \href{http://adsabs.harvard.edu/abs/2011A%26A...533A..69C}{\adsurllinklabel}
{}

\bibitem[{{Carretta} {et~al.}(2009){Carretta}, {Bragaglia}, {Gratton},
  {Lucatello}, {Catanzaro}, {Leone}, {Bellazzini}, {Claudi}, {D'Orazi},
  {Momany}, {Ortolani}, {Pancino}, {Piotto}, {Recio-Blanco}, \&
  {Sabbi}}]{carretta;et-al_2009a}
{Carretta}, E., {et~al.} 2009, \aap, 505, 117
 \href{http://adsabs.harvard.edu/abs/2009A%26A...505..117C}{\adsurllinklabel}
{}

\bibitem[{{Carretta} {et~al.}(2010){Carretta}, {Bragaglia}, {Gratton},
  {Lucatello}, {Bellazzini}, {Catanzaro}, {Leone}, {Momany}, {Piotto}, \&
  {D'Orazi}}]{carretta;et-al_2010}
---. 2010, \aap, 520, A95
 \href{http://adsabs.harvard.edu/abs/2010A%26A...520A..95C}{\adsurllinklabel}
{}

\bibitem[{{Casagrande} {et~al.}(2010){Casagrande}, {Ram{\'{\i}}rez},
  {Mel{\'e}ndez}, {Bessell}, \& {Asplund}}]{casagrande;et-al_2010}
{Casagrande}, L., {Ram{\'{\i}}rez}, I., {Mel{\'e}ndez}, J., {Bessell}, M., \&
  {Asplund}, M. 2010, \aap, 512, A54
 \href{http://adsabs.harvard.edu/abs/2010A%26A...512A..54C}{\adsurllinklabel}
{}

\bibitem[{{Casetti-Dinescu} {et~al.}(2009){Casetti-Dinescu}, {Girard},
  {Majewski}, {Vivas}, {Wilhelm}, {Carlin}, {Beers}, \& {van
  Altena}}]{Casetti-Dinescu;et-al_2009}
{Casetti-Dinescu}, D.~I., {Girard}, T.~M., {Majewski}, S.~R., {Vivas}, A.~K.,
  {Wilhelm}, R., {Carlin}, J.~L., {Beers}, T.~C., \& {van Altena}, W.~F. 2009,
  \apjl, 701, L29
 \href{http://adsabs.harvard.edu/abs/2009ApJ...701L..29C}{\adsurllinklabel}
{}

\bibitem[{{Casey} {et~al.}(2013{\natexlab{a}}){Casey}, {Da Costa}, {Keller}, \&
  {Maunder}}]{casey;et-al_2013a}
{Casey}, A.~R., {Da Costa}, G., {Keller}, S.~C., \& {Maunder}, E.
  2013{\natexlab{a}}, \apj, 764, 39
 \href{http://adsabs.harvard.edu/abs/2013ApJ...764...39C}{\adsurllinklabel}
{}

\bibitem[{{Casey} {et~al.}(2013{\natexlab{b}}){Casey}, {Keller}, {Da Costa},
  {Frebel}, \& {Maunder}}]{casey;et-al_2013c}
{Casey}, A.~R., {Keller}, S., {Da Costa}, G., {Frebel}, A., \& {Maunder}, E.
  2013{\natexlab{b}}, ArXiv e-prints
 \href{http://adsabs.harvard.edu/abs/2013arXiv1309.3563C}{\adsurllinklabel}
{}

\bibitem[{{Casey} {et~al.}(2013{\natexlab{c}}){Casey}, {Keller}, {Alves-Brito},
  {Frebel}, {Da Costa}, {Karakas}, {Yong}, {Schlaufman}, {Jacobson}, {Yu}, \&
  {Fishlock}}]{casey;et-al_2013b}
{Casey}, A.~R., {et~al.} 2013{\natexlab{c}}, ArXiv e-prints
 \href{http://adsabs.harvard.edu/abs/2013arXiv1309.3562C}{\adsurllinklabel}
{}

\bibitem[{{Castelli} \& {Kurucz}(2003)}]{castelli;kurucz_2003}
{Castelli}, F., \& {Kurucz}, R.~L. 2003, in IAU Symposium, Vol. 210, Modelling
  of Stellar Atmospheres, ed. N.~{Piskunov}, W.~W. {Weiss}, \& D.~F. {Gray}, 20
 \href{http://adsabs.harvard.edu/abs/2003IAUS..210P.A20C}{\adsurllinklabel}
{}

\bibitem[{{Castelli} \& {Kurucz}(2004)}]{castelli;kurucz_2004}
{Castelli}, F., \& {Kurucz}, R.~L. 2004, ArXiv Astrophysics e-prints 0405087
 \href{http://adsabs.harvard.edu/abs/2004astro.ph..5087C}{\adsurllinklabel}
{}

\bibitem[{{Cayrel} {et~al.}(2004){Cayrel}, {Depagne}, {Spite}, {Hill}, {Spite},
  {Fran{\c c}ois}, {Plez}, {Beers}, {Primas}, {Andersen}, {Barbuy},
  {Bonifacio}, {Molaro}, \& {Nordstr{\"o}m}}]{cayrel;et-al_2004}
{Cayrel}, R., {et~al.} 2004, \aap, 416, 1117
 \href{http://adsabs.harvard.edu/abs/2004A%26A...416.1117C}{\adsurllinklabel}
{}

\bibitem[{{Cayrel de Strobel} \& {Spite}(1988)}]{cayrel}
{Cayrel de Strobel}, G., \& {Spite}, M., eds. 1988, IAU Symposium, Vol. 132,
  {The impact of very high S/N spectroscopy on stellar physics: proceedings of
  the 132nd Symposium of the International Astronomical Union held in Paris,
  France, June 29-July 3, 1987.}
 \href{http://adsabs.harvard.edu/abs/1988IAUS..132.....C}{\adsurllinklabel}
{}

\bibitem[{{Chou} {et~al.}(2007){Chou}, {Majewski}, {Cunha}, {Smith},
  {Patterson}, {Mart{\'{\i}}nez-Delgado}, {Law}, {Crane}, {Mu{\~n}oz}, {Garcia
  L{\'o}pez}, {Geisler}, \& {Skrutskie}}]{Chou;et-al_2007}
{Chou}, M., {et~al.} 2007, \apj, 670, 346
 \href{http://adsabs.harvard.edu/abs/2007ApJ...670..346C}{\adsurllinklabel}
{}

\bibitem[{{Cohen} {et~al.}(2011){Cohen}, {Huang}, \&
  {Kirby}}]{cohen;et-al_2011}
{Cohen}, J.~G., {Huang}, W., \& {Kirby}, E.~N. 2011, \apj, 740, 60
 \href{http://adsabs.harvard.edu/abs/2011ApJ...740...60C}{\adsurllinklabel}
{}

\bibitem[{{Cohen} {et~al.}(2010){Cohen}, {Kirby}, {Simon}, \&
  {Geha}}]{cohen;et-al_2010}
{Cohen}, J.~G., {Kirby}, E.~N., {Simon}, J.~D., \& {Geha}, M. 2010, \apj, 725,
  288
 \href{http://adsabs.harvard.edu/abs/2010ApJ...725..288C}{\adsurllinklabel}
{}

\bibitem[{{Coleman} {et~al.}(2005){Coleman}, {Da Costa}, {Bland-Hawthorn}, \&
  {Freeman}}]{coleman_fornax}
{Coleman}, M.~G., {Da Costa}, G.~S., {Bland-Hawthorn}, J., \& {Freeman}, K.~C.
  2005, \aj, 129, 1443
 \href{http://adsabs.harvard.edu/abs/2005AJ....129.1443C}{\adsurllinklabel}
{}

\bibitem[{{Coleman} {et~al.}(2007){Coleman}, {De Jong}, {Martin}, {Rix},
  {Sand}, {Bell}, {Pogge}, {Thompson}, {Hippelein}, {Giallongo}, {Ragazzoni},
  {DiPaola}, {Farinato}, {Smareglia}, {Testa}, {Bechtold}, {Hill}, {Garnavich},
  \& {Green}}]{coleman;et-al_2007}
{Coleman}, M.~G., {et~al.} 2007, ArXiv e-prints
 \href{http://adsabs.harvard.edu/abs/2007arXiv0706.1669C}{\adsurllinklabel}
{}

\bibitem[{{Creze} {et~al.}(1998){Creze}, {Chereul}, {Bienayme}, \&
  {Pichon}}]{creze;et-al_1998}
{Creze}, M., {Chereul}, E., {Bienayme}, O., \& {Pichon}, C. 1998, \aap, 329,
  920
 \href{http://adsabs.harvard.edu/abs/1998A%26A...329..920C}{\adsurllinklabel}
{}

\bibitem[{{Cristallo} {et~al.}(2009){Cristallo}, {Straniero}, {Gallino},
  {Piersanti}, {Dom{\'{\i}}nguez}, \& {Lederer}}]{cristallo;et-al_2009}
{Cristallo}, S., {Straniero}, O., {Gallino}, R., {Piersanti}, L.,
  {Dom{\'{\i}}nguez}, I., \& {Lederer}, M.~T. 2009, \apj, 696, 797
 \href{http://adsabs.harvard.edu/abs/2009ApJ...696..797C}{\adsurllinklabel}
{}

\bibitem[{{Da Costa} \& {Coleman}(2008)}]{da_costa;coleman_2008}
{Da Costa}, G.~S., \& {Coleman}, M.~G. 2008, \aj, 136, 506
 \href{http://adsabs.harvard.edu/abs/2008AJ....136..506D}{\adsurllinklabel}
{}

\bibitem[{{Da Costa} {et~al.}(2009){Da Costa}, {Held}, {Saviane}, \&
  {Gullieuszik}}]{da_costa;et-al_2009}
{Da Costa}, G.~S., {Held}, E.~V., {Saviane}, I., \& {Gullieuszik}, M. 2009,
  \apj, 705, 1481
 \href{http://adsabs.harvard.edu/abs/2009ApJ...705.1481D}{\adsurllinklabel}
{}

\bibitem[{{Da Costa} {et~al.}(2013){Da Costa}, {Norris}, \&
  {Yong}}]{da_costa;et-al_2013}
{Da Costa}, G.~S., {Norris}, J.~E., \& {Yong}, D. 2013, \apj, 769, 8
 \href{http://adsabs.harvard.edu/abs/2013ApJ...769....8D}{\adsurllinklabel}
{}

\bibitem[{{de Blok} \& {Walter}(2000)}]{de_block;et-al_2000}
{de Blok}, W.~J.~G., \& {Walter}, F. 2000, \apjl, 537, L95
 \href{http://adsabs.harvard.edu/abs/2000ApJ...537L..95D}{\adsurllinklabel}
{}

\bibitem[{{de Marchi} {et~al.}(1999){de Marchi}, {Leibundgut}, {Paresce}, \&
  {Pulone}}]{de_marchi;et-al_1999}
{de Marchi}, G., {Leibundgut}, B., {Paresce}, F., \& {Pulone}, L. 1999, \aap,
  343, L9
 \href{http://adsabs.harvard.edu/abs/1999A%26A...343L...9D}{\adsurllinklabel}
{}

\bibitem[{{De Propris} {et~al.}(2010){De Propris}, {Harrison}, \&
  {Mares}}]{de_propris;et-al_2010}
{De Propris}, R., {Harrison}, C.~D., \& {Mares}, P.~J. 2010, \apj, 719, 1582
 \href{http://adsabs.harvard.edu/abs/2010ApJ...719.1582D}{\adsurllinklabel}
{}

\bibitem[{{De Silva} {et~al.}(2013){De Silva}, {D'Orazi}, {Melo}, {Torres},
  {Gieles}, {Quast}, \& {Sterzik}}]{de_silva;et-al_2013}
{De Silva}, G.~M., {D'Orazi}, V., {Melo}, C., {Torres}, C.~A.~O., {Gieles}, M.,
  {Quast}, G.~R., \& {Sterzik}, M. 2013, ArXiv e-prints
 \href{http://adsabs.harvard.edu/abs/2013arXiv1301.5967D}{\adsurllinklabel}
{}

\bibitem[{{De Silva} {et~al.}(2007){De Silva}, {Freeman}, {Bland-Hawthorn},
  {Asplund}, \& {Bessell}}]{de_silva;et-al_2007}
{De Silva}, G.~M., {Freeman}, K.~C., {Bland-Hawthorn}, J., {Asplund}, M., \&
  {Bessell}, M.~S. 2007, \aj, 133, 694
 \href{http://adsabs.harvard.edu/abs/2007AJ....133..694D}{\adsurllinklabel}
{}

\bibitem[{{de Vaucouleurs}(1970{\natexlab{a}})}]{de_vaucouleurs_1970b}
{de Vaucouleurs}, G. 1970{\natexlab{a}}, in IAU Symposium, Vol.~38, The Spiral
  Structure of our Galaxy, ed. W.~{Becker} \& G.~I. {Kontopoulos}, 18
 \href{http://adsabs.harvard.edu/abs/1970IAUS...38...18D}{\adsurllinklabel}
{}

\bibitem[{{de Vaucouleurs}(1970{\natexlab{b}})}]{de_vaucouleurs_1970a}
{de Vaucouleurs}, G. 1970{\natexlab{b}}, Science, 167, 1203
 \href{http://adsabs.harvard.edu/abs/1970Sci...167.1203D}{\adsurllinklabel}
{}

\bibitem[{{Dehnen}(2000)}]{dehnen_2000}
{Dehnen}, W. 2000, \aj, 119, 800
 \href{http://adsabs.harvard.edu/abs/2000AJ....119..800D}{\adsurllinklabel}
{}

\bibitem[{{Deutsch}(1994)}]{Deutsch_1994}
{Deutsch}, E.~W. 1994, \pasp, 106, 1134
 \href{http://adsabs.harvard.edu/abs/1994PASP..106.1134D}{\adsurllinklabel}
{}

\bibitem[{{Diemand} {et~al.}(2008){Diemand}, {Kuhlen}, {Madau}, {Zemp},
  {Moore}, {Potter}, \& {Stadel}}]{diemand;et-al_2008}
{Diemand}, J., {Kuhlen}, M., {Madau}, P., {Zemp}, M., {Moore}, B., {Potter},
  D., \& {Stadel}, J. 2008, \nat, 454, 735
 \href{http://adsabs.harvard.edu/abs/2008Natur.454..735D}{\adsurllinklabel}
{}

\bibitem[{{Dinescu}(2002)}]{dinescu_2002}
{Dinescu}, D.~I. 2002, in Astronomical Society of the Pacific Conference
  Series, Vol. 265, Omega Centauri, A Unique Window into Astrophysics, ed.
  F.~{van Leeuwen}, J.~D. {Hughes}, \& G.~{Piotto}, 365
 \href{http://adsabs.harvard.edu/abs/2002ASPC..265..365D}{\adsurllinklabel}
{}

\bibitem[{{Dotter} {et~al.}(2008){Dotter}, {Chaboyer}, {Jevremovi{\'c}},
  {Kostov}, {Baron}, \& {Ferguson}}]{Dotter;et-al_2008}
{Dotter}, A., {Chaboyer}, B., {Jevremovi{\'c}}, D., {Kostov}, V., {Baron}, E.,
  \& {Ferguson}, J.~W. 2008, \apjs, 178, 89
 \href{http://adsabs.harvard.edu/abs/2008ApJS..178...89D}{\adsurllinklabel}
{}

\bibitem[{{Downes} {et~al.}(2004){Downes}, {Margon}, {Anderson}, {Harris},
  {Knapp}, {Schroeder}, {Schneider}, {York}, {Pier}, \&
  {Brinkmann}}]{Downes;et-al_2004}
{Downes}, R.~A., {et~al.} 2004, \aj, 127, 2838
 \href{http://adsabs.harvard.edu/abs/2004AJ....127.2838D}{\adsurllinklabel}
{}

\bibitem[{{Drake} {et~al.}(2013){Drake}, {Catelan}, {Djorgovski}, {Torrealba},
  {Graham}, {Mahabal}, {Prieto}, {Donalek}, {Williams}, {Larson},
  {Christensen}, \& {Beshore}}]{drake;et-al_2013}
{Drake}, A.~J., {et~al.} 2013, ArXiv e-prints
 \href{http://adsabs.harvard.edu/abs/2013arXiv1301.6168D}{\adsurllinklabel}
{}

\bibitem[{{Duffau} {et~al.}(2006){Duffau}, {Zinn}, {Vivas}, {Carraro},
  {M{\'e}ndez}, {Winnick}, \& {Gallart}}]{Duffau;et-al_2006}
{Duffau}, S., {Zinn}, R., {Vivas}, A.~K., {Carraro}, G., {M{\'e}ndez}, R.~A.,
  {Winnick}, R., \& {Gallart}, C. 2006, \apjl, 636, L97
 \href{http://adsabs.harvard.edu/abs/2006ApJ...636L..97D}{\adsurllinklabel}
{}

\bibitem[{{Edelsohn} \& {Elmegreen}(1997)}]{edelsohn_1997}
{Edelsohn}, D.~J., \& {Elmegreen}, B.~G. 1997, \mnras, 290, 7
 \href{http://adsabs.harvard.edu/abs/1997MNRAS.290....7E}{\adsurllinklabel}
{}

\bibitem[{{Eggen} {et~al.}(1962){Eggen}, {Lynden-Bell}, \&
  {Sandage}}]{eggen;et-al_1962}
{Eggen}, O.~J., {Lynden-Bell}, D., \& {Sandage}, A.~R. 1962, \apj, 136, 748
 \href{http://adsabs.harvard.edu/abs/1962ApJ...136..748E}{\adsurllinklabel}
{}

\bibitem[{{Eidelman} {et~al.}(2004){Eidelman}, {Hayes}, {Olive},
  {Aguilar-Benitez}, {Amsler}, {Asner}, {Babu}, {Barnett}, {Beringer},
  {Burchat}, {Carone}, {Caso}, {Conforto}, {Dahl}, {D'Ambrosio}, {Doser},
  {Feng}, {Gherghetta}, {Gibbons}, {Goodman}, {Grab}, {Groom}, {Gurtu},
  {Hagiwara}, {Hern{\'a}ndez-Rey}, {Hikasa}, {Honscheid}, {Jawahery}, {Kolda},
  {Kwon}, {Mangano}, {Manohar}, {March-Russell}, {Masoni}, {Miquel},
  {M{\"o}nig}, {Murayama}, {Nakamura}, {Navas}, {Pape}, {Patrignani}, {Piepke},
  {Raffelt}, {Roos}, {Tanabashi}, {Terning}, {T{\"o}rnqvist}, {Trippe},
  {Vogel}, {Wohl}, {Workman}, {Yao}, {Zyla}, {Armstrong}, {Gee}, {Harper},
  {Lugovsky}, {Lugovsky}, {Lugovsky}, {Rom}, {Artuso}, {Barberio}, {Battaglia},
  {Bichsel}, {Biebel}, {Bloch}, {Cahn}, {Casper}, {Cattai}, {Chivukula},
  {Cowan}, {Damour}, {Desler}, {Dobbs}, {Drees}, {Edwards}, {Edwards},
  {Elvira}, {Erler}, {Ezhela}, {Fetscher}, {Fields}, {Foster}, {Froidevaux},
  {Fukugita}, {Gaisser}, {Garren}, {Gerber}, {Gerbier}, {Gilman}, {Haber},
  {Hagmann}, {Hewett}, {Hinchliffe}, {Hogan}, {H{\"o}hler}, {Igo-Kemenes},
  {Jackson}, {Johnson}, {Karlen}, {Kayser}, {Kirkby}, {Klein}, {Kleinknecht},
  {Knowles}, {Kreitz}, {Kuyanov}, {Lahav}, {Langacker}, {Liddle}, {Littenberg},
  {Manley}, {Martin}, {Narain}, {Nason}, {Nir}, {Peacock}, {Quinn}, {Raby},
  {Ratcliff}, {Razuvaev}, {Renk}, {Rolandi}, {Ronan}, {Rosenberg}, {Sachrajda},
  {Sakai}, {Sanda}, {Sarkar}, {Schmitt}, {Schneider}, {Scott}, {Seligman},
  {Shaevitz}, {Sj{\"o}strand}, {Smoot}, {Spanier}, {Spieler}, {Spooner},
  {Srednicki}, {Stahl}, {Stanev}, {Suzuki}, {Tkachenko}, {Trilling},
  {Valencia}, {van Bibber}, {Vincter}, {Ward}, {Webber}, {Whalley},
  {Wolfenstein}, {Womersley}, {Woody}, {Zenin}, \& {Zhu}}]{2004PhLB..592....1P}
{Eidelman}, S., {et~al.} 2004, Physics Letters B, 592, 1
 \href{http://adsabs.harvard.edu/abs/2004PhLB..592....1P}{\adsurllinklabel}
{}

\bibitem[{{Essig} {et~al.}(2010){Essig}, {Sehgal}, {Strigari}, {Geha}, \&
  {Simon}}]{essig}
{Essig}, R., {Sehgal}, N., {Strigari}, L.~E., {Geha}, M., \& {Simon}, J.~D.
  2010, \prd, 82, 123503
 \href{http://adsabs.harvard.edu/abs/2010PhRvD..82l3503E}{\adsurllinklabel}
{}

\bibitem[{{Fellhauer} {et~al.}(2006){Fellhauer}, {Belokurov}, {Evans},
  {Wilkinson}, {Zucker}, {Gilmore}, {Irwin}, {Bramich}, {Vidrih}, {Wyse},
  {Beers}, \& {Brinkmann}}]{Fellhauer;et-al_2006}
{Fellhauer}, M., {et~al.} 2006, \apj, 651, 167
 \href{http://adsabs.harvard.edu/abs/2006ApJ...651..167F}{\adsurllinklabel}
{}

\bibitem[{{Fellhauer} {et~al.}(2007){Fellhauer}, {Evans}, {Belokurov},
  {Zucker}, {Yanny}, {Wilkinson}, {Gilmore}, {Irwin}, {Bramich}, {Vidrih},
  {Hewett}, \& {Beers}}]{fellhaur;et-al_2007}
---. 2007, \mnras, 375, 1171
 \href{http://adsabs.harvard.edu/abs/2007MNRAS.375.1171F}{\adsurllinklabel}
{}

\bibitem[{{Fitzpatrick}(1999)}]{fitzpatrick_1999}
{Fitzpatrick}, E.~L. 1999, \pasp, 111, 63
 \href{http://adsabs.harvard.edu/abs/1999PASP..111...63F}{\adsurllinklabel}
{}

\bibitem[{{Francois} {et~al.}(1988){Francois}, {Spite}, \&
  {Spite}}]{francois;et-al_1988}
{Francois}, P., {Spite}, M., \& {Spite}, F. 1988, \aap, 191, 267
 \href{http://adsabs.harvard.edu/abs/1988A%26A...191..267F}{\adsurllinklabel}
{}

\bibitem[{{Frebel} {et~al.}(2013){Frebel}, {Casey}, {Jacobson}, \&
  {Yu}}]{frebel;et-al_2013}
{Frebel}, A., {Casey}, A.~R., {Jacobson}, H.~R., \& {Yu}, Q. 2013, \apj, 769,
  57
 \href{http://adsabs.harvard.edu/abs/2013ApJ...769...57F}{\adsurllinklabel}
{}

\bibitem[{{Frebel} {et~al.}(2010){Frebel}, {Simon}, {Geha}, \&
  {Willman}}]{frebel;et-al_2010}
{Frebel}, A., {Simon}, J.~D., {Geha}, M., \& {Willman}, B. 2010, \apj, 708, 560
 \href{http://adsabs.harvard.edu/abs/2010ApJ...708..560F}{\adsurllinklabel}
{}

\bibitem[{{Freeman} \& {Bland-Hawthorn}(2002)}]{freeman;bland-hawthorn_2002}
{Freeman}, K., \& {Bland-Hawthorn}, J. 2002, \araa, 40, 487
 \href{http://adsabs.harvard.edu/abs/2002ARA%26A..40..487F}{\adsurllinklabel}
{}

\bibitem[{{Freeman} {et~al.}(2013){Freeman}, {Ness}, {Wylie-de-Boer},
  {Athanassoula}, {Bland-Hawthorn}, {Asplund}, {Lewis}, {Yong}, {Lane}, {Kiss},
  \& {Ibata}}]{argosI}
{Freeman}, K., {et~al.} 2013, \mnras, 428, 3660
 \href{http://adsabs.harvard.edu/abs/2013MNRAS.428.3660F}{\adsurllinklabel}
{}

\bibitem[{{Freeman}(1970)}]{freeman_1970}
{Freeman}, K.~C. 1970, \apj, 160, 811
 \href{http://adsabs.harvard.edu/abs/1970ApJ...160..811F}{\adsurllinklabel}
{}

\bibitem[{{Freeman}(2012)}]{hermes}
{Freeman}, K.~C. 2012, in Astronomical Society of the Pacific Conference
  Series, Vol. 458, Galactic Archaeology: Near-Field Cosmology and the
  Formation of the Milky Way, ed. W.~{Aoki}, M.~{Ishigaki}, T.~{Suda},
  T.~{Tsujimoto}, \& N.~{Arimoto}, 393
 \href{http://adsabs.harvard.edu/abs/2012ASPC..458..393F}{\adsurllinklabel}
{}

\bibitem[{{Frischknecht} {et~al.}(2012){Frischknecht}, {Hirschi}, \&
  {Thielemann}}]{frischknecht;et-al_2012}
{Frischknecht}, U., {Hirschi}, R., \& {Thielemann}, F.-K. 2012, \aap, 538, 2
 \href{http://adsabs.harvard.edu/abs/2012A%26A...538L...2F}{\adsurllinklabel}
{}

\bibitem[{{Fuhrmann}(2011)}]{fuhrmann;et-al_2011}
{Fuhrmann}, K. 2011, \mnras, 414, 2893
 \href{http://adsabs.harvard.edu/abs/2011MNRAS.414.2893F}{\adsurllinklabel}
{}

\bibitem[{{Fulbright}(2000)}]{fulbright_2000}
{Fulbright}, J.~P. 2000, \aj, 120, 1841
 \href{http://adsabs.harvard.edu/abs/2000AJ....120.1841F}{\adsurllinklabel}
{}

\bibitem[{{Fux}(2001)}]{fux_2001}
{Fux}, R. 2001, \aap, 373, 511
 \href{http://adsabs.harvard.edu/abs/2001A%26A...373..511F}{\adsurllinklabel}
{}

\bibitem[{{Gao} \& {Theuns}(2007)}]{gao;theuns_2007}
{Gao}, L., \& {Theuns}, T. 2007, Science, 317, 1527
 \href{http://adsabs.harvard.edu/abs/2007Sci...317.1527G}{\adsurllinklabel}
{}

\bibitem[{{Garc{\'{\i}}a P{\'e}rez} {et~al.}(2006){Garc{\'{\i}}a P{\'e}rez},
  {Asplund}, {Primas}, {Nissen}, \& {Gustafsson}}]{perez;et-al_2006}
{Garc{\'{\i}}a P{\'e}rez}, A.~E., {Asplund}, M., {Primas}, F., {Nissen}, P.~E.,
  \& {Gustafsson}, B. 2006, \aap, 451, 621
 \href{http://adsabs.harvard.edu/abs/2006A%26A...451..621G}{\adsurllinklabel}
{}

\bibitem[{{Geha} {et~al.}(2009){Geha}, {Willman}, {Simon}, {Strigari}, {Kirby},
  {Law}, \& {Strader}}]{geha;et-al_2009}
{Geha}, M., {Willman}, B., {Simon}, J.~D., {Strigari}, L.~E., {Kirby}, E.~N.,
  {Law}, D.~R., \& {Strader}, J. 2009, \apj, 692, 1464
 \href{http://adsabs.harvard.edu/abs/2009ApJ...692.1464G}{\adsurllinklabel}
{}

\bibitem[{{Gilmore} \& {Reid}(1983)}]{gilmore;reid_1983}
{Gilmore}, G., \& {Reid}, N. 1983, \mnras, 202, 1025
 \href{http://adsabs.harvard.edu/abs/1983MNRAS.202.1025G}{\adsurllinklabel}
{}

\bibitem[{{Gilmore} {et~al.}(2012){Gilmore}, {Randich}, {Asplund}, {Binney},
  {Bonifacio}, {Drew}, {Feltzing}, {Ferguson}, {Jeffries}, {Micela},
  {Negueruela}, {Prusti}, {Rix}, {Vallenari}, {Alfaro}, {Allende-Prieto},
  {Babusiaux}, {Bensby}, {Blomme}, {Bragaglia}, {Flaccomio}, {Fran{\c c}ois},
  {Irwin}, {Koposov}, {Korn}, {Lanzafame}, {Pancino}, {Paunzen},
  {Recio-Blanco}, {Sacco}, {Smiljanic}, {Van Eck}, \& {Walton}}]{gaia-eso}
{Gilmore}, G., {et~al.} 2012, The Messenger, 147, 25
 \href{http://adsabs.harvard.edu/abs/2012Msngr.147...25G}{\adsurllinklabel}
{}

\bibitem[{{Gilmore} {et~al.}(2013){Gilmore}, {Koposov}, {Norris}, {Monaco},
  {Yong}, {Wyse}, {Belokurov}, {Geisler}, {Evans}, {Fellhauer}, {Gieren},
  {Irwin}, {Walker}, {Wilkinson}, \& {Zucker}}]{gilmore}
---. 2013, The Messenger, 151, 25
 \href{http://adsabs.harvard.edu/abs/2013Msngr.151...25G}{\adsurllinklabel}
{}

\bibitem[{{Girardi} {et~al.}(2004){Girardi}, {Grebel}, {Odenkirchen}, \&
  {Chiosi}}]{girardi;et-al_2004}
{Girardi}, L., {Grebel}, E.~K., {Odenkirchen}, M., \& {Chiosi}, C. 2004, \aap,
  422, 205
 \href{http://adsabs.harvard.edu/abs/2004A%26A...422..205G}{\adsurllinklabel}
{}

\bibitem[{{G{\'o}mez} {et~al.}(2013){G{\'o}mez}, {Minchev}, {O'Shea}, {Beers},
  {Bullock}, \& {Purcell}}]{gomez;et-al_2013}
{G{\'o}mez}, F.~A., {Minchev}, I., {O'Shea}, B.~W., {Beers}, T.~C., {Bullock},
  J.~S., \& {Purcell}, C.~W. 2013, \mnras, 429, 159
 \href{http://adsabs.harvard.edu/abs/2013MNRAS.429..159G}{\adsurllinklabel}
{}

\bibitem[{{G{\'o}mez} {et~al.}(2012{\natexlab{a}}){G{\'o}mez}, {Minchev},
  {Villalobos}, {O'Shea}, \& {Williams}}]{gomez;et-al_2012}
{G{\'o}mez}, F.~A., {Minchev}, I., {Villalobos}, {\'A}., {O'Shea}, B.~W., \&
  {Williams}, M.~E.~K. 2012{\natexlab{a}}, \mnras, 419, 2163
 \href{http://adsabs.harvard.edu/abs/2012MNRAS.419.2163G}{\adsurllinklabel}
{}

\bibitem[{{G{\'o}mez} {et~al.}(2012{\natexlab{b}}){G{\'o}mez}, {Minchev},
  {O'Shea}, {Lee}, {Beers}, {An}, {Bullock}, {Purcell}, \&
  {Villalobos}}]{gomez;et-al_2012b}
{G{\'o}mez}, F.~A., {et~al.} 2012{\natexlab{b}}, \mnras, 423, 3727
 \href{http://adsabs.harvard.edu/abs/2012MNRAS.423.3727G}{\adsurllinklabel}
{}

\bibitem[{{Goswami} {et~al.}(2010){Goswami}, {Karinkuzhi}, \&
  {Shantikumar}}]{Goswami;et-al_2010}
{Goswami}, A., {Karinkuzhi}, D., \& {Shantikumar}, N.~S. 2010, \mnras, 402,
  1111
 \href{http://adsabs.harvard.edu/abs/2010MNRAS.402.1111G}{\adsurllinklabel}
{}

\bibitem[{{Gratton} \& {Sneden}(1991)}]{gratton_sneden_1991}
{Gratton}, R.~G., \& {Sneden}, C. 1991, \aap, 241, 501
 \href{http://cdsads.u-strasbg.fr/abs/1991A%26A...241..501G}{\adsurllinklabel}
{}

\bibitem[{{Gratton} {et~al.}(2000){Gratton}, {Sneden}, {Carretta}, \&
  {Bragaglia}}]{gratton;et-al_2000}
{Gratton}, R.~G., {Sneden}, C., {Carretta}, E., \& {Bragaglia}, A. 2000, \aap,
  354, 169
 \href{http://adsabs.harvard.edu/abs/2000A%26A...354..169G}{\adsurllinklabel}
{}

\bibitem[{{Green} \& {Morrison}(1993)}]{green;morrison_1993}
{Green}, E.~M., \& {Morrison}, H.~L. 1993, in Astronomical Society of the
  Pacific Conference Series, Vol.~48, The Globular Cluster-Galaxy Connection,
  ed. G.~H. {Smith} \& J.~P. {Brodie}, 318
 \href{http://adsabs.harvard.edu/abs/1993ASPC...48..318G}{\adsurllinklabel}
{}

\bibitem[{{Green} {et~al.}(1994){Green}, {Margon}, {Anderson}, \&
  {Cook}}]{Green;et-al_1994}
{Green}, P.~J., {Margon}, B., {Anderson}, S.~F., \& {Cook}, K.~H. 1994, \apj,
  434, 319
 \href{http://adsabs.harvard.edu/abs/1994ApJ...434..319G}{\adsurllinklabel}
{}

\bibitem[{{Green} {et~al.}(1992){Green}, {Margon}, {Anderson}, \&
  {MacConnell}}]{Green;et-al_1992}
{Green}, P.~J., {Margon}, B., {Anderson}, S.~F., \& {MacConnell}, D.~J. 1992,
  \apj, 400, 659
 \href{http://adsabs.harvard.edu/abs/1992ApJ...400..659G}{\adsurllinklabel}
{}

\bibitem[{{Grevesse} \& {Sauval}(1998)}]{grevesse;sauval_1998}
{Grevesse}, N., \& {Sauval}, A.~J. 1998, \ssr, 85, 161
 \href{http://adsabs.harvard.edu/abs/1998SSRv...85..161G}{\adsurllinklabel}
{}

\bibitem[{{Grillmair}(2006)}]{grillmair_2006}
{Grillmair}, C.~J. 2006, \apjl, 645, 37
 \href{http://adsabs.harvard.edu/abs/2006ApJ...645L..37G}{\adsurllinklabel}
{}

\bibitem[{{Grillmair}(2009)}]{grillmair_2009}
---. 2009, \apj, 693, 1118
 \href{http://adsabs.harvard.edu/abs/2009ApJ...693.1118G}{\adsurllinklabel}
{}

\bibitem[{{Grillmair} \&
  {Dionatos}(2006{\natexlab{a}})}]{grillmair_dionatos_2006a}
{Grillmair}, C.~J., \& {Dionatos}, O. 2006{\natexlab{a}}, \apjl, 641, 37
 \href{http://adsabs.harvard.edu/abs/2006ApJ...641L..37G}{\adsurllinklabel}
{}

\bibitem[{{Grillmair} \&
  {Dionatos}(2006{\natexlab{b}})}]{grillmair_dionatos_2006b}
---. 2006{\natexlab{b}}, \apjl, 643, 17
 \href{http://adsabs.harvard.edu/abs/2006ApJ...643L..17G}{\adsurllinklabel}
{}

\bibitem[{{Grillmair} {et~al.}(1995){Grillmair}, {Freeman}, {Irwin}, \&
  {Quinn}}]{grillmair;et-al_1995}
{Grillmair}, C.~J., {Freeman}, K.~C., {Irwin}, M., \& {Quinn}, P.~J. 1995, \aj,
  109, 2553
 \href{http://adsabs.harvard.edu/abs/1995AJ....109.2553G}{\adsurllinklabel}
{}

\bibitem[{{Grillmair} \& {Johnson}(2006)}]{grillmair_johnson_2006}
{Grillmair}, C.~J., \& {Johnson}, R. 2006, \apjl, 639, 17
 \href{http://adsabs.harvard.edu/abs/2006ApJ...639L..17G}{\adsurllinklabel}
{}

\bibitem[{{Gustafsson} {et~al.}(2008){Gustafsson}, {Edvardsson}, {Eriksson},
  {J{\o}rgensen}, {Nordlund}, \& {Plez}}]{marcs_models}
{Gustafsson}, B., {Edvardsson}, B., {Eriksson}, K., {J{\o}rgensen}, U.~G.,
  {Nordlund}, {\AA}., \& {Plez}, B. 2008, \aap, 486, 951
 \href{http://adsabs.harvard.edu/abs/2008A%26A...486..951G}{\adsurllinklabel}
{}

\bibitem[{{Guth}(1997)}]{guth_1997}
{Guth}, A.~H., ed. 1997, {The inflationary universe. The quest for a new theory
  of cosmic origins}
 \href{http://adsabs.harvard.edu/abs/1997iuqn.conf.....G}{\adsurllinklabel}
{}

\bibitem[{{Harris}(1996)}]{harris_1996}
{Harris}, W.~E. 1996, \aj, 112, 1487
 \href{http://adsabs.harvard.edu/abs/1996AJ....112.1487H}{\adsurllinklabel}
{}

\bibitem[{{Helmi}(2004)}]{Helmi_2004}
{Helmi}, A. 2004, \apjl, 610, L97
 \href{http://adsabs.harvard.edu/abs/2004ApJ...610L..97H}{\adsurllinklabel}
{}

\bibitem[{{Helmi}(2008)}]{helmi_2008}
---. 2008, \aapr, 15, 145
 \href{http://adsabs.harvard.edu/abs/2008A%26ARv..15..145H}{\adsurllinklabel}
{}

\bibitem[{{Helmi} \& {White}(1999)}]{Helmi;White_1999}
{Helmi}, A., \& {White}, S.~D.~M. 1999, \mnras, 307, 495
 \href{http://adsabs.harvard.edu/abs/1999MNRAS.307..495H}{\adsurllinklabel}
{}

\bibitem[{{Henden} {et~al.}(2012){Henden}, {Levine}, {Terrell}, {Smith}, \&
  {Welch}}]{henden;et-al_2012}
{Henden}, A.~A., {Levine}, S.~E., {Terrell}, D., {Smith}, T.~C., \& {Welch}, D.
  2012, Journal of the American Association of Variable Star Observers
  (JAAVSO), 40, 430
 \href{http://adsabs.harvard.edu/abs/2012JAVSO..40..430H}{\adsurllinklabel}
{}

\bibitem[{{Hinkle} {et~al.}(2003){Hinkle}, {Wallace}, {Livingston}, {Ayres},
  {Harmer}, \& {Valenti}}]{Hinkle;et-al_2003}
{Hinkle}, K., {Wallace}, L., {Livingston}, W., {Ayres}, T., {Harmer}, D., \&
  {Valenti}, J. 2003, in The Future of Cool-Star Astrophysics: 12th Cambridge
  Workshop on Cool Stars, Stellar Systems, and the Sun, ed. A.~{Brown}, G.~M.
  {Harper}, \& T.~R. {Ayres}, Vol.~12, 851
 \href{http://adsabs.harvard.edu/abs/2003csss...12..851H}{\adsurllinklabel}
{}

\bibitem[{{Hodapp} {et~al.}(2004){Hodapp}, {Kaiser}, {Aussel}, {Burgett},
  {Chambers}, {Chun}, {Dombeck}, {Douglas}, {Hafner}, {Heasley}, {Hoblitt},
  {Hude}, {Isani}, {Jedicke}, {Jewitt}, {Laux}, {Luppino}, {Lupton}, {Maberry},
  {Magnier}, {Mannery}, {Monet}, {Morgan}, {Onaka}, {Price}, {Ryan},
  {Siegmund}, {Szapudi}, {Tonry}, {Wainscoat}, \&
  {Waterson}}]{hodapp;et-al_2004}
{Hodapp}, K.~W., {et~al.} 2004, Astronomische Nachrichten, 325, 636
 \href{http://adsabs.harvard.edu/abs/2004AN....325..636H}{\adsurllinklabel}
{}

\bibitem[{{Ibata} {et~al.}(2001){Ibata}, {Lewis}, {Irwin}, {Totten}, \&
  {Quinn}}]{ibata;et-al_2001}
{Ibata}, R., {Lewis}, G.~F., {Irwin}, M., {Totten}, E., \& {Quinn}, T. 2001,
  \apj, 551, 294
 \href{http://adsabs.harvard.edu/abs/2001ApJ...551..294I}{\adsurllinklabel}
{}

\bibitem[{{Ibata} {et~al.}(1994){Ibata}, {Gilmore}, \&
  {Irwin}}]{ibata;et-al_1994}
{Ibata}, R.~A., {Gilmore}, G., \& {Irwin}, M.~J. 1994, \nat, 370, 194
 \href{http://adsabs.harvard.edu/abs/1994Natur.370..194I}{\adsurllinklabel}
{}

\bibitem[{{Ibata} \& {Lewis}(1998)}]{Ibata;Lewis_1998}
{Ibata}, R.~A., \& {Lewis}, G.~F. 1998, \apj, 500, 575
 \href{http://adsabs.harvard.edu/abs/1998ApJ...500..575I}{\adsurllinklabel}
{}

\bibitem[{{Irwin} \& {Hatzidimitriou}(1995)}]{irwin_1995}
{Irwin}, M., \& {Hatzidimitriou}, D. 1995, \mnras, 277, 1354
 \href{http://adsabs.harvard.edu/abs/1995MNRAS.277.1354I}{\adsurllinklabel}
{}

\bibitem[{{Ishigaki} {et~al.}(2013){Ishigaki}, {Aoki}, \&
  {Chiba}}]{ishigaki;et-al_2013}
{Ishigaki}, M.~N., {Aoki}, W., \& {Chiba}, M. 2013, \apj, 771, 67
 \href{http://adsabs.harvard.edu/abs/2013ApJ...771...67I}{\adsurllinklabel}
{}

\bibitem[{{Ivezi{\'c}} {et~al.}(2000){Ivezi{\'c}}, {Goldston}, {Finlator},
  {Knapp}, {Yanny}, {McKay}, {Amrose}, {Krisciunas}, {Willman}, {Anderson},
  {Schaber}, {Erb}, {Logan}, {Stubbs}, {Chen}, {Neilsen}, {Uomoto}, {Pier},
  {Fan}, {Gunn}, {Lupton}, {Rockosi}, {Schlegel}, {Strauss}, {Annis},
  {Brinkmann}, {Csabai}, {Doi}, {Fukugita}, {Hennessy}, {Hindsley}, {Margon},
  {Munn}, {Newberg}, {Schneider}, {Smith}, {Szokoly}, {Thakar}, {Vogeley},
  {Waddell}, {Yasuda}, \& {York}}]{Ivezic;et-al_2000}
{Ivezi{\'c}}, {\v Z}., {et~al.} 2000, \aj, 120, 963
 \href{http://adsabs.harvard.edu/abs/2000AJ....120..963I}{\adsurllinklabel}
{}

\bibitem[{{Jester} {et~al.}(2005){Jester}, {Schneider}, {Richards}, {Green},
  {Schmidt}, {Hall}, {Strauss}, {Vanden Berk}, {Stoughton}, {Gunn},
  {Brinkmann}, {Kent}, {Smith}, {Tucker}, \& {Yanny}}]{Jester;et-al_2005}
{Jester}, S., {et~al.} 2005, \aj, 130, 873
 \href{http://adsabs.harvard.edu/abs/2005AJ....130..873J}{\adsurllinklabel}
{}

\bibitem[{{Jin} \& {Lynden-Bell}(2007)}]{jin;lynden-bell_2007}
{Jin}, S., \& {Lynden-Bell}, D. 2007, \mnras, 378, L64
 \href{http://adsabs.harvard.edu/abs/2007MNRAS.378L..64J}{\adsurllinklabel}
{}

\bibitem[{{Johnson} \& {Pilachowski}(2010)}]{johnson_pilachowski_2010}
{Johnson}, C.~I., \& {Pilachowski}, C.~A. 2010, \apj, 722, 1373
 \href{http://esoads.eso.org/abs/2010ApJ...722.1373J}{\adsurllinklabel}
{}

\bibitem[{{Johnson}(2002)}]{johnson_2002}
{Johnson}, J.~A. 2002, \apjs, 139, 219
 \href{http://adsabs.harvard.edu/abs/2002ApJS..139..219J}{\adsurllinklabel}
{}

\bibitem[{{Johnston}(1998)}]{johnston_1998}
{Johnston}, K.~V. 1998, \apj, 495, 297
 \href{http://adsabs.harvard.edu/abs/1998ApJ...495..297J}{\adsurllinklabel}
{}

\bibitem[{{Johnston} {et~al.}(2005){Johnston}, {Law}, \&
  {Majewski}}]{Johnston;et-al_2005}
{Johnston}, K.~V., {Law}, D.~R., \& {Majewski}, S.~R. 2005, \apj, 619, 800
 \href{http://adsabs.harvard.edu/abs/2005ApJ...619..800J}{\adsurllinklabel}
{}

\bibitem[{{Jordi} {et~al.}(2006){Jordi}, {Grebel}, \&
  {Ammon}}]{jordi;et-al_2006}
{Jordi}, K., {Grebel}, E.~K., \& {Ammon}, K. 2006, \aap, 460, 339
 \href{http://adsabs.harvard.edu/abs/2006A%26A...460..339J}{\adsurllinklabel}
{}

\bibitem[{{Juri{\'c}} {et~al.}(2008){Juri{\'c}}, {Ivezi{\'c}}, {Brooks},
  {Lupton}, {Schlegel}, {Finkbeiner}, {Padmanabhan}, {Bond}, {Sesar},
  {Rockosi}, {Knapp}, {Gunn}, {Sumi}, {Schneider}, {Barentine}, {Brewington},
  {Brinkmann}, {Fukugita}, {Harvanek}, {Kleinman}, {Krzesinski}, {Long},
  {Neilsen}, {Nitta}, {Snedden}, \& {York}}]{Juric;et-al_2008}
{Juri{\'c}}, M., {et~al.} 2008, \apj, 673, 864
 \href{http://adsabs.harvard.edu/abs/2008ApJ...673..864J}{\adsurllinklabel}
{}

\bibitem[{{Kafle} {et~al.}(2012){Kafle}, {Sharma}, {Lewis}, \&
  {Bland-Hawthorn}}]{kafle;et-al_2012}
{Kafle}, P.~R., {Sharma}, S., {Lewis}, G.~F., \& {Bland-Hawthorn}, J. 2012,
  \apj, 761, 98
 \href{http://adsabs.harvard.edu/abs/2012ApJ...761...98K}{\adsurllinklabel}
{}

\bibitem[{{Kang} \& {Lee}(2012)}]{tame}
{Kang}, W., \& {Lee}, S.-G. 2012, \mnras, 425, 3162
 \href{http://adsabs.harvard.edu/abs/2012MNRAS.425.3162K}{\adsurllinklabel}
{}

\bibitem[{{Karakas} \& {Lattanzio}(2007)}]{karakas;lattanzio_2007}
{Karakas}, A., \& {Lattanzio}, J.~C. 2007, Proc. Astron. Soc. Aust., 24, 103
 \href{http://adsabs.harvard.edu/abs/2007PASA...24..103K}{\adsurllinklabel}
{}

\bibitem[{{Karakas}(2010)}]{karakas_2010}
{Karakas}, A.~I. 2010, \mnras, 403, 1413
 \href{http://adsabs.harvard.edu/abs/2010MNRAS.403.1413K}{\adsurllinklabel}
{}

\bibitem[{{Karakas} \& {Lattanzio}(2003)}]{karakas;lattanzio_2003}
{Karakas}, A.~I., \& {Lattanzio}, J.~C. 2003, Publications of the Astron. Soc.
  of Australia, 20, 279
 \href{http://adsabs.harvard.edu/abs/2003PASA...20..279K}{\adsurllinklabel}
{}

\bibitem[{{Karakas} {et~al.}(2006){Karakas}, {Lugaro}, {Wiescher},
  {G{\"o}rres}, \& {Ugalde}}]{karakas;et-al_2006}
{Karakas}, A.~I., {Lugaro}, M.~A., {Wiescher}, M., {G{\"o}rres}, J., \&
  {Ugalde}, C. 2006, \apj, 643, 471
 \href{http://adsabs.harvard.edu/abs/2006ApJ...643..471K}{\adsurllinklabel}
{}

\bibitem[{{Keller} {et~al.}(2008){Keller}, {Murphy}, {Prior}, {Da Costa}, \&
  {Schmidt}}]{keller;et-al_2008}
{Keller}, S.~C., {Murphy}, S., {Prior}, S., {Da Costa}, G., \& {Schmidt}, B.
  2008, \apj, 678, 851
 \href{http://adsabs.harvard.edu/abs/2008ApJ...678..851K}{\adsurllinklabel}
{}

\bibitem[{{Keller} {et~al.}(2010){Keller}, {Yong}, \& {Da
  Costa}}]{keller;et-al_2010}
{Keller}, S.~C., {Yong}, D., \& {Da Costa}, G.~S. 2010, \apj, 720, 940
 \href{http://adsabs.harvard.edu/abs/2010ApJ...720..940K}{\adsurllinklabel}
{}

\bibitem[{{Keller} {et~al.}(2007){Keller}, {Schmidt}, {Bessell}, {Conroy},
  {Francis}, {Granlund}, {Kowald}, {Oates}, {Martin-Jones}, {Preston},
  {Tisserand}, {Vaccarella}, \& {Waterson}}]{keller;et-al_2007}
{Keller}, S.~C., {et~al.} 2007, Publications of the Astronomical Society of
  Australia, 24, 1
 \href{http://adsabs.harvard.edu/abs/2007PASA...24....1K}{\adsurllinklabel}
{}

\bibitem[{{Kerr} \& {Lynden-Bell}(1986)}]{kerr;lynden-bell_1986}
{Kerr}, F.~J., \& {Lynden-Bell}, D. 1986, \mnras, 221, 1023
 \href{http://adsabs.harvard.edu/abs/1986MNRAS.221.1023K}{\adsurllinklabel}
{}

\bibitem[{{Kerzendorf} {et~al.}(2012){Kerzendorf}, {Yong}, {Schmidt}, {Simon},
  {Jeffery}, {Anderson}, {Podsiadlowski}, {Gal-Yam}, {Silverman}, {Filippenko},
  {Nomoto}, {Murphy}, {Bessell}, {Venn}, \& {Foley}}]{kerzendorf}
{Kerzendorf}, W.~E., {et~al.} 2012, ArXiv e-prints
 \href{http://adsabs.harvard.edu/abs/2012arXiv1210.2713K}{\adsurllinklabel}
{}

\bibitem[{{Kirby} {et~al.}(2011){Kirby}, {Lanfranchi}, {Simon}, {Cohen}, \&
  {Guhathakurta}}]{kirby;et-al_2011}
{Kirby}, E.~N., {Lanfranchi}, G.~A., {Simon}, J.~D., {Cohen}, J.~G., \&
  {Guhathakurta}, P. 2011, \apj, 727, 78
 \href{http://adsabs.harvard.edu/abs/2011ApJ...727...78K}{\adsurllinklabel}
{}

\bibitem[{{Klypin} {et~al.}(1999){Klypin}, {Kravtsov}, {Valenzuela}, \&
  {Prada}}]{klypin;et-al_1999}
{Klypin}, A., {Kravtsov}, A.~V., {Valenzuela}, O., \& {Prada}, F. 1999, \apj,
  522, 82
 \href{http://adsabs.harvard.edu/abs/1999ApJ...522...82K}{\adsurllinklabel}
{}

\bibitem[{{Kobayashi} {et~al.}(2011){Kobayashi}, {Karakas}, \&
  {Umeda}}]{kobayashi;et-al_2011}
{Kobayashi}, C., {Karakas}, A.~I., \& {Umeda}, H. 2011, \mnras, 414, 3231
 \href{http://adsabs.harvard.edu/abs/2011MNRAS.414.3231K}{\adsurllinklabel}
{}

\bibitem[{{Kobayashi} \& {Nakasato}(2011)}]{kobayashi;nakasato_2011}
{Kobayashi}, C., \& {Nakasato}, N. 2011, \apj, 729, 16
 \href{http://adsabs.harvard.edu/abs/2011ApJ...729...16K}{\adsurllinklabel}
{}

\bibitem[{{Kochanek}(1996)}]{kochanek_1996}
{Kochanek}, C.~S. 1996, \apj, 466, 638
 \href{http://adsabs.harvard.edu/abs/1996ApJ...466..638K}{\adsurllinklabel}
{}

\bibitem[{{Koposov} {et~al.}(2007){Koposov}, {de Jong}, {Belokurov}, {Rix},
  {Zucker}, {Evans}, {Gilmore}, {Irwin}, \& {Bell}}]{koposov;et-al_2008}
{Koposov}, S., {et~al.} 2007, \apj, 669, 337
 \href{http://adsabs.harvard.edu/abs/2007ApJ...669..337K}{\adsurllinklabel}
{}

\bibitem[{{Koposov} {et~al.}(2010){Koposov}, {Rix}, \&
  {Hogg}}]{koposov;et-al_2010}
{Koposov}, S.~E., {Rix}, H.-W., \& {Hogg}, D.~W. 2010, \apj, 712, 260
 \href{http://adsabs.harvard.edu/abs/2010ApJ...712..260K}{\adsurllinklabel}
{}

\bibitem[{{Kordopatis} {et~al.}(2013){Kordopatis}, {Gilmore}, {Steinmetz},
  {Boeche}, {Seabroke}, {Siebert}, {Zwitter}, {Binney}, {de Laverny},
  {Recio-Blanco}, {Williams}, {Piffl}, {Enke}, {Roeser}, {Bijaoui}, {Wyse},
  {Freeman}, {Munari}, {Carillo}, {Anguiano}, {Burton}, {Campbell}, {Cass},
  {Fiegert}, {Hartley}, {Parker}, {Reid}, {Ritter}, {Russell}, {Stupart},
  {Watson}, {Bienayme}, {Bland-Hawthorn}, {Gerhard}, {Gibson}, {Grebel},
  {Helmi}, {Navarro}, {Conrad}, {Famaey}, {Faure}, {Just}, {Kos}, {Matijevic},
  {McMillan}, {Minchev}, {Scholz}, {Sharma}, {Siviero}, {de Boer}, \&
  {Zerjal}}]{kordopatis}
{Kordopatis}, G., {et~al.} 2013, ArXiv e-prints
 \href{http://adsabs.harvard.edu/abs/2013arXiv1309.4284K}{\adsurllinklabel}
{}

\bibitem[{{Kravtsov} {et~al.}(2004){Kravtsov}, {Gnedin}, \&
  {Klypin}}]{kravtsov;et-al_2004}
{Kravtsov}, A.~V., {Gnedin}, O.~Y., \& {Klypin}, A.~A. 2004, \apj, 609, 482
 \href{http://adsabs.harvard.edu/abs/2004ApJ...609..482K}{\adsurllinklabel}
{}

\bibitem[{{Kurucz} \& {Bell}(1995)}]{sneden}
{Kurucz}, R., \& {Bell}, B. 1995, Atomic Line Data (R.L.~Kurucz and B.~Bell)
  Kurucz CD-ROM No.~23.~Cambridge, Mass.: Smithsonian Astrophysical
  Observatory, 1995., 23
{}
{}

\bibitem[{Kurucz \& {Bell}(1995)}]{Kurucz;Bell_1995}
Kurucz, R.~L., \& {Bell}, B. 1995, Atomic Line Data, Kurucz CD-ROM No. 23,
  (Smithsonian Astrophysical Observatory, Cambridge, MA)
{}
{}

\bibitem[{{Law} {et~al.}(2005){Law}, {Johnston}, \&
  {Majewski}}]{Law;et-al_2005}
{Law}, D.~R., {Johnston}, K.~V., \& {Majewski}, S.~R. 2005, \apj, 619, 807
 \href{http://adsabs.harvard.edu/abs/2005ApJ...619..807L}{\adsurllinklabel}
{}

\bibitem[{{Law} \& {Majewski}(2010)}]{Law;Majewski_2010}
{Law}, D.~R., \& {Majewski}, S.~R. 2010, \apj, 714, 229
 \href{http://adsabs.harvard.edu/abs/2010ApJ...714..229L}{\adsurllinklabel}
{}

\bibitem[{{Law} {et~al.}(2009){Law}, {Majewski}, \&
  {Johnston}}]{Law;et-al_2009}
{Law}, D.~R., {Majewski}, S.~R., \& {Johnston}, K.~V. 2009, \apjl, 703, L67
 \href{http://adsabs.harvard.edu/abs/2009ApJ...703L..67L}{\adsurllinklabel}
{}

\bibitem[{{Lawler} {et~al.}(2001){Lawler}, {Bonvallet}, \&
  {Sneden}}]{Lawler;et-al_2001}
{Lawler}, J.~E., {Bonvallet}, G., \& {Sneden}, C. 2001, \apj, 556, 452
 \href{http://adsabs.harvard.edu/abs/2001ApJ...556..452L}{\adsurllinklabel}
{}

\bibitem[{{Lee} {et~al.}(2008){Lee}, {Beers}, {Sivarani}, {Allende Prieto},
  {Koesterke}, {Wilhelm}, {Re Fiorentin}, {Bailer-Jones}, {Norris}, {Rockosi},
  {Yanny}, {Newberg}, {Covey}, {Zhang}, \& {Luo}}]{sspp}
{Lee}, Y.~S., {et~al.} 2008, \aj, 136, 2022
 \href{http://adsabs.harvard.edu/abs/2008AJ....136.2022L}{\adsurllinklabel}
{}

\bibitem[{{Leon} {et~al.}(2000){Leon}, {Meylan}, \& {Combes}}]{leon;et-al_2000}
{Leon}, S., {Meylan}, G., \& {Combes}, F. 2000, \aap, 359, 907
 \href{http://adsabs.harvard.edu/abs/2000A%26A...359..907L}{\adsurllinklabel}
{}

\bibitem[{{Letarte} {et~al.}(2010){Letarte}, {Hill}, {Tolstoy}, {Jablonka},
  {Shetrone}, {Venn}, {Spite}, {Irwin}, {Battaglia}, {Helmi}, {Primas},
  {Fran{\c c}ois}, {Kaufer}, {Szeifert}, {Arimoto}, \&
  {Sadakane}}]{letarte;et-al_2010}
{Letarte}, B., {et~al.} 2010, \aap, 523, A17
 \href{http://adsabs.harvard.edu/abs/2010A%26A...523A..17L}{\adsurllinklabel}
{}

\bibitem[{{Liddle} \& {Lyth}(2000)}]{liddle;lyth_2000}
{Liddle}, A.~R., \& {Lyth}, D.~H. 2000, {Cosmological Inflation and Large-Scale
  Structure}
 \href{http://adsabs.harvard.edu/abs/2000cils.book.....L}{\adsurllinklabel}
{}

\bibitem[{{Lineweaver}(1999)}]{lineweaver_1999}
{Lineweaver}, C.~H. 1999, Science, 284, 1503
 \href{http://adsabs.harvard.edu/abs/1999Sci...284.1503L}{\adsurllinklabel}
{}

\bibitem[{{Loeb}(2010)}]{loeb_2010}
{Loeb}, A. 2010, {How Did the First Stars and Galaxies Form?}
 \href{http://adsabs.harvard.edu/abs/2010hdfs.book.....L}{\adsurllinklabel}
{}

\bibitem[{{Lopez-Corredoira} {et~al.}(2012){Lopez-Corredoira}, {Moitinho},
  {Zaggia}, {Momany}, {Carraro}, {Hammersley}, {Cabrera-Lavers}, \&
  {Vazquez}}]{lopez-corredoira;et-al_2012}
{Lopez-Corredoira}, M., {Moitinho}, A., {Zaggia}, S., {Momany}, Y., {Carraro},
  G., {Hammersley}, P.~L., {Cabrera-Lavers}, A., \& {Vazquez}, R.~A. 2012,
  ArXiv e-prints
 \href{http://adsabs.harvard.edu/abs/2012arXiv1207.2749L}{\adsurllinklabel}
{}

\bibitem[{{Madau} {et~al.}(2008){Madau}, {Diemand}, \&
  {Kuhlen}}]{madau;et-al_2008}
{Madau}, P., {Diemand}, J., \& {Kuhlen}, M. 2008, \apj, 679, 1260
 \href{http://adsabs.harvard.edu/abs/2008ApJ...679.1260M}{\adsurllinklabel}
{}

\bibitem[{{Magic} {et~al.}(2013){Magic}, {Collet}, {Asplund}, {Trampedach},
  {Hayek}, {Chiavassa}, {Stein}, \& {Nordlund}}]{stagger_grid}
{Magic}, Z., {Collet}, R., {Asplund}, M., {Trampedach}, R., {Hayek}, W.,
  {Chiavassa}, A., {Stein}, R.~F., \& {Nordlund}, {\AA}. 2013, \aap, 557, A26
 \href{http://adsabs.harvard.edu/abs/2013A%26A...557A..26M}{\adsurllinklabel}
{}

\bibitem[{{Magrini} {et~al.}(2013){Magrini}, {Randich}, {Friel}, {Spina},
  {Jacobson}, {Cantat-Gaudin}, {Donati}, {Baglioni}, {Maiorca}, {Bragaglia},
  {Sordo}, \& {Vallenari}}]{fama}
{Magrini}, L., {et~al.} 2013, ArXiv e-prints
 \href{http://adsabs.harvard.edu/abs/2013arXiv1307.2367M}{\adsurllinklabel}
{}

\bibitem[{{Majewski} {et~al.}(2012){Majewski}, {Nidever}, {Smith}, {Damke},
  {Kunkel}, {Patterson}, {Bizyaev}, \& {Garc{\'{\i}}a
  P{\'e}rez}}]{majewski;et-al_2012}
{Majewski}, S.~R., {Nidever}, D.~L., {Smith}, V.~V., {Damke}, G.~J., {Kunkel},
  W.~E., {Patterson}, R.~J., {Bizyaev}, D., \& {Garc{\'{\i}}a P{\'e}rez}, A.~E.
  2012, \apjl, 747, L37
 \href{http://adsabs.harvard.edu/abs/2012ApJ...747L..37M}{\adsurllinklabel}
{}

\bibitem[{{Majewski} {et~al.}(1999){Majewski}, {Siegel}, {Kunkel}, {Reid},
  {Johnston}, {Thompson}, {Landolt}, \& {Palma}}]{Majewski;et-al_1999}
{Majewski}, S.~R., {Siegel}, M.~H., {Kunkel}, W.~E., {Reid}, I.~N., {Johnston},
  K.~V., {Thompson}, I.~B., {Landolt}, A.~U., \& {Palma}, C. 1999, \aj, 118,
  1709
 \href{http://adsabs.harvard.edu/abs/1999AJ....118.1709M}{\adsurllinklabel}
{}

\bibitem[{{Majewski} {et~al.}(2003){Majewski}, {Skrutskie}, {Weinberg}, \&
  {Ostheimer}}]{Majewski;et-al_2003}
{Majewski}, S.~R., {Skrutskie}, M.~F., {Weinberg}, M.~D., \& {Ostheimer}, J.~C.
  2003, \apj, 599, 1082
 \href{http://adsabs.harvard.edu/abs/2003ApJ...599.1082M}{\adsurllinklabel}
{}

\bibitem[{{Marino} {et~al.}(2009){Marino}, {Milone}, {Piotto}, {Villanova},
  {Bedin}, {Bellini}, \& {Renzini}}]{marino;et-al_2009}
{Marino}, A.~F., {Milone}, A.~P., {Piotto}, G., {Villanova}, S., {Bedin},
  L.~R., {Bellini}, A., \& {Renzini}, A. 2009, \aap, 505, 1099
 \href{http://adsabs.harvard.edu/abs/2009A%26A...505.1099M}{\adsurllinklabel}
{}

\bibitem[{{Marino} {et~al.}(2011){Marino}, {Sneden}, {Kraft}, {Wallerstein},
  {Norris}, {da Costa}, {Milone}, {Ivans}, {Gonzalez}, {Fulbright}, {Hilker},
  {Piotto}, {Zoccali}, \& {Stetson}}]{marino;et-al_2011}
{Marino}, A.~F., {et~al.} 2011, \aap, 532, 8
 \href{http://adsabs.harvard.edu/abs/2011A%26A...532A...8M}{\adsurllinklabel}
{}

\bibitem[{{Martin} {et~al.}(2004){Martin}, {Ibata}, {Bellazzini}, {Irwin},
  {Lewis}, \& {Dehnen}}]{martin;et-al_2004}
{Martin}, N.~F., {Ibata}, R.~A., {Bellazzini}, M., {Irwin}, M.~J., {Lewis},
  G.~F., \& {Dehnen}, W. 2004, \mnras, 348, 12
 \href{http://adsabs.harvard.edu/abs/2004MNRAS.348...12M}{\adsurllinklabel}
{}

\bibitem[{{Mart{\'{\i}}nez-Delgado} {et~al.}(2004){Mart{\'{\i}}nez-Delgado},
  {G{\'o}mez-Flechoso}, {Aparicio}, \& {Carrera}}]{Martinez-Delgado;et-al_2004}
{Mart{\'{\i}}nez-Delgado}, D., {G{\'o}mez-Flechoso}, M.~{\'A}., {Aparicio}, A.,
  \& {Carrera}, R. 2004, \apj, 601, 242
 \href{http://adsabs.harvard.edu/abs/2004ApJ...601..242M}{\adsurllinklabel}
{}

\bibitem[{{Mateo}(1998)}]{mateo_1998}
{Mateo}, M.~L. 1998, \araa, 36, 435
 \href{http://adsabs.harvard.edu/abs/1998ARA%26A..36..435M}{\adsurllinklabel}
{}

\bibitem[{{Meyer}(1994)}]{meyer_1994}
{Meyer}, B.~S. 1994, \araa, 32, 153
 \href{http://adsabs.harvard.edu/abs/1994ARA%26A..32..153M}{\adsurllinklabel}
{}

\bibitem[{{Mihalas} \& {Binney}(1981)}]{mihalas;binney_1981}
{Mihalas}, D., \& {Binney}, J. 1981, Science, 214, 829
 \href{http://adsabs.harvard.edu/abs/1981Sci...214..829M}{\adsurllinklabel}
{}

\bibitem[{{Minchev} {et~al.}(2009){Minchev}, {Quillen}, {Williams}, {Freeman},
  {Nordhaus}, {Siebert}, \& {Bienaym{\'e}}}]{minchev;et-al_2009}
{Minchev}, I., {Quillen}, A.~C., {Williams}, M., {Freeman}, K.~C., {Nordhaus},
  J., {Siebert}, A., \& {Bienaym{\'e}}, O. 2009, \mnras, 396, L56
 \href{http://adsabs.harvard.edu/abs/2009MNRAS.396L..56M}{\adsurllinklabel}
{}

\bibitem[{{Miyamoto} \& {Nagai}(1975)}]{miyamoto;nagai_1975}
{Miyamoto}, M., \& {Nagai}, R. 1975, \pasj, 27, 533
 \href{http://adsabs.harvard.edu/abs/1975PASJ...27..533M}{\adsurllinklabel}
{}

\bibitem[{{Mo} {et~al.}(2010){Mo}, {van den Bosch}, \& {White}}]{mo;et-al_2010}
{Mo}, H., {van den Bosch}, F.~C., \& {White}, S. 2010, {Galaxy Formation and
  Evolution}
 \href{http://adsabs.harvard.edu/abs/2010gfe..book.....M}{\adsurllinklabel}
{}

\bibitem[{{Momany} {et~al.}(2006){Momany}, {Zaggia}, {Gilmore}, {Piotto},
  {Carraro}, {Bedin}, \& {de Angeli}}]{momany_2006}
{Momany}, Y., {Zaggia}, S., {Gilmore}, G., {Piotto}, G., {Carraro}, G.,
  {Bedin}, L.~R., \& {de Angeli}, F. 2006, \aap, 451, 515
 \href{http://adsabs.harvard.edu/abs/2006A%26A...451..515M}{\adsurllinklabel}
{}

\bibitem[{{Momany} {et~al.}(2004){Momany}, {Zaggia}, {Bonifacio}, {Piotto}, {De
  Angeli}, {Bedin}, \& {Carraro}}]{momany;et-al_2004}
{Momany}, Y., {Zaggia}, S.~R., {Bonifacio}, P., {Piotto}, G., {De Angeli}, F.,
  {Bedin}, L.~R., \& {Carraro}, G. 2004, \aap, 421, L29
 \href{http://adsabs.harvard.edu/abs/2004A%26A...421L..29M}{\adsurllinklabel}
{}

\bibitem[{{Monaco} {et~al.}(2005){Monaco}, {Bellazzini}, {Bonifacio},
  {Ferraro}, {Marconi}, {Pancino}, {Sbordone}, \& {Zaggia}}]{Monaco;et-al_2005}
{Monaco}, L., {Bellazzini}, M., {Bonifacio}, P., {Ferraro}, F.~R., {Marconi},
  G., {Pancino}, E., {Sbordone}, L., \& {Zaggia}, S. 2005, \aap, 441, 141
 \href{http://adsabs.harvard.edu/abs/2005A%26A...441..141M}{\adsurllinklabel}
{}

\bibitem[{{Moore} {et~al.}(1999){Moore}, {Ghigna}, {Governato}, {Lake},
  {Quinn}, {Stadel}, \& {Tozzi}}]{moore;et-al_1999}
{Moore}, B., {Ghigna}, S., {Governato}, F., {Lake}, G., {Quinn}, T., {Stadel},
  J., \& {Tozzi}, P. 1999, \apjl, 524, L19
 \href{http://adsabs.harvard.edu/abs/1999ApJ...524L..19M}{\adsurllinklabel}
{}

\bibitem[{{Morrison}(1993)}]{morrison_1993}
{Morrison}, H.~L. 1993, \aj, 106, 578
 \href{http://adsabs.harvard.edu/abs/1993AJ....106..578M}{\adsurllinklabel}
{}

\bibitem[{{Mu{\~n}oz} {et~al.}(2006){Mu{\~n}oz}, {Majewski}, {Zaggia},
  {Kunkel}, {Frinchaboy}, {Nidever}, {Crnojevic}, {Patterson}, {Crane},
  {Johnston}, {Sohn}, {Bernstein}, \& {Shectman}}]{munoz_2006}
{Mu{\~n}oz}, R.~R., {et~al.} 2006, \apj, 649, 201
 \href{http://adsabs.harvard.edu/abs/2006ApJ...649..201M}{\adsurllinklabel}
{}

\bibitem[{{Mucciarelli} {et~al.}(2012){Mucciarelli}, {Bellazzini}, {Ibata},
  {Merle}, {Chapman}, {Dalessandro}, \& {Sollima}}]{mucciarelli;et-al_2012}
{Mucciarelli}, A., {Bellazzini}, M., {Ibata}, R., {Merle}, T., {Chapman},
  S.~C., {Dalessandro}, E., \& {Sollima}, A. 2012, \mnras, 426, 2889
 \href{http://adsabs.harvard.edu/abs/2012MNRAS.426.2889M}{\adsurllinklabel}
{}

\bibitem[{{Munari} {et~al.}(2005){Munari}, {Sordo}, {Castelli}, \&
  {Zwitter}}]{munari;et-al_2005}
{Munari}, U., {Sordo}, R., {Castelli}, F., \& {Zwitter}, T. 2005, \aap, 442,
  1127
 \href{http://adsabs.harvard.edu/abs/2005A%26A...442.1127M}{\adsurllinklabel}
{}

\bibitem[{{Navarro} {et~al.}(1997){Navarro}, {Frenk}, \&
  {White}}]{navarro;et-al_1997}
{Navarro}, J.~F., {Frenk}, C.~S., \& {White}, S.~D.~M. 1997, \apj, 490, 493
 \href{http://adsabs.harvard.edu/abs/1997ApJ...490..493N}{\adsurllinklabel}
{}

\bibitem[{Nelder \& Mead(1965)}]{nelder-mead}
Nelder, J.~A., \& Mead, R. 1965, The Computer Journal, 7, 308
 \href{http://adsabs.harvard.edu/abs/2002A%26A...385..951B}{\adsurllinklabel}
{}

\bibitem[{{Ness} {et~al.}(2012){Ness}, {Freeman}, {Athanassoula},
  {Wylie-De-Boer}, {Bland-Hawthorn}, {Lewis}, {Yong}, {Asplund}, {Lane},
  {Kiss}, \& {Ibata}}]{ness;et-al_2012}
{Ness}, M., {et~al.} 2012, \apj, 756, 22
 \href{http://adsabs.harvard.edu/abs/2012ApJ...756...22N}{\adsurllinklabel}
{}

\bibitem[{{Ness} {et~al.}(2013{\natexlab{a}}){Ness}, {Freeman}, {Athanassoula},
  {Wylie-de-Boer}, {Bland-Hawthorn}, {Asplund}, {Lewis}, {Yong}, {Lane}, \&
  {Kiss}}]{argosII}
---. 2013{\natexlab{a}}, \mnras, 430, 836
 \href{http://adsabs.harvard.edu/abs/2013MNRAS.430..836N}{\adsurllinklabel}
{}

\bibitem[{{Ness} {et~al.}(2013{\natexlab{b}}){Ness}, {Freeman}, {Athanassoula},
  {Wylie-de-Boer}, {Bland-Hawthorn}, {Asplund}, {Lewis}, {Yong}, {Lane},
  {Kiss}, \& {Ibata}}]{argosIII}
---. 2013{\natexlab{b}}, \mnras, 432, 2092
 \href{http://adsabs.harvard.edu/abs/2013MNRAS.432.2092N}{\adsurllinklabel}
{}

\bibitem[{{Newberg} {et~al.}(2010){Newberg}, {Willett}, {Yanny}, \&
  {Xu}}]{newberg;et-al_2010}
{Newberg}, H.~J., {Willett}, B.~A., {Yanny}, B., \& {Xu}, Y. 2010, \apj, 711,
  32
 \href{http://adsabs.harvard.edu/abs/2010ApJ...711...32N}{\adsurllinklabel}
{}

\bibitem[{{Newberg} {et~al.}(2007){Newberg}, {Yanny}, {Cole}, {Beers}, {Re
  Fiorentin}, {Schneider}, \& {Wilhelm}}]{Newberg;et-al_2007}
{Newberg}, H.~J., {Yanny}, B., {Cole}, N., {Beers}, T.~C., {Re Fiorentin}, P.,
  {Schneider}, D.~P., \& {Wilhelm}, R. 2007, \apj, 668, 221
 \href{http://adsabs.harvard.edu/abs/2007ApJ...668..221N}{\adsurllinklabel}
{}

\bibitem[{{Newberg} {et~al.}(2009){Newberg}, {Yanny}, \&
  {Willett}}]{newberg;et-al_2009}
{Newberg}, H.~J., {Yanny}, B., \& {Willett}, B.~A. 2009, \apjl, 700, L61
 \href{http://adsabs.harvard.edu/abs/2009ApJ...700L..61N}{\adsurllinklabel}
{}

\bibitem[{{Newberg} {et~al.}(2002){Newberg}, {Yanny}, {Rockosi}, {Grebel},
  {Rix}, {Brinkmann}, {Csabai}, {Hennessy}, {Hindsley}, {Ibata}, {Ivezi{\'c}},
  {Lamb}, {Nash}, {Odenkirchen}, {Rave}, {Schneider}, {Smith}, {Stolte}, \&
  {York}}]{Newberg;et-al_2002}
{Newberg}, H.~J., {et~al.} 2002, \apj, 569, 245
 \href{http://adsabs.harvard.edu/abs/2002ApJ...569..245N}{\adsurllinklabel}
{}

\bibitem[{{Newberg} {et~al.}(2003){Newberg}, {Yanny}, {Grebel}, {Hennessy},
  {Ivezi{\'c}}, {Martinez-Delgado}, {Odenkirchen}, {Rix}, {Brinkmann}, {Lamb},
  {Schneider}, \& {York}}]{Newberg;et-al_2003}
---. 2003, \apjl, 596, L191
 \href{http://adsabs.harvard.edu/abs/2003ApJ...596L.191N}{\adsurllinklabel}
{}

\bibitem[{{Nichols} {et~al.}(2012){Nichols}, {Lin}, \&
  {Bland-Hawthorn}}]{nichols;et-al_2012}
{Nichols}, M., {Lin}, D., \& {Bland-Hawthorn}, J. 2012, \apj, 748, 149
 \href{http://adsabs.harvard.edu/abs/2012ApJ...748..149N}{\adsurllinklabel}
{}

\bibitem[{{Niederste-Ostholt} {et~al.}(2009){Niederste-Ostholt}, {Belokurov},
  {Evans}, {Gilmore}, {Wyse}, \& {Norris}}]{ostholt}
{Niederste-Ostholt}, M., {Belokurov}, V., {Evans}, N.~W., {Gilmore}, G.,
  {Wyse}, R.~F.~G., \& {Norris}, J.~E. 2009, \mnras, 398, 1771
 \href{http://adsabs.harvard.edu/abs/2009MNRAS.398.1771N}{\adsurllinklabel}
{}

\bibitem[{{Nissen} {et~al.}(2000){Nissen}, {Chen}, {Schuster}, \&
  {Zhao}}]{nissen;et-al_2000}
{Nissen}, P.~E., {Chen}, Y.~Q., {Schuster}, W.~J., \& {Zhao}, G. 2000, \aap,
  353, 722
 \href{http://cdsads.u-strasbg.fr/abs/2000A%26A...353..722N}{\adsurllinklabel}
{}

\bibitem[{{Nissen} \& {Schuster}(1997)}]{nissen;schuster_1997}
{Nissen}, P.~E., \& {Schuster}, W.~J. 1997, \aap, 326, 751
 \href{http://adsabs.harvard.edu/abs/1997A%26A...326..751N}{\adsurllinklabel}
{}

\bibitem[{{Nissen} \& {Schuster}(2010)}]{nissen;schuster_2010}
---. 2010, \aap, 511, L10
 \href{http://adsabs.harvard.edu/abs/2010A%26A...511L..10N}{\adsurllinklabel}
{}

\bibitem[{{Nissen} \& {Schuster}(2011)}]{nissen;schuster_2011}
---. 2011, \aap, 530, 15
 \href{http://adsabs.harvard.edu/abs/2011A%26A...530A..15N}{\adsurllinklabel}
{}

\bibitem[{{Nordstr{\"o}m} {et~al.}(2004){Nordstr{\"o}m}, {Mayor}, {Andersen},
  {Holmberg}, {Pont}, {J{\o}rgensen}, {Olsen}, {Udry}, \&
  {Mowlavi}}]{nordstrom;et-al_2004}
{Nordstr{\"o}m}, B., {et~al.} 2004, \aap, 418, 989
 \href{http://adsabs.harvard.edu/abs/2004A%26A...418..989N}{\adsurllinklabel}
{}

\bibitem[{{Norris} \& {Da Costa}(1995)}]{norris;da_costa_1995}
{Norris}, J.~E., \& {Da Costa}, G.~S. 1995, \apjl, 441, L81
 \href{http://adsabs.harvard.edu/abs/1995ApJ...441L..81N}{\adsurllinklabel}
{}

\bibitem[{{Norris} {et~al.}(1996){Norris}, {Ryan}, \&
  {Beers}}]{norris;et-al_1996}
{Norris}, J.~E., {Ryan}, S.~G., \& {Beers}, T.~C. 1996, \apjs, 107, 391
 \href{http://adsabs.harvard.edu/abs/1996ApJS..107..391N}{\adsurllinklabel}
{}

\bibitem[{{Norris} {et~al.}(2001){Norris}, {Ryan}, \&
  {Beers}}]{norris;et-al_2001}
---. 2001, \apj, 561, 1034
 \href{http://adsabs.harvard.edu/abs/2001ApJ...561.1034N}{\adsurllinklabel}
{}

\bibitem[{{Norris} {et~al.}(2010){Norris}, {Wyse}, {Gilmore}, {Yong}, {Frebel},
  {Wilkinson}, {Belokurov}, \& {Zucker}}]{norris;et-al_2010}
{Norris}, J.~E., {Wyse}, R.~F.~G., {Gilmore}, G., {Yong}, D., {Frebel}, A.,
  {Wilkinson}, M.~I., {Belokurov}, V., \& {Zucker}, D.~B. 2010, \apj, 723, 1632
 \href{http://adsabs.harvard.edu/abs/2010ApJ...723.1632N}{\adsurllinklabel}
{}

\bibitem[{{Odenkirchen} {et~al.}(2009){Odenkirchen}, {Grebel}, {Kayser}, {Rix},
  \& {Dehnen}}]{odenkirchen;et-al_2009}
{Odenkirchen}, M., {Grebel}, E.~K., {Kayser}, A., {Rix}, H.-W., \& {Dehnen}, W.
  2009, \aj, 137, 3378
 \href{http://adsabs.harvard.edu/abs/2009AJ....137.3378O}{\adsurllinklabel}
{}

\bibitem[{{Odenkirchen} {et~al.}(2003){Odenkirchen}, {Grebel}, {Dehnen}, {Rix},
  {Yanny}, {Newberg}, {Rockosi}, {Mart{\'{\i}}nez-Delgado}, {Brinkmann}, \&
  {Pier}}]{odenkirchen;et-al_2003}
{Odenkirchen}, M., {et~al.} 2003, \aj, 126, 2385
 \href{http://adsabs.harvard.edu/abs/2003AJ....126.2385O}{\adsurllinklabel}
{}

\bibitem[{{Okamoto} {et~al.}(2008){Okamoto}, {Arimoto}, {Yamada}, \&
  {Onodera}}]{okamoto;et-al_2008}
{Okamoto}, S., {Arimoto}, N., {Yamada}, Y., \& {Onodera}, M. 2008, \aap, 487,
  103
 \href{http://adsabs.harvard.edu/abs/2008A%26A...487..103O}{\adsurllinklabel}
{}

\bibitem[{{Oswalt} {et~al.}(1996){Oswalt}, {Smith}, {Wood}, \&
  {Hintzen}}]{oswalt_1996}
{Oswalt}, T.~D., {Smith}, J.~A., {Wood}, M.~A., \& {Hintzen}, P. 1996, \nat,
  382, 692
 \href{http://adsabs.harvard.edu/abs/1996Natur.382..692O}{\adsurllinklabel}
{}

\bibitem[{{Pakzad} {et~al.}(2004){Pakzad}, {Majewski}, {Frinchaboy}, {Hummels},
  {Ivezic}, {Johnston}, {Law}, {Patterson}, {Prada}, \&
  {Skrutskie}}]{pakzad;et-al_2004}
{Pakzad}, S.~L., {et~al.} 2004, in Bulletin of the American Astronomical
  Society, Vol.~36, American Astronomical Society Meeting Abstracts, 142.05
 \href{http://adsabs.harvard.edu/abs/2004AAS...20514205P}{\adsurllinklabel}
{}

\bibitem[{{Palma} {et~al.}(2003){Palma}, {Majewski}, {Siegel}, {Patterson},
  {Ostheimer}, \& {Link}}]{palma;et-al_2003}
{Palma}, C., {Majewski}, S.~R., {Siegel}, M.~H., {Patterson}, R.~J.,
  {Ostheimer}, J.~C., \& {Link}, R. 2003, \aj, 125, 1352
 \href{http://adsabs.harvard.edu/abs/2003AJ....125.1352P}{\adsurllinklabel}
{}

\bibitem[{{Peebles}(1971)}]{peebles_1971}
{Peebles}, P.~J.~E. 1971, {Physical cosmology}
 \href{http://adsabs.harvard.edu/abs/1971phco.book.....P}{\adsurllinklabel}
{}

\bibitem[{{Peebles}(1974)}]{peebles_1974}
---. 1974, \apjl, 189, L51
 \href{http://adsabs.harvard.edu/abs/1974ApJ...189L..51P}{\adsurllinklabel}
{}

\bibitem[{{Perryman} {et~al.}(2001){Perryman}, {de Boer}, {Gilmore}, {H{\o}g},
  {Lattanzi}, {Lindegren}, {Luri}, {Mignard}, {Pace}, \& {de Zeeuw}}]{perryman}
{Perryman}, M.~A.~C., {et~al.} 2001, \aap, 369, 339
 \href{http://adsabs.harvard.edu/abs/2001A%26A...369..339P}{\adsurllinklabel}
{}

\bibitem[{{Phelps} {et~al.}(1994){Phelps}, {Janes}, \&
  {Montgomery}}]{phelps;et-al_1994}
{Phelps}, R.~L., {Janes}, K.~A., \& {Montgomery}, K.~A. 1994, \aj, 107, 1079
 \href{http://adsabs.harvard.edu/abs/1994AJ....107.1079P}{\adsurllinklabel}
{}

\bibitem[{{Plez} {et~al.}(2008){Plez}, {Masseron}, \& {Van
  Eck}}]{plez;et-al_2008}
{Plez}, B., {Masseron}, T., \& {Van Eck}, S. 2008, in Cool Stars, Stellar
  Systems and the Sun, ASP Conference Series
{}
{}

\bibitem[{{Prantzos} {et~al.}(2007){Prantzos}, {Charbonnel}, \&
  {Iliadis}}]{prantzos;et-al_2007}
{Prantzos}, N., {Charbonnel}, C., \& {Iliadis}, C. 2007, \aap, 470, 179
 \href{http://adsabs.harvard.edu/abs/2007A%26A...470..179P}{\adsurllinklabel}
{}

\bibitem[{{Press} \& {Schechter}(1974)}]{press;schechter_1974}
{Press}, W.~H., \& {Schechter}, P. 1974, \apj, 187, 425
 \href{http://adsabs.harvard.edu/abs/1974ApJ...187..425P}{\adsurllinklabel}
{}

\bibitem[{{Preston} {et~al.}(1991){Preston}, {Shectman}, \&
  {Beers}}]{preston;et-al_1991}
{Preston}, G.~W., {Shectman}, S.~A., \& {Beers}, T.~C. 1991, \apj, 375, 121
 \href{http://adsabs.harvard.edu/abs/1991ApJ...375..121P}{\adsurllinklabel}
{}

\bibitem[{{Price-Whelan} \& {Johnston}(2013)}]{adrn_2013}
{Price-Whelan}, A.~M., \& {Johnston}, K.~V. 2013, ArXiv e-prints
 \href{http://adsabs.harvard.edu/abs/2013arXiv1308.2670P}{\adsurllinklabel}
{}

\bibitem[{{Prior} {et~al.}(2009{\natexlab{a}}){Prior}, {Da Costa}, \&
  {Keller}}]{Prior;et-al_2009b}
{Prior}, S.~L., {Da Costa}, G.~S., \& {Keller}, S.~C. 2009{\natexlab{a}}, \apj,
  704, 1327
 \href{http://adsabs.harvard.edu/abs/2009ApJ...704.1327P}{\adsurllinklabel}
{}

\bibitem[{{Prior} {et~al.}(2009{\natexlab{b}}){Prior}, {Da Costa}, {Keller}, \&
  {Murphy}}]{Prior;et-al_2009a}
{Prior}, S.~L., {Da Costa}, G.~S., {Keller}, S.~C., \& {Murphy}, S.~J.
  2009{\natexlab{b}}, \apj, 691, 306
 \href{http://adsabs.harvard.edu/abs/2009ApJ...691..306P}{\adsurllinklabel}
{}

\bibitem[{{Purcell} {et~al.}(2011){Purcell}, {Bullock}, {Tollerud}, {Rocha}, \&
  {Chakrabarti}}]{purcell;et-al_2011}
{Purcell}, C.~W., {Bullock}, J.~S., {Tollerud}, E.~J., {Rocha}, M., \&
  {Chakrabarti}, S. 2011, \nat, 477, 301
 \href{http://adsabs.harvard.edu/abs/2011Natur.477..301P}{\adsurllinklabel}
{}

\bibitem[{{Putman} {et~al.}(1998){Putman}, {Gibson}, {Staveley-Smith}, {Banks},
  {Barnes}, {Bhatal}, {Disney}, {Ekers}, {Freeman}, {Haynes}, {Henning},
  {Jerjen}, {Kilborn}, {Koribalski}, {Knezek}, {Malin}, {Mould}, {Oosterloo},
  {Price}, {Ryder}, {Sadler}, {Stewart}, {Stootman}, {Vaile}, {Webster}, \&
  {Wright}}]{putman_1998}
{Putman}, M.~E., {et~al.} 1998, \nat, 394, 752
 \href{http://adsabs.harvard.edu/abs/1998Natur.394..752P}{\adsurllinklabel}
{}

\bibitem[{{Qian} \& {Wasserburg}(2001)}]{qian;wasserburg}
{Qian}, Y.-Z., \& {Wasserburg}, G.~J. 2001, \apj, 559, 925
 \href{http://adsabs.harvard.edu/abs/2001ApJ...559..925Q}{\adsurllinklabel}
{}

\bibitem[{{Racine} \& {Harris}(1975)}]{photometry_for_ngc2419}
{Racine}, R., \& {Harris}, W.~E. 1975, \apj, 196, 413
 \href{http://adsabs.harvard.edu/abs/1975ApJ...196..413R}{\adsurllinklabel}
{}

\bibitem[{{Ram{\'{\i}}rez} \& {Mel{\'e}ndez}(2005)}]{ramirez;melendez_2005}
{Ram{\'{\i}}rez}, I., \& {Mel{\'e}ndez}, J. 2005, \apj, 626, 465
 \href{http://adsabs.harvard.edu/abs/2005ApJ...626..465R}{\adsurllinklabel}
{}

\bibitem[{{Ram{\'{\i}}rez} {et~al.}(2001){Ram{\'{\i}}rez}, {Cohen}, {Buss}, \&
  {Briley}}]{ewdet}
{Ram{\'{\i}}rez}, S.~V., {Cohen}, J.~G., {Buss}, J., \& {Briley}, M.~M. 2001,
  \aj, 122, 1429
 \href{http://adsabs.harvard.edu/abs/2001AJ....122.1429R}{\adsurllinklabel}
{}

\bibitem[{{Read} {et~al.}(2008){Read}, {Lake}, {Agertz}, \&
  {Debattista}}]{read;et-al_2008}
{Read}, J.~I., {Lake}, G., {Agertz}, O., \& {Debattista}, V.~P. 2008, \mnras,
  389, 1041
 \href{http://adsabs.harvard.edu/abs/2008MNRAS.389.1041R}{\adsurllinklabel}
{}

\bibitem[{{Recio-Blanco} {et~al.}(2006){Recio-Blanco}, {Bijaoui}, \& {de
  Laverny}}]{matisse}
{Recio-Blanco}, A., {Bijaoui}, A., \& {de Laverny}, P. 2006, \mnras, 370, 141
 \href{http://adsabs.harvard.edu/abs/2006MNRAS.370..141R}{\adsurllinklabel}
{}

\bibitem[{{Reddy} {et~al.}(2006){Reddy}, {Lambert}, \& {Allende
  Prieto}}]{reddy;et-al_2006}
{Reddy}, B.~E., {Lambert}, D.~L., \& {Allende Prieto}, C. 2006, \mnras, 367,
  1329
 \href{http://adsabs.harvard.edu/abs/2006MNRAS.367.1329R}{\adsurllinklabel}
{}

\bibitem[{{Reddy} {et~al.}(2003){Reddy}, {Tomkin}, {Lambert}, \& {Allende
  Prieto}}]{reddy;et-al_2003}
{Reddy}, B.~E., {Tomkin}, J., {Lambert}, D.~L., \& {Allende Prieto}, C. 2003,
  \mnras, 340, 304
 \href{http://adsabs.harvard.edu/abs/2003MNRAS.340..304R}{\adsurllinklabel}
{}

\bibitem[{{Rees}(1986)}]{rees_1986}
{Rees}, M.~J. 1986, \mnras, 222, 27P
 \href{http://adsabs.harvard.edu/abs/1986MNRAS.222P..27R}{\adsurllinklabel}
{}

\bibitem[{{Reynolds}(1989)}]{reynolds;et-al_1989}
{Reynolds}, R.~J. 1989, \apjl, 339, L29
 \href{http://adsabs.harvard.edu/abs/1989ApJ...339L..29R}{\adsurllinklabel}
{}

\bibitem[{{Robin} {et~al.}(2003){Robin}, {Reyl{\'e}}, {Derri{\`e}re}, \&
  {Picaud}}]{robin;et-al_2003}
{Robin}, A.~C., {Reyl{\'e}}, C., {Derri{\`e}re}, S., \& {Picaud}, S. 2003,
  \aap, 409, 523
 \href{http://adsabs.harvard.edu/abs/2003A%26A...409..523R}{\adsurllinklabel}
{}

\bibitem[{{Rocha-Pinto} {et~al.}(2000){Rocha-Pinto}, {Scalo}, {Maciel}, \&
  {Flynn}}]{rocha-pinto;et-al_2000}
{Rocha-Pinto}, H.~J., {Scalo}, J., {Maciel}, W.~J., \& {Flynn}, C. 2000, \aap,
  358, 869
 \href{http://adsabs.harvard.edu/abs/2000A%26A...358..869R}{\adsurllinklabel}
{}

\bibitem[{{Roederer} {et~al.}(2010){Roederer}, {Sneden}, {Thompson}, {Preston},
  \& {Shectman}}]{roederer;et-al_2010}
{Roederer}, I.~U., {Sneden}, C., {Thompson}, I.~B., {Preston}, G.~W., \&
  {Shectman}, S.~A. 2010, \apj, 711, 573
 \href{http://adsabs.harvard.edu/abs/2010ApJ...711..573R}{\adsurllinklabel}
{}

\bibitem[{{Roeser} {et~al.}(2010{\natexlab{a}}){Roeser}, {Demleitner}, \&
  {Schilbach}}]{roser;et-al_2010}
{Roeser}, S., {Demleitner}, M., \& {Schilbach}, E. 2010{\natexlab{a}}, \aj,
  139, 2440
 \href{http://adsabs.harvard.edu/abs/2010AJ....139.2440R}{\adsurllinklabel}
{}

\bibitem[{{Roeser} {et~al.}(2010{\natexlab{b}}){Roeser}, {Demleitner}, \&
  {Schilbach}}]{roeser;et-al_2010}
---. 2010{\natexlab{b}}, \aj, 139, 2440
 \href{http://adsabs.harvard.edu/abs/2010AJ....139.2440R}{\adsurllinklabel}
{}

\bibitem[{{Rubin} \& {Ford}(1970)}]{rubin;et-al_1970}
{Rubin}, V.~C., \& {Ford}, Jr., W.~K. 1970, \apj, 159, 379
 \href{http://adsabs.harvard.edu/abs/1970ApJ...159..379R}{\adsurllinklabel}
{}

\bibitem[{{Rubin} {et~al.}(1980){Rubin}, {Ford}, \&
  {.~Thonnard}}]{rubin;et-al_1980}
{Rubin}, V.~C., {Ford}, W.~K.~J., \& {.~Thonnard}, N. 1980, \apj, 238, 471
 \href{http://adsabs.harvard.edu/abs/1980ApJ...238..471R}{\adsurllinklabel}
{}

\bibitem[{{Ruhland} {et~al.}(2011){Ruhland}, {Bell}, {Rix}, \&
  {Xue}}]{Ruhland;et-al_2011}
{Ruhland}, C., {Bell}, E.~F., {Rix}, H.-W., \& {Xue}, X.-X. 2011, \apj, 731,
  119
 \href{http://adsabs.harvard.edu/abs/2011ApJ...731..119R}{\adsurllinklabel}
{}

\bibitem[{{Rutledge} {et~al.}(1997){Rutledge}, {Hesser}, \&
  {Stetson}}]{Rutledge;Hesser;Stetson_1997}
{Rutledge}, G.~A., {Hesser}, J.~E., \& {Stetson}, P.~B. 1997, \pasp, 109, 907
 \href{http://adsabs.harvard.edu/abs/1997PASP..109..907R}{\adsurllinklabel}
{}

\bibitem[{{Ryan} \& {Norris}(1991)}]{Ryan;Norris_1991}
{Ryan}, S.~G., \& {Norris}, J.~E. 1991, \aj, 101, 1865
 \href{http://adsabs.harvard.edu/abs/1991AJ....101.1865R}{\adsurllinklabel}
{}

\bibitem[{{Sales} {et~al.}(2008){Sales}, {Helmi}, {Starkenburg}, {Morrison},
  {Engle}, {Harding}, {Mateo}, {Olszewski}, \& {Sivarani}}]{sales;et-al_2008}
{Sales}, L.~V., {et~al.} 2008, \mnras, 389, 1391
 \href{http://adsabs.harvard.edu/abs/2008MNRAS.389.1391S}{\adsurllinklabel}
{}

\bibitem[{{Saviane} {et~al.}(2012){Saviane}, {da Costa}, {Held}, {Sommariva},
  {Gullieuszik}, {Barbuy}, \& {Ortolani}}]{saviane;et-al_2012}
{Saviane}, I., {da Costa}, G.~S., {Held}, E.~V., {Sommariva}, V.,
  {Gullieuszik}, M., {Barbuy}, B., \& {Ortolani}, S. 2012, \aap, 540, A27
 \href{http://adsabs.harvard.edu/abs/2012A%26A...540A..27S}{\adsurllinklabel}
{}

\bibitem[{{Schlafly} \& {Finkbeiner}(2011)}]{Schlafly;Finkbeiner_2011}
{Schlafly}, E.~F., \& {Finkbeiner}, D.~P. 2011, \apj, 737, 103
 \href{http://adsabs.harvard.edu/abs/2011ApJ...737..103S}{\adsurllinklabel}
{}

\bibitem[{{Schlegel} {et~al.}(1998){Schlegel}, {Finkbeiner}, \&
  {Davis}}]{schlegel;et-al_1998}
{Schlegel}, D.~J., {Finkbeiner}, D.~P., \& {Davis}, M. 1998, \apj, 500, 525
 \href{http://adsabs.harvard.edu/abs/1998ApJ...500..525S}{\adsurllinklabel}
{}

\bibitem[{{Sch{\"o}nrich}(2012)}]{schonrich;et-al_2012}
{Sch{\"o}nrich}, R. 2012, \mnras, 427, 274
 \href{http://adsabs.harvard.edu/abs/2012MNRAS.427..274S}{\adsurllinklabel}
{}

\bibitem[{{Searle} \& {Zinn}(1978)}]{searle;zinn_1978}
{Searle}, L., \& {Zinn}, R. 1978, \apj, 225, 357
 \href{http://adsabs.harvard.edu/abs/1978ApJ...225..357S}{\adsurllinklabel}
{}

\bibitem[{{Sesar} {et~al.}(2013){Sesar}, {Grillmair}, {Cohen}, {Bellm},
  {Bhalerao}, {Levitan}, {Laher}, {Ofek}, {Surace}, {Tang}, {Waszczak},
  {Kulkarni}, \& {Prince}}]{sesar;et-al_2013}
{Sesar}, B., {et~al.} 2013, ArXiv e-prints
 \href{http://adsabs.harvard.edu/abs/2013arXiv1308.0857S}{\adsurllinklabel}
{}

\bibitem[{{Sharma} {et~al.}(2011){Sharma}, {Bland-Hawthorn}, {Johnston}, \&
  {Binney}}]{sharma;et-al_2011}
{Sharma}, S., {Bland-Hawthorn}, J., {Johnston}, K.~V., \& {Binney}, J. 2011,
  \apj, 730, 3
 \href{http://adsabs.harvard.edu/abs/2011ApJ...730....3S}{\adsurllinklabel}
{}

\bibitem[{{Shetrone} {et~al.}(2003){Shetrone}, {Venn}, {Tolstoy}, {Primas},
  {Hill}, \& {Kaufer}}]{shetrone;et-al_2003}
{Shetrone}, M., {Venn}, K.~A., {Tolstoy}, E., {Primas}, F., {Hill}, V., \&
  {Kaufer}, A. 2003, \aj, 125, 684
 \href{http://adsabs.harvard.edu/abs/2003AJ....125..684S}{\adsurllinklabel}
{}

\bibitem[{{Shetrone}(1996)}]{shetrone_1996}
{Shetrone}, M.~D. 1996, \aj, 112, 1517
 \href{http://cdsads.u-strasbg.fr/abs/1996AJ....112.1517S}{\adsurllinklabel}
{}

\bibitem[{{Siebert} {et~al.}(2011){Siebert}, {Williams}, {Siviero}, {Reid},
  {Boeche}, {Steinmetz}, {Fulbright}, {Munari}, {Zwitter}, {Watson}, {Wyse},
  {de Jong}, {Enke}, {Anguiano}, {Burton}, {Cass}, {Fiegert}, {Hartley},
  {Ritter}, {Russel}, {Stupar}, {Bienaym{\'e}}, {Freeman}, {Gilmore}, {Grebel},
  {Helmi}, {Navarro}, {Binney}, {Bland-Hawthorn}, {Campbell}, {Famaey},
  {Gerhard}, {Gibson}, {Matijevi{\v c}}, {Parker}, {Seabroke}, {Sharma},
  {Smith}, \& {Wylie-de Boer}}]{siebert;et-al_2011}
{Siebert}, A., {et~al.} 2011, \aj, 141, 187
 \href{http://adsabs.harvard.edu/abs/2011AJ....141..187S}{\adsurllinklabel}
{}

\bibitem[{{Siegel} {et~al.}(2007){Siegel}, {Dotter}, {Majewski}, {Sarajedini},
  {Chaboyer}, {Nidever}, {Anderson}, {Mar{\'{\i}}n-Franch}, {Rosenberg},
  {Bedin}, {Aparicio}, {King}, {Piotto}, \& {Reid}}]{Siegel;et-al_2007}
{Siegel}, M.~H., {et~al.} 2007, \apjl, 667, L57
 \href{http://adsabs.harvard.edu/abs/2007ApJ...667L..57S}{\adsurllinklabel}
{}

\bibitem[{{Simmerer} {et~al.}(2013){Simmerer}, {Ivans}, {Filler}, {Francois},
  {Charbonnel}, {Monier}, \& {James}}]{simmerer;et-al_2013}
{Simmerer}, J., {Ivans}, I.~I., {Filler}, D., {Francois}, P., {Charbonnel}, C.,
  {Monier}, R., \& {James}, G. 2013, \apjl, 764, L7
 \href{http://adsabs.harvard.edu/abs/2013ApJ...764L...7S}{\adsurllinklabel}
{}

\bibitem[{{Simon} \& {Geha}(2007)}]{simon;et-al_2007}
{Simon}, J.~D., \& {Geha}, M. 2007, \apj, 670, 313
 \href{http://adsabs.harvard.edu/abs/2007ApJ...670..313S}{\adsurllinklabel}
{}

\bibitem[{{Simon} {et~al.}(2011){Simon}, {Geha}, {Minor}, {Martinez}, {Kirby},
  {Bullock}, {Kaplinghat}, {Strigari}, {Willman}, {Choi}, {Tollerud}, \&
  {Wolf}}]{simon;et-al_2011}
{Simon}, J.~D., {et~al.} 2011, \apj, 733, 46
 \href{http://adsabs.harvard.edu/abs/2011ApJ...733...46S}{\adsurllinklabel}
{}

\bibitem[{{Sirko} {et~al.}(2004){Sirko}, {Goodman}, {Knapp}, {Brinkmann},
  {Ivezi{\'c}}, {Knerr}, {Schlegel}, {Schneider}, \& {York}}]{Sirko;et-al_2004}
{Sirko}, E., {et~al.} 2004, \aj, 127, 899
 \href{http://adsabs.harvard.edu/abs/2004AJ....127..899S}{\adsurllinklabel}
{}

\bibitem[{{Skrutskie} {et~al.}(2006){Skrutskie}, {Cutri}, {Stiening},
  {Weinberg}, {Schneider}, {Carpenter}, {Beichman}, {Capps}, {Chester},
  {Elias}, {Huchra}, {Liebert}, {Lonsdale}, {Monet}, {Price}, {Seitzer},
  {Jarrett}, {Kirkpatrick}, {Gizis}, {Howard}, {Evans}, {Fowler}, {Fullmer},
  {Hurt}, {Light}, {Kopan}, {Marsh}, {McCallon}, {Tam}, {Van Dyk}, \&
  {Wheelock}}]{skrutskie;et-al_2006}
{Skrutskie}, M.~F., {et~al.} 2006, \aj, 131, 1163
 \href{http://adsabs.harvard.edu/abs/2006AJ....131.1163S}{\adsurllinklabel}
{}

\bibitem[{{Sluis} \& {Arnold}(1998)}]{sluis;arnold_1998}
{Sluis}, A.~P.~N., \& {Arnold}, R.~A. 1998, \mnras, 297, 732
 \href{http://adsabs.harvard.edu/abs/1998MNRAS.297..732S}{\adsurllinklabel}
{}

\bibitem[{{Smith} {et~al.}(2007){Smith}, {Draine}, {Dale}, {Moustakas},
  {Kennicutt}, {Helou}, {Armus}, {Roussel}, {Sheth}, {Bendo}, {Buckalew},
  {Calzetti}, {Engelbracht}, {Gordon}, {Hollenbach}, {Li}, {Malhotra},
  {Murphy}, \& {Walter}}]{smith;et-al_2007}
{Smith}, J.~D.~T., {et~al.} 2007, \apj, 656, 770
 \href{http://adsabs.harvard.edu/abs/2007ApJ...656..770S}{\adsurllinklabel}
{}

\bibitem[{{Smith} {et~al.}(2000){Smith}, {Suntzeff}, {Cunha}, {Gallino},
  {Busso}, {Lambert}, \& {Straniero}}]{smith;et-al_2000}
{Smith}, V.~V., {Suntzeff}, N.~B., {Cunha}, K., {Gallino}, R., {Busso}, M.,
  {Lambert}, D.~L., \& {Straniero}, O. 2000, \aj, 119, 1239
 \href{http://adsabs.harvard.edu/abs/2000AJ....119.1239S}{\adsurllinklabel}
{}

\bibitem[{{Sneden}(1973)}]{sneden_1973}
{Sneden}, C.~A. 1973, PhD thesis, The University of Texas at Austin.
 \href{http://adsabs.harvard.edu/abs/1973PhDT.......180S}{\adsurllinklabel}
{}

\bibitem[{{Sobeck} {et~al.}(2011){Sobeck}, {Kraft}, {Sneden}, {Preston},
  {Cowan}, {Smith}, {Thompson}, {Shectman}, \& {Burley}}]{sobeck;et-al_2011}
{Sobeck}, J.~S., {et~al.} 2011, \aj, 141, 175
 \href{http://adsabs.harvard.edu/abs/2011AJ....141..175S}{\adsurllinklabel}
{}

\bibitem[{{Sohn} {et~al.}(2007){Sohn}, {Majewski}, {Mu{\~n}oz}, {Kunkel},
  {Johnston}, {Ostheimer}, {Guhathakurta}, {Patterson}, {Siegel}, \&
  {Cooper}}]{sohn;et-al_2007}
{Sohn}, S.~T., {et~al.} 2007, \apj, 663, 960
 \href{http://adsabs.harvard.edu/abs/2007ApJ...663..960S}{\adsurllinklabel}
{}

\bibitem[{{Sousa}(2012)}]{ares}
{Sousa}, S.~G. 2012, {ARES: Automatic Routine for Line Equivalent Widths in
  Stellar Spectra}, astrophysics Source Code Library
 \href{http://adsabs.harvard.edu/abs/2012ascl.soft05009S}{\adsurllinklabel}
{}

\bibitem[{{Springel} {et~al.}(2005){Springel}, {White}, {Jenkins}, {Frenk},
  {Yoshida}, {Gao}, {Navarro}, {Thacker}, {Croton}, {Helly}, {Peacock}, {Cole},
  {Thomas}, {Couchman}, {Evrard}, {Colberg}, \& {Pearce}}]{springel;et-al_2005}
{Springel}, V., {et~al.} 2005, \nat, 435, 629
 \href{http://adsabs.harvard.edu/abs/2005Natur.435..629S}{\adsurllinklabel}
{}

\bibitem[{{Stanford} {et~al.}(2010){Stanford}, {Da Costa}, \&
  {Norris}}]{stanford;et-al_2010}
{Stanford}, L.~M., {Da Costa}, G.~S., \& {Norris}, J.~E. 2010, \apj, 714, 1001
 \href{http://adsabs.harvard.edu/abs/2010ApJ...714.1001S}{\adsurllinklabel}
{}

\bibitem[{{Starkenburg} {et~al.}(2009){Starkenburg}, {Helmi}, {Morrison},
  {Harding}, {van Woerden}, {Mateo}, {Olszewski}, {Sivarani}, {Norris},
  {Freeman}, {Shectman}, {Dohm-Palmer}, {Frey}, \&
  {Oravetz}}]{starkenburg;et-al_2009}
{Starkenburg}, E., {et~al.} 2009, \apj, 698, 567
 \href{http://adsabs.harvard.edu/abs/2009ApJ...698..567S}{\adsurllinklabel}
{}

\bibitem[{{Starkenburg} {et~al.}(2010){Starkenburg}, {Hill}, {Tolstoy},
  {Gonz{\'a}lez Hern{\'a}ndez}, {Irwin}, {Helmi}, {Battaglia}, {Jablonka},
  {Tafelmeyer}, {Shetrone}, {Venn}, \& {de Boer}}]{starkenburg;et-al_2010}
---. 2010, \aap, 513, A34
 \href{http://adsabs.harvard.edu/abs/2010A%26A...513A..34S}{\adsurllinklabel}
{}

\bibitem[{{Steinmetz} {et~al.}(2006){Steinmetz}, {Zwitter}, {Siebert},
  {Watson}, {Freeman}, {Munari}, {Campbell}, {Williams}, {Seabroke}, {Wyse},
  {Parker}, {Bienaym{\'e}}, {Roeser}, {Gibson}, {Gilmore}, {Grebel}, {Helmi},
  {Navarro}, {Burton}, {Cass}, {Dawe}, {Fiegert}, {Hartley}, {Russell},
  {Saunders}, {Enke}, {Bailin}, {Binney}, {Bland-Hawthorn}, {Boeche}, {Dehnen},
  {Eisenstein}, {Evans}, {Fiorucci}, {Fulbright}, {Gerhard}, {Jauregi}, {Kelz},
  {Mijovi{\'c}}, {Minchev}, {Parmentier}, {Pe{\~n}arrubia}, {Quillen}, {Read},
  {Ruchti}, {Scholz}, {Siviero}, {Smith}, {Sordo}, {Veltz}, {Vidrih}, {von
  Berlepsch}, {Boyle}, \& {Schilbach}}]{steinmetz;et-al_2006}
{Steinmetz}, M., {et~al.} 2006, \aj, 132, 1645
 \href{http://adsabs.harvard.edu/abs/2006AJ....132.1645S}{\adsurllinklabel}
{}

\bibitem[{{Stetson} \& {Pancino}(2008)}]{daospec}
{Stetson}, P.~B., \& {Pancino}, E. 2008, \pasp, 120, 1332
 \href{http://adsabs.harvard.edu/abs/2008PASP..120.1332S}{\adsurllinklabel}
{}

\bibitem[{{Tolstoy} {et~al.}(2009){Tolstoy}, {Hill}, \&
  {Tosi}}]{tolstoy;et-al_2009}
{Tolstoy}, E., {Hill}, V., \& {Tosi}, M. 2009, \araa, 47, 371
 \href{http://adsabs.harvard.edu/abs/2009ARA%26A..47..371T}{\adsurllinklabel}
{}

\bibitem[{{Tolstoy} {et~al.}(2001){Tolstoy}, {Irwin}, {Cole}, {Pasquini},
  {Gilmozzi}, \& {Gallagher}}]{Tolstoy;et-al_2001}
{Tolstoy}, E., {Irwin}, M.~J., {Cole}, A.~A., {Pasquini}, L., {Gilmozzi}, R.,
  \& {Gallagher}, J.~S. 2001, \mnras, 327, 918
 \href{http://adsabs.harvard.edu/abs/2001MNRAS.327..918T}{\adsurllinklabel}
{}

\bibitem[{{Tolstoy} {et~al.}(2003){Tolstoy}, {Venn}, {Shetrone}, {Primas},
  {Hill}, {Kaufer}, \& {Szeifert}}]{tolstoy;et-al_2003}
{Tolstoy}, E., {Venn}, K.~A., {Shetrone}, M., {Primas}, F., {Hill}, V.,
  {Kaufer}, A., \& {Szeifert}, T. 2003, \aj, 125, 707
 \href{http://adsabs.harvard.edu/abs/2003AJ....125..707T}{\adsurllinklabel}
{}

\bibitem[{{Tonry} \& {Davis}(1979)}]{tonry;davis_1978}
{Tonry}, J., \& {Davis}, M. 1979, \aj, 84, 1511
 \href{http://adsabs.harvard.edu/abs/1979AJ.....84.1511T}{\adsurllinklabel}
{}

\bibitem[{{Totten} \& {Irwin}(1998)}]{Totten;Irwin_1998}
{Totten}, E.~J., \& {Irwin}, M.~J. 1998, \mnras, 294, 1
 \href{http://adsabs.harvard.edu/abs/1998MNRAS.294....1T}{\adsurllinklabel}
{}

\bibitem[{{Travaglio} {et~al.}(2004){Travaglio}, {Gallino}, {Arnone}, {Cowan},
  {Jordan}, \& {Sneden}}]{travaglio;et-al_2004}
{Travaglio}, C., {Gallino}, R., {Arnone}, E., {Cowan}, J., {Jordan}, F., \&
  {Sneden}, C. 2004, \apj, 601, 864
 \href{http://adsabs.harvard.edu/abs/2004ApJ...601..864T}{\adsurllinklabel}
{}

\bibitem[{{Tsuchiya} {et~al.}(2003){Tsuchiya}, {Dinescu}, \&
  {Korchagin}}]{tsuchiya_2003}
{Tsuchiya}, T., {Dinescu}, D.~I., \& {Korchagin}, V.~I. 2003, \apjl, 589, L29
 \href{http://adsabs.harvard.edu/abs/2003ApJ...589L..29T}{\adsurllinklabel}
{}

\bibitem[{{Tsuchiya} {et~al.}(2004){Tsuchiya}, {Korchagin}, \&
  {Dinescu}}]{tsuchiya_2004}
{Tsuchiya}, T., {Korchagin}, V.~I., \& {Dinescu}, D.~I. 2004, \mnras, 350, 1141
 \href{http://adsabs.harvard.edu/abs/2004MNRAS.350.1141T}{\adsurllinklabel}
{}

\bibitem[{{Tsujimoto} {et~al.}(1998){Tsujimoto}, {Miyamoto}, \&
  {Yoshii}}]{Tsujimoto;et-al_1998}
{Tsujimoto}, T., {Miyamoto}, M., \& {Yoshii}, Y. 1998, \apjl, 492, L79
 \href{http://adsabs.harvard.edu/abs/1998ApJ...492L..79T}{\adsurllinklabel}
{}

\bibitem[{{Urquhart} {et~al.}(2014){Urquhart}, {Figura}, {Moore}, {Hoare},
  {Lumsden}, {Mottram}, {Thompson}, \& {Oudmaijer}}]{urquhart;et-al_2014}
{Urquhart}, J.~S., {Figura}, C.~C., {Moore}, T.~J.~T., {Hoare}, M.~G.,
  {Lumsden}, S.~L., {Mottram}, J.~C., {Thompson}, M.~A., \& {Oudmaijer}, R.~D.
  2014, \mnras, 437, 1791
 \href{http://adsabs.harvard.edu/abs/2014MNRAS.437.1791U}{\adsurllinklabel}
{}

\bibitem[{{Uttenthaler} {et~al.}(2011){Uttenthaler}, {van Stiphout}, {Voet},
  {van Winckel}, {van Eck}, {Jorissen}, {Kerschbaum}, {Raskin}, {Prins},
  {Pessemier}, {Waelkens}, {Fr{\'e}mat}, {Hensberge}, {Dumortier}, \&
  {Lehmann}}]{uttenthaler;et-al_2011}
{Uttenthaler}, S., {et~al.} 2011, \aap, 531, A88
 \href{http://adsabs.harvard.edu/abs/2011A%26A...531A..88U}{\adsurllinklabel}
{}

\bibitem[{{Valenti} \& {Piskunov}(1996)}]{sme}
{Valenti}, J.~A., \& {Piskunov}, N. 1996, \aaps, 118, 595
 \href{http://adsabs.harvard.edu/abs/1996A%26AS..118..595V}{\adsurllinklabel}
{}

\bibitem[{{Van Eck} \& {Jorissen}(1999)}]{van_eck_jorrissen_1999}
{Van Eck}, S., \& {Jorissen}, A. 1999, \aap, 345, 127
 \href{http://adsabs.harvard.edu/abs/1999A%26A...345..127V}{\adsurllinklabel}
{}

\bibitem[{{Vargas} {et~al.}(2013){Vargas}, {Geha}, {Kirby}, \&
  {Simon}}]{vargas;et-al_2013}
{Vargas}, L.~C., {Geha}, M., {Kirby}, E.~N., \& {Simon}, J.~D. 2013, \apj, 767,
  134
 \href{http://adsabs.harvard.edu/abs/2013ApJ...767..134V}{\adsurllinklabel}
{}

\bibitem[{{Venn} {et~al.}(2004){Venn}, {Irwin}, {Shetrone}, {Tout}, {Hill}, \&
  {Tolstoy}}]{venn;et-al_2004}
{Venn}, K.~A., {Irwin}, M., {Shetrone}, M.~D., {Tout}, C.~A., {Hill}, V., \&
  {Tolstoy}, E. 2004, \aj, 128, 1177
 \href{http://adsabs.harvard.edu/abs/2004AJ....128.1177V}{\adsurllinklabel}
{}

\bibitem[{{Venn} {et~al.}(2006){Venn}, {Irwin}, {Shetrone}, {Tout}, {Hill}, \&
  {Tolstoy}}]{venn;et-al_2006}
---. 2006, \aj, 132, 1726
 \href{http://adsabs.harvard.edu/abs/2006AJ....132.1726V}{\adsurllinklabel}
{}

\bibitem[{{Ventura} {et~al.}(2011){Ventura}, {Carini}, \&
  {D'Antona}}]{ventura;et-al_2011}
{Ventura}, P., {Carini}, R., \& {D'Antona}, F. 2011, \mnras, 415, 3865
 \href{http://adsabs.harvard.edu/abs/2011MNRAS.415.3865V}{\adsurllinklabel}
{}

\bibitem[{{Villalobos} \& {Helmi}(2008)}]{villalobos;helmi_2008}
{Villalobos}, {\'A}., \& {Helmi}, A. 2008, \mnras, 391, 1806
 \href{http://adsabs.harvard.edu/abs/2008MNRAS.391.1806V}{\adsurllinklabel}
{}

\bibitem[{{Vivas} \& {Zinn}(2006)}]{vivas;zinn_2006}
{Vivas}, A.~K., \& {Zinn}, R. 2006, \aj, 132, 714
 \href{http://adsabs.harvard.edu/abs/2006AJ....132..714V}{\adsurllinklabel}
{}

\bibitem[{{Vivas} {et~al.}(2005){Vivas}, {Zinn}, \&
  {Gallart}}]{Vivas;et-al_2005}
{Vivas}, A.~K., {Zinn}, R., \& {Gallart}, C. 2005, \aj, 129, 189
 \href{http://adsabs.harvard.edu/abs/2005AJ....129..189V}{\adsurllinklabel}
{}

\bibitem[{{Vivas} {et~al.}(2001){Vivas}, {Zinn}, {Andrews}, {Bailyn}, {Baltay},
  {Coppi}, {Ellman}, {Girard}, {Rabinowitz}, {Schaefer}, {Shin}, {Snyder},
  {Sofia}, {van Altena}, {Abad}, {Bongiovanni}, {Brice{\~n}o}, {Bruzual},
  {Della Prugna}, {Herrera}, {Magris}, {Mateu}, {Pacheco}, {S{\'a}nchez},
  {S{\'a}nchez}, {Schenner}, {Stock}, {Vicente}, {Vieira}, {Ferr{\'{\i}}n},
  {Hernandez}, {Gebhard}, {Honeycutt}, {Mufson}, {Musser}, \&
  {Rengstorf}}]{Vivas;et-al_2001}
{Vivas}, A.~K., {et~al.} 2001, \apjl, 554, L33
 \href{http://adsabs.harvard.edu/abs/2001ApJ...554L..33V}{\adsurllinklabel}
{}

\bibitem[{{Waters} \& {Hollek}(2013)}]{robospect}
{Waters}, C.~Z., \& {Hollek}, J.~K. 2013, ArXiv e-prints
 \href{http://adsabs.harvard.edu/abs/2013arXiv1308.0757W}{\adsurllinklabel}
{}

\bibitem[{{Watkins} {et~al.}(2009){Watkins}, {Evans}, {Belokurov}, {Smith},
  {Hewett}, {Bramich}, {Gilmore}, {Irwin}, {Vidrih}, {Wyrzykowski}, \&
  {Zucker}}]{Watkins;et-al_2009}
{Watkins}, L.~L., {et~al.} 2009, \mnras, 398, 1757
 \href{http://adsabs.harvard.edu/abs/2009MNRAS.398.1757W}{\adsurllinklabel}
{}

\bibitem[{{White} \& {Rees}(1978)}]{white;rees_1978}
{White}, S.~D.~M., \& {Rees}, M.~J. 1978, \mnras, 183, 341
 \href{http://adsabs.harvard.edu/abs/1978MNRAS.183..341W}{\adsurllinklabel}
{}

\bibitem[{{Widrow} {et~al.}(2012){Widrow}, {Gardner}, {Yanny}, {Dodelson}, \&
  {Chen}}]{widrow;et-al_2012}
{Widrow}, L.~M., {Gardner}, S., {Yanny}, B., {Dodelson}, S., \& {Chen}, H.-Y.
  2012, \apjl, 750, L41
 \href{http://adsabs.harvard.edu/abs/2012ApJ...750L..41W}{\adsurllinklabel}
{}

\bibitem[{{Williams} {et~al.}(2011){Williams}, {Steinmetz}, {Sharma},
  {Bland-Hawthorn}, {de Jong}, {Seabroke}, {Helmi}, {Freeman}, {Binney},
  {Minchev}, {Bienaym{\'e}}, {Campbell}, {Fulbright}, {Gibson}, {Gilmore},
  {Grebel}, {Munari}, {Navarro}, {Parker}, {Reid}, {Siebert}, {Siviero},
  {Watson}, {Wyse}, \& {Zwitter}}]{williams;et-al_2011}
{Williams}, M.~E.~K., {et~al.} 2011, \apj, 728, 102
 \href{http://adsabs.harvard.edu/abs/2011ApJ...728..102W}{\adsurllinklabel}
{}

\bibitem[{{Woosley} \& {Weaver}(1995)}]{woosley;weaver_1995}
{Woosley}, S.~E., \& {Weaver}, T.~A. 1995, \apjs, 101, 181
 \href{http://adsabs.harvard.edu/abs/1995ApJS..101..181W}{\adsurllinklabel}
{}

\bibitem[{{Wylie-de Boer} {et~al.}(2010){Wylie-de Boer}, {Freeman}, \&
  {Williams}}]{wylie-de-boer;et-al_2010}
{Wylie-de Boer}, E., {Freeman}, K., \& {Williams}, M. 2010, \aj, 139, 636
 \href{http://adsabs.harvard.edu/abs/2010AJ....139..636W}{\adsurllinklabel}
{}

\bibitem[{{Wylie-de Boer} {et~al.}(2012){Wylie-de Boer}, {Freeman}, {Williams},
  {Steinmetz}, {Munari}, \& {Keller}}]{wylie-de-boer;et-al_2012}
{Wylie-de Boer}, E., {Freeman}, K., {Williams}, M., {Steinmetz}, M., {Munari},
  U., \& {Keller}, S. 2012, \apj, 755, 35
 \href{http://adsabs.harvard.edu/abs/2012ApJ...755...35W}{\adsurllinklabel}
{}

\bibitem[{{Xue} {et~al.}(2008){Xue}, {Rix}, {Zhao}, {Re Fiorentin}, {Naab},
  {Steinmetz}, {van den Bosch}, {Beers}, {Lee}, {Bell}, {Rockosi}, {Yanny},
  {Newberg}, {Wilhelm}, {Kang}, {Smith}, \& {Schneider}}]{xue;et-al_2008}
{Xue}, X.~X., {et~al.} 2008, \apj, 684, 1143
 \href{http://adsabs.harvard.edu/abs/2008ApJ...684.1143X}{\adsurllinklabel}
{}

\bibitem[{{Yanny} {et~al.}(2009){Yanny}, {Newberg}, {Johnson}, {Lee}, {Beers},
  {Bizyaev}, {Brewington}, {Fiorentin}, {Harding}, {Malanushenko},
  {Malanushenko}, {Oravetz}, {Pan}, {Simmons}, \& {Snedden}}]{Yanny;et-al_2009}
{Yanny}, B., {et~al.} 2009, \apj, 700, 1282
 \href{http://adsabs.harvard.edu/abs/2009ApJ...700.1282Y}{\adsurllinklabel}
{}

\bibitem[{{Yong} {et~al.}(2006){Yong}, {Aoki}, \& {Lambert}}]{yong;et-al_2006}
{Yong}, D., {Aoki}, W., \& {Lambert}, D.~L. 2006, \apj, 638, 1018
 \href{http://adsabs.harvard.edu/abs/2006ApJ...638.1018Y}{\adsurllinklabel}
{}

\bibitem[{{Yong} {et~al.}(2005){Yong}, {Carney}, \& {Teixera de
  Almeida}}]{yong;et-al_2005}
{Yong}, D., {Carney}, B.~W., \& {Teixera de Almeida}, M.~L. 2005, \aj, 130, 597
 \href{http://adsabs.harvard.edu/abs/2005AJ....130..597Y}{\adsurllinklabel}
{}

\bibitem[{{Yong} {et~al.}(2008){Yong}, {Mel{\'e}ndez}, {Cunha}, {Karakas},
  {Norris}, \& {Smith}}]{yong}
{Yong}, D., {Mel{\'e}ndez}, J., {Cunha}, K., {Karakas}, A.~I., {Norris}, J.~E.,
  \& {Smith}, V.~V. 2008, \apj, 689, 1020
 \href{http://adsabs.harvard.edu/abs/2008ApJ...689.1020Y}{\adsurllinklabel}
{}

\bibitem[{{Yong} {et~al.}(2013){Yong}, {Norris}, {Bessell}, {Christlieb},
  {Asplund}, {Beers}, {Barklem}, {Frebel}, \& {Ryan}}]{yong;et-al_2013}
{Yong}, D., {et~al.} 2013, \apj, 762, 26
 \href{http://adsabs.harvard.edu/abs/2013ApJ...762...26Y}{\adsurllinklabel}
{}

\bibitem[{{York} {et~al.}(2000){York}, {Adelman}, {Anderson}, {Anderson},
  {Annis}, {Bahcall}, {Bakken}, {Barkhouser}, {Bastian}, {Berman}, {Boroski},
  {Bracker}, {Briegel}, {Briggs}, {Brinkmann}, {Brunner}, {Burles}, {Carey},
  {Carr}, {Castander}, {Chen}, {Colestock}, {Connolly}, {Crocker}, {Csabai},
  {Czarapata}, {Davis}, {Doi}, {Dombeck}, {Eisenstein}, {Ellman}, {Elms},
  {Evans}, {Fan}, {Federwitz}, {Fiscelli}, {Friedman}, {Frieman}, {Fukugita},
  {Gillespie}, {Gunn}, {Gurbani}, {de Haas}, {Haldeman}, {Harris}, {Hayes},
  {Heckman}, {Hennessy}, {Hindsley}, {Holm}, {Holmgren}, {Huang}, {Hull},
  {Husby}, {Ichikawa}, {Ichikawa}, {Ivezi{\'c}}, {Kent}, {Kim}, {Kinney},
  {Klaene}, {Kleinman}, {Kleinman}, {Knapp}, {Korienek}, {Kron}, {Kunszt},
  {Lamb}, {Lee}, {Leger}, {Limmongkol}, {Lindenmeyer}, {Long}, {Loomis},
  {Loveday}, {Lucinio}, {Lupton}, {MacKinnon}, {Mannery}, {Mantsch}, {Margon},
  {McGehee}, {McKay}, {Meiksin}, {Merelli}, {Monet}, {Munn}, {Narayanan},
  {Nash}, {Neilsen}, {Neswold}, {Newberg}, {Nichol}, {Nicinski}, {Nonino},
  {Okada}, {Okamura}, {Ostriker}, {Owen}, {Pauls}, {Peoples}, {Peterson},
  {Petravick}, {Pier}, {Pope}, {Pordes}, {Prosapio}, {Rechenmacher}, {Quinn},
  {Richards}, {Richmond}, {Rivetta}, {Rockosi}, {Ruthmansdorfer}, {Sandford},
  {Schlegel}, {Schneider}, {Sekiguchi}, {Sergey}, {Shimasaku}, {Siegmund},
  {Smee}, {Smith}, {Snedden}, {Stone}, {Stoughton}, {Strauss}, {Stubbs},
  {SubbaRao}, {Szalay}, {Szapudi}, {Szokoly}, {Thakar}, {Tremonti}, {Tucker},
  {Uomoto}, {Vanden Berk}, {Vogeley}, {Waddell}, {Wang}, {Watanabe},
  {Weinberg}, {Yanny}, \& {Yasuda}}]{York;et-al_2000}
{York}, D.~G., {et~al.} 2000, \aj, 120, 1579
 \href{http://adsabs.harvard.edu/abs/2000AJ....120.1579Y}{\adsurllinklabel}
{}

\bibitem[{{Yoshida} {et~al.}(2004){Yoshida}, {Bromm}, \&
  {Hernquist}}]{yoshida;et-al_2004}
{Yoshida}, N., {Bromm}, V., \& {Hernquist}, L. 2004, \apj, 605, 579
 \href{http://adsabs.harvard.edu/abs/2004ApJ...605..579Y}{\adsurllinklabel}
{}

\bibitem[{{Zhao} {et~al.}(2006){Zhao}, {Zhao}, {Chen}, {Shi}, {Liu}, \&
  {Zhang}}]{zhao}
{Zhao}, J.-K., {Zhao}, G., {Chen}, Y.-Q., {Shi}, J.-R., {Liu}, Y.-J., \&
  {Zhang}, J.-Y. 2006, Chinese Journal of Astronomy and Astrophysics, 6, 689
 \href{http://adsabs.harvard.edu/abs/2006ChJAA...6..689Z}{\adsurllinklabel}
{}

\bibitem[{{Zinn} {et~al.}(2004){Zinn}, {Vivas}, {Gallart}, \&
  {Winnick}}]{Zinn;et-al_2004}
{Zinn}, R., {Vivas}, A.~K., {Gallart}, C., \& {Winnick}, R. 2004, in
  Astronomical Society of the Pacific Conference Series, Vol. 327, Satellites
  and Tidal Streams, ed. {F.~Prada, D.~Martinez Delgado, \& T.~J.~Mahoney}, 92
 \href{http://adsabs.harvard.edu/abs/2004ASPC..327...92Z}{\adsurllinklabel}
{}

\bibitem[{{Zucker} {et~al.}(2006){Zucker}, {Belokurov}, {Evans}, {Kleyna},
  {Irwin}, {Wilkinson}, {Fellhauer}, {Bramich}, {Gilmore}, {Newberg}, {Yanny},
  {Smith}, {Hewett}, {Bell}, {Rix}, {Gnedin}, {Vidrih}, {Wyse}, {Willman},
  {Grebel}, {Schneider}, {Beers}, {Kniazev}, {Barentine}, {Brewington},
  {Brinkmann}, {Harvanek}, {Kleinman}, {Krzesinski}, {Long}, {Nitta}, \&
  {Snedden}}]{zucker;et-al_2006}
{Zucker}, D.~B., {et~al.} 2006, \apjl, 650, L41
 \href{http://adsabs.harvard.edu/abs/2006ApJ...650L..41Z}{\adsurllinklabel}
{}

\bibitem[{{Zwitter} {et~al.}(2008){Zwitter}, {Siebert}, {Munari}, {Freeman},
  {Siviero}, {Watson}, {Fulbright}, {Wyse}, {Campbell}, {Seabroke}, {Williams},
  {Steinmetz}, {Bienaym{\'e}}, {Gilmore}, {Grebel}, {Helmi}, {Navarro},
  {Anguiano}, {Boeche}, {Burton}, {Cass}, {Dawe}, {Fiegert}, {Hartley},
  {Russell}, {Veltz}, {Bailin}, {Binney}, {Bland-Hawthorn}, {Brown}, {Dehnen},
  {Evans}, {Re Fiorentin}, {Fiorucci}, {Gerhard}, {Gibson}, {Kelz}, {Kujken},
  {Matijevi{\v c}}, {Minchev}, {Parker}, {Pe{\~n}arrubia}, {Quillen}, {Read},
  {Reid}, {Roeser}, {Ruchti}, {Scholz}, {Smith}, {Sordo}, {Tolstoi},
  {Tomasella}, {Vidrih}, \& {Wylie-de Boer}}]{zwitter;et-al_2008}
{Zwitter}, T., {et~al.} 2008, \aj, 136, 421
 \href{http://adsabs.harvard.edu/abs/2008AJ....136..421Z}{\adsurllinklabel}
{}

\bibitem[{{Zwitter} {et~al.}(2010){Zwitter}, {Matijevi{\v c}}, {Breddels},
  {Smith}, {Helmi}, {Munari}, {Bienaym{\'e}}, {Binney}, {Bland-Hawthorn},
  {Boeche}, {Brown}, {Campbell}, {Freeman}, {Fulbright}, {Gibson}, {Gilmore},
  {Grebel}, {Navarro}, {Parker}, {Seabroke}, {Siebert}, {Siviero}, {Steinmetz},
  {Watson}, {Williams}, \& {Wyse}}]{zwitter;et-al_2010}
---. 2010, \aap, 522, A54
 \href{http://adsabs.harvard.edu/abs/2010A%26A...522A..54Z}{\adsurllinklabel}
{}

\end{thebibliography}

\cleardoublepage
 \phantomsection
 %
 %
\begin{appendices}

\chapter{Equivalent Widths for Orphan Stream Stars}
\label{app:orphan_ews}

\setlength\tabcolsep{2pt}
\begin{landscape}
\begin{center}
\begin{longtable}{cccrcccccccccccc}
\\
\hline
\hline
Wavelength & Species & $\chi$ & $\log{gf}$ & HD 41667 & HD 44007 & HD 76932 & HD 122563 & HD 136316 & HD 141531 & HD 142948 & OSS 3 & OSS 6 & OSS 8 & OSS 14 & OSS 18 \\
({\AA}) & & (eV) & & (m{\AA}) & (m{\AA}) & (m{\AA}) & (m{\AA}) & (m{\AA}) & (m{\AA}) & (m{\AA}) & (m{\AA}) & (m{\AA}) & (m{\AA}) & (m{\AA}) & (m{\AA}) \\
\hline
\endfirsthead
\hline
Wavelength & Species & $\chi$ & $\log{gf}$ & HD 41667 & HD 44007 & HD 76932 & HD 122563 & HD 136316 & HD 141531 & HD 142948 & OSS 3 & OSS 6 & OSS 8 & OSS 14 & OSS 18 \\
({\AA}) & & (eV) & & (m{\AA}) & (m{\AA}) & (m{\AA}) & (m{\AA}) & (m{\AA}) & (m{\AA}) & (m{\AA}) & (m{\AA}) & (m{\AA}) & (m{\AA}) & (m{\AA}) & (m{\AA}) \\
\hline
\endhead
\hline
\multicolumn{16}{r}{Continued..}
\endfoot
\endlastfoot

 6300.30 &       O I &      0.00 &   $-$9.72 &   \nodata &     14.79 &   \nodata &      6.77 &   \nodata &     38.16 &     35.37 &   \nodata &      9.72 &   \nodata &   \nodata &   \nodata \\
 6363.78 &       O I &      0.02 &  $-$10.19 &     11.35 &   \nodata &   \nodata &   \nodata &     11.58 &     14.74 &     12.70 &   \nodata &   \nodata &   \nodata &   \nodata &   \nodata \\
 5688.19 &      Na I &      2.11 &   $-$0.42 &     71.85 &     30.87 &     57.44 &   \nodata &     29.30 &     38.07 &   \nodata &     55.17 &     24.02 &     20.39 &   \nodata &     71.19 \\
 6154.23 &      Na I &      2.10 &   $-$1.53 &     10.66 &   \nodata &      7.75 &   \nodata &   \nodata &      5.23 &     28.38 &   \nodata &   \nodata &   \nodata &   \nodata &   \nodata \\
 6160.75 &      Na I &      2.10 &   $-$1.23 &     16.55 &   \nodata &     13.38 &   \nodata &      4.58 &      6.98 &     43.40 &     17.39 &   \nodata &   \nodata &   \nodata &   \nodata \\
 3829.35 &      Mg I &      2.71 &   $-$0.21 &   \nodata &   \nodata &   \nodata &    190.12 &   \nodata &   \nodata &   \nodata &   \nodata &   \nodata &   \nodata &    162.57 &   \nodata \\
 3832.30 &      Mg I &      2.71 &      0.27 &   \nodata &   \nodata &   \nodata &    226.99 &   \nodata &   \nodata &   \nodata &   \nodata &   \nodata &   \nodata &    186.18 &   \nodata \\
 3986.75 &      Mg I &      4.35 &   $-$1.03 &   \nodata &   \nodata &    132.19 &     34.70 &   \nodata &    116.98 &    181.70 &    104.81 &   \nodata &     93.49 &   \nodata &   \nodata \\
\hline
\end{longtable}
\end{center}
\end{landscape}

\chapter{Equivalent Widths for Aquarius Stream Stars}
\label{app:aquarius_ews}

The EWs of atomic absorption lines for standard and program stars presented in Chapter \ref{chap:aquarius} are tabulated here. Some of the stars adopted as field standards in Chapter \ref{chap:orphan_II} were also employed as comparison stars in Chapter \ref{chap:aquarius}. The input data and atomic lines utilised for analysis in each study are different, however there are some atomic lines that are common to both studies. The EWs for atomic lines common to both studies for a given star agree excellently, but are not guaranteed to be precisely equal. Improvements have been made to the SMH analysis software over time, yielding EWs that may differ up to a few per cent. Comparisons between existing automatic and manual measurements were made after every software iteration to ensure consistent, sane measurements. Furthermore, we reiterate that every profile fitted to atomic absorption lines -- in both Chapter \ref{chap:orphan_II} and Chapter \ref{chap:aquarius} -- have been repeatedly examined by eye for quality assurance.

\setlength\tabcolsep{2pt}
\begin{landscape}
\begin{center}
\begin{longtable}{cccrccccccccccc}
\\
\hline
\hline
Wavelength & Species & $\chi$ & $\log{gf}$ & HD 41667 & HD 44007 & HD 76932 & HD 136316 & HD 141531 & HD 142948 & J221821 & C222531 & J223504 & J223811 & C230626 \\
({\AA}) & & (eV) & & (m{\AA}) & (m{\AA}) & (m{\AA}) & (m{\AA}) & (m{\AA}) & (m{\AA}) & (m{\AA}) & (m{\AA}) & (m{\AA}) & (m{\AA}) & (m{\AA})  \\
\hline
\endfirsthead
\hline
Wavelength & Species & $\chi$ & $\log{gf}$ & HD 41667 & HD 44007 & HD 76932 & HD 136316 & HD 141531 & HD 142948 & J221821 & C222531 & J223504 & J223811 & C230626 \\
({\AA}) & & (eV) & & (m{\AA}) & (m{\AA}) & (m{\AA}) & (m{\AA}) & (m{\AA}) & (m{\AA}) & (m{\AA}) & (m{\AA}) & (m{\AA}) & (m{\AA}) & (m{\AA})  \\
\hline
\endhead
\hline
\multicolumn{15}{r}{Continued..}
\endfoot
\endlastfoot

 6300.30 &       O I &      0.00 &     $-$9.72 &   \nodata &     13.47 &   \nodata &   \nodata &     35.88 &     33.62 &     38.48 &     45.37 &     32.85 &     17.87 &     66.93 \\
 6363.78 &       O I &      0.02 &    $-$10.19 &     10.62 &   \nodata &   \nodata &     11.63 &     13.34 &     11.83 &     12.96 &     19.47 &     18.83 &   \nodata &     28.31 \\
 5688.19 &      Na I &      2.11 &     $-$0.42 &     74.23 &     29.51 &     56.11 &     25.72 &   \nodata &    115.40 &     48.97 &   \nodata &    131.49 &     38.41 &   \nodata \\
 6154.23 &      Na I &      2.10 &     $-$1.53 &      9.88 &   \nodata &      6.92 &   \nodata &      3.98 &     27.13 &   \nodata &     24.06 &     48.49 &   \nodata &     38.89 \\
 6160.75 &      Na I &      2.10 &     $-$1.23 &     12.51 &      4.33 &     12.33 &      4.70 &      6.19 &     41.94 &   \nodata &     37.72 &     65.72 &   \nodata &     58.49 \\
 6318.72 &      Mg I &      5.11 &     $-$1.97 &     26.57 &      9.70 &     18.59 &   \nodata &     10.21 &     47.81 &     14.69 &   \nodata &     62.79 &      8.88 &     47.94 \\
 6319.24 &      Mg I &      5.11 &     $-$2.22 &     18.07 &   \nodata &     12.42 &      8.23 &   \nodata &   \nodata &      5.51 &     30.84 &   \nodata &   \nodata &   \nodata \\
 6965.41 &      Mg I &      5.75 &     $-$1.51 &     15.59 &   \nodata &      6.23 &   \nodata &      9.44 &     33.18 &   \nodata &   \nodata &     59.22 &   \nodata &   \nodata \\
\hline
\end{longtable}
\end{center}
\end{landscape}




\end{appendices}
\end{document}